%% file: DetectorPerformance.tex
\documentclass[11pt,a4paper]{scrartcl}
\pdfoutput=1
\usepackage{CLICdp}

\usepackage{CLICdp_definitions}

\usepackage{todonotes}

\usepackage{multirow}
\usepackage[nolist]{acronym}

\input{include/packages.tex}

\input{include/definitions.tex}
\input{include/abbreviations.tex}

\title{A detector for CLIC\@: main parameters and performance}

\clicdpnote{2018}{005}  

\date{\formatdate{17}{12}{2018}}

\onbehalfof{the CLICdp Collaboration} 

\abstract{Together with the recent CLIC detector model CLICdet a new
software suite was introduced for the simulation and reconstruction of 
events in this detector. This note gives a brief introduction to CLICdet
and describes the CLIC experimental conditions at \SI{380}{GeV} and \SI{3}{TeV},
including beam-induced backgrounds. The simulation and reconstruction tools
are introduced, and the physics performance
obtained is described in terms
of single particles, particles in jets, jet energy resolution and flavour tagging.
The performance of the very forward electromagnetic calorimeters is also discussed.
}

\titlecomment{}

\graphicspath{{./logos/}{./figures/}}

\begin{document}


\input{AuthorList/authorslist}
\titlepage
\clearpage
\tableofcontents
\clearpage

\setcounter{footnote}{1}
\renewcommand*{\thefootnote}{\arabic{footnote}}

\section{Introduction}
A state-of-the-art detector, built using cutting-edge technology and optimised through simulation, is
crucial to exploit the physics potential of CLIC\@. Two detector models were previously defined, based
on concepts for the ILC detectors and adapted for the higher centre-of-mass energies at CLIC\@. The
CLIC\_ILD and CLIC\_SiD models were used for physics studies in the CDR~\cite{cdrvol2}.
Based on the lessons learnt for the CDR as well as the experience from several additional
optimisation studies, a new model, dubbed CLICdet, has been designed for the forthcoming physics
benchmark studies. The CLICdet model is described in detail in~\cite{CLICdet_note_2017}. A summary of the main parameters
of CLICdet is given in \cref{parameters}. \Cref{clic_beam_and_bg} gives an overview of CLIC
experimental conditions linked to the characteristics of the beam and of the detector requirements.

In parallel to the development of a new detector design and its hardware technologies, the software chain for simulation and reconstruction
has been re-designed. The simulated model of CLICdet has been implemented using the DD4hep detector description toolkit. 
The performances of the simulated CLICdet model with the new software chain have been assessed in terms of
lower level physics observables, including flavour tagging and jet energy resolution.
Performance results for CLIC operation at \SI{380}{GeV} and \SI{3}{TeV} are shown in \cref{performances}.

\section{CLICdet Layout and Main Parameters}\label{parameters}
\subsection{Overview}

\begin{figure}[htbp]
  \centering
  \includegraphics[width=0.6\textwidth]{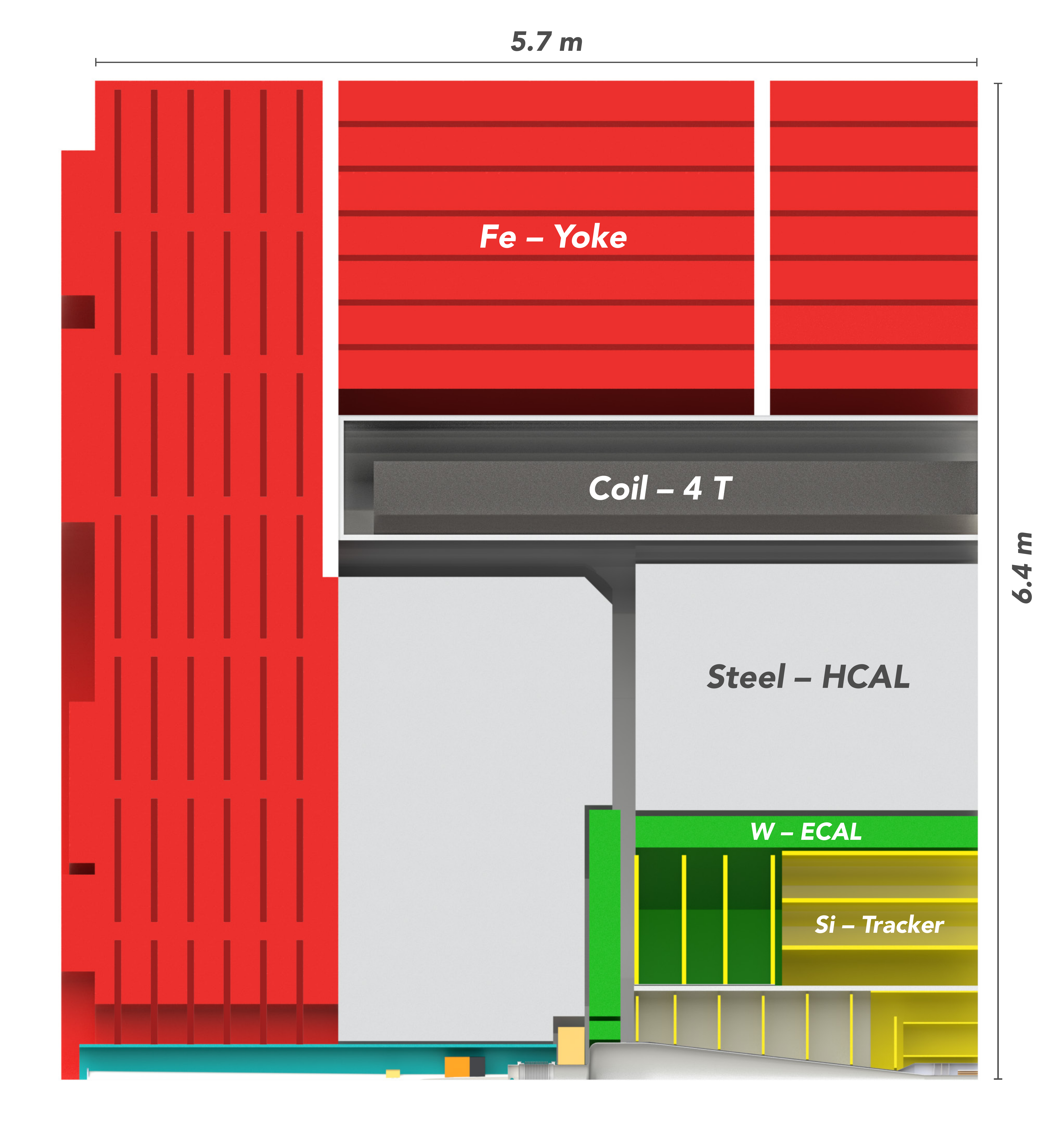}%
  \caption{Longitudinal cross section showing a quadrant of CLICdet (side view)~\cite{CLICdet_note_2017}. The structures shown on the left of the image (i.e.\ outside of the yoke) represent the end coils.}
  \label{quarter_view}
\end{figure}

\begin{figure}[htbp]
  \centering
  \includegraphics[width=0.50\textwidth]{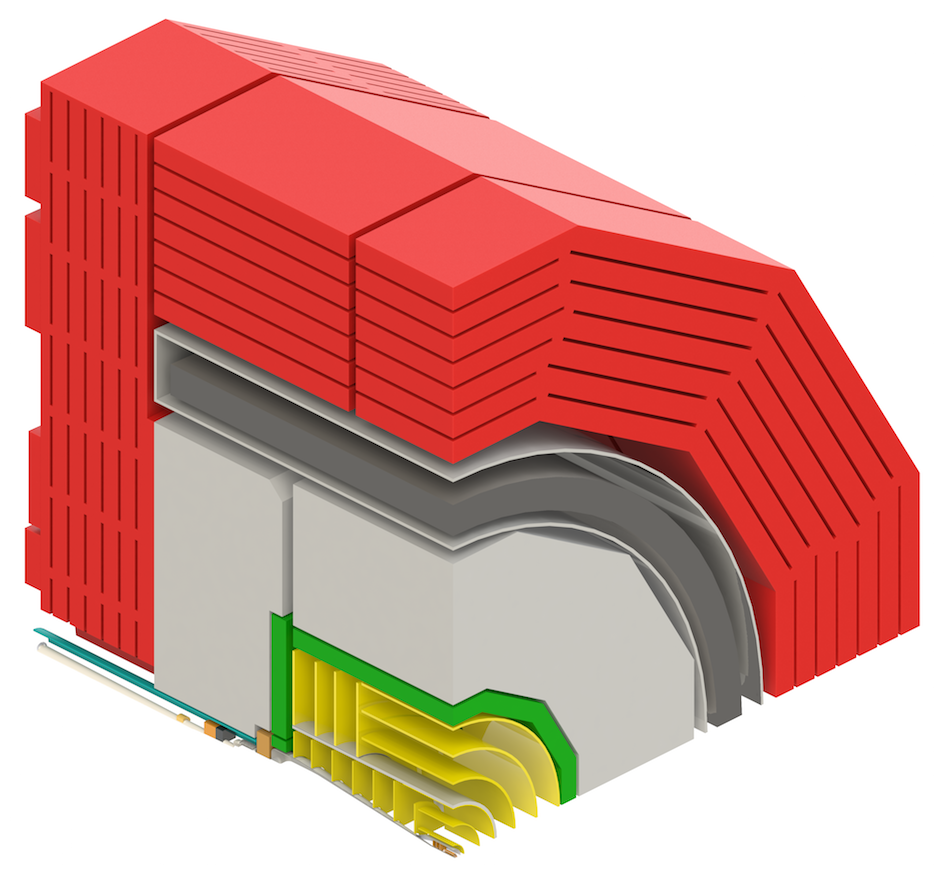}
  \caption{Isometric view of CLICdet.}\label{isometric}
\end{figure}

The CLICdet layout follows the typical collider detector scheme of a vertex detector surrounding the beryllium beam pipe, a large tracker volume with barrel and disks of silicon sensors, and an ECAL and HCAL, all embedded
inside a superconducting solenoid providing a \SI{4}{T} field. The surrounding iron yoke is interleaved with muon chambers needed for a clean muon identification in complex events. A quarter-view of the cross section of CLICdet is shown in \cref{quarter_view}, and an isometric view is presented in \cref{isometric}. Key parameters of the CLICdet model are compared to CLIC\_ILD and CLIC\_SiD in \cref{table.overall}.

An important change with respect to the CDR detector models concerns the location of the \ac{QD0}: to improve the angular coverage of the HCAL endcap to reach smaller polar angles, 
this quadrupole is moved outside of the detector into the tunnel. Nevertheless, the \ac{QD0} must be as close as possible to
the \ac{IP}. The overall length of CLICdet has been minimised by reducing the thickness of the iron yoke endcaps.
The missing iron is compensated by a set of end coils, shown schematically in \cref{quarter_view}.

\begin{figure}[htbp]
  \centering
  \includegraphics[width=0.90\textwidth]{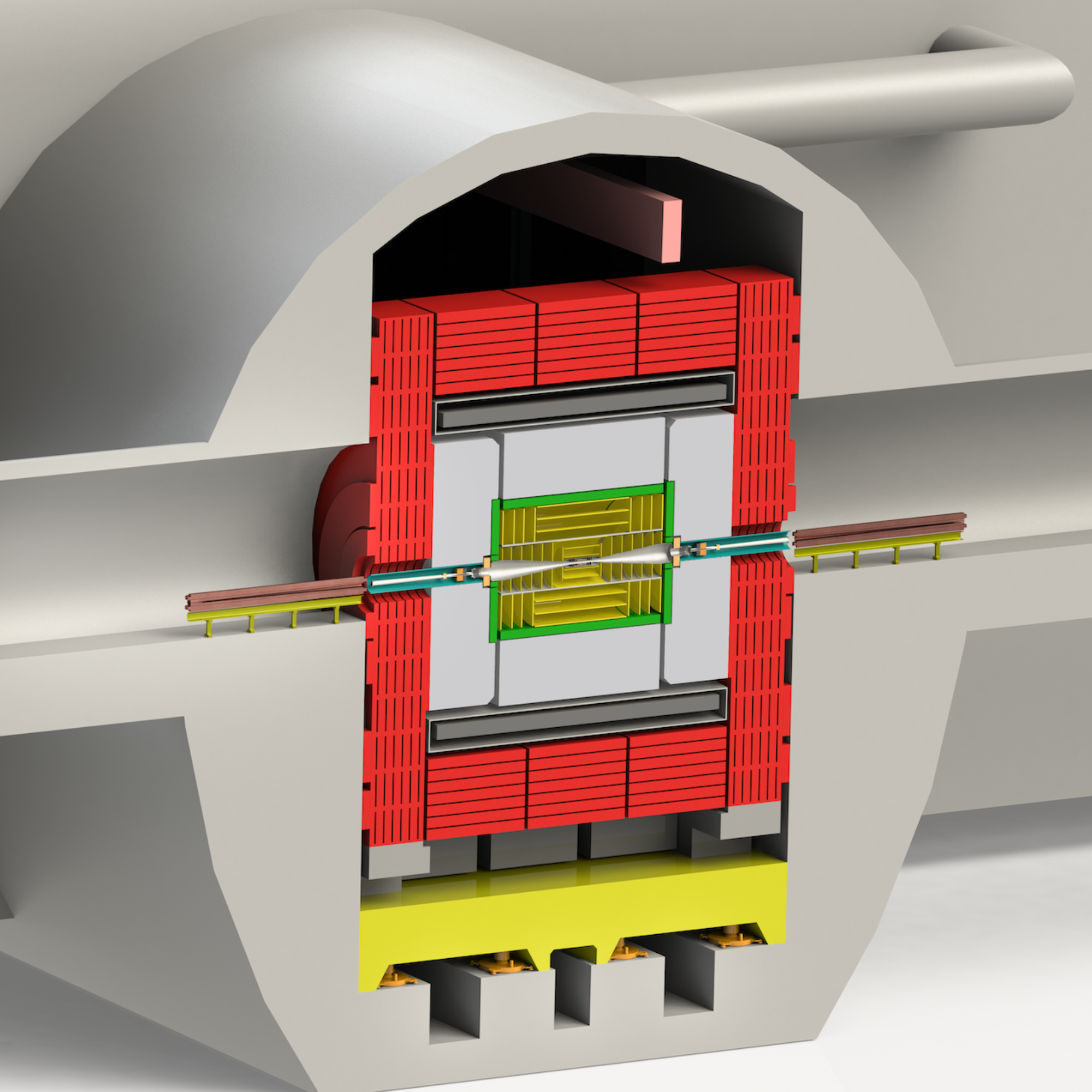}%
  \caption{Cut view of CLICdet on the interaction point. The \ac{QD0} quadrupoles in the accelerator tunnel are visible on both sides of the detector.}\label{on_beam}
\end{figure}

\cref{on_beam} shows a vertical cut-view of CLICdet installed on the interaction point, together with a section of accelerator tunnel on either side.
The \ac{QD0} quadrupoles, located just outside of the detector at $\lstar=\SI{6}{m}$ are also shown. \lstar{} is the
distance from the downstream end of \ac{QD0} to the interaction point.
The next magnetic elements, further upstream, are located outside the tunnel section covered by the drawing.

\begin{table}[hbtp]
\caption{\label{table_key_parameters} 
Comparison of  key parameters of the different CLIC detector concepts.
CLIC\_ILD and CLIC\_SiD values are taken from the CDR~\cite{cdrvol2}. 
The inner radius of the electromagnetic calorimeter (ECAL) is given by the smallest distance of the
calorimeter to the main detector axis. For the hadronic calorimeter (HCAL), materials are given separately for the barrel and the endcap.\label{table.overall}}
\centering
\begin{tabular}{l l l l l}
    \toprule
    Concept & CLICdet & \clicild  & \clicsid \\
    \midrule
Be vacuum pipe inner radius [mm] &29.4&29.4 &24.5 \\
Be vacuum pipe wall thickness [mm] &0.6&0.6&0.5\\
  Vertex technology                          & Silicon      & Silicon      & Silicon      \\
Vertex inner radius [mm] & 31 & 31  & 27       \\    
Tracker technology& Silicon & TPC/Silicon  & Silicon  \\
Tracker half length [m] & 2.2 &  2.3  & 1.5  \\
Tracker outer radius [m] & 1.5 & 1.8 & 1.3 \\
  ECAL technology                            & Silicon      & Silicon      & Silicon      \\
 ECAL absorber  & W & W   & W    \\
    ECAL radiation lengths & 22 & 23  & 23      \\
    ECAL barrel $r_{\min}$ [m] & 1.5 & 1.8  & 1.3      \\
    ECAL barrel $\Delta r$ [mm] & 202 & 172  & 139      \\
ECAL endcap $z_{\min}$ [m] & 2.31 & 2.45 &1.66  \\
ECAL endcap $\Delta z$ [mm] & 202  &  172 & 139 \\
  HCAL technology                            & Scintillator & Scintillator & Scintillator \\
    HCAL absorber barrel / endcap  & Fe / Fe & W / Fe  & W / Fe   \\
    HCAL nuclear interaction lengths & 7.5 & 7.5  & 7.5      \\
    HCAL barrel $r_{\min}$ [m] &1.74  & 2.06 & 1.45      \\
    HCAL barrel $\Delta r$ [mm] & 1590 & 1238  & 1177     \\
HCAL endcap $z_{\min}$ [m] &2.45  & 2.65 &1.80  \\
HCAL endcap $\Delta z$ [mm] & 1590  &  1590 & 1595 \\
Solenoid field [T] & 4 & 4 & 5        \\
    Solenoid length [m] & 8.3 & 8.3  & 6.5      \\
    Solenoid bore radius [m] & 3.5  & 3.4  & 2.7      \\
Yoke with muon system barrel $r_{\min}$ [m] &4.46&4.40 &3.91 \\
Yoke with muon system barrel $\Delta r$ [mm] & 1.99& 2.59& 3.09\\
Yoke with muon system endcap $z_{\min}$ [m] &4.18& 4.24&3.40 \\
Yoke with muon system endcap $\Delta z$ [mm] &1.52& 1.96& 2.81\\
    Overall height [m]  & 12.9 & 14.0  & 14.0     \\
    Overall length [m]  & 11.4 & 12.8 & 12.8     \\
Overall weight [t] & 8100 & 10800 & 12500 \\
    \bottomrule
\end{tabular}
\end{table}

\clearpage
\subsection{Vertex and Tracker}
\label{sec:tracker}

The vertex detector in CLICdet, similarly to the CDR detector models, consists of a cylindrical barrel
detector closed off in the forward directions by ``disks''. The layout is based on double layers, i.e.\ two
sensitive layers fixed on one support structure, in both barrel and forward region. The barrel consists of
three double layers. In the forward region, the three ``disks'' are split up in 8 segments which are arranged
to create a ``spiral''. This spiral geometry allows efficient air-flow cooling of the vertex detector. The
air-flow imposes that both spirals have the same sense of rotation -- this leads to an asymmetric layout
of the vertex ``disks''. The vertex detector is built from modules of \SI{50}{\micron} thin silicon pixel detectors
(plus an additional \SI{50}{\micron} thick ASIC)  with a pixel size
of \SI{25 x 25}{\micron\squared}.

\begin{figure}[bp]
  \centering
  \includegraphics[scale=0.7]{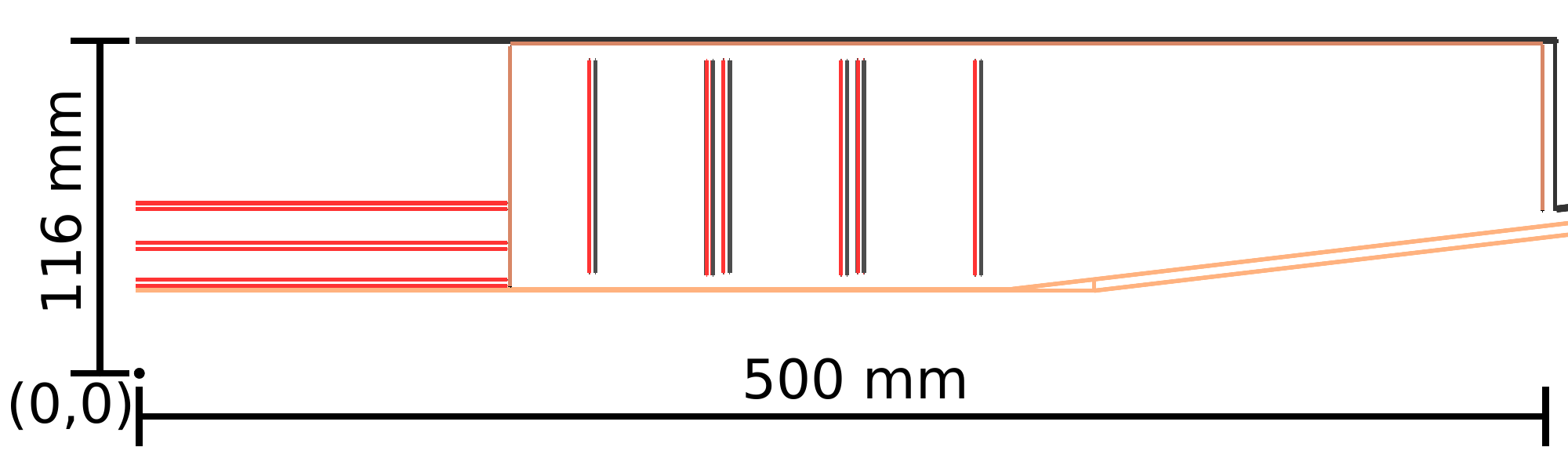}
  \caption{Sketch of the barrel and forward vertex detector region in the $X-Z$ plane. The vacuum tube is shown in yellow, sensors in red, support structures in black and cables in brown. Note that
the spiral structure of the vertex forward disks is not visible in this cut view.
 However, due to the choice of the cut, the two middle disks appear ``double'' due to the small azimuthal overlap between modules.}\label{VertexXZ}
  \includegraphics[scale=0.7]{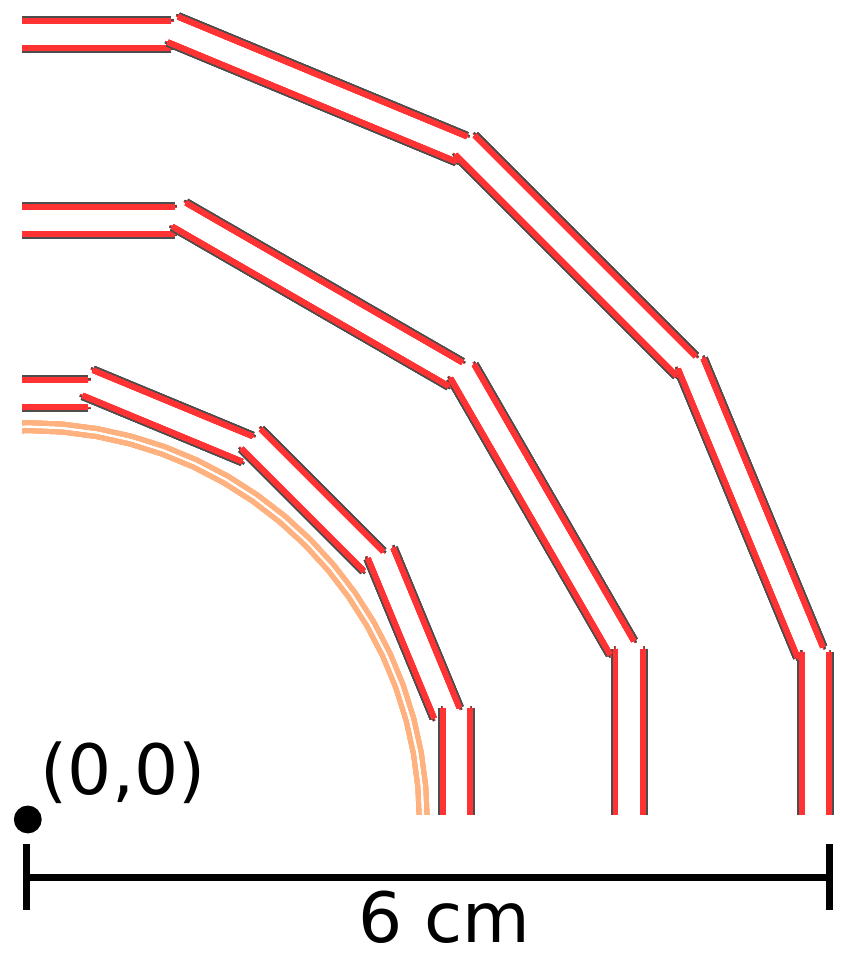}
  \caption{Sketch of the barrel vertex detector region in the $X-Y$ plane; colour coding as in \cref{VertexXZ}.}\label{VertexXY}
\end{figure}

\cref{VertexXZ} shows the arrangement of the three vertex barrel layers and the
forward spirals in the $X-Z$ plane, together with the vacuum tube and additional material representing supports
and a surrounding air-guiding cylinder. An $X-Y$ section through the vertex barrel layers is shown in \cref{VertexXY},
indicating the arrangement of modules as currently implemented in the simulation model of CLICdet. \Cref{sec:appendix}
contains further images indicating polar angles relevant for the efficiency studies.

\begin{figure}[bpt]
  \centering
  \includegraphics[scale=0.4]{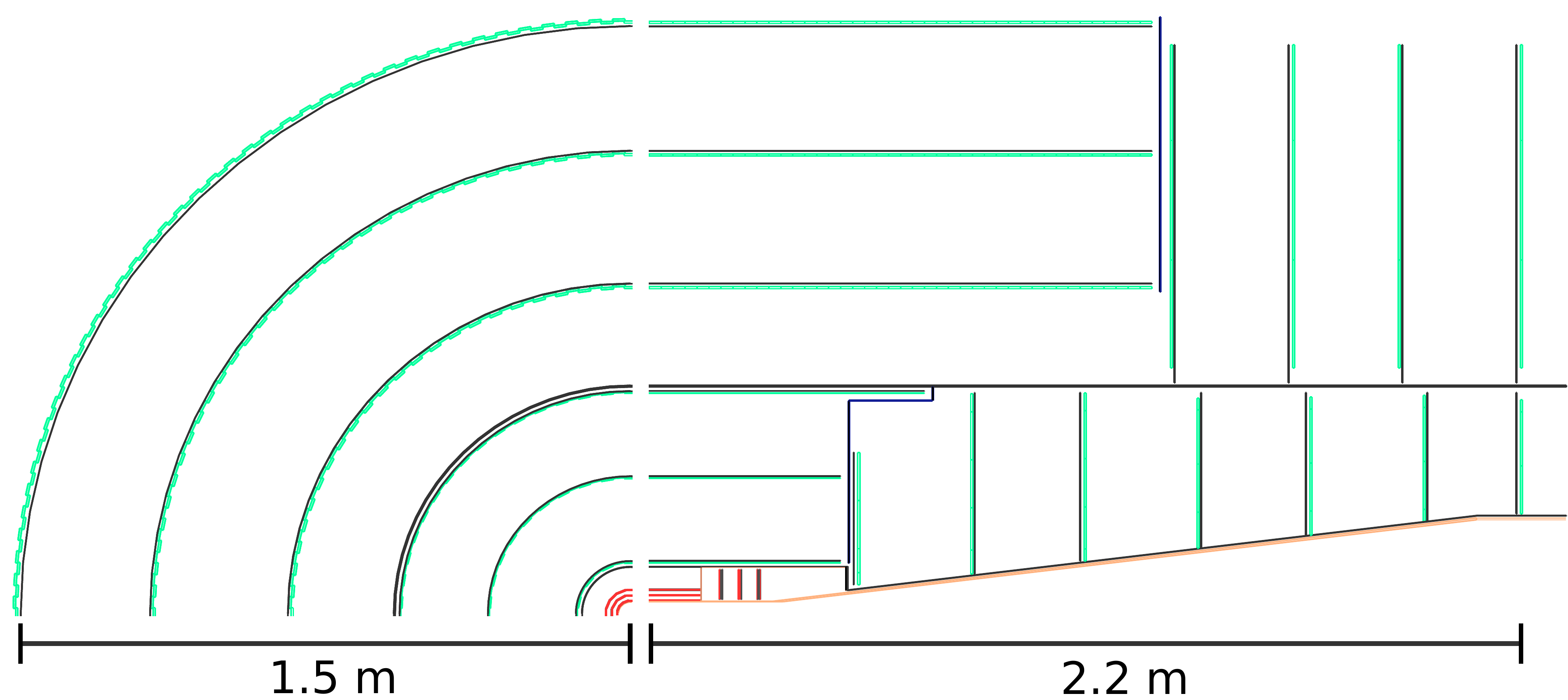}
  \caption{Layout of the tracking system in the $X-Y$  plane (left) and the $X-Z$  plane (right). Tracker sensors are shown in green, support material in black. The blue lines represent additional material (e.g.\ cables), which has only been added in critical regions.} \label{TrackerLayout}
\end{figure}

The all-silicon tracker volume has
a radius of \SI{1.5}{m} and a half-length of \SI{2.2}{m}. The main support tube, among other things needed to carry the weight of the vacuum tube and the vertex detector,
has an inner and outer radius of
\SI{0.575}{m} and \SI{0.600}{m}, respectively, and a half-length of \SI{2.25}{m}. This support tube effectively divides the
tracker volume into two regions: the ``Inner Tracker'' and ``Outer Tracker''. The Inner Tracker contains
three tracker barrel layers and, on each side of the barrel, seven inner tracker disks.
The Outer Tracker is built from three large barrel layers complemented on either side by four
outer tracker disks. The layout of the CLICdet tracker as implemented in the simulation model is shown in \cref{TrackerLayout}.
The sensors envisaged have a thickness of \SI{200}{\micron} including electronics and, in the simulation model, are assembled in modules either \SI{15x15}{mm\squared} or \SI{30x30}{mm\squared}.

In reconstruction, the hits are smeared with Gaussian distributions in order to represent the single point resolution.
A certain degree of charge sharing/cluster size is assumed when estimating resolutions from a given pixel/strip size -- verifying these assumptions is part of an ongoing R\&D programme.
The $\sigma$ values as used in reconstruction are given in \cref{tab:resolutions}.

\begin{table}[bt]
  \centering
  \caption{\label{tab:resolutions}Vertex and tracker pixels/strips, and assumed single point resolutions. The smaller numbers for each of the strips refer to the direction in the bending plane.
Note that the tracker numbers in the second column are driven by occupancy studies~\cite{Nurnberg_Dannheim_2017}, while the resolutions in the third column are the values currently used in the
reconstruction software.}
\begin{tabular}{l *2{l@{$\:\times\,$}r}} \toprule
  \tabt{Subdetector}        & \tabtt{Layout sizes} & \tabtt{Resolutions}                                   \\ \midrule
  Vertex (Barrel and Disks) & \SI{25}{\micron}     & \SI{25}{\micron} & \SI{3}{\micron} & \SI{3}{\micron}  \\
  Inner Tracker Disk 1      & \SI{25}{\micron}     & \SI{25}{\micron} & \SI{5}{\micron} & \SI{5}{\micron}  \\
  Inner Tracker Disks 2--7  & \SI{50}{\micron}     & \SI{1}{mm}       & \SI{7}{\micron} & \SI{90}{\micron} \\
  Outer Tracker Disks       & \SI{50}{\micron}     & \SI{10}{mm}      & \SI{7}{\micron} & \SI{90}{\micron} \\
  Inner Tracker Barrel 1--2 & \SI{50}{\micron}     & \SI{1}{mm}       & \SI{7}{\micron} & \SI{90}{\micron} \\
  Inner Tracker Barrel 3    & \SI{50}{\micron}     & \SI{5}{mm}       & \SI{7}{\micron} & \SI{90}{\micron} \\
  Outer Tracker Barrel 1--3 & \SI{50}{\micron}     & \SI{10}{mm}      & \SI{7}{\micron} & \SI{90}{\micron} \\ \bottomrule
  \end{tabular}
\end{table}

\begin{figure}[htbp]
         \centering
\includegraphics[scale=0.492]{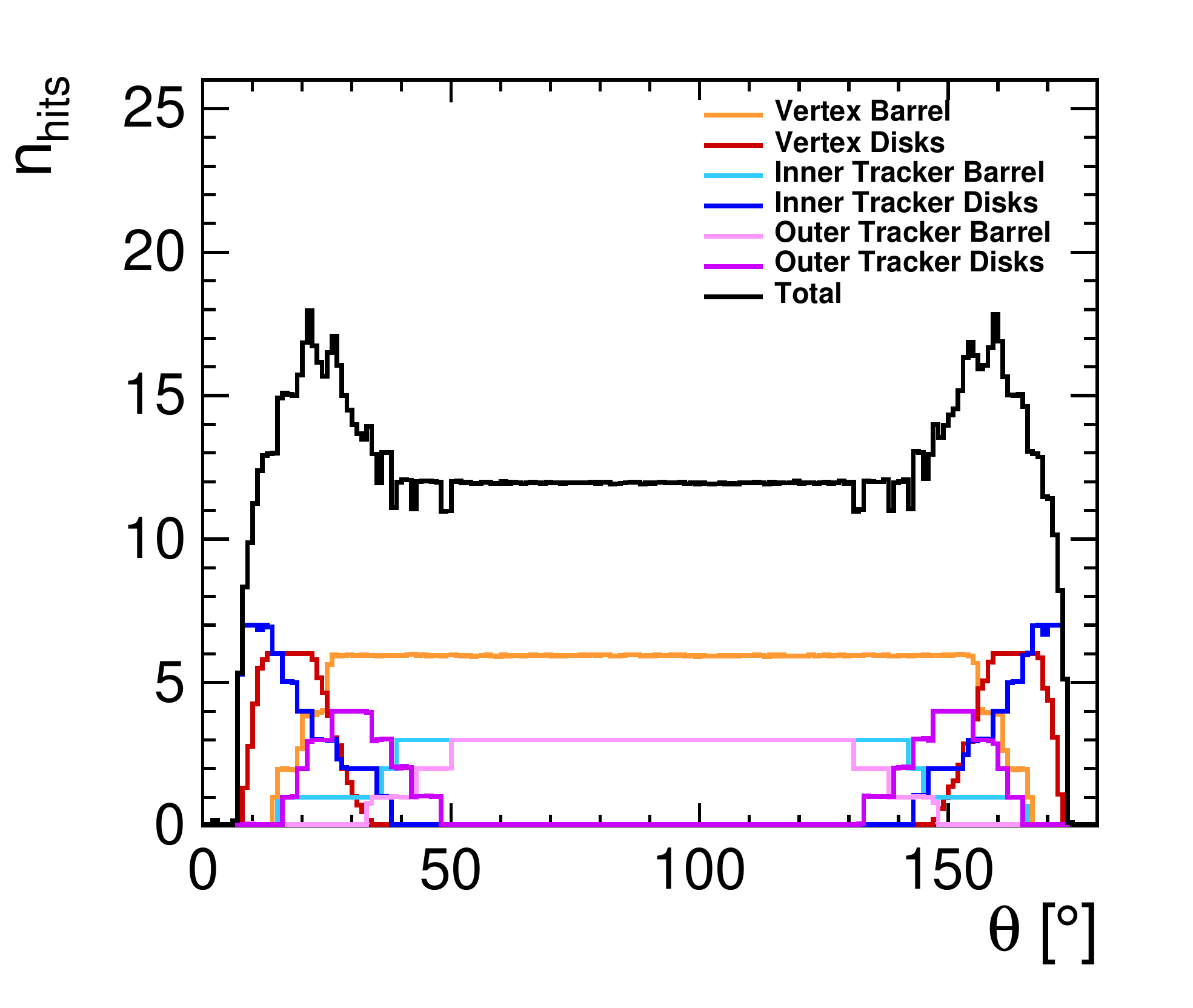}
         \caption{The coverage of the tracking systems with respect to the polar angle $\theta$~\cite{CLICdet_note_2017}.
 Shown is the mean number of hits created by \SI{500}{GeV} muons in full simulation. Only primary muon hits are taken into consideration (hits from secondary particles are ignored).
At least eight hits are measured for all tracks with a polar angle down to about 8\degrees.}
        \label{TrackerHits}
\end{figure}

\begin{figure}[htbp]
  \renewcommand{\thesubfigure}{(\lr{subfigure})}
  \centering
  \begin{subfigure}{.5\textwidth}
    \centering
    \includegraphics[width=\linewidth]{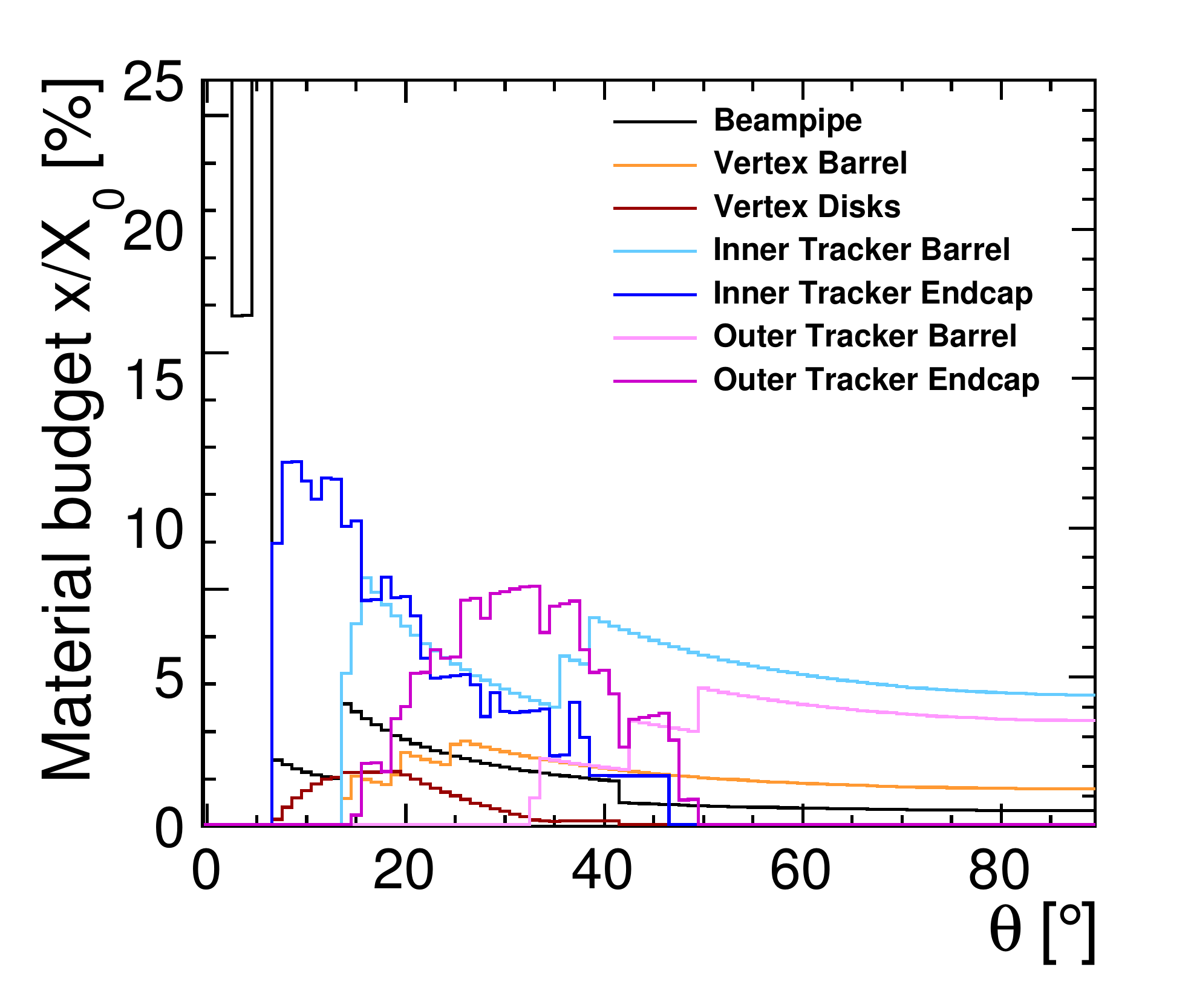}%
    \phantomsubcaption\label{fig:MatBudgetSub}
  \end{subfigure}%
  \begin{subfigure}{.5\textwidth}
    \centering
    \includegraphics[width=\linewidth]{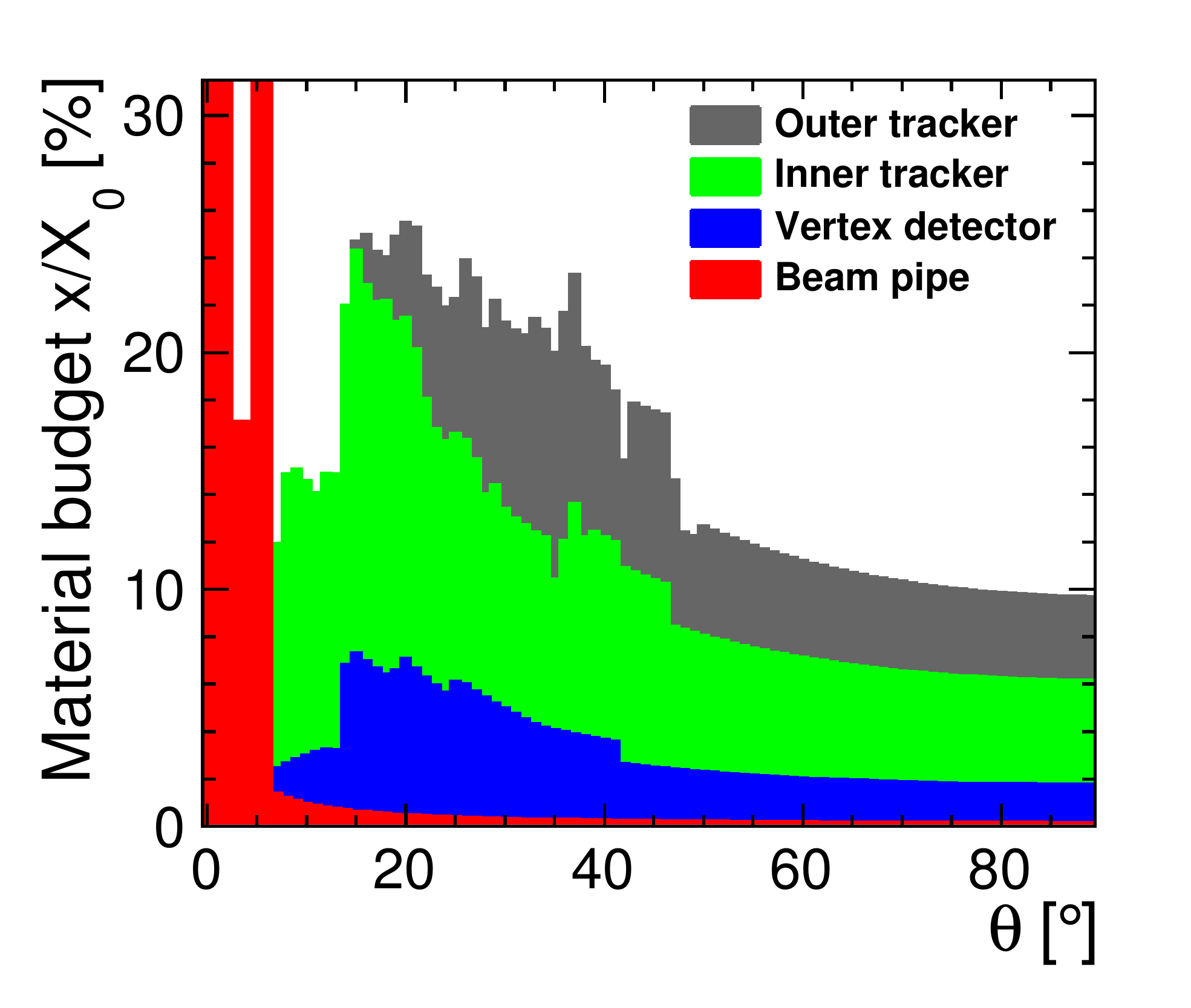}%
    \phantomsubcaption\label{fig:MatBudgetStacked}
  \end{subfigure}
  \vspace{-3mm}
  \caption{Material budget as a function of the polar angle and averaged over azimuthal angles, distinguished for the beam pipe and subdetectors in the tracking system~\subref{fig:MatBudgetSub} and stacked material budget of the different regions inside the tracking system~\subref{fig:MatBudgetStacked}. Contributions from sensitive layers, cables, supports and cooling are included in the respective regions.}\label{fig:MatBudget}
\end{figure}

When compared to CLIC\_SiD, CLICdet has a much larger tracking system, in particular extending
the forward region acceptance. The number of expected hits in CLICdet as a function of the polar angle $\theta$ is
shown in \cref{TrackerHits}.
The total material budget considering all elements up to the calorimeters is presented in \cref{fig:MatBudget}.

\subsection{Calorimetry}
Calorimetry at CLIC is designed according to requirements given by the particle flow paradigm. 
An additional design criterion is good photon energy resolution over a wide energy range.
The ECAL and HCAL barrel of CLICdet are arranged in dodecagons around the tracker volume. The endcap calorimeters
are arranged to provide good coverage in the transition region, and maximum coverage to small polar angles. The overall dimensions of the calorimeters are given in \cref{table.overall}.

The ECAL is a highly granular array of 40 layers of silicon sensors and tungsten plates. The \SI{1.9}{mm} tungsten plates, together with sensors and readout, add up to 22 radiation lengths.
The lateral segmentation of the \SI{300}{\micron} thick sensors is chosen to be \SI{5x5}{mm\squared}. The present layout of the ECAL allows for excellent energy resolution (e.g.\ for high energy photons),
see \cref{sec:single_particle}.

The HCAL is built from 60 layers of plastic scintillator tiles, read out by silicon photomultipliers, interleaved with \SI{20}{mm} thick steel absorber plates.
The scintillator tiles are \SI{3}{mm} thick and have lateral dimensions of \SI{3x3}{cm\squared}. Together with tracker and ECAL, and using the Pandora particle-flow analysis software~\cite{Marshall:2015rfaPandoraSDK,Thomson:2009rp,Marshall:2012ryPandoraPFA},
jet energy resolutions in the order of 4\% to 5\% are obtained (see \cref{sec:jets_E}).

\begin{figure}[htbp]
         \centering
\includegraphics[scale=0.492]{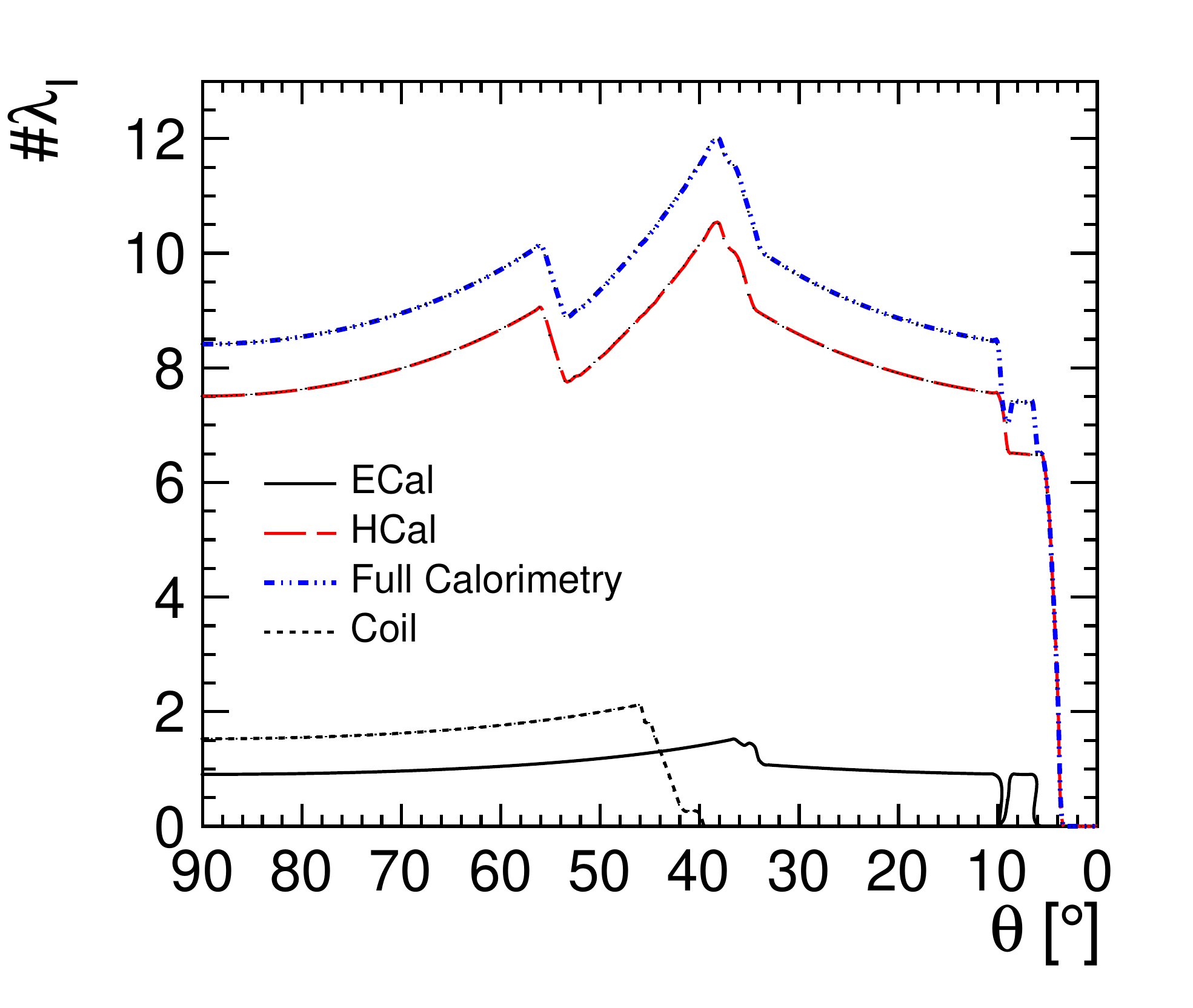}
         \caption{Nuclear interaction lengths \nuclen{} in the calorimeters with respect to the polar angle $\theta$~\cite{CLICdet_note_2017}.
The interaction length corresponding to the material of the superconducting coil is shown for completeness.}
        \label{int_length}
\end{figure}

Earlier studies had shown that about 7.5 nuclear interaction lengths (\nuclen{}) of depth are needed in the HCAL, in addition to the 1~\nuclen{} from the ECAL~\cite{cdrvol2}.
The integrated thickness of the CLICdet calorimeters, in terms of nuclear interaction lengths \nuclen, is shown in
\cref{int_length}.

\subsection{Muon Detector System}

\begin{figure}[tb]
  \centering
  \includegraphics[scale=0.3]{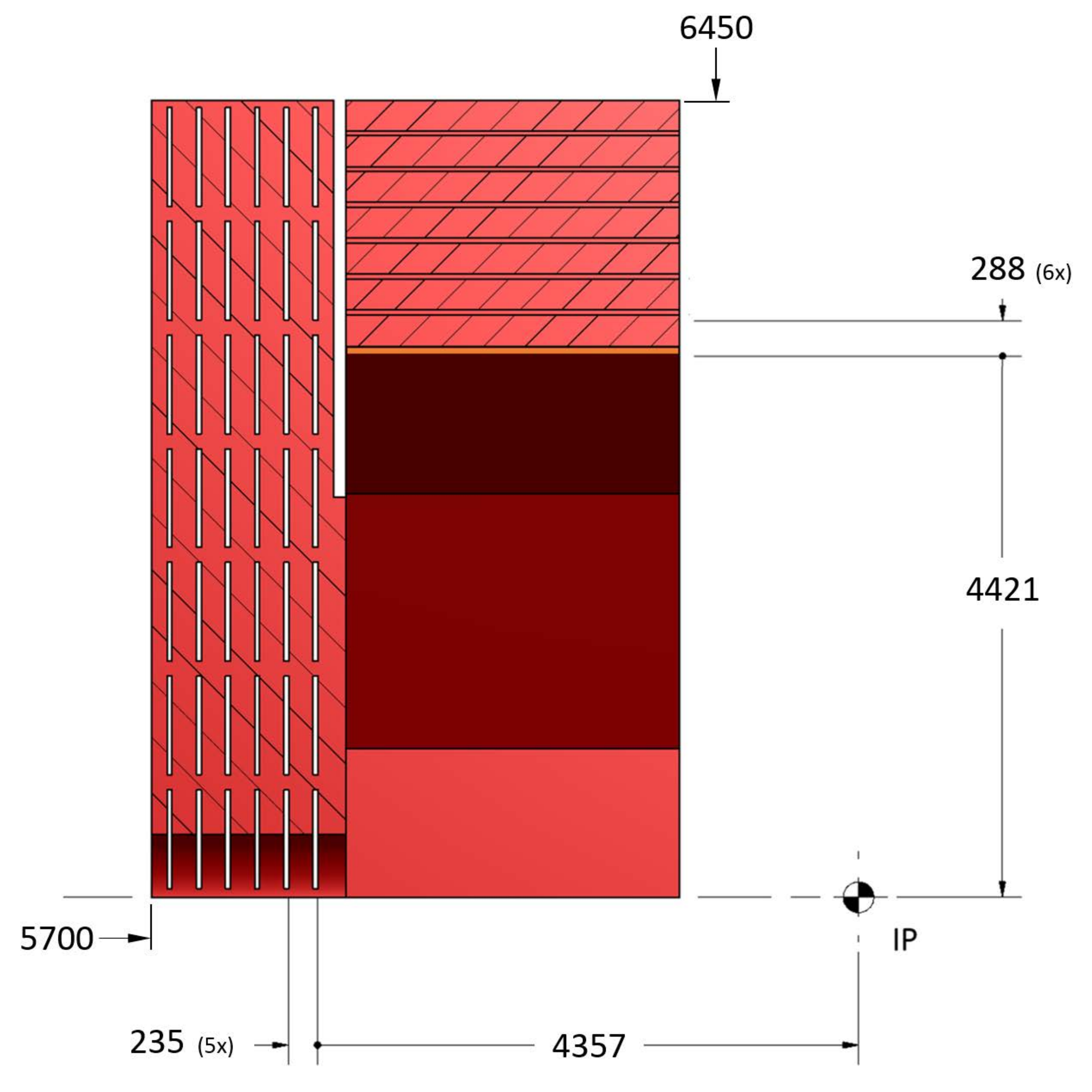}
  \caption{Schematic cross section of the muon system layout in the yoke of CLICdet~\cite{CLICdet_note_2017}.}\label{fig:muons}
\end{figure}

The muon system contains 6 layers of detectors interleaved with the yoke steel plates.
In the barrel, a 7$^\mathrm{th}$ layer as close as possible to the coil is foreseen, which can also act as tail catcher for the calorimeter system.
The layout of the muon system is shown in \cref{fig:muons}.

As in the CDR, the muon detection layers are proposed to be built as RPCs with cells of \SI{30x30}{mm\squared}
(alternatively, crossed scintillator bars could be envisaged). The free space between yoke steel layers is
\SI{40}{mm}, which is considered generous given present-day technologies for building RPCs.

\subsection{Very Forward Calorimeters LumiCal and BeamCal}
\begin{figure}[tb]
  \centering
  \includegraphics[width=0.8\textwidth,trim=0 30 0 30,clip]{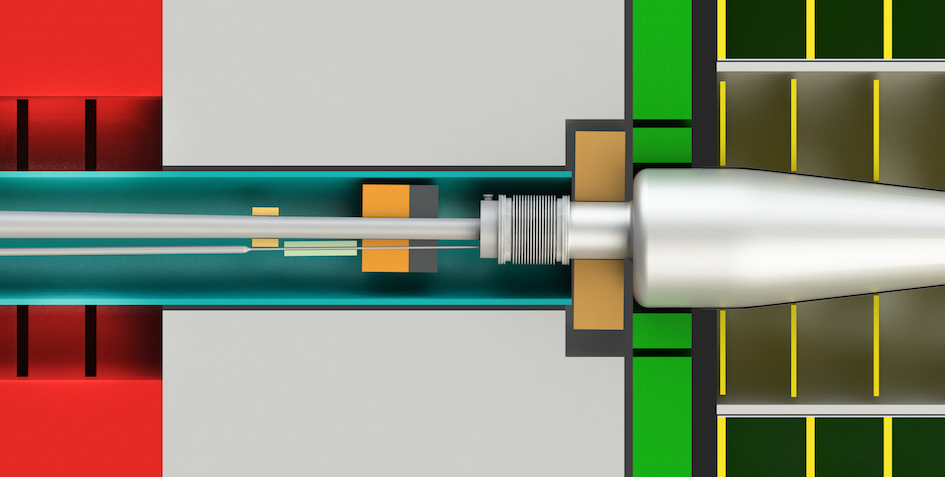}
  \caption{Layout of the forward region in CLICdet (top view)~\cite{CLICdet_note_2017}. The LumiCal is shown in
    khaki, the BeamCal in orange. Downstream of LumiCal are the bellows of the vacuum system. A graphite cylinder is installed upstream of BeamCal -- this \ign{allows one to considerably reduce} backscattering from BeamCal into the central detector region.}\label{very_forward_layout}
\end{figure}

Two smaller electromagnetic calorimeters close the very forward angular region of CLICdet: LumiCal, covering an angular range from \SI{39}{mrad} to \SI{134}{mrad}, and BeamCal,
covering from \SI{10}{mrad} to \SI{46}{mrad}. The layout of the very forward region is shown schematically in \cref{very_forward_layout}. Both calorimeters are built from 40 layers of \SI{3.5}{mm} thick tungsten plates, interleaved with sensors.
The readout electronics is located at the periphery of these calorimeters. LumiCal sensors will be \SI{300}{\micron} thick silicon pads, in a layout optimised for high precision luminosity measurements using Bhabha events. The LumiCal sensor pads have a radial size of \SI{3.75}{\milli\meter} and an azimuthal size of \ang{7.5}.
In the BeamCal, the sensors must be more radiation tolerant, but at the moment silicon sensors are used in the simulation model. The BeamCal cell sizes are constant in radius and $R\phi$ with about
\SI{8x8}{\square\milli\meter}.

\clearpage
\section{Summary of CLIC Experimental Conditions and Detector Requirements}
\label{clic_beam_and_bg}

The design requirements for a detector at CLIC have been described in detail in the CDR~\cite[Chapter 2]{cdrvol2}.
An updated summary, using beam optics parameters for the new default $\lstar=\SI{6}{m}$, and with emphasis on the first stage of CLIC at \SI{380}{GeV} centre-of-mass energy, is given here.

\subsection{The CLIC Beam}

The main parameters of the CLIC beam of relevance to the physics reach, to the beam--beam backgrounds at the \ac{IP} and thus the detector design
are summarised in \cref{tab:clicBeam} for the first CLIC energy stage, \SI{380}{GeV}, and the ultimate high energy stage at \SI{3}{TeV}~\cite{CLIC_staging}.

The time structure of the CLIC beam, with bunch trains colliding every \SI{20}{ms}, is very similar at \SI{380}{GeV} and at \SI{3}{TeV}\@. Bunches inside the trains are separated by
\SI{0.5}{ns}. The number of bunches per train is slightly larger at \SI{380}{GeV}, and the number of particles per bunch is significantly larger at \SI{380}{GeV} than at \SI{3}{TeV}.

\begin{figure}[htbp]
\begin{subfigure}{.495\textwidth}
  \centering
  \includegraphics[width=\linewidth]{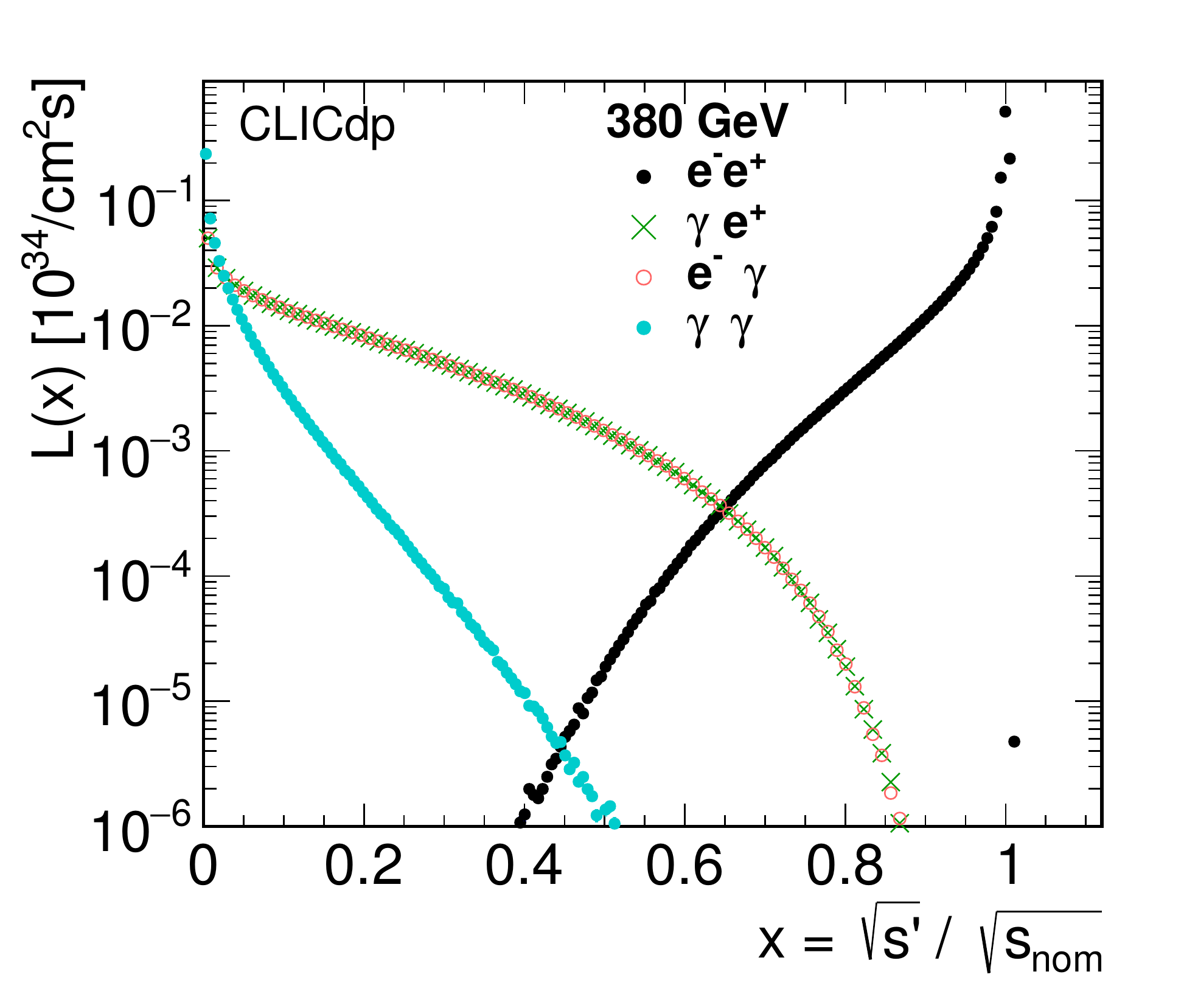}
\end{subfigure}
\begin{subfigure}{.495\textwidth}
  \centering
  \includegraphics[width=\linewidth]{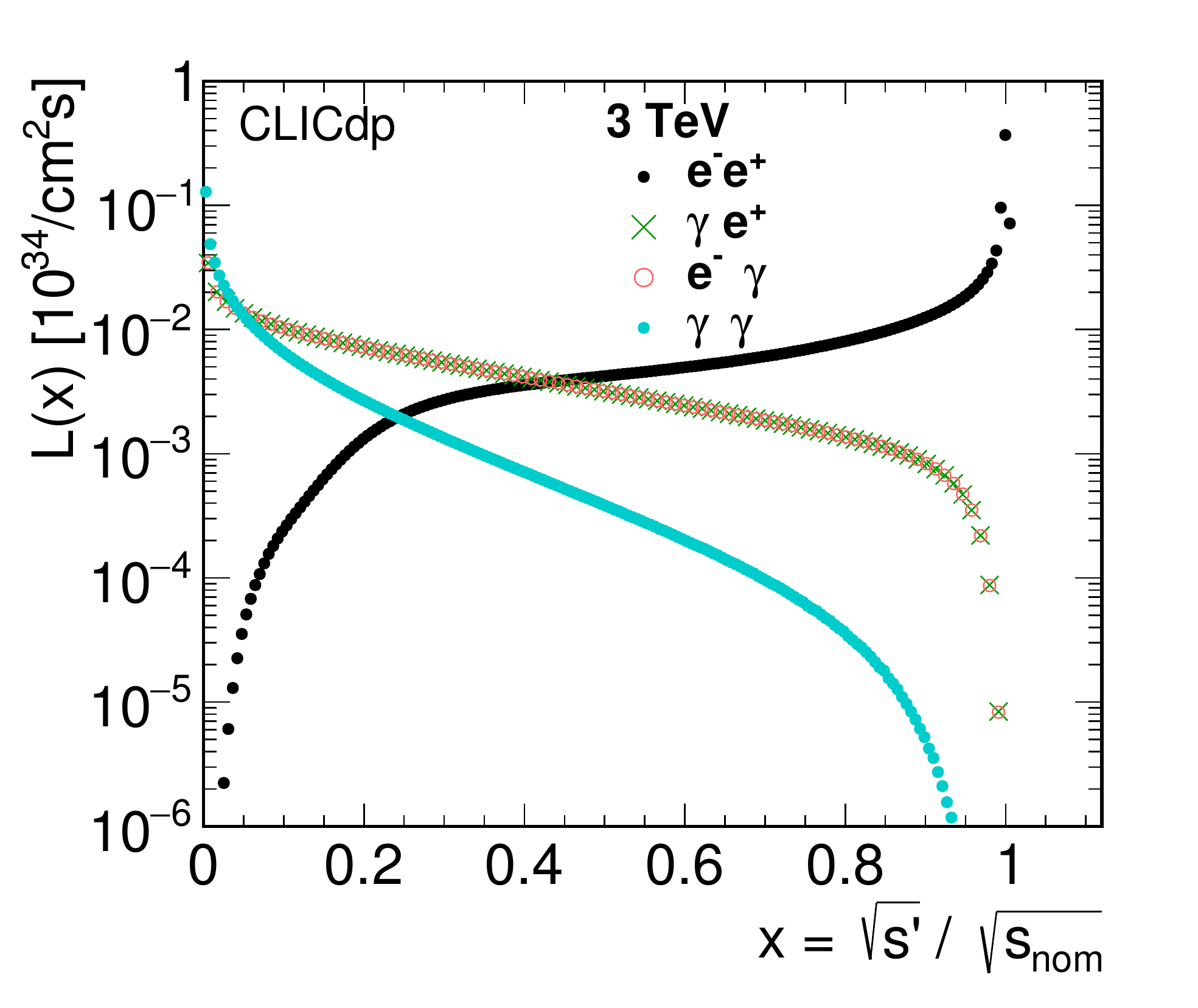}
\end{subfigure}
\vspace{-3mm}
\caption{Luminosity distributions at \SI{380}{GeV} (left) and at \SI{3}{TeV} (right).}\label{fig:lumispectra}
\end{figure}

The beam--beam effects strongly vary with increasing centre-of-mass energy of CLIC\@.
This fact manifests itself e.g.\ in the difference of the number of coherent and incoherent pairs produced,
as well as the number of hadronic events produced by gamma-gamma interactions. As described in~\cite{cdrvol2}, this leads also to very different luminosity spectra
at different energies, as illustrated in \cref{fig:lumispectra}: while there is a strong tail to lower energies at \SI{3}{TeV} CLIC, the tail at \SI{380}{GeV} is much less prominent.
The fraction of luminosity in different regions of the luminosity spectrum, both for \SI{380}{GeV} and \SI{3}{TeV}, is given in \cref{tab:lumiSpectrum}.

\begin{table}[!hbt]
  \centering
  \begin{threeparttable}
  \captionsetup{width=\linewidth}
  \caption{The main parameters of the CLIC machine and background rates at the interaction point.
    The listed variables are:
    $\theta_\mathrm{c}$, the horizontal crossing angle of the beams at the \ac{IP}\@;
    $f_{\mathrm{rep}}$, the repetition frequency\@;
    $n_\mathrm{b}$, the number of bunches per bunch train\@;
    $\Delta t$, the separation between bunches in a train\@;
    $N$, the number of particles per bunch\@;
    $\sigma_\mathrm{x}$, $\sigma_\mathrm{y}$, and $\sigma_\mathrm{z}$, the bunch dimensions at the \ac{IP}\@;
    $\beta_x$ and $\beta_y$, the beta functions at the \ac{IP}\@;
    \lstar{}, the distance from the last quadrupole to the \ac{IP}\@;
    ${\cal{L}}$, the design luminosity\@;
    ${\cal{L}}_{0.01}$, the luminosity with $\rootsprime>0.99\roots$\@;
    $\Delta E/E$, the average fraction of energy lost through beamstrahlung\@;
    $n_\upgamma$, the average number of beamstrahlung photons per beam
    particle\@;
    $N_{\mathrm{coh}}$, the number of coherent pair particles per
    \acf{BX}\@;
    $E_{\mathrm{coh}}$, the total energy of coherent pair particles per \ac{BX}\@;
    $N_{\mathrm{incoh}}$, the number of incoherent pair particles per \ac{BX}\@;
    $E_{\mathrm{incoh}}$, the total energy of incoherent pair particles per \ac{BX}\@;
    and,  $n_{\mathrm{Had}}$, the number of \gghadron{} events per \ac{BX}
    for a $\upgamma\upgamma$ centre-of-mass energy threshold of $W_{\upgamma\upgamma} > \SI{2}{GeV}$.  
    The background rates and energy releases are quoted excluding
    safety factors representing the simulation uncertainties.
    \label{tab:clicBeam}}
  \begin{tabular}{c c c c}
    \toprule
    Parameter              & Units               & $\roots=\SI{380}{GeV}$   & $\roots=\SI{3}{TeV}$     \\  \midrule
    $\theta_\mathrm{c}$    &      mrad&       16.5   &        20          \\
    $f_{\mathrm{rep}}$     & Hz                  & 50                 & 50                 \\
    $n_{\mathrm{b}}$       &                     & 352                & 312                \\
    $\Delta t$             & ns                  & 0.5                & 0.5                \\
    $N$                    &                     & $5.2 \cdot 10^9$    & $3.72 \cdot 10^9$    \\
    $\sigma_x$             & nm                  & $\approx 149$                & $\approx 45$        \\
    $\sigma_y$             & nm                  & $\approx 2.9$               & $\approx 1$         \\
    $\sigma_z$             & \micron{}           & 70                 & 44                 \\
    $\beta_x$           & mm                  & 8                  & 7                  \\
    $\beta_y$           & mm                  & 0.1                & 0.12               \\  
    {L$^*$}   &     m            &  6                &  6         \\
    ${\cal{L}}$            & $\mathrm{ cm^{-2}s^{-1}}$ & $1.5 \cdot 10^{34}$ & $5.9 \cdot 10^{34}$ \\
    ${\cal{L}}_{0.01}$     & $\mathrm{ cm^{-2}s^{-1}}$ & $0.9 \cdot 10^{34}$ & $2.0 \cdot 10^{34}$ \\
    $n_\upgamma$           &                     & 1.4                & 2.0                \\
    {$\Delta E/E$}         &                     & {$0.08$}           & {$0.25$}           \\ \midrule
    {$N_{\mathrm{coh}}$}   &                     & {$\approx 0$}  & {$6.1 \cdot 10^8$}  \\
    {$E_{\mathrm{coh}}$}   & TeV                 & {$\approx 0$}  & {$2.1 \cdot 10^8$}  \\ \midrule
    {$N_{\mathrm{incoh}}$} &                     & {$4.6 \cdot 10^4$}    & {$2.8 \cdot 10^5$}    \\
    {$E_{\mathrm{incoh}}$} & TeV                 & $2.1 \cdot 10^2$    & $2.1 \cdot 10^4$    \\ \midrule
    {$n_{\mathrm{Had}}$} ($W_{\upgamma\upgamma}>$2~GeV)   &                     & {0.17}             & {3.1}              \\\bottomrule
    & & &  \\ 
  \end{tabular}

\end{threeparttable}
\end{table}

\begin{table}[h]
  \centering
  \caption{\label{tab:lumiSpectrum} Fraction of luminosity above $\rootsprime/\roots$.}
  \begin{tabular}{c  *2{c}} \toprule
    \tabt{Fraction \rootsprime/\roots} & \tabt{\SI{380}{GeV}} & \tabt{\SI{3}{TeV}} \\ \midrule
    $>0.99$                            & 60\%           & 36\%         \\
    $>0.90$                            & 90\%           & 57\%         \\
    $>0.80$                            & 97,6\%         & 69\%         \\
    $>0.70$                            & 99.5\%         & 76.8\%       \\
    $>0.50$                            & 99.99\%        & 88.6\%       \\ \bottomrule
  \end{tabular}
\end{table}


\subsection{Beam-Induced Backgrounds}
\label{sec:beam-induc-backgr}

All the sources of beam related backgrounds, in particular for the \SI{3}{TeV} stage of CLIC, have been discussed in detail in the CDR~\cite[Section 2.1.2]{cdrvol2}.
Comparisons of background particle distributions at \SI{380}{GeV}, direct hits in all subdetectors as well as indirect hits from secondary and  backscattered particles are
given in the following. 

Background hits from coherent pairs, which are produced in large quantities and have high energy, are
avoided by the design of the interaction region, and in particular the shape of the outgoing beam pipe: a cone with
opening angle \SI{10}{mrad} allows sufficient space for the coherent pairs to leave the detector region. The
post-collision line is designed to transport these particles to the beam dump, together with the 
spent beam and the beamstrahlung photons.

Backscattering from the post-collision line and the main CLIC beam dumps, \SI{315}{m} downstream of the \ac{IP},
 has been investigated~\cite{Appleby:1403134}. The average flux of backscattered photons and neutrons hitting the detector area
was found to be negligible.

Beam halo muons can be largely suppressed by optimising the beam delivery system, and in particular the
collimation system. The expected level of halo muons traversing the detector should be easily handled by
CLICdet, mainly due to the high granularity and timing resolution of the subdetector systems~\cite{cdrvol2}.

\begin{figure}[hbt]
\centering
\begin{subfigure}{.495\textwidth}
  \centering
  \includegraphics[width=\linewidth]{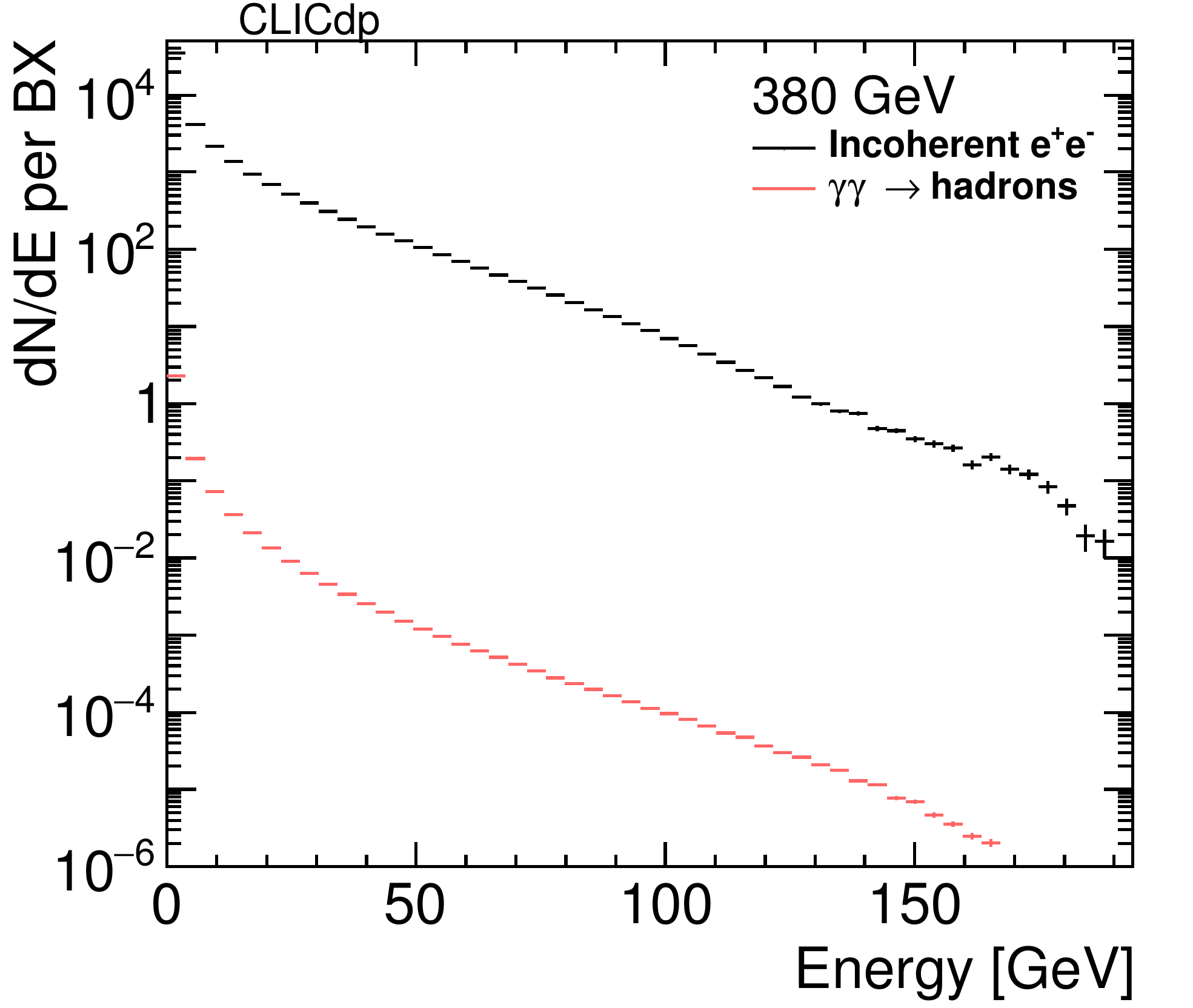}\label{fig:background_380gev}
\end{subfigure}
\begin{subfigure}{.495\textwidth}
  \centering
  \includegraphics[width=\linewidth]{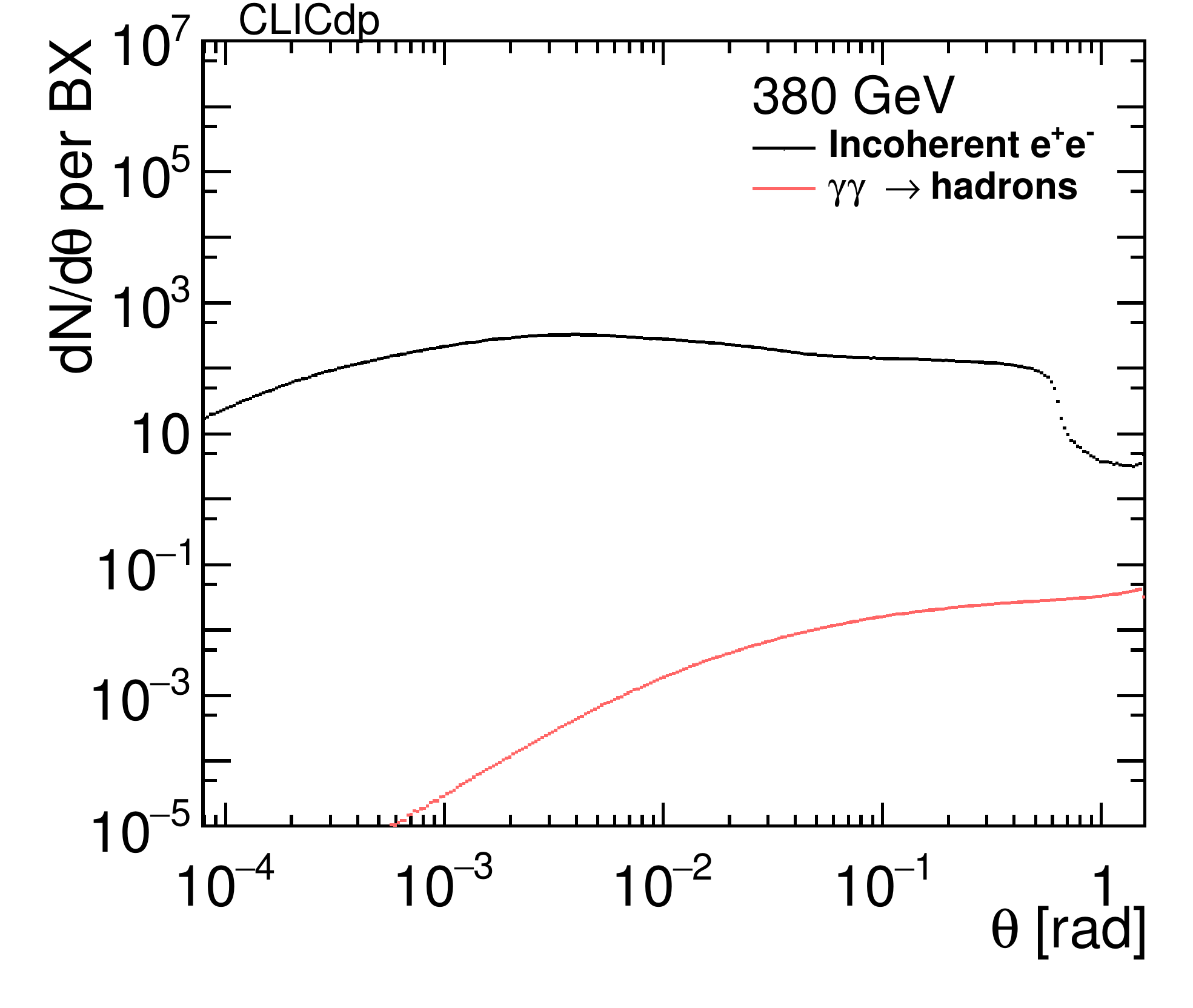}\label{fig:background_3tev}
\end{subfigure}
\caption{Energy distribution (left) and polar angle distribution (right) per \ac{BX} of beam-induced backgrounds. Both
  figures are for CLIC at \SI{380}{GeV}\@. Generated particle distributions are shown, no cuts other than the
  \SI{2}{GeV} centre-of-mass threshold for \gghad{} are applied.}\label{fig:background}
\end{figure}

Incoherent pairs, produced from the interaction of real or virtual photons with individual particles of the oncoming beam,
can be produced at larger angles than coherent pairs, and are potentially a significant source of background hits, 
in particular in the vertex detector. The energy and angular distribution of the incoherent pairs at \SI{380}{GeV} centre-of-mass energy
are shown in \cref{fig:background}.

The interaction of real or virtual photons from the colliding beams can produce hadronic final states. These \gghadron{}
interactions can produce particles at a large angle to the beam line. The energy and angular distributions of these
background particles are also shown in \cref{fig:background}. The impact of backgrounds is discussed in \cref{sec:impact}.

\subsection{Overview of Requirements for Physics Reconstruction}
\label{physicsRequirements}

The detector requirements, in particular for a \SI{3}{TeV} CLIC collider, have been described in detail in the CDR~\cite[Section 2.2]{cdrvol2}.
Summarising the findings, from the perspective of the likely physics measurements at CLIC the detector requirements are:
\begin{itemize}
 \item jet energy resolution of $\sigma_E/E \lesssim 5\text{--}3.5\%$ for light quark jet energies in the
 range \SI{50}{GeV}--\SI{1}{TeV}\@;
 \item track momentum resolution of $\sigma_{\pT}/\pT^2 \lesssim \SI{2e-5}{\per\gev}$ for high momentum tracks;
 \item transverse impact parameter resolution $\sigma_{d_{0}}(p, \theta) = \sqrt{a^2+b^2\cdot\si{\gev\squared}/(p^2 \sin^3(\theta))}$
   with $a \lesssim 5~\micron$, $b \lesssim 15~\micron$;
 \item lepton identification efficiency: $>95\%$ over the full range of energies; 
 \item detector coverage for electrons down to very small angles. 
\end{itemize}


\subsection{Impact of Backgrounds on the Detector Requirements}
\label{sec:impact}
The main beam-related backgrounds in the CLIC detector are from incoherent pairs, and particles from \gghadron events.
As discussed in the CDR, particles from incoherent pairs are the dominant backgrounds in the vertex and the
very forward region. The particles from \gghad{} are less forward-peaked and the dominant source
of background in the silicon tracker and the calorimeters. As shown in \cref{tab:clicBeam}, the
number of background particles varies strongly with the CLIC centre-of-mass energy.
 
A detailed optimisation of the position of BeamCal, and the openings to allow for the incoming and outgoing
beam pipes, had been performed at the time of the CDR and was not repeated for CLICdet. In the following,
results from full simulation studies, thus including all multiple- and back-scattering effects, are demonstrating
the impact of these beam-related backgrounds. 

\subsubsection{Impact on Vertex and Tracking Detectors}
The dense core of particles from the incoherent pair background, spiralling near the beam axis due to the \SI{4}{T} solenoid field,
must not intercept any material of the detector. As shown in the CDR, at the \SI{3}{TeV} stage of CLIC this imposes an inner radius
of the beryllium beam pipe of \SI{29.4}{mm}.

\begin{figure}[htbp]
  \renewcommand{\thesubfigure}{(\lr{subfigure})}
  \centering
  \begin{subfigure}{.49\textwidth}
    \centering
    \includegraphics[width=\linewidth]{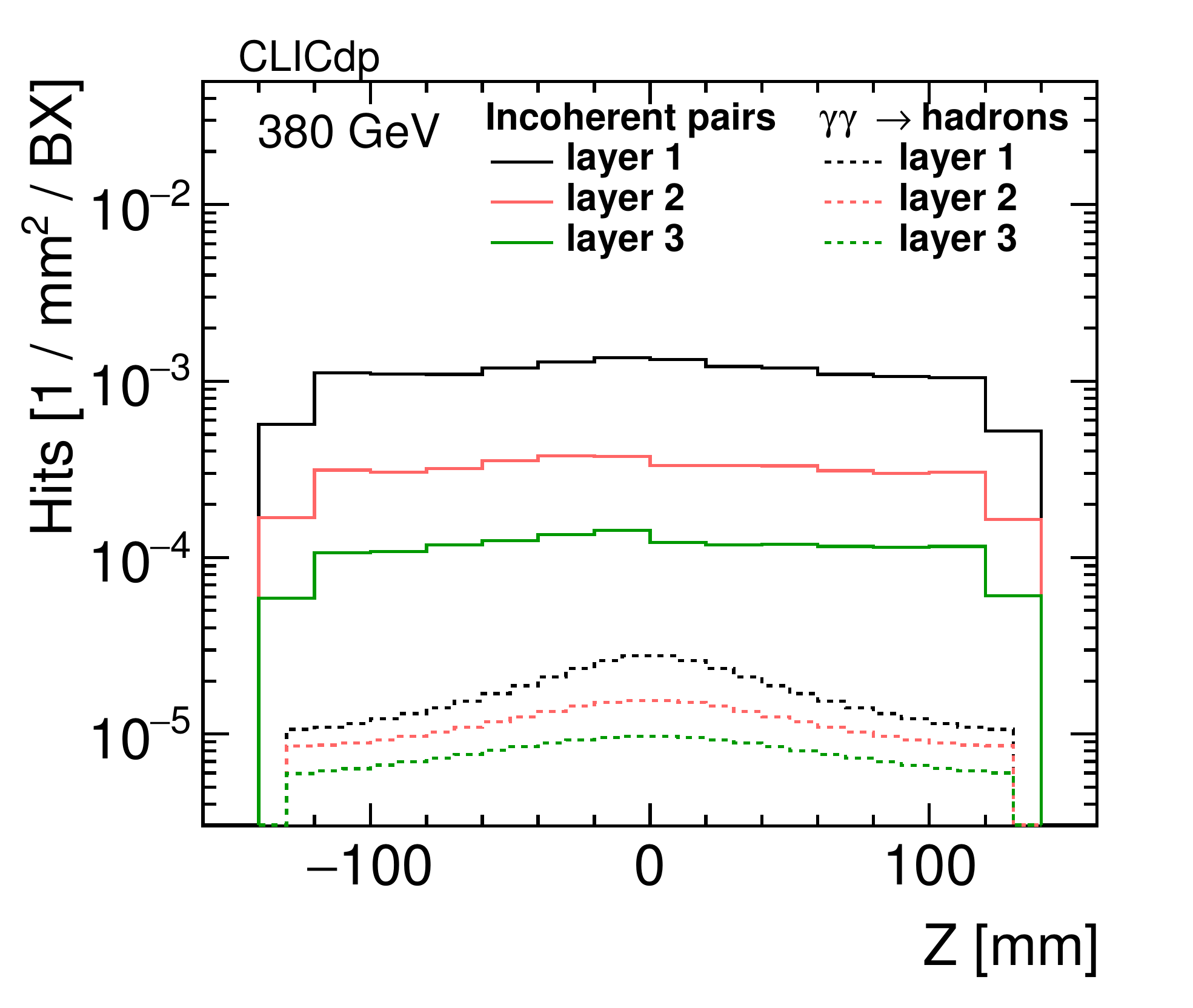}%
    \phantomsubcaption\label{fig:vtx_barrel_hits_380}
  \end{subfigure}%
  \begin{subfigure}{.49\textwidth}
    \centering
    \includegraphics[width=\linewidth]{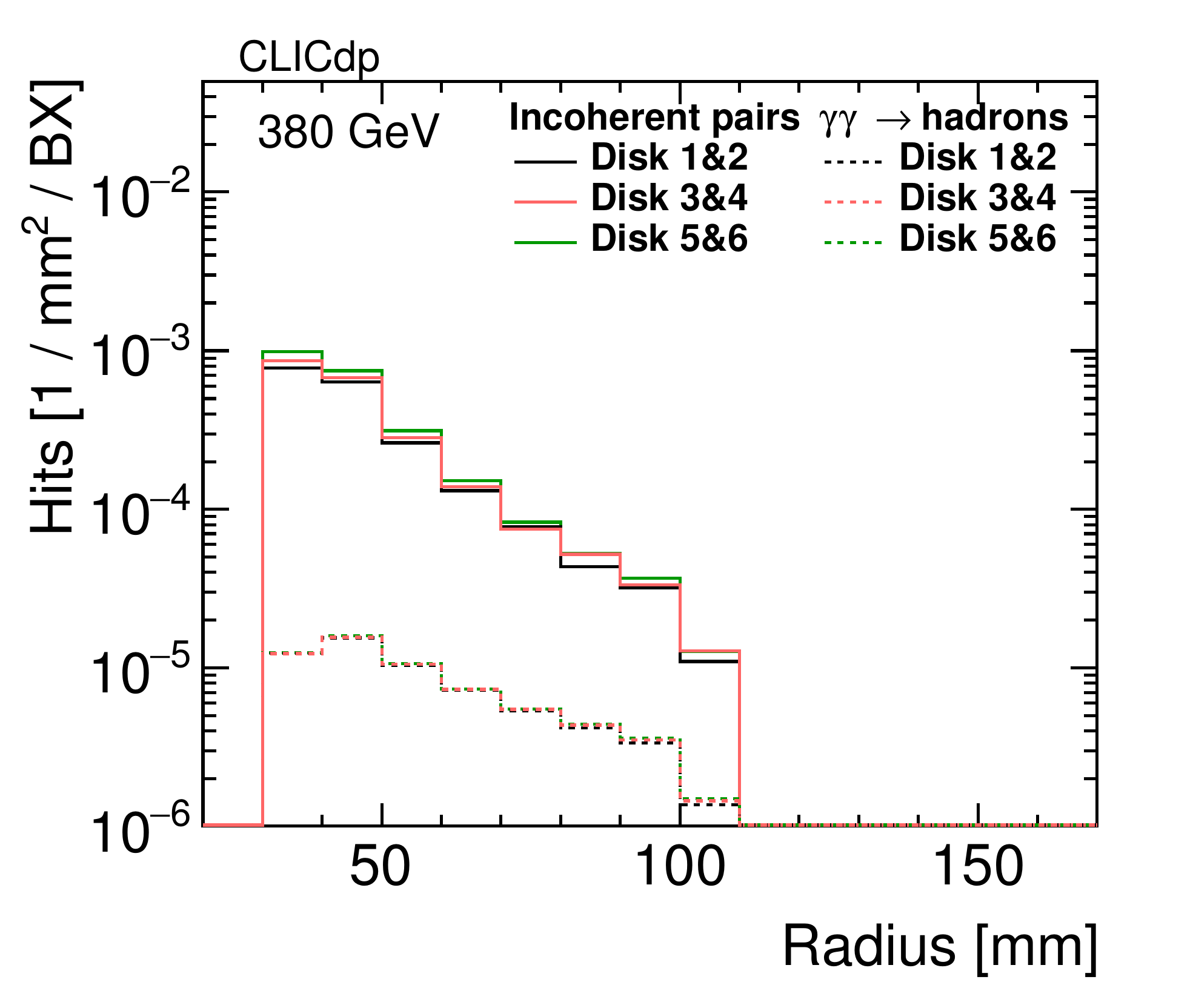}%
    \phantomsubcaption\label{fig:vtx_disks_hits_380}
  \end{subfigure}
  \vspace{-2mm}
  \caption{Hit densities per bunch crossing in the vertex barrel~\subref{fig:vtx_barrel_hits_380}
    and disks~\subref{fig:vtx_disks_hits_380} from incoherent electron--positron pairs and \gghad{} at
    \SI{380}{GeV}\@. Safety factors are not included.} \label{fig:vtx_hits_380}
\end{figure}

\begin{figure}[htbp]
  \renewcommand{\thesubfigure}{(\lr{subfigure})}
  \centering
  \begin{subfigure}{.49\textwidth}
    \centering
    \includegraphics[width=\linewidth]{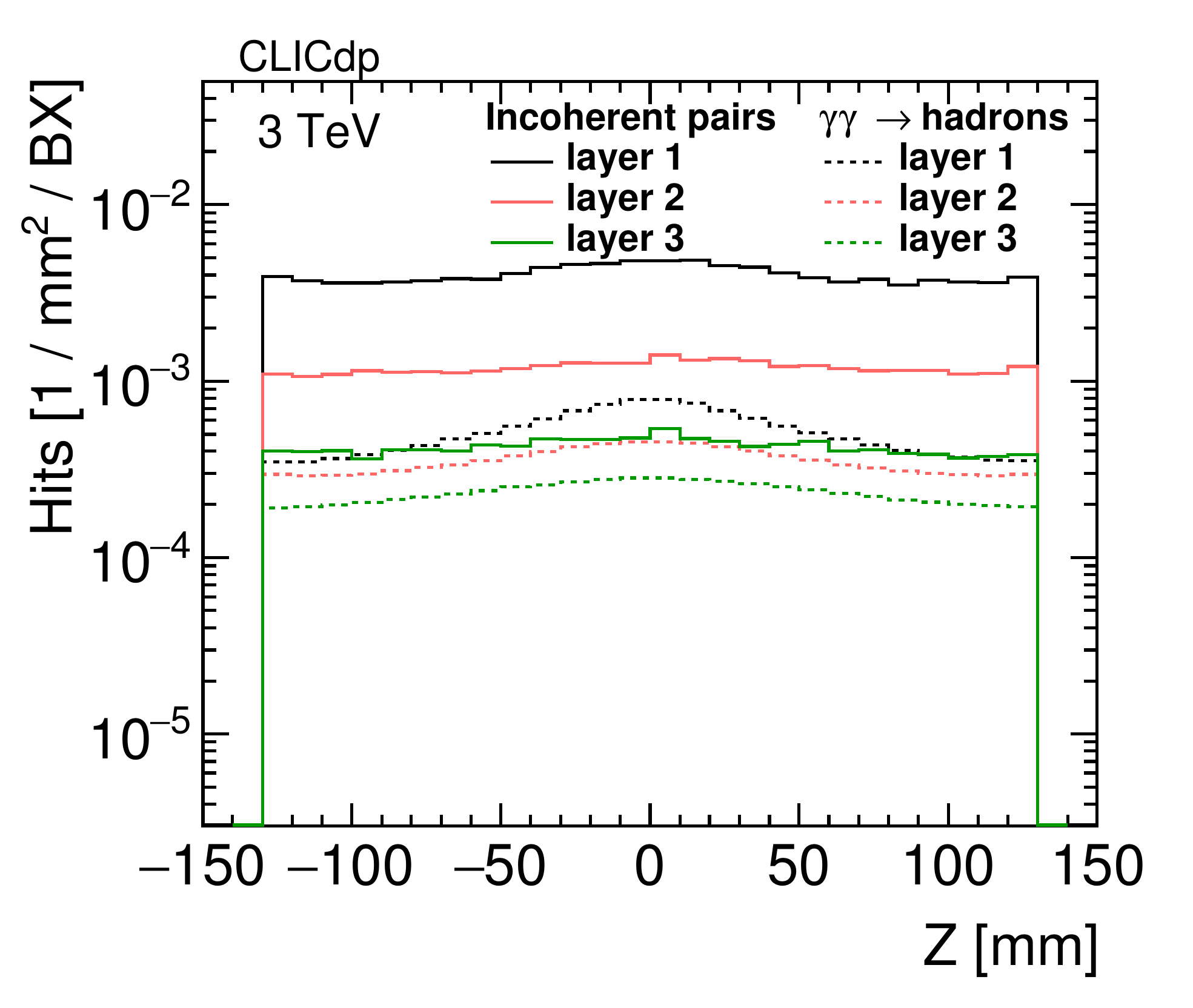}%
    \phantomsubcaption\label{fig:vtx_barrel_hits_3tev}
  \end{subfigure}%
  \begin{subfigure}{.49\textwidth}
    \centering
    \includegraphics[width=\linewidth]{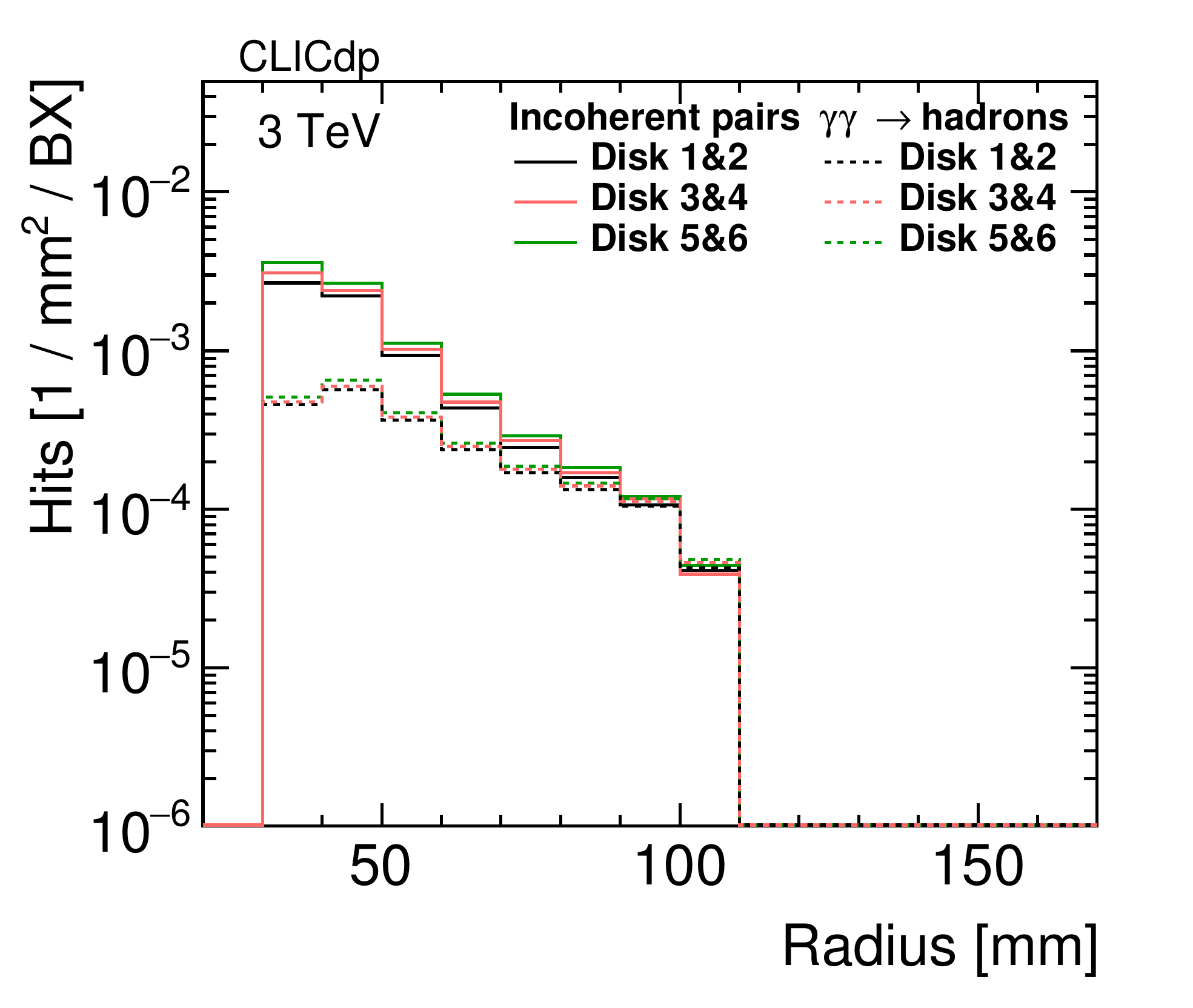}%
    \phantomsubcaption\label{fig:vtx_disks_hits_3tev}
  \end{subfigure}
  \vspace{-2mm}
  \caption{Hit densities per bunch crossing in the vertex barrel~\subref{fig:vtx_barrel_hits_3tev}
    and disks~\subref{fig:vtx_disks_hits_3tev} detector from incoherent electron--positron pairs and \gghad{} at
    \SI{3}{TeV}\@. Safety factors are not included.}\label{fig:vtx_hits_3tev}
\end{figure}

\begin{figure}[htbp]
  \renewcommand{\thesubfigure}{(\lr{subfigure})}
  \centering
  \begin{subfigure}{.49\textwidth}
    \centering
    \includegraphics[width=\linewidth]{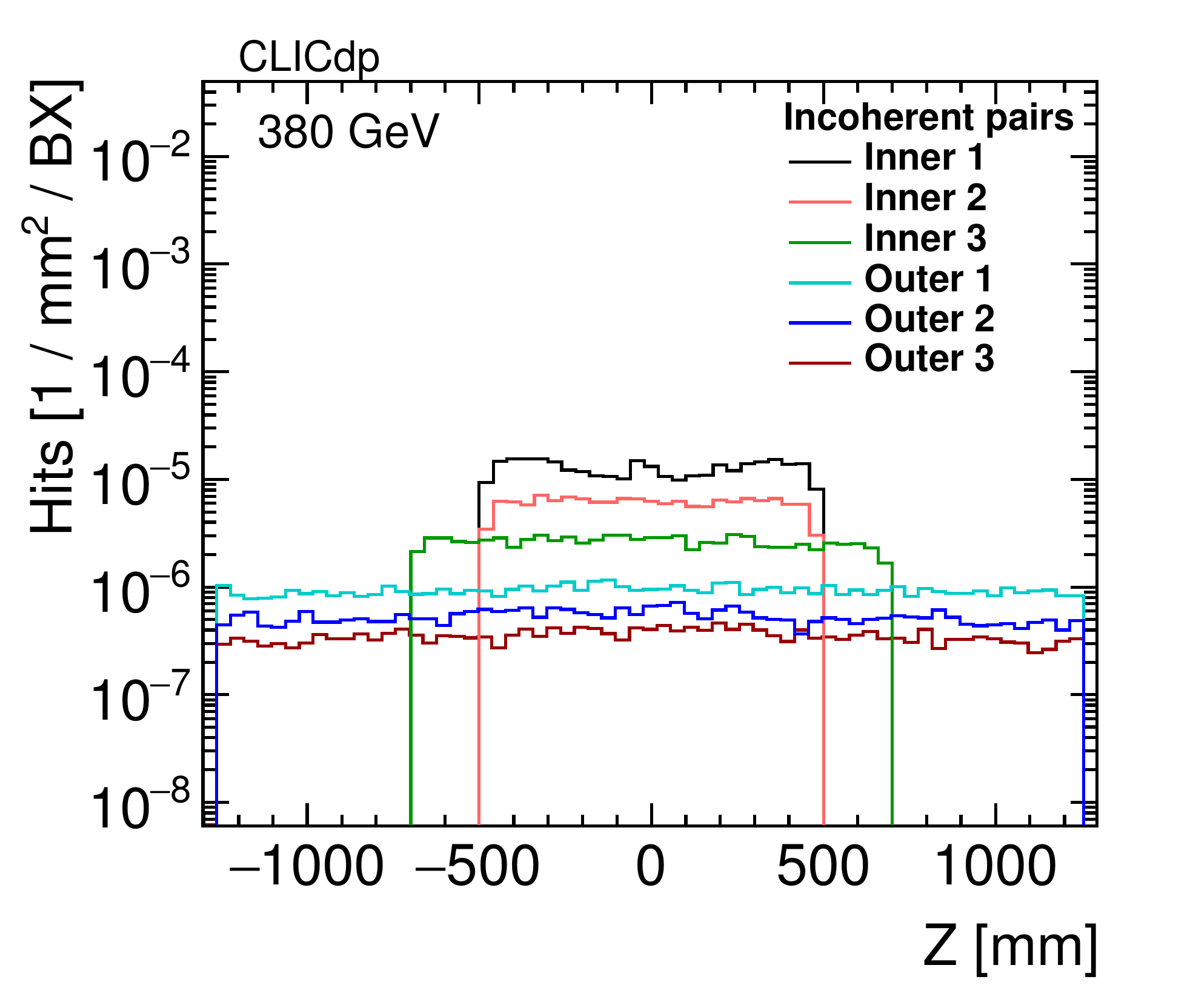}%
    \phantomsubcaption\label{fig:trk_barrel_hits_pairs_380gev}
  \end{subfigure}%
  \begin{subfigure}{.49\textwidth}
    \centering
    \includegraphics[width=\linewidth]{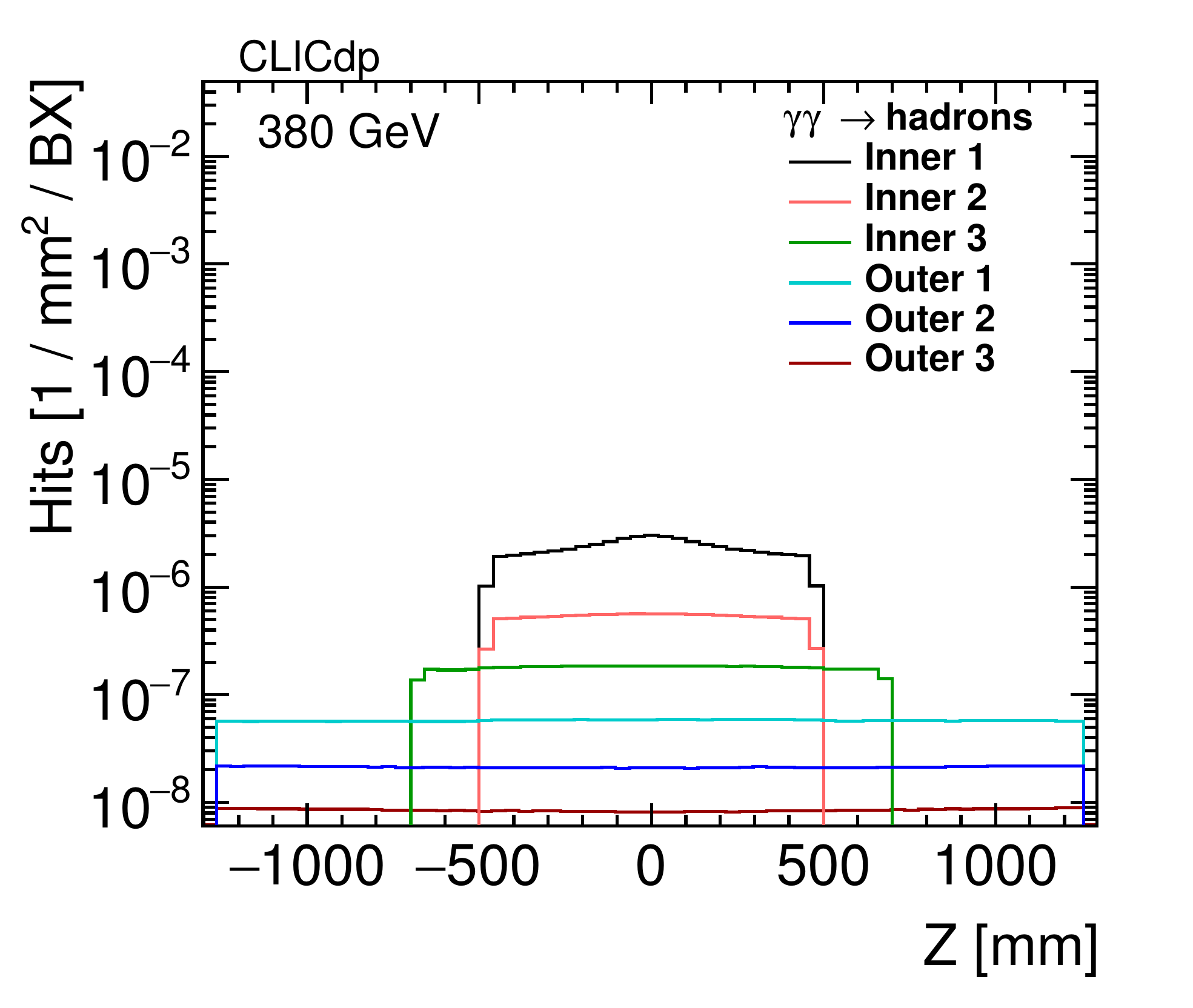}%
    \phantomsubcaption\label{fig:trk_barrel_hits_gghad_380gev}
  \end{subfigure}
  \vspace{-2mm}
  \caption{Hit densities per bunch crossing in the tracker barrel layers from incoherent
    electron--positron pairs~\subref{fig:trk_barrel_hits_pairs_380gev} and \gghad~\subref{fig:trk_barrel_hits_gghad_380gev}
    at \SI{380}{GeV}\@. Safety factors are not included.}
\end{figure}

\begin{figure}[htbp]
  \renewcommand{\thesubfigure}{(\lr{subfigure})}
  \centering
  \begin{subfigure}{.49\textwidth}
    \centering
    \includegraphics[width=\linewidth]{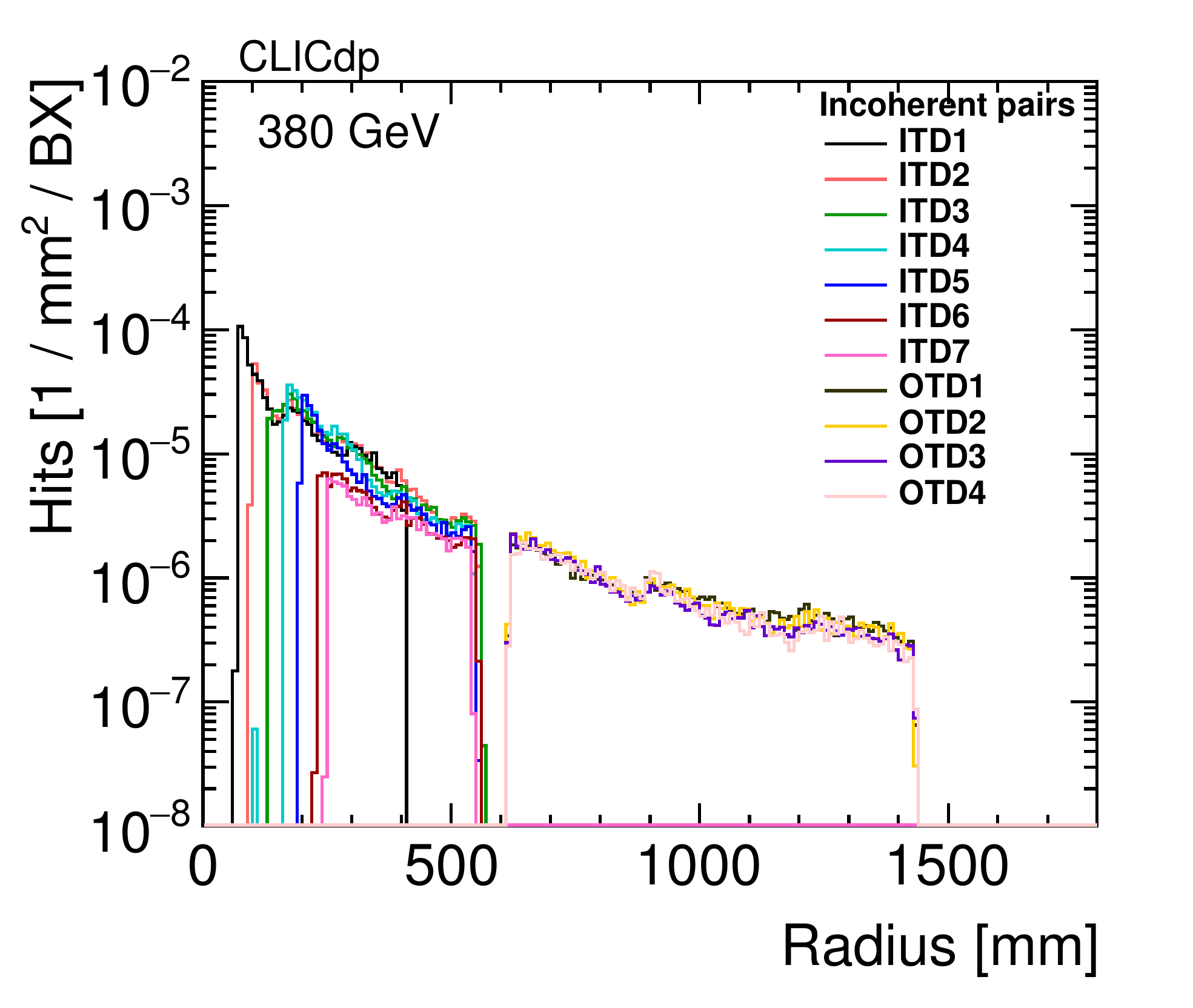}%
    \phantomsubcaption\label{fig:trk_disk_hits_pairs_380gev}
  \end{subfigure}%
  \begin{subfigure}{.49\textwidth}
    \centering
    \includegraphics[width=\linewidth]{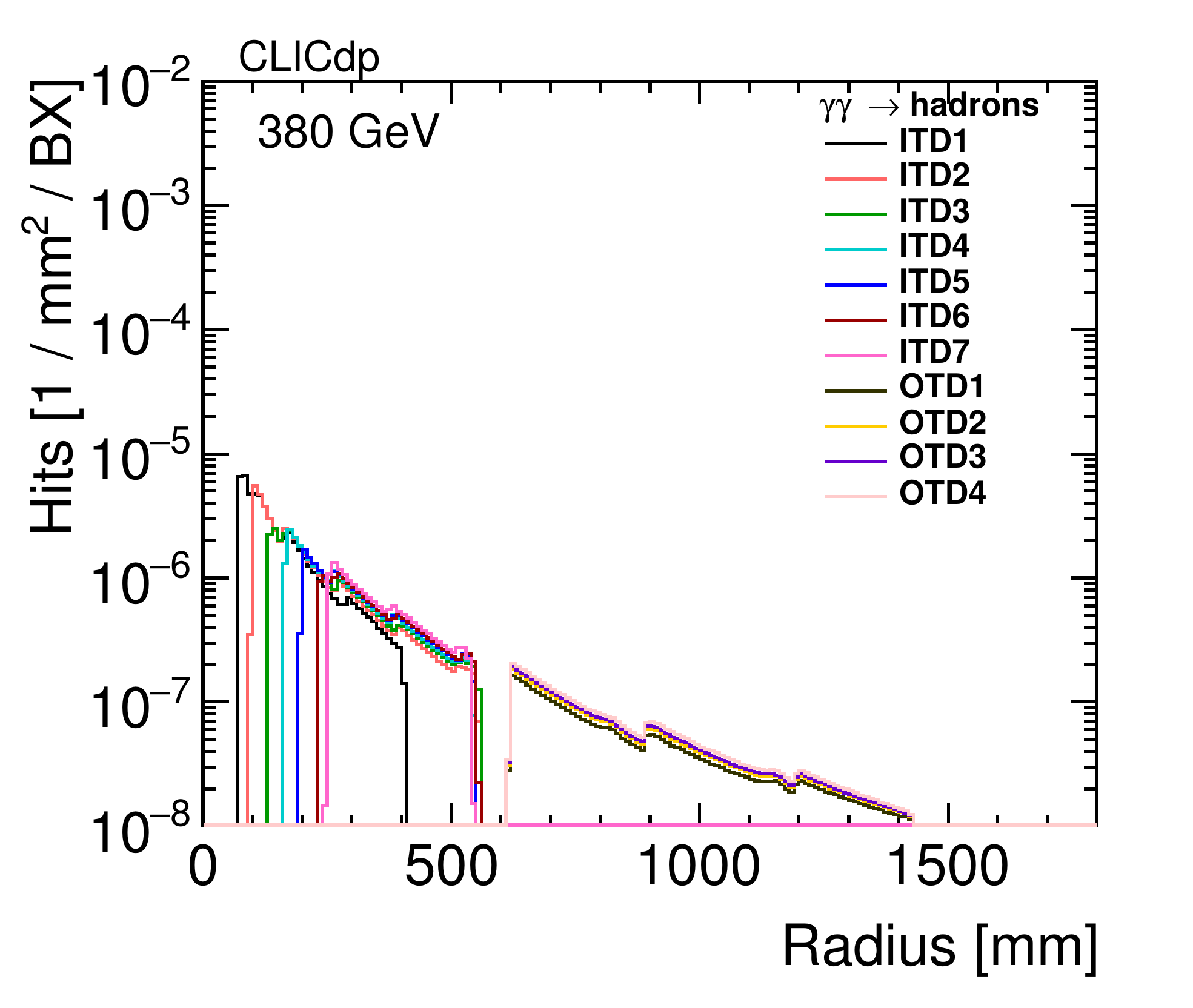}%
    \phantomsubcaption\label{fig:trk_disk_hits_gghad_380gev}
  \end{subfigure}
  \vspace{-2mm}
  \caption{Hit densities per bunch crossing in the tracker disks from incoherent electron--positron
    pairs~\subref{fig:trk_disk_hits_pairs_380gev} and \gghad~\subref{fig:trk_disk_hits_gghad_380gev} at
    \SI{380}{GeV}. Safety factors are not included.}
\end{figure}

\begin{figure}[htbp]
  \renewcommand{\thesubfigure}{(\lr{subfigure})}
  \centering
  \begin{subfigure}{.49\textwidth}
    \centering
    \includegraphics[width=\linewidth]{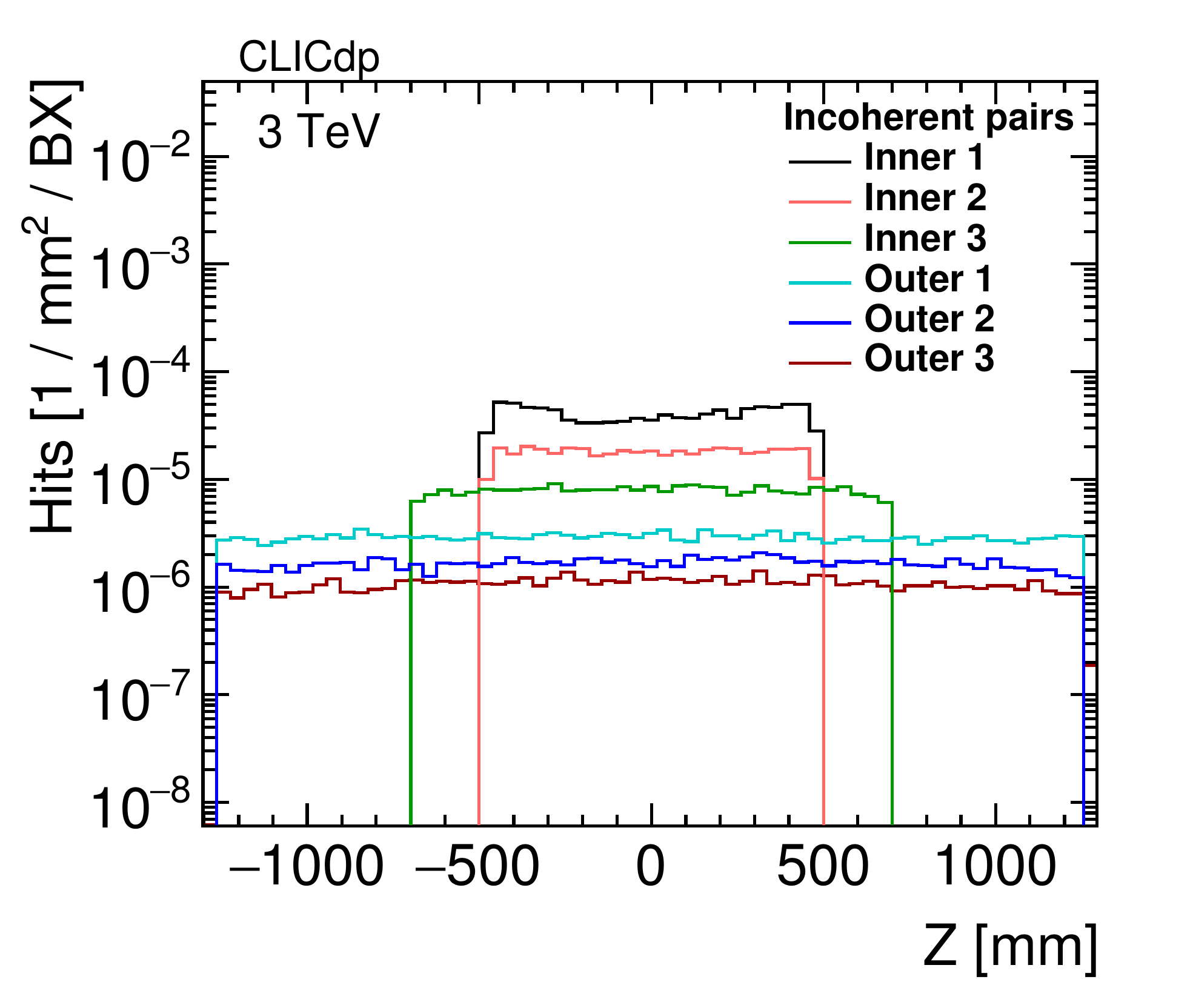}%
    \phantomsubcaption\label{fig:trk_barrel_hits_pairs_3tev}
  \end{subfigure}%
  \begin{subfigure}{.49\textwidth}
    \centering
    \includegraphics[width=\linewidth]{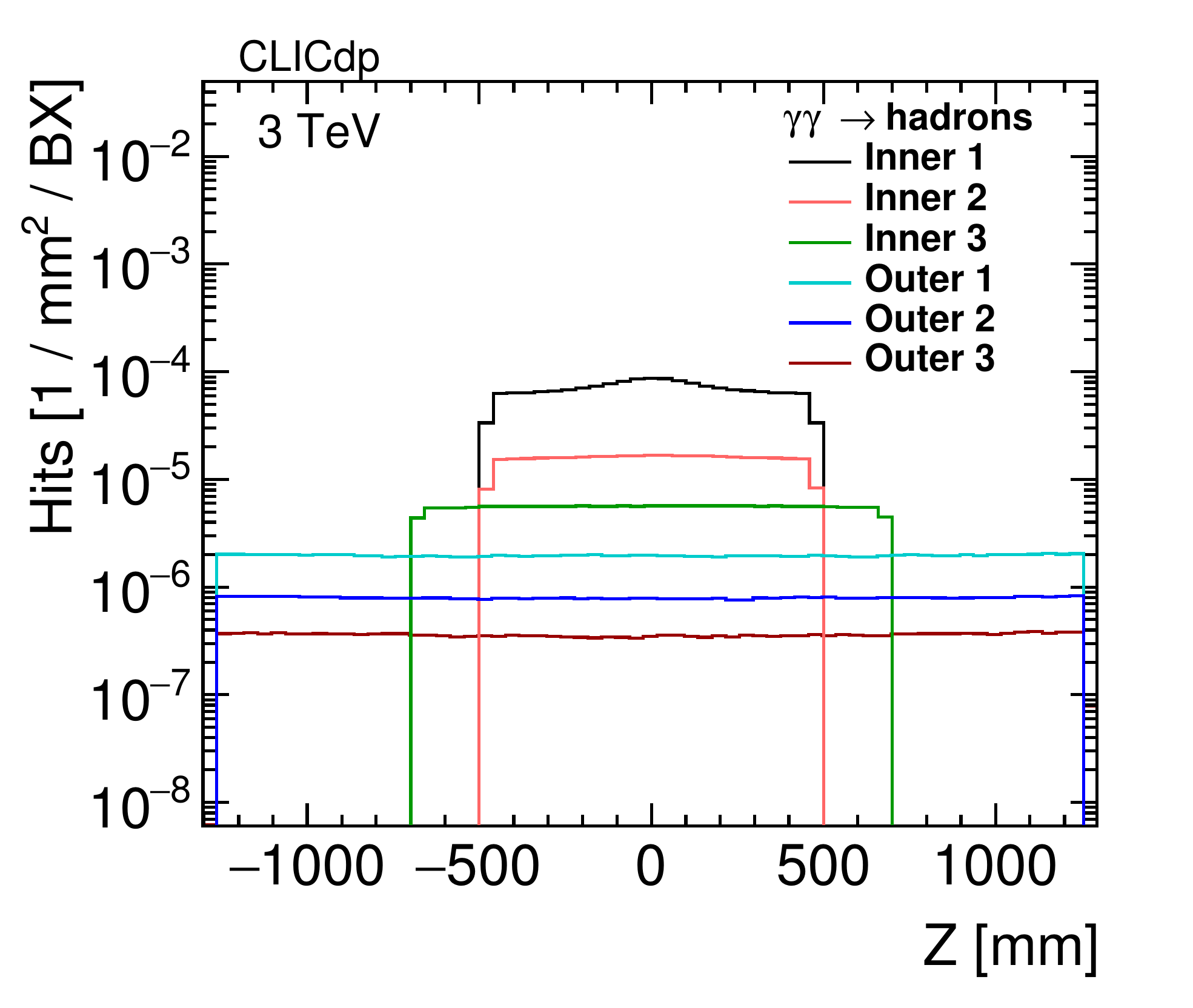}%
    \phantomsubcaption\label{fig:trk_barrel_hits_gghad_3tev}
  \end{subfigure}
  \vspace{-2mm}
  \caption{Hit densities per bunch crossing in the tracker barrel layers from incoherent
    electron--positron pairs~\subref{fig:trk_barrel_hits_pairs_3tev} and \gghad~\subref{fig:trk_barrel_hits_gghad_3tev}
    at \SI{3}{TeV}. Safety factors are not included.}
\end{figure}

\begin{figure}[htbp]
  \renewcommand{\thesubfigure}{(\lr{subfigure})}
  \centering
  \begin{subfigure}{.49\textwidth}
    \centering
    \includegraphics[width=\linewidth]{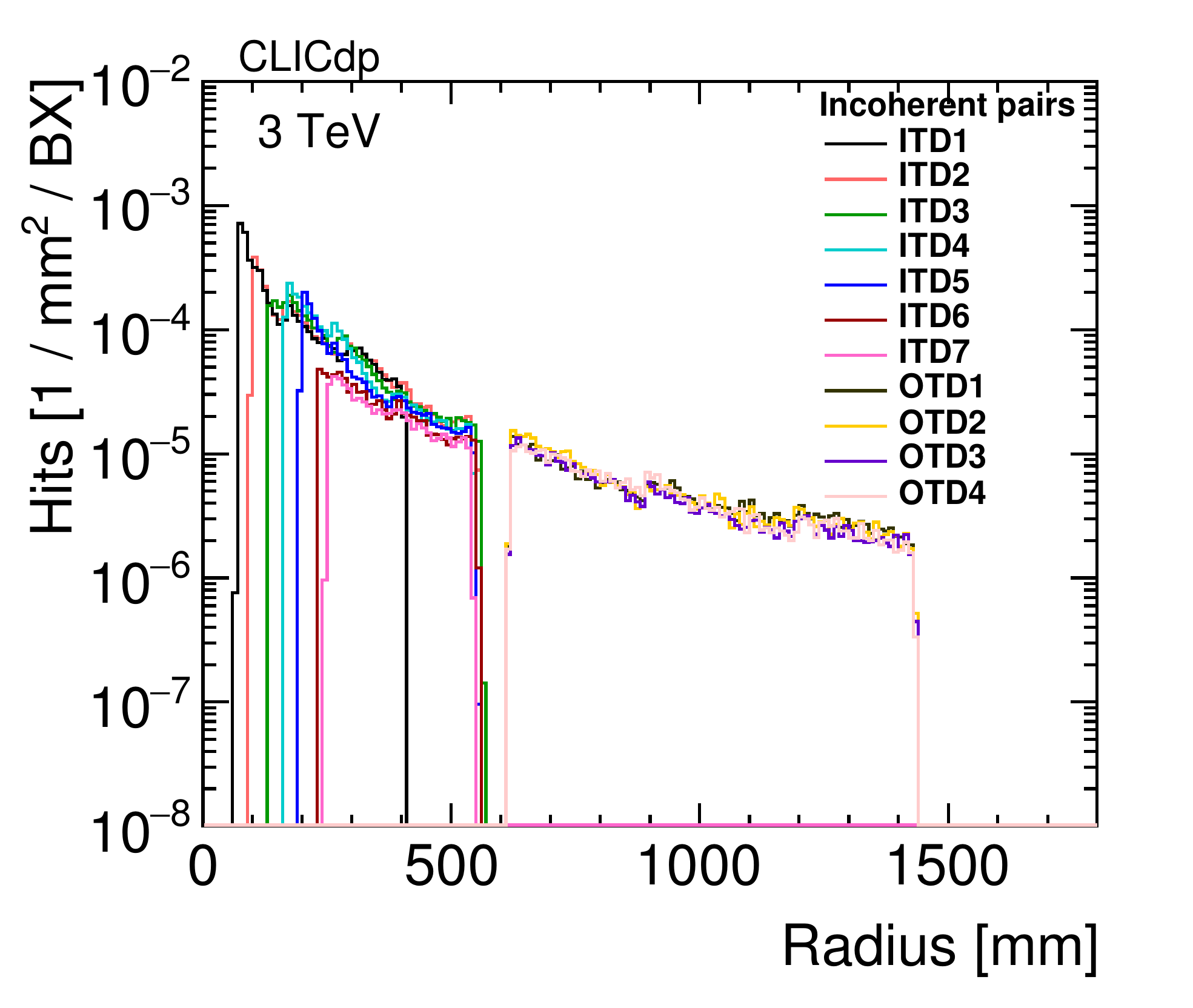}%
    \phantomsubcaption\label{fig:trk_disk_hits_pairs_3tev}
  \end{subfigure}%
  \begin{subfigure}{.49\textwidth}
    \centering
    \includegraphics[width=\linewidth]{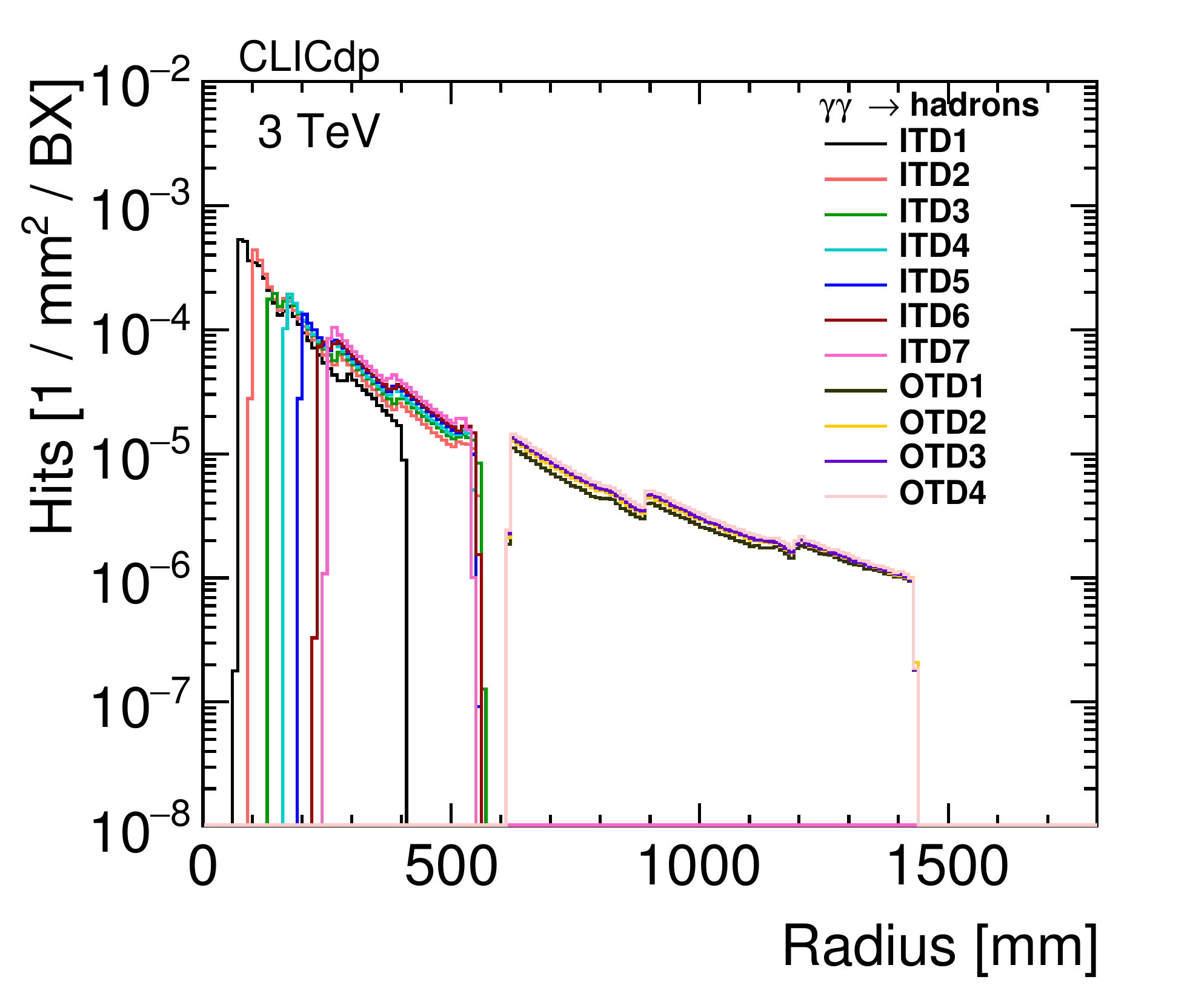}%
    \phantomsubcaption\label{fig:trk_disk_hits_gghad_3tev}
  \end{subfigure}
  \vspace{-2mm}
  \caption{Hit densities per bunch crossing in the tracker disks from incoherent electron--positron
    pairs~\subref{fig:trk_disk_hits_pairs_3tev} and \gghad~\subref{fig:trk_disk_hits_gghad_3tev} at
    \SI{3}{TeV}. Safety factors are not included.}\label{fig:trk_disk_hits_3tev}
\end{figure}

In the silicon vertex and tracker sensor layers of CLICdet, hits caused by direct and backscattering particles from incoherent pairs and \gghad{} add up to significant
occupancies, as shown in \crefrange{fig:vtx_hits_380}{fig:trk_disk_hits_3tev}.
The goal at CLIC is to keep occupancies below 3\% per bunch train, including
safety factors of 5 for incoherent pairs, and 2 for \gghadron{} events\footnote{For the definition of occupancy see~\cite[Equation 2]{Nurnberg_Dannheim_2017}}.
The pixel sizes of the vertex detector and the cell sizes in the tracker disks and barrel layers are chosen so that the occupancies are
below this limit, as described for \SI{3}{TeV} CLIC in~\cite{Nurnberg_Dannheim_2017}.

As shown in the figures  below, and as expected from the total numbers of background particles given in \cref{tab:clicBeam}, the hit density and rates from
beam induced background particles are considerably lower at \SI{380}{GeV}\@. The cell sizes in the CLICdet silicon tracker at \SI{3}{TeV}, given in~\cite{Nurnberg_Dannheim_2017}, can therefore safely be used for all CLIC energy stages.

\clearpage
\subsubsection{Backgrounds in ECAL and HCAL}

\begin{figure}[hbt]
\centering
\begin{subfigure}{.495\textwidth}
  \centering
  \includegraphics[width=\linewidth]{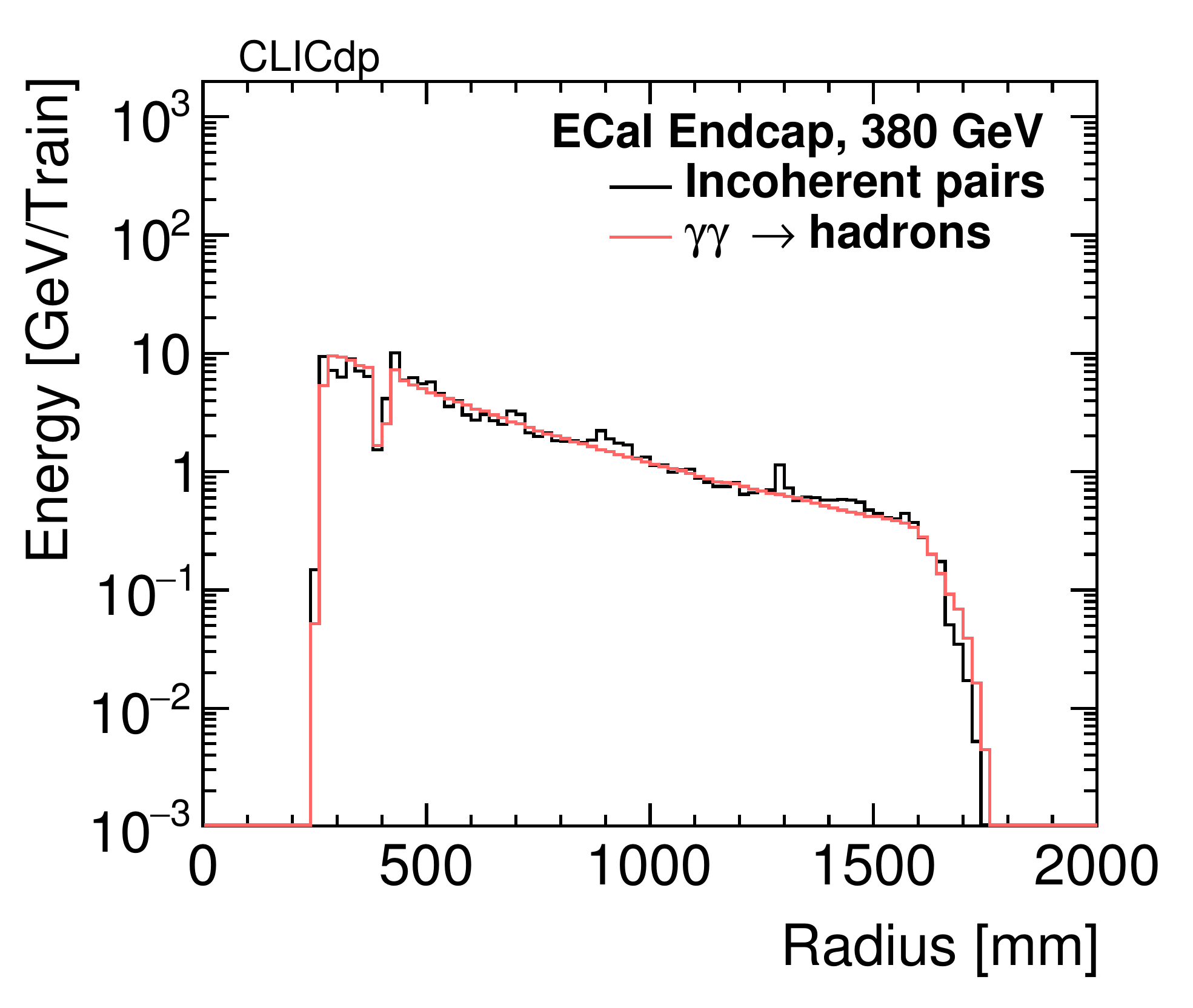}
\end{subfigure}
\begin{subfigure}{.495\textwidth}
  \centering
  \includegraphics[width=\linewidth]{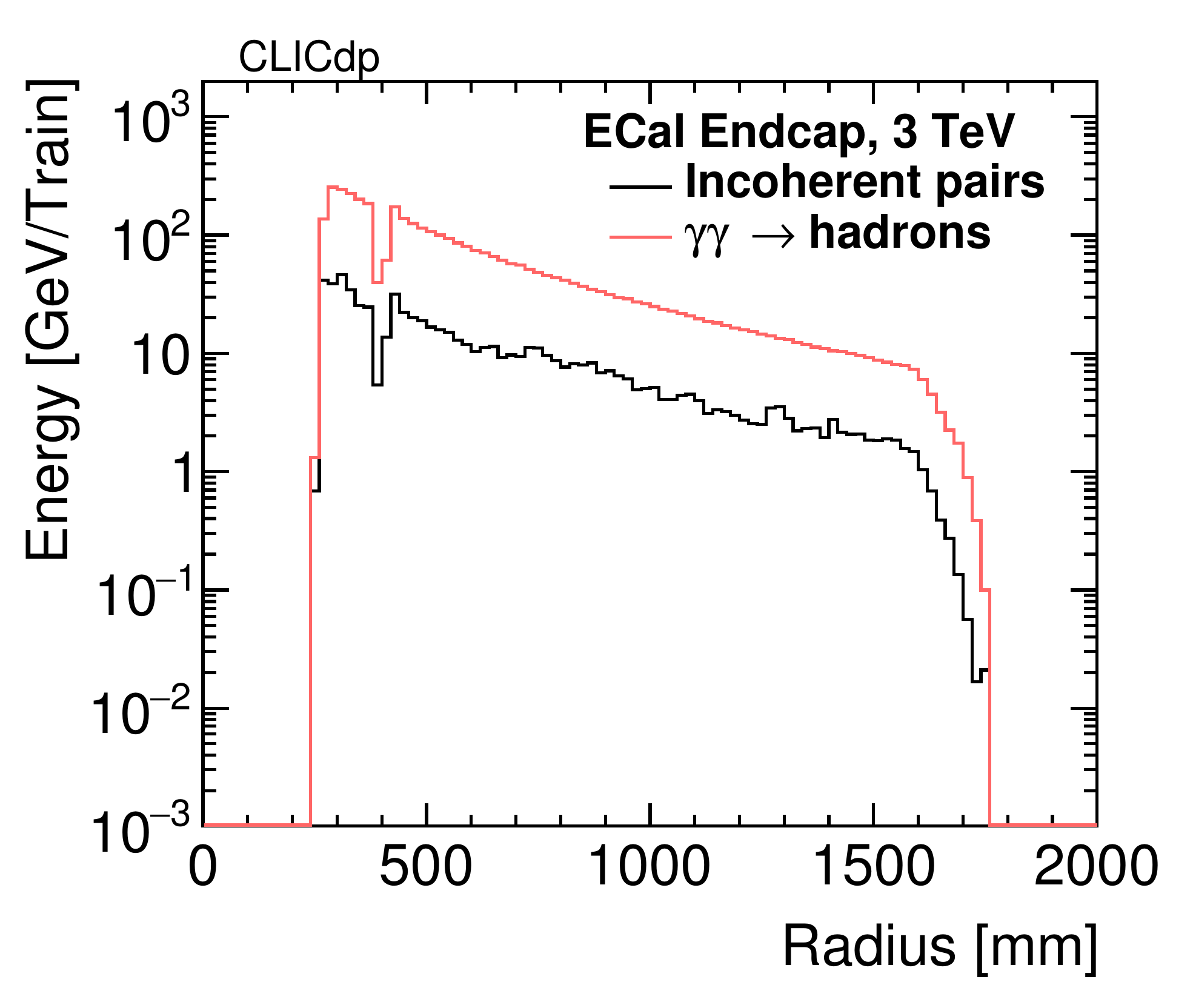}
\end{subfigure}
\caption{Radial distribution of the calorimetric energy deposition in the ECAL endcap of CLICdet, for \SI{380}{GeV} (left) and for \SI{3}{TeV} (right), for an entire bunch train, within \SI{200}{ns} from the start
  of the train. 
Safety factors representing the simulation uncertainties are not included. The bin width is \SI{20}{mm}. The dip visible around $R=\SI{400}{mm}$
corresponds to the space between ECAL `plug' and ECAL endcap, which is \SI{30}{mm} wide.
 \label{fig:ECALBackEnergy}}

\end{figure}

\begin{figure}[hbt]
\centering
\begin{subfigure}{.495\textwidth}
  \centering
  \includegraphics[width=\linewidth]{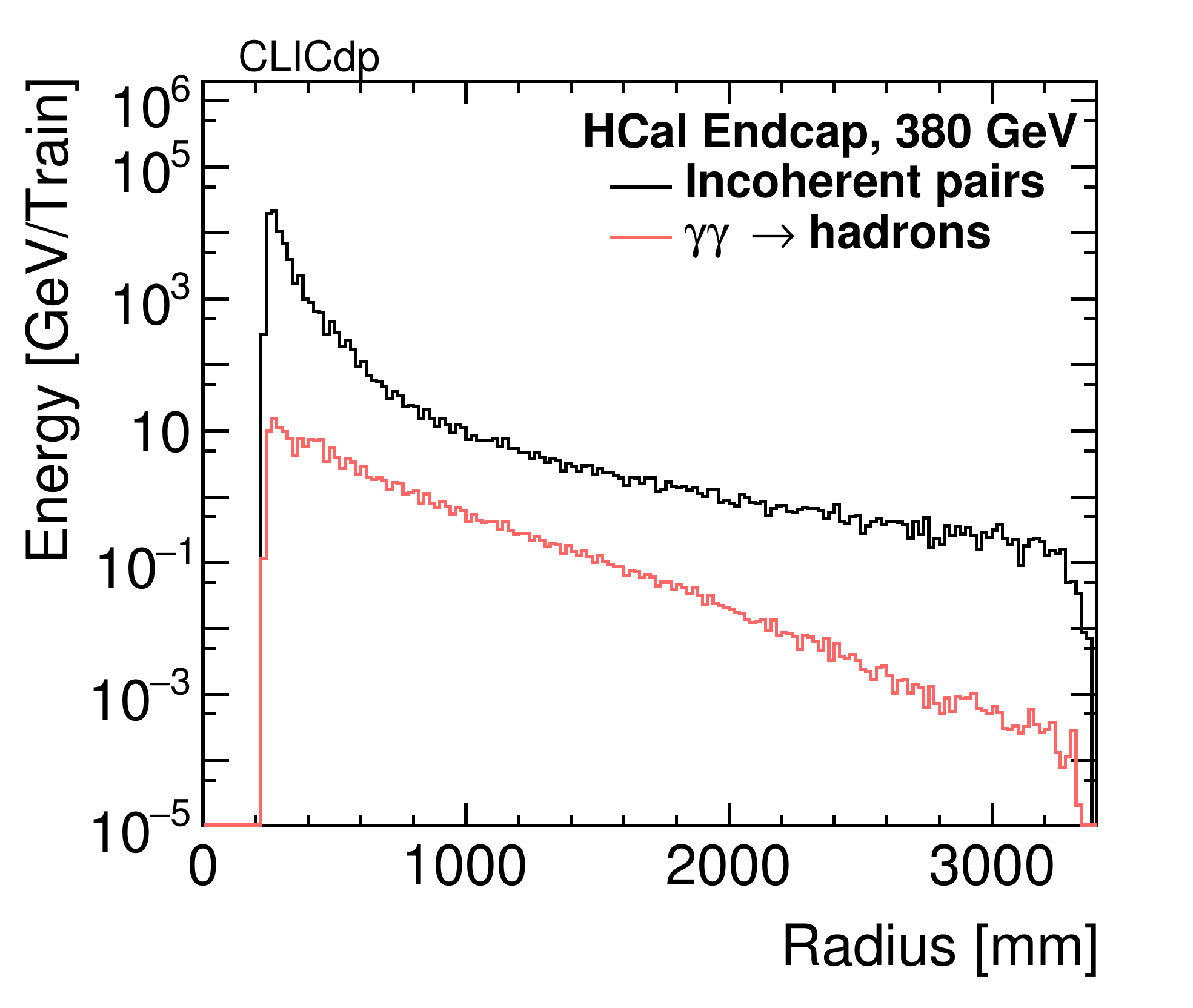}
\end{subfigure}
\begin{subfigure}{.495\textwidth}
  \centering
  \includegraphics[width=\linewidth]{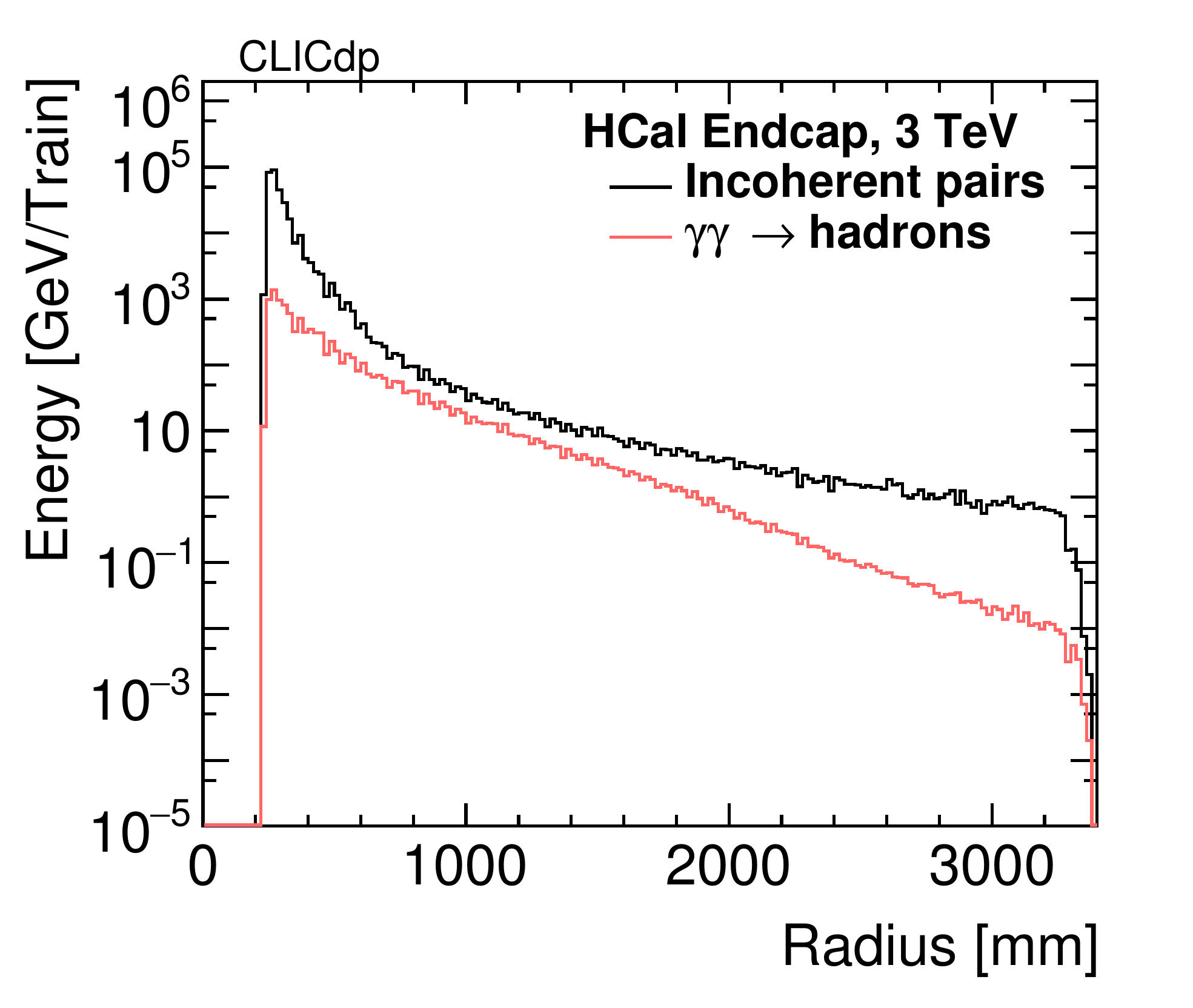}
\end{subfigure}
\caption{Radial distribution of the calorimetric energy deposition in the HCAL endcap of CLICdet, for \SI{380}{GeV} (left) and for \SI{3}{TeV} (right), for an entire bunch train, within \SI{200}{ns} from the start of the train.
Safety factors representing the simulation uncertainties are not included. The bin width is \SI{20}{mm}.
 \label{fig:HCALBackEnergy}}
\end{figure}

The distribution of deposited energy from incoherent pairs and \gghad{}, including backscattering, has been studied for both the ECAL and the HCAL\@.
The radial calorimetric energy distributions from both types of background in the ECAL endcap are shown in \cref{fig:ECALBackEnergy} and for the HCAL
endcap in \cref{fig:HCALBackEnergy}. The general features found in the CDR~\cite[Section 2.4.3]{cdrvol2} are confirmed for CLICdet. The energy deposition and occupancy
are found to increase significantly at the lowest radii of the HCAL endcap, now at $R=\SI{250}{mm}$, instead of \SI{400}{mm} for \clicild.

\cref{tab:clicBackgrounds} summarises the simulated background conditions
in CLICdet calorimeters for an entire CLIC bunch train. The total calorimetric energy 
deposition is large compared to the centre-of-mass energy and implies strict 
requirements on the timing resolution of CLIC calorimeters. Even excluding the HCAL contribution 
from the incoherent pair background, the overall energy deposited in the CLIC ECAL and HCAL detectors corresponds
to more than \SI{28}{TeV} per bunch train at \SI{3}{TeV}\@. This is predominantly forward peaked, but nevertheless poses a serious challenge to
the design of a detector at CLIC\@. The deposited energies without applied calibration factors can be found in \cref{tab:clicBackgroundsRaw} in \cref{sec:appendixBKG}.

\newcommand{\sr}[1]{\sisetup{round-precision=#1,round-mode=figures}}
\begin{table}[tpb]
\centering
  \caption{Energy from beam-induced backgrounds in the CLICdet calorimeters. The
    numbers correspond to the background for an entire CLIC bunch train and  nominal background rates.
    Safety factors representing the simulation uncertainties are not included.
    \label{tab:clicBackgrounds}}
\begin{threeparttable}
  \begin{tabular}{l *4{S[table-format=5.5,round-mode=figures,round-precision=2]} }
    \toprule
    Energy stage          & \multicolumn{2}{c}{\SI{380}{GeV}} & \multicolumn{2}{c}{\SI{3}{TeV}}                             \\\cmidrule(r){2-3}\cmidrule(l){4-5}
    Subdetector           & \tabt{Incoherent pairs}           & \tabt{\gghad{}} & \tabt{Incoherent pairs} & \tabt{\gghad{}} \\
                          & \tabt{[TeV]}                      & \tabt{[TeV]}    & \tabt{[TeV]}            & \tabt{[TeV]}    \\\midrule
    ECAL barrel           &    0.1283671                      & 0.0767185       &     0.4942848           &  1.8710751      \\
    ECAL endcaps\tnote{a} &    0.3991715                      & 0.3386558       &     1.4055785           &  9.0326494      \\
    HCAL barrel           &  \sr{1}0.0026036                  & \sr{1}0.0088904 &     0.0109936           &  0.2446977      \\
    HCAL endcaps          &  \sr{3}154.0601200                & 0.3726348       & \sr{3}631.9245800       & 16.7018610      \\\midrule
    ECAL\&HCAL            &  \sr{3}154.59026                  & 0.7968995       & \sr{3}633.83544         & 27.850283       \\\midrule
    LumiCal               &    5.6446434                      & 0.3727381       &    23.2832810           & 15.8697500      \\
    BeamCal               & \sr{3}6372.3532000                & 0.6568639       & \sr{3}31504.7520000     & 62.9328480      \\\bottomrule
  \end{tabular}
\begin{tablenotes}
\item[a] Including the ECAL plugs
\end{tablenotes}
\end{threeparttable}
\end{table}

\begin{figure}[hbt]
\centering
\begin{subfigure}{.495\textwidth}
  \centering
  \includegraphics[width=\linewidth]{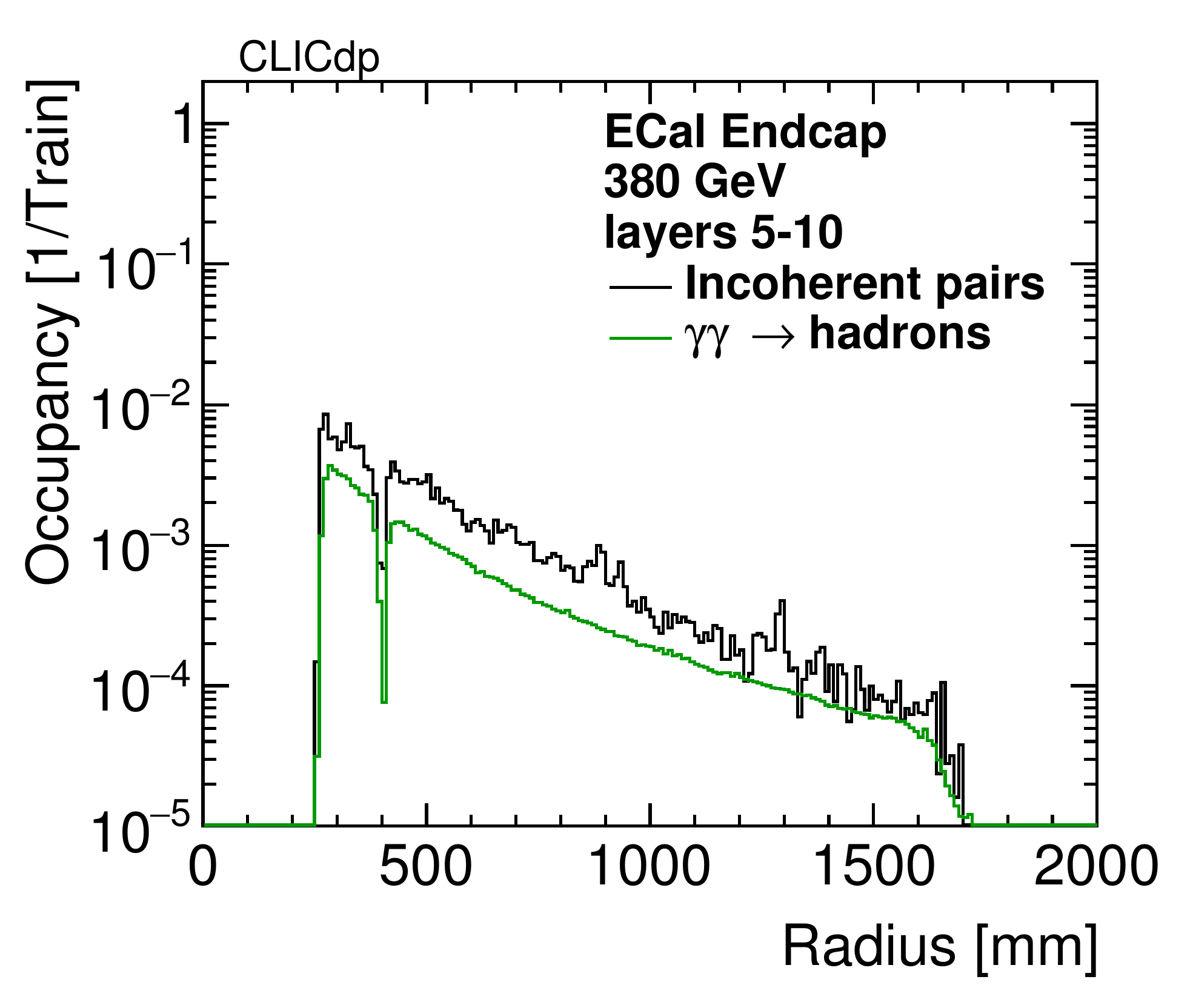}
\end{subfigure}
\begin{subfigure}{.495\textwidth}
  \centering
\includegraphics[width=\linewidth]{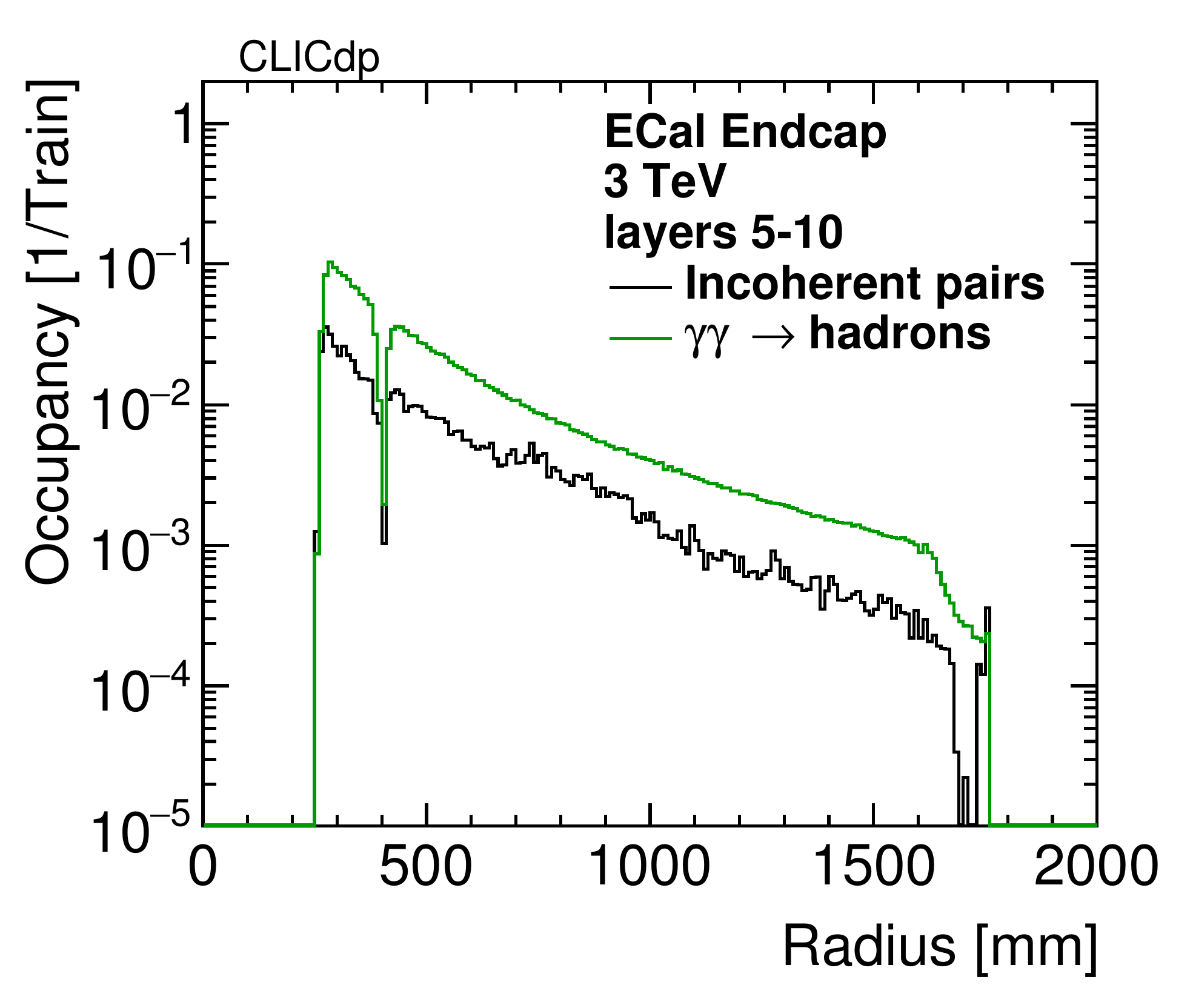}
\end{subfigure}
\caption{Average cell occupancy in the ECAL endcaps of CLICdet, at \SI{380}{GeV} (left) and \SI{3}{TeV}~(right).
 The average is given for layers 5--10 which broadly correspond to maximum energy deposit of typical
 electromagnetic showers. The results are obtained for nominal background
 rates, excluding safety factors representing the simulation uncertainties.
  \label{fig:ECALBackOccupancy}}
\end{figure}

\begin{figure}[hbt]
\centering
\begin{subfigure}{.495\textwidth}
  \centering
  \includegraphics[width=\linewidth]{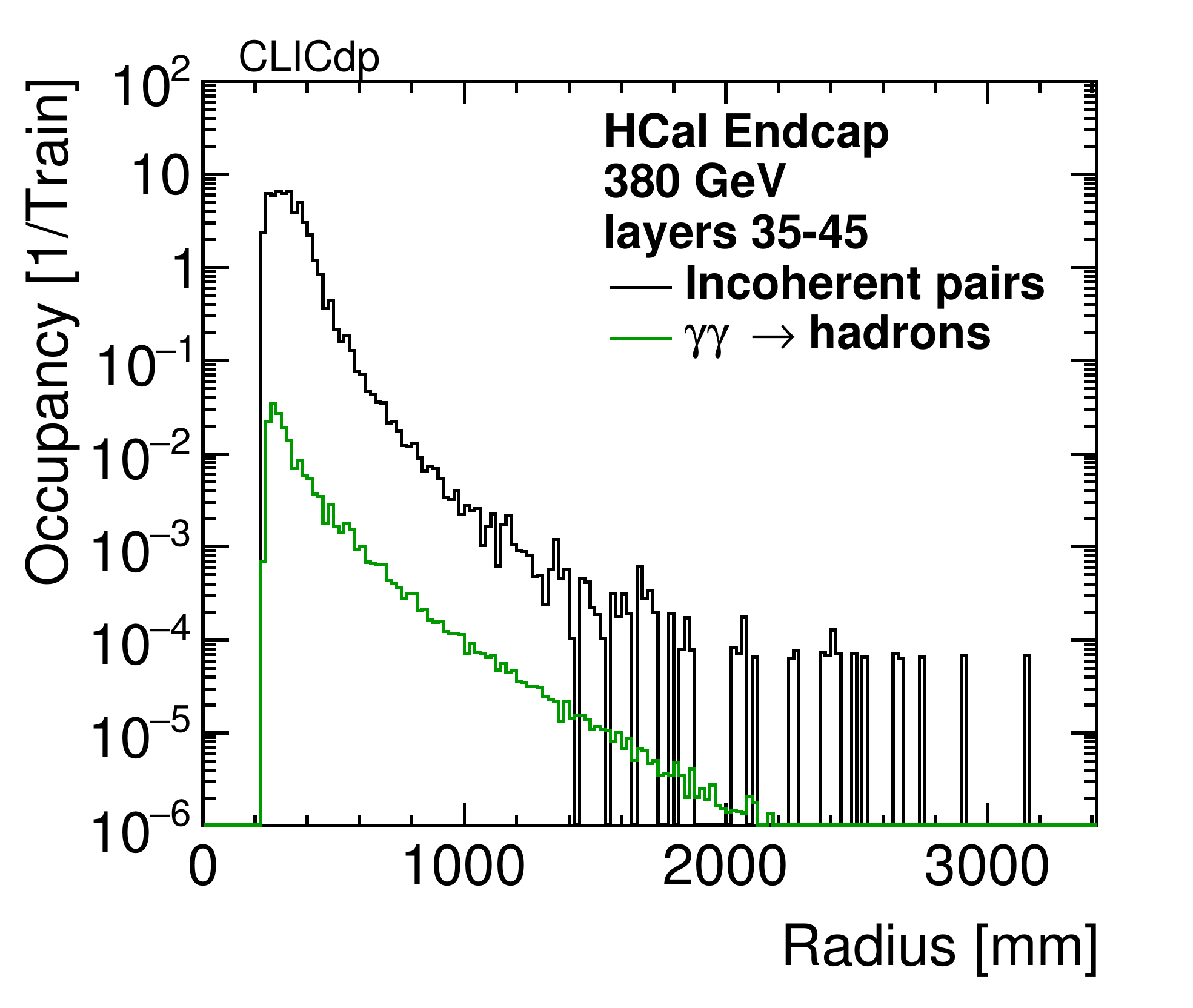}
\end{subfigure}
\begin{subfigure}{.495\textwidth}
  \centering
 \includegraphics[width=\linewidth]{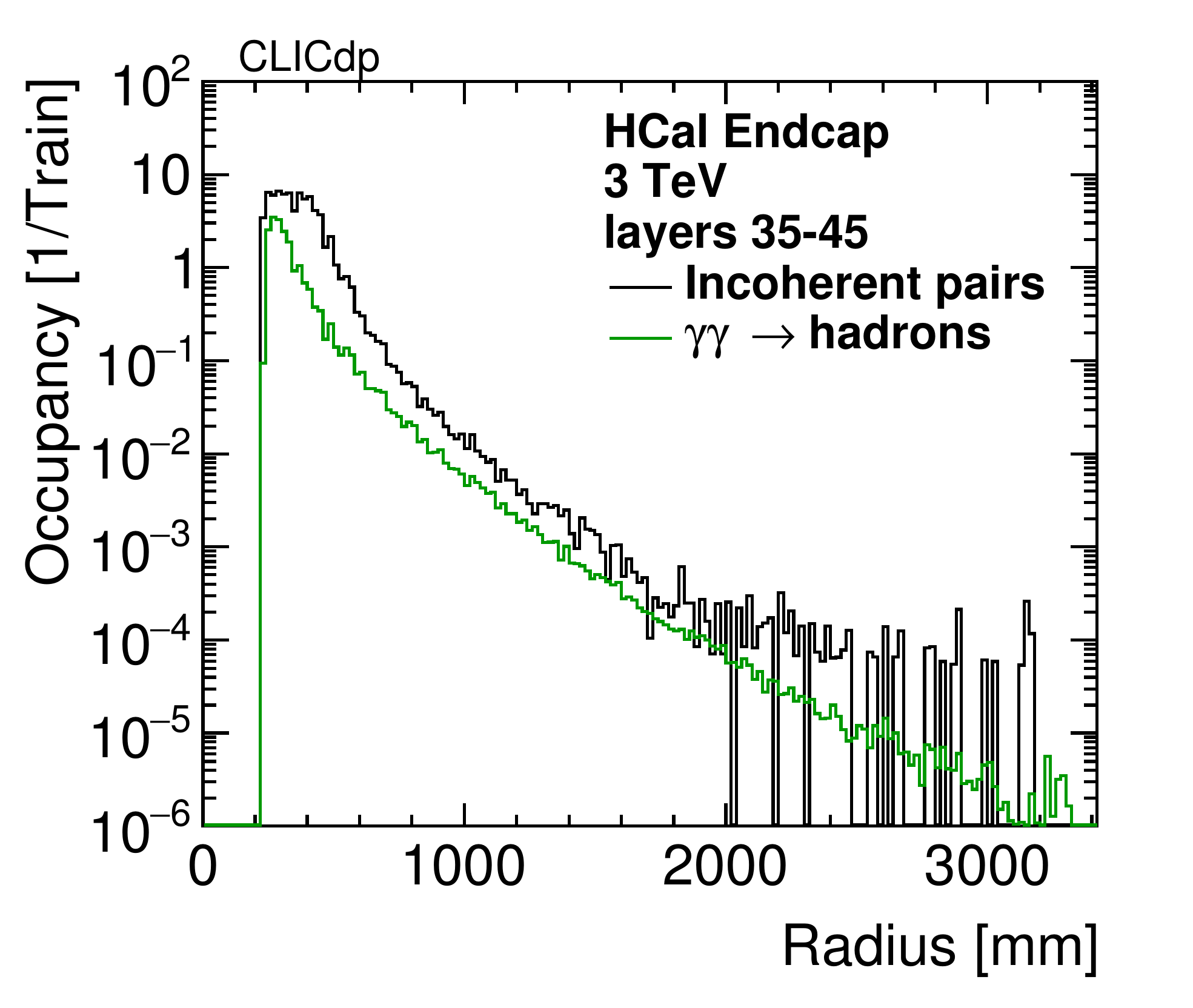}
\end{subfigure}
 \caption{
Average cell occupancy in the HCAL endcaps of CLICdet, at \SI{380}{GeV} (left) and at \SI{3}{TeV}~(right).
The average is quoted for layers 35--45 where the maximum activity
 from the neutron background is observed. The results are obtained for nominal background
 rates, excluding safety factors representing the simulation uncertainties.
  \label{fig:HCALBackOccupancy}}
\end{figure}

Another  important consideration is the level of occupancy per calorimeter cell.
In CLICdet, the ECAL silicon cells are \SI{5x5}{mm\squared}, while the scintillator tiles in the HCAL are \SI{30x30}{mm\squared}.
For the occupancy calculation the time window of \SI{200}{ns} from the start of the bunch train was divided into eight \SI{25}{ns} time windows.
The mean number of hits above the threshold of 0.3 minimum ionising particle equivalents is shown for the ECAL in
\cref{fig:ECALBackOccupancy} and for the HCAL in \cref{fig:HCALBackOccupancy}.

The occupancies in the ECAL are acceptable, both at \SI{380}{GeV} and at \SI{3}{TeV}\@. In the HCAL endcap at small radii, occupancies exceeding 0.1 per train are observed.
This is clearly too high, and studies are on-going to reduce these occupancies which are mainly stemming from neutrons produced by incoherent pair particles in BeamCal,
which is located inside the HCAL endcap opening. One possibility could be improved shielding, along the lines of what has been studied previously~\cite{vanDam_Sailer_2014}.
Other options include the use of a different active material less sensitive to neutrons, or increasing the transverse segmentation to resolve the occupancy issue.


\clearpage
\subsubsection{Backgrounds in LumiCal and BeamCal}

Due to the angular dependence of the incoherent pairs the very forward
calorimeters -- the LumiCal and the BeamCal -- receive larger background
contributions from incoherent pairs than from \gghadron{} events.
See \cref{tab:clicBackgrounds} for the background energies in the forward calorimeters.

\begin{figure}[tbp]
  \begin{subfigure}{\subfigwidth}
    \centering
    \includegraphics[width=1.0\textwidth]{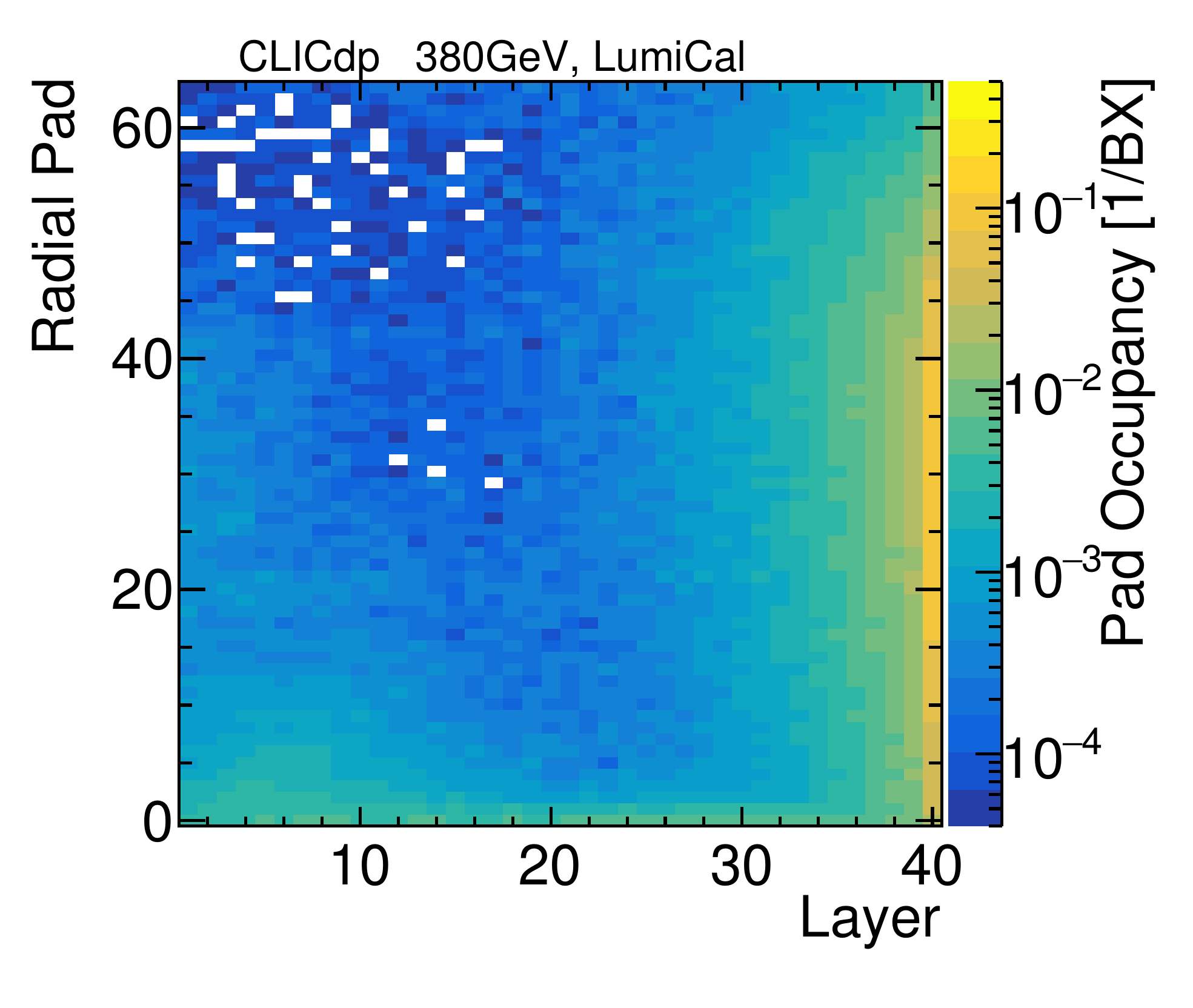}%
  \end{subfigure}
  \begin{subfigure}{\subfigwidth}
    \centering
    \includegraphics[width=1.0\textwidth]{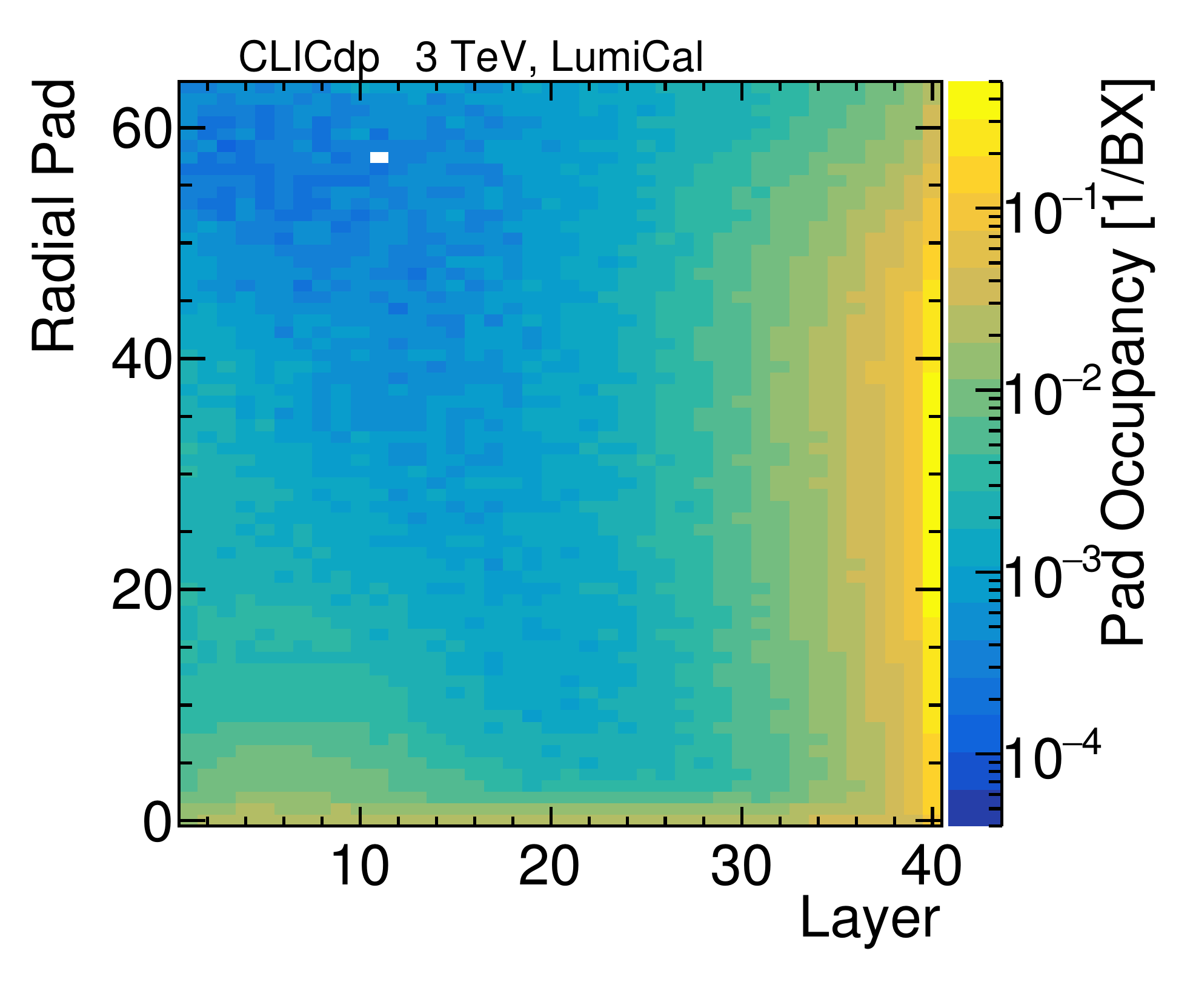}%
  \end{subfigure}
  \vspace{-5mm}
  \caption{Occupancy per bunch crossing per pad in LumiCal, from incoherent pairs, for \SI{380}{GeV} (left) and
    \SI{3}{TeV} (right) collisions.\label{fig:lcal_occ}}
\end{figure}

\Cref{fig:lcal_occ} shows the occupancies per bunch crossing in the
LumiCal from incoherent pairs at \SI{380}{GeV} and \SI{3}{TeV}\@. The occupancies are
averaged over all pads with the same radius. The largest occupancy of the
LumiCal is in its last layer and is caused by particles scattering back from the
BeamCal. Similarly the innermost pads suffer from backscattered particles. In
the largest part of the LumiCal the occupancy is below 1\%.

\begin{figure}[tbp]
  \begin{subfigure}{\subfigwidth}
    \centering
    \includegraphics[width=1.0\textwidth]{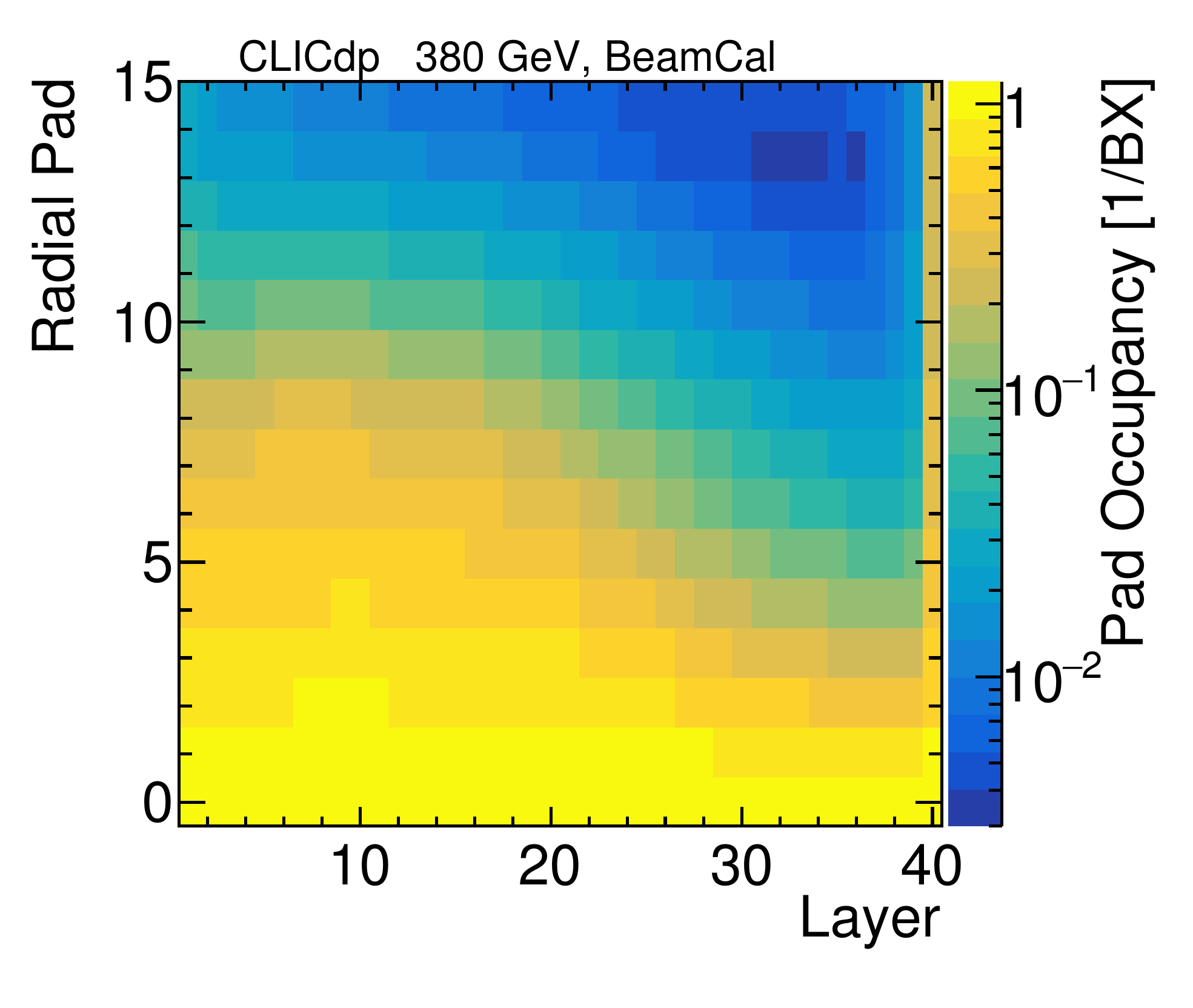}%
  \end{subfigure}
  \begin{subfigure}{\subfigwidth}
    \centering
    \includegraphics[width=1.0\textwidth]{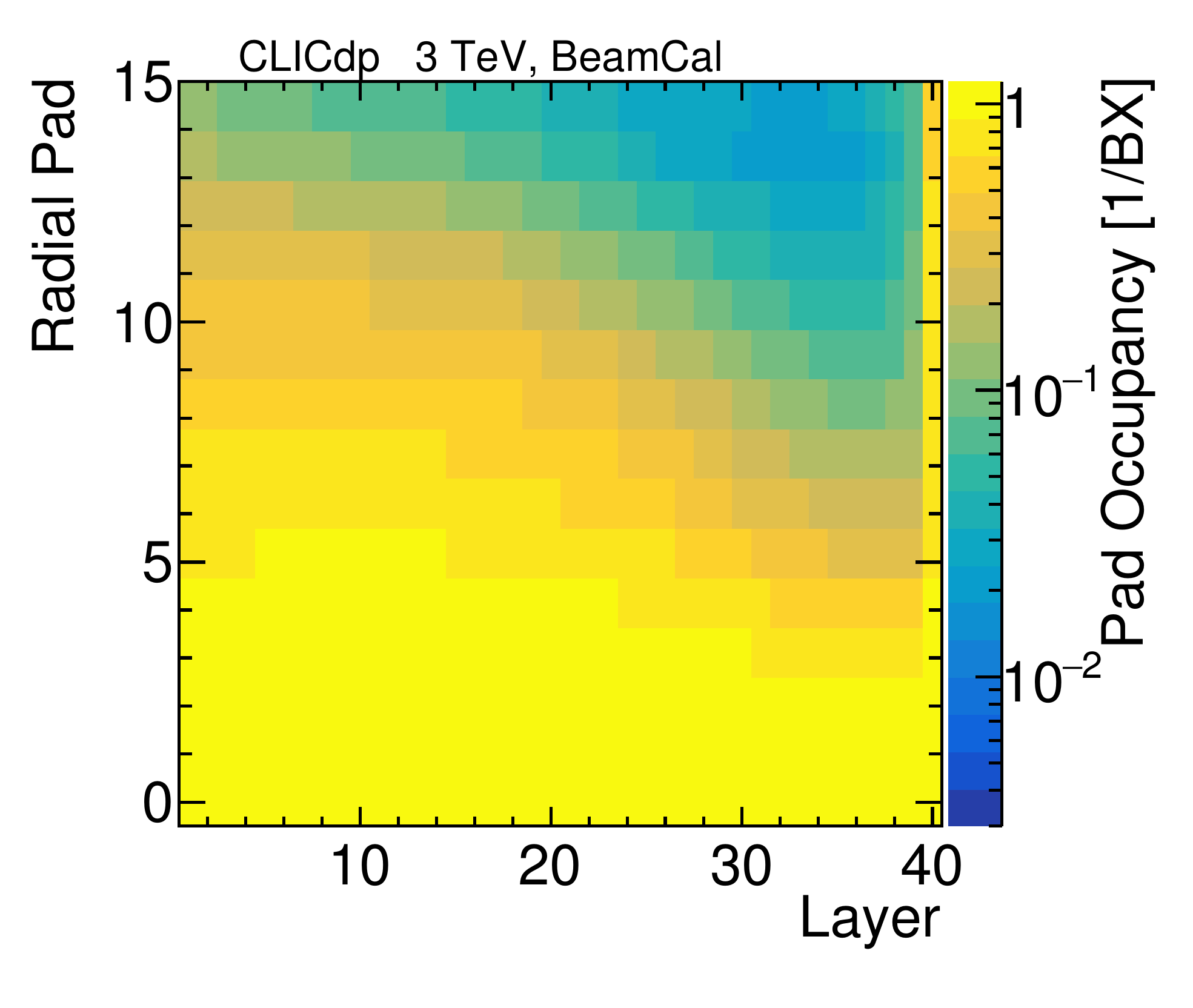}%
  \end{subfigure}
  \vspace{-5mm}
  \caption{Occupancy per bunch crossing per pad in the BeamCal, from incoherent pairs, for \SI{380}{GeV}~(left) and \SI{3}{TeV} (right) collisions.\label{fig:bcal_occ}}
\end{figure}

The BeamCal occupancies at \SI{380}{GeV} and \SI{3}{TeV} are shown in
\cref{fig:bcal_occ}. Both at \SI{380}{GeV} and at \SI{3}{TeV}, the pads
of the BeamCal with the smallest
radii with respect to the outgoing beam axis
will receive an energy deposit in each bunch crossing. This limits the electron identification efficiency
in BeamCal at the smallest polar angles as described in \cref{sec:VF_perf}.

\subsection{Overview of Detector Timing Requirements at CLIC}
\label{sec:overv-detect-timing}
The timing requirements and the impact of timing cuts at CLIC are described in detail in the CDR~\cite[Section 2.5]{cdrvol2}.
For CLICdet, these requirements remain unchanged, with one notable difference: since the HCAL absorber
material is now steel in both barrel and endcap -- instead of tungsten in the barrel and steel in the endcap -- the timing requirements are
the same for the entire HCAL\@.
The time windows and required single hit time resolutions are summarised in \cref{tab:timingcuts}.

\begin{table}[tbp]
\centering
  \caption{Assumed time windows used for the event reconstruction and the required single hit time resolutions.
    \label{tab:timingcuts}}
  \begin{tabular}{l c c }
    \toprule
    Subdetector       & Reconstruction window & Hit resolution                               \\\midrule
    ECAL              & \SI{10}{ns}           & \SI{1}{ns}                                   \\
    HCAL  	      & \SI{10}{ns}           & \SI{1}{ns}                                   \\
    Silicon detectors & \SI{10}{ns}           & \SI[parse-numbers=false]{10 / \sqrt{12}}{ns} \\\bottomrule
  \end{tabular}
\end{table}

\subsection{A Detector at CLIC for \SI[detect-all]{380}{GeV}, \SI[detect-all]{1.5}{TeV} and \SI[detect-all]{3}{TeV}}

As illustrated in the previous sections, the CLIC beam conditions and parameters (see \cref{tab:clicBeam}) and the resulting beam-induced background conditions
are rather different for the different CLIC energy stages. A priori, this would allow one to consider a somewhat different detector layout for each of the 
CLIC stages. However, from a practical point of view, it is very likely that the calorimeters, the solenoid, the yoke and  muon systems 
but also the outer tracker will remain unchanged.

On the other hand, the different crossing angle alone imposes a change of the vacuum pipes, which in turn implies replacing the BeamCal when
moving from the initial \SI{380}{GeV} stage to the higher centre-of-mass energies. The diameter of the incoming and outgoing beam pipes must also be adjusted.

In addition, because of the reduced number and lower \pT{} of the incoherent pairs produced at the \SI{380}{GeV} stage,
the radius of the central, cylindrical beryllium beam pipe can be \SI{6}{mm} smaller~\cite[Section 12.4.7]{cdrvol2}. This, in turn, allows one to move the first vertex barrel detector
layer to a smaller radius and the positions of the remaining vertex barrel layers can be re-optimised. The need for also adapting the inner tracker layout
still remains to be studied.

At the present stage of the CLIC detector and physics studies, it was decided to use CLICdet, i.e.\ the detector with a  layout optimised for the \SI{3}{TeV} stage, for all energy stages.
This is mainly motivated by the need to optimise the use of resources in the studies.
 
\section{Physics Performance}
\label{performances}

\input{./include/software}


\subsection{Performance for Lower Level Physics Observables}

\subsubsection{Single Particle Performances}
\label{sec:single_particle}

\paragraph{Impact-Parameter, Angular and Momentum Resolution}

To identify heavy-flavour quark states and tau-leptons with high efficiency, a precise measurement of the impact parameter point and of the charge of the tracks originating from the secondary vertex is required.
 Monte Carlo simulations show that these goals can be met with a constant term in the transverse impact-parameter resolution of $a \simeq \SI{5}{\micron}$ and a multiple-scattering term of $b \simeq \SI{15}{\micron}$, using the parametrization:

\begin{equation}
\sigma_{d_{0}}(p, \theta) = \sqrt{a^2+b^2 \cdot \si{\gev\squared}/(p^2 \sin^3(\theta))}.
\label{eq:d0}
\end{equation}

\begin{figure}[tbp]
  \renewcommand{\thesubfigure}{(\lr{subfigure})}
  \centering
  \begin{subfigure}{.5\textwidth}
    \centering
    \includegraphics[width=\linewidth]{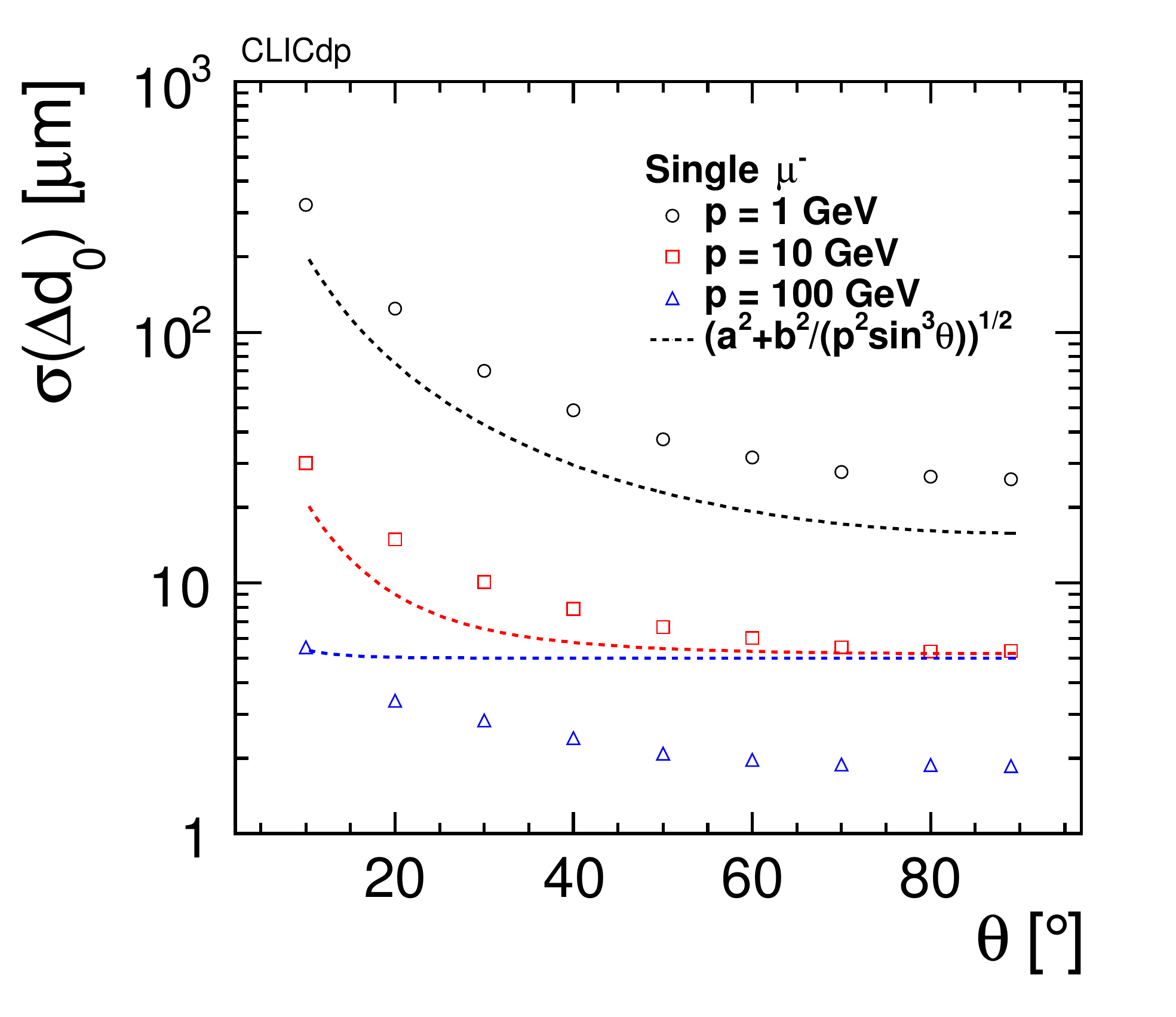}%
    \phantomsubcaption\label{fig:d0res}
  \end{subfigure}%
  \begin{subfigure}{.5\textwidth}
    \centering
    \includegraphics[width=\linewidth]{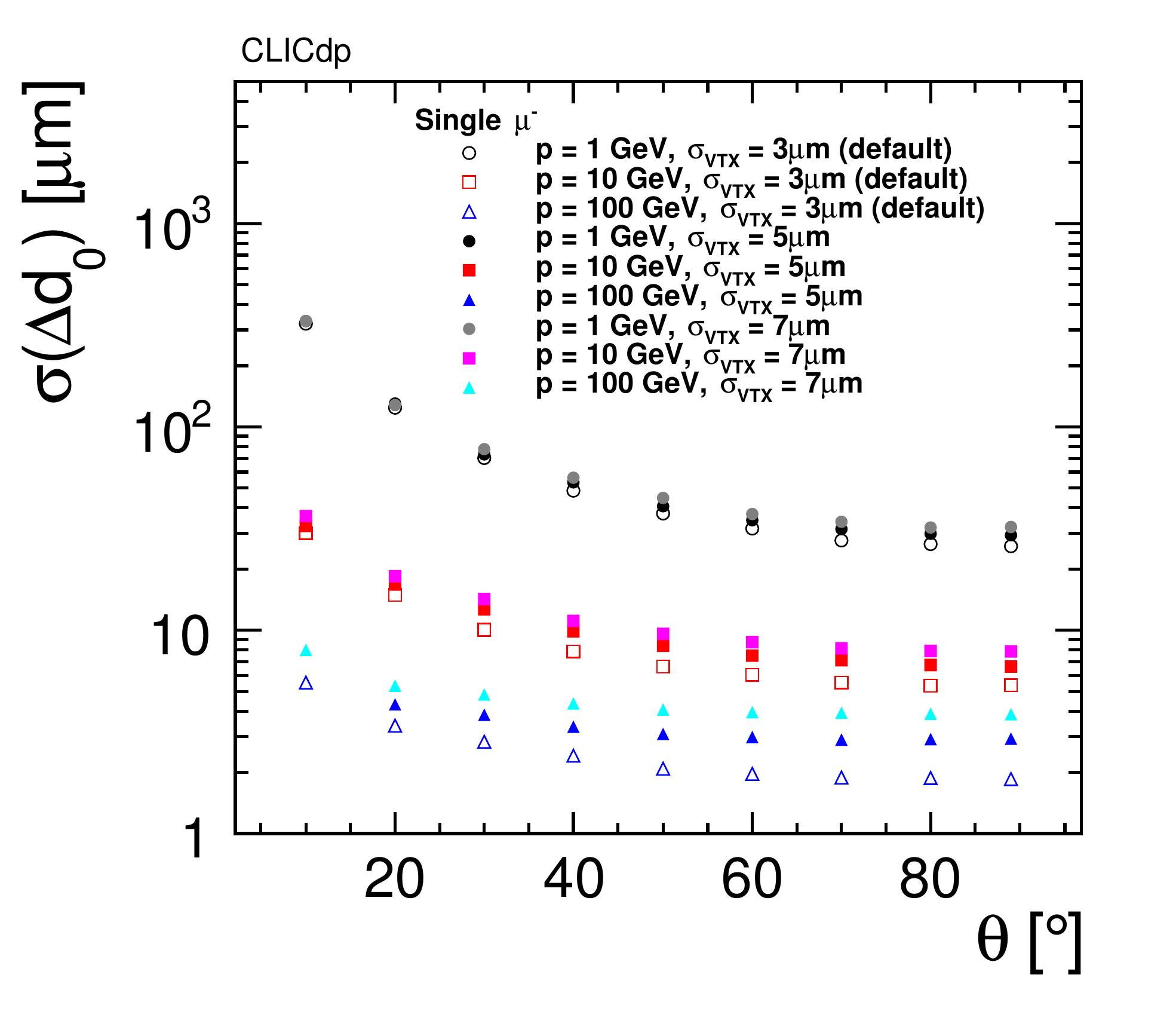}%
    \phantomsubcaption\label{fig:d0res_spr}
  \end{subfigure}
  \vspace{-5mm}
  \caption{Transverse impact-parameter resolutions as a function of the polar angle for muons with momenta of \SIlist{1;10;100}{GeV}, obtained with the baseline \SI{3}{\micron} single point resolution~\subref{fig:d0res} and different single point resolutions~\subref{fig:d0res_spr} in the vertex detector.}
  \label{fig:d0_res}
\end{figure}

\cref{fig:d0res} shows the transverse impact-parameter resolutions obtained with the baseline configuration for CLICdet
(single point resolution in the vertex detector of \SI{3}{\micron})
for isolated muon tracks with momenta of \SIlist{1;10;100}{GeV}, as a function of the polar angle.
Each data point corresponds to 10\,000 single muons simulated at fixed energy and polar angle.
For each dataset, the resolution is calculated as the width of the Gaussian fit of the residual distributions. Residuals are obtained as difference between the reconstructed and simulated parameters per track.
Also shown, for reference, are three dashed lines corresponding to \cref{eq:d0} for the three energies,
with the target values of $a \simeq \SI{5}{\micron}$ and $b \simeq \SI{15}{\micron}$.
The resolution for \SI{100}{GeV} muons is well below the high-momentum limit of \SI{5}{\micron}.
At \SI{10}{GeV}, the effect of an increased material becomes visible in the forward region. The same effect impacts the \SI{1}{GeV} result for all angles.
Note that the results for CLICdet at low momenta are slightly worse than those obtained in the CDR~\cite{cdrvol2}, due to a more realistic material budget in CLICdet.

The dependence of the $d_{0}$ resolution on the pixel size has been studied by varying the single point resolution for the vertex detector layers from the baseline value of \SI{3}{\micron} (which would correspond to a pixel size of \SI{10}{\micron} if no charge-sharing is assumed)
to \SI{5}{\micron} and \SI{7}{\micron}
(corresponding to pixel sizes of 17 and \SI{24}{\micron} respectively). The resulting resolutions are shown in \cref{fig:d0res_spr}.
The single point resolution dominates at higher energies, especially in the central region, where a change from \SI{3}{\micron} to \SI{5}{\micron} results in an increase by approximately 50\%. However, even in the worst scenario of \SI{7}{\micron} single point resolution,
the $d_{0}$ resolution for \SI{100}{GeV} tracks does not exceed the target value for the high-momentum limit of $a \simeq \SI{5}{\micron}$. For the \SI{10}{GeV} tracks, on the other hand, with the $d_{0}$ resolution already at the limit for the default \SI{3}{\micron},
any increase in single point resolution is detrimental. 
For \SI{1}{GeV} muons, where multiple scattering dominates, the effect of a single point resolution increase from \SI{3}{\micron} to \SI{5}{\micron} amounts to at most 8\%.

\begin{figure}[tbp]
  \renewcommand{\thesubfigure}{(\lr{subfigure})}
  \centering
  \begin{subfigure}{.5\textwidth}
    \centering
    \includegraphics[width=\linewidth]{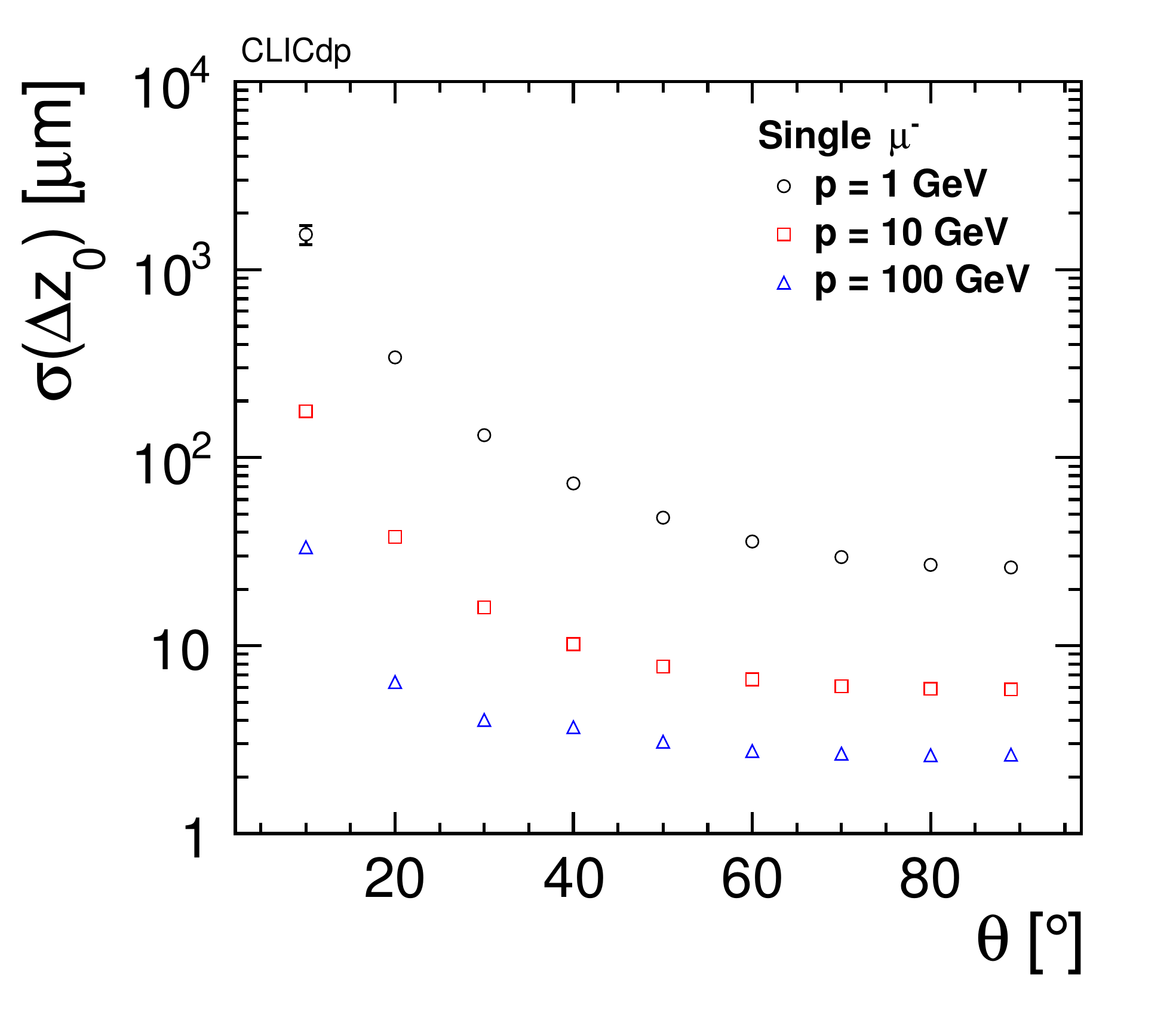}%
    \phantomsubcaption\label{fig:z0res}
  \end{subfigure}%
  \begin{subfigure}{.5\textwidth}
    \centering
    \includegraphics[width=\linewidth]{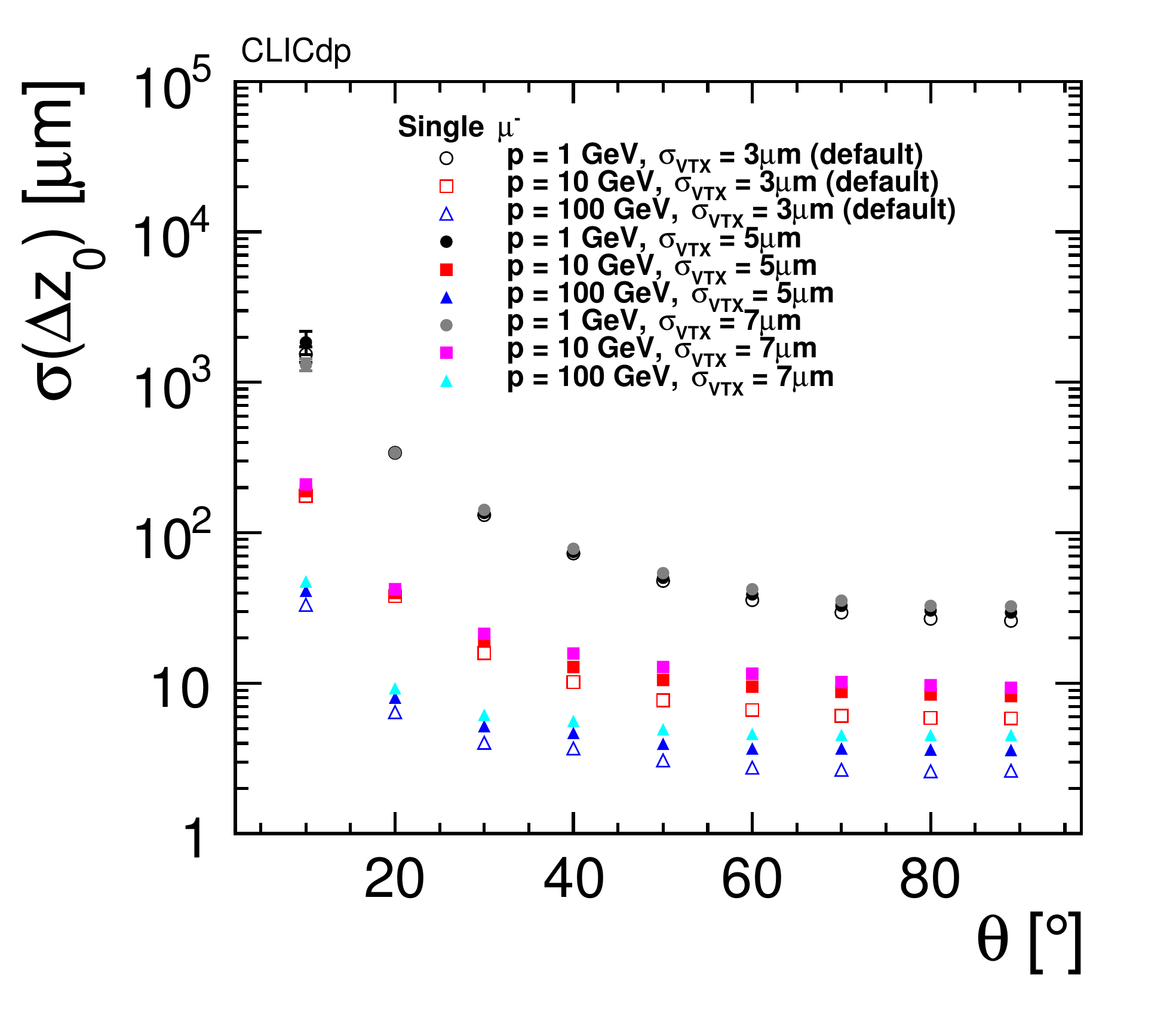}%
    \phantomsubcaption\label{fig:z0res_spr}
  \end{subfigure}
  \vspace{-5mm}
  \caption{Longitudinal impact-parameter resolutions as a function of the polar angle for muons with momenta of
    \SIlist{1;10;100}{GeV}, obtained with the baseline \SI{3}{\micron} single point resolution~\subref{fig:z0res} and
    different single point resolutions~\subref{fig:z0res_spr} in the vertex detector.}
  \label{fig:z0_res}
\end{figure}

Similarly, the longitudinal impact-parameter resolution for isolated muon tracks with momenta of \SIlist{1;10;100}{GeV} is shown as a function of the polar angle in \cref{fig:z0res}.
The achieved resolution for high-energy muons at all polar angles is smaller than the longitudinal bunch length of \SI{44}{\micron} at \SI{3}{TeV} centre-of-mass energy. A minimum of approximately 1.\SI{5}{\micron} is reached for \SI{100}{GeV} muons at 90\degrees.
As for the transverse impact-parameter resolution, the variation of the single point resolution of the vertex detector, shown in \cref{fig:z0res_spr}, affects mostly the $z_{0}$ resolution of higher-energy muons.

\begin{figure}[tbp]
  \renewcommand{\thesubfigure}{(\lr{subfigure})}
  \centering
  \begin{subfigure}{.5\textwidth}
    \centering
    \includegraphics[width=\linewidth]{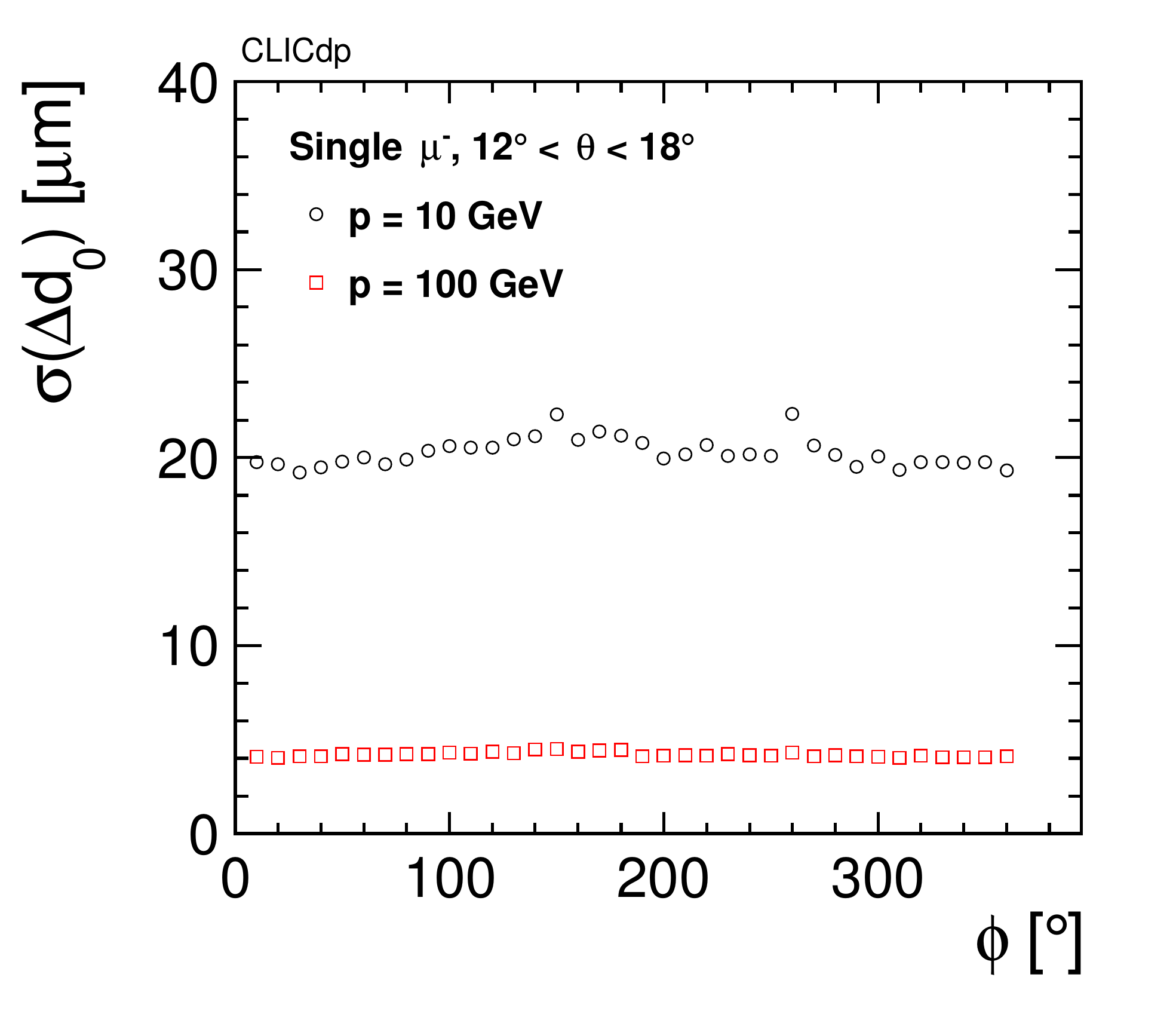}%
    \phantomsubcaption\label{fig:d0res_phi}
  \end{subfigure}%
  \begin{subfigure}{.5\textwidth}
    \centering
    \includegraphics[width=\linewidth]{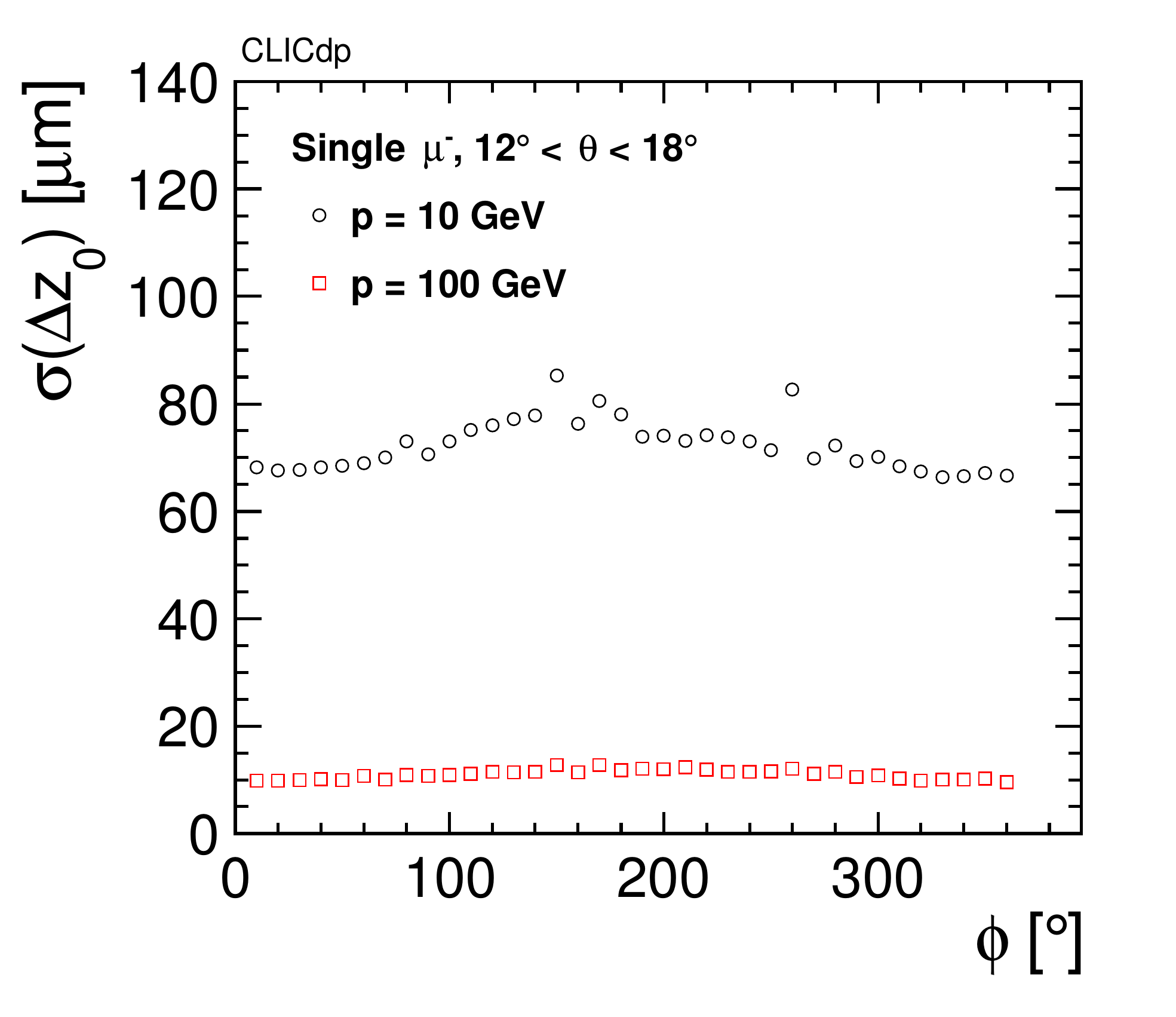}%
    \phantomsubcaption\label{fig:z0res_phi}
  \end{subfigure}
  \vspace{-5mm}
  \caption{Transverse~\subref{fig:d0res_phi} and longitudinal~\subref{fig:z0res_phi} impact-parameter resolutions as a function of the azimuthal angle in the region $12\degrees < \theta < 18\degrees$ for muons with momenta of \SI{10}{GeV} and \SI{100}{GeV}.}
  \label{fig:d0z0_res_phi}
\end{figure}

The $\phi$ dependence of transverse and longitudinal impact-parameter resolutions has been investigated in the region $12\degrees < \theta < 18\degrees$ for muons with momenta of \SI{10}{GeV} and \SI{100}{GeV}. As shown in \cref{fig:d0z0_res_phi}, the angular dependence is negligible for the highest-energy muons, but it is visible for the lower-energy ones, which, due to multiple scattering and higher curvature, may hit non-equidistant sensors in the different spiral-shaped layers. This results in a non-uniform impact parameter resolution. The variation for the \SI{10}{GeV} muons, however, amounts to at most 15\% for both $d_{0}$ and $z_{0}$, thus validating the robustness of the spiral geometry of the vertex disks.

\begin{figure}[tbp]
  \renewcommand{\thesubfigure}{(\lr{subfigure})}
  \centering
  \begin{subfigure}{.5\textwidth}
    \centering
    \includegraphics[width=\linewidth]{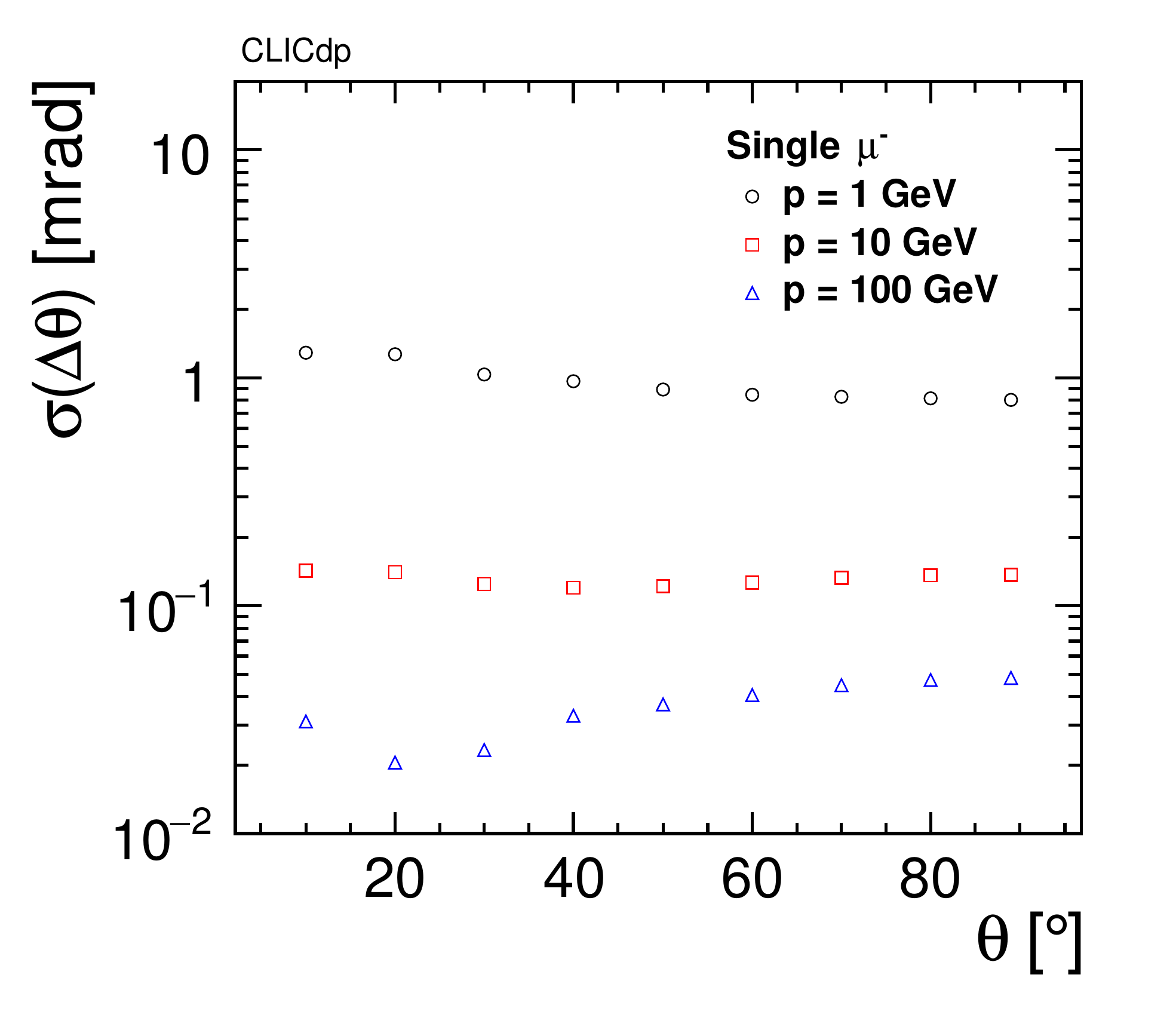}%
    \phantomsubcaption\label{fig:theta_res}
  \end{subfigure}%
  \begin{subfigure}{.5\textwidth}
    \centering
    \includegraphics[width=\linewidth]{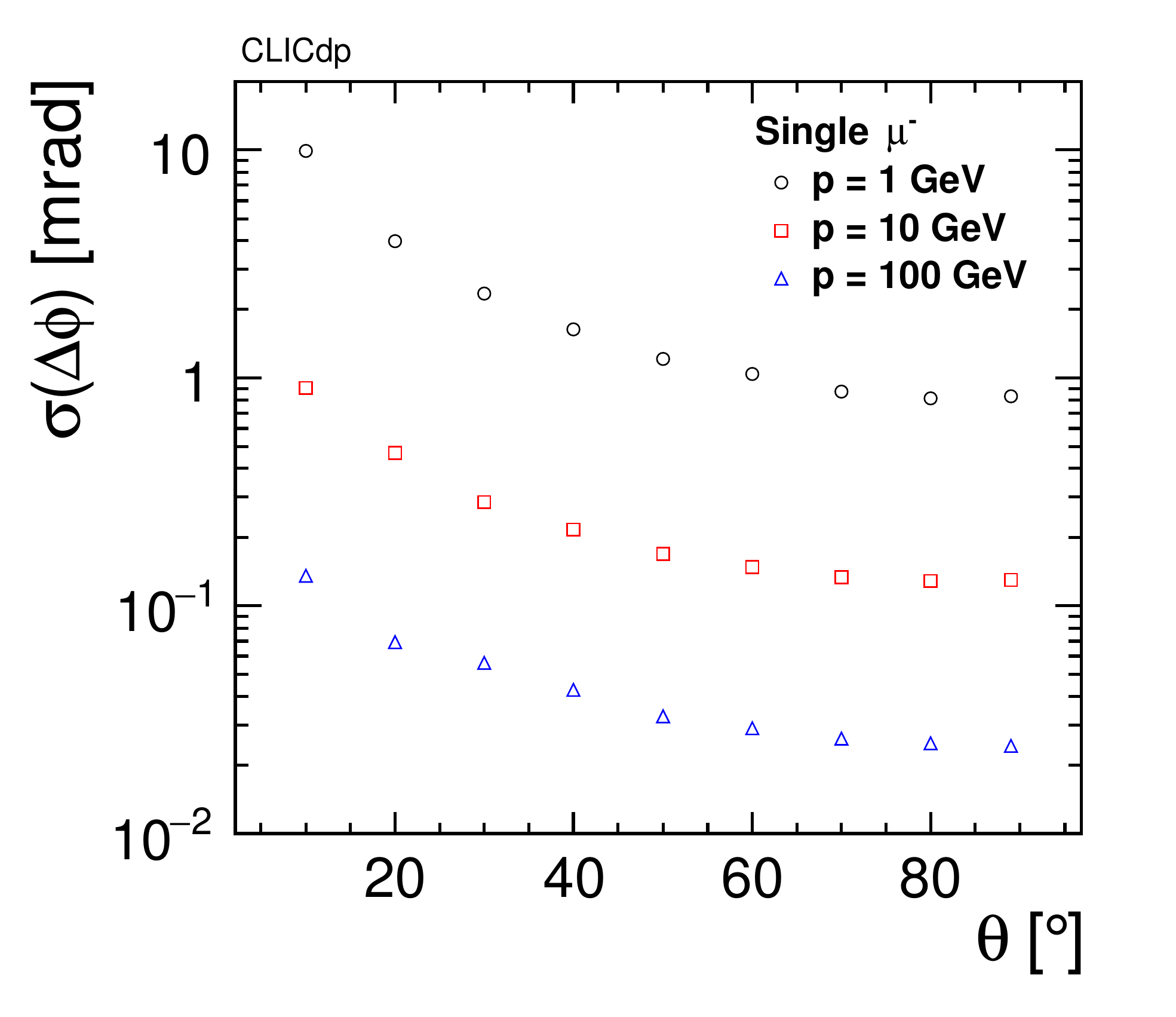}%
    \phantomsubcaption\label{fig:phi_res}
  \end{subfigure}
  \vspace{-5mm}
  \caption{Polar angle~\subref{fig:theta_res} and azimuthal angle~\subref{fig:phi_res} resolutions as a function of the polar angle for muons with momenta of \SIlist{1;10;100}{GeV}\@.}
  \label{fig:angular_res}
\end{figure}

\cref{fig:angular_res} shows the polar angle resolution (left) and the azimuthal angle resolution (right), both as a function of the polar angle $\theta$, for isolated muon tracks with momenta of \SIlist{1;10;100}{GeV}\@.
The polar angle resolution follows different trends according to the muon energy. For \SI{1}{GeV} muons, it decreases slightly while going from the forward to the transition region and levels up in the central region. For \SI{10}{GeV}  muons, the dependence on polar angle is negligible. On the contrary, a visible dependence is observed for \SI{100}{GeV} muons: smaller resolutions are obtained in the transition regions, where a higher number of measurements (traversed layers) is available, while resolutions in the barrel are limited by single point resolution.
The azimuthal angle resolution follows a similar trend for different muon energies, decreasing monotonically while going from the forward to the central region. For high-energy muons, it reaches a minimum of \SI{0.025}{mrad} in the barrel. The same value is obtained for the polar angle resolution in the transition region.

\begin{figure}[tbp]
  \renewcommand{\thesubfigure}{(\lr{subfigure})}
  \centering
  \begin{subfigure}{.5\textwidth}
    \centering
    \includegraphics[width=\linewidth]{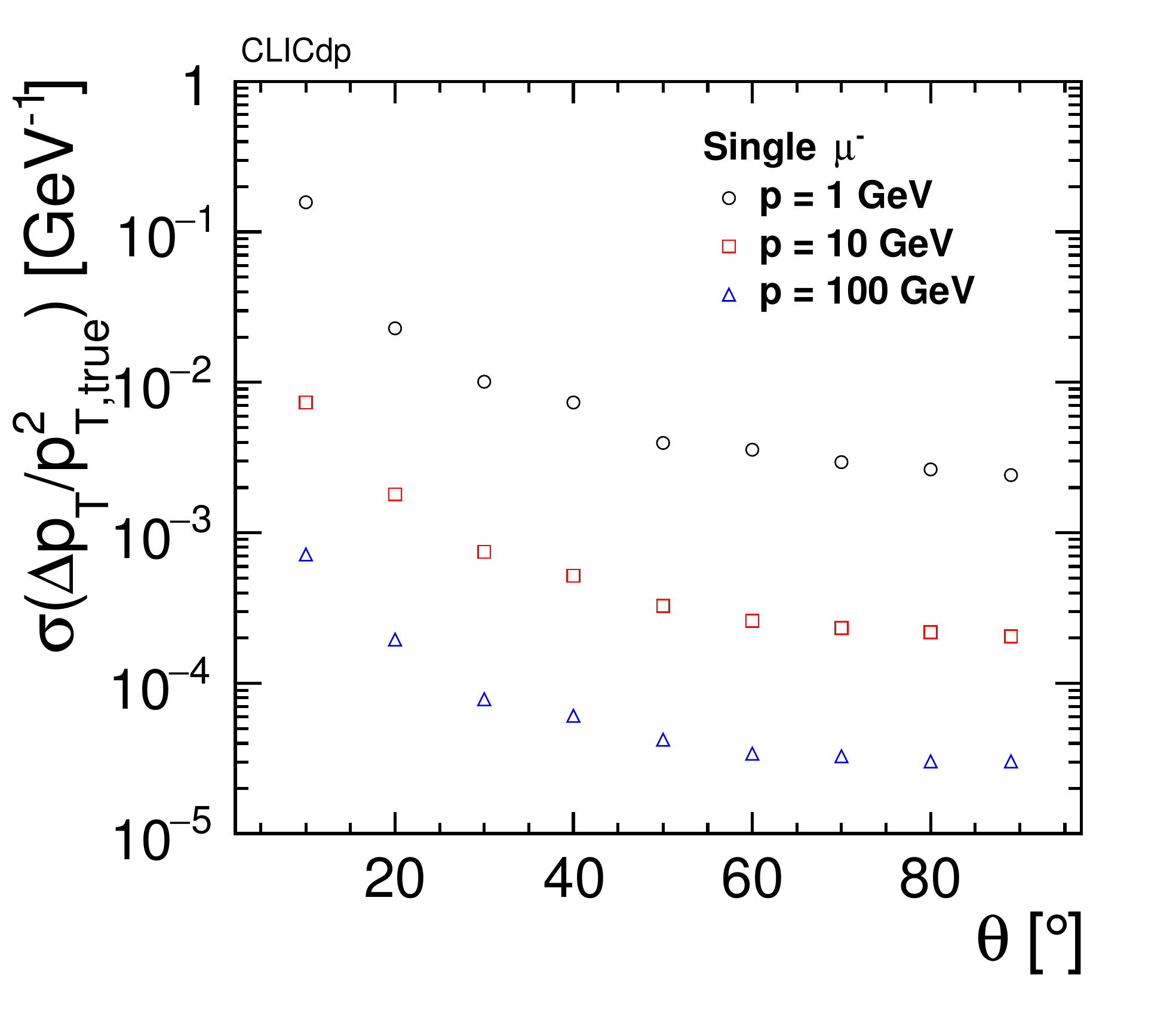}%
    \phantomsubcaption\label{fig:mom_res_theta}
  \end{subfigure}%
  \begin{subfigure}{.5\textwidth}
    \centering
    \includegraphics[width=\linewidth]{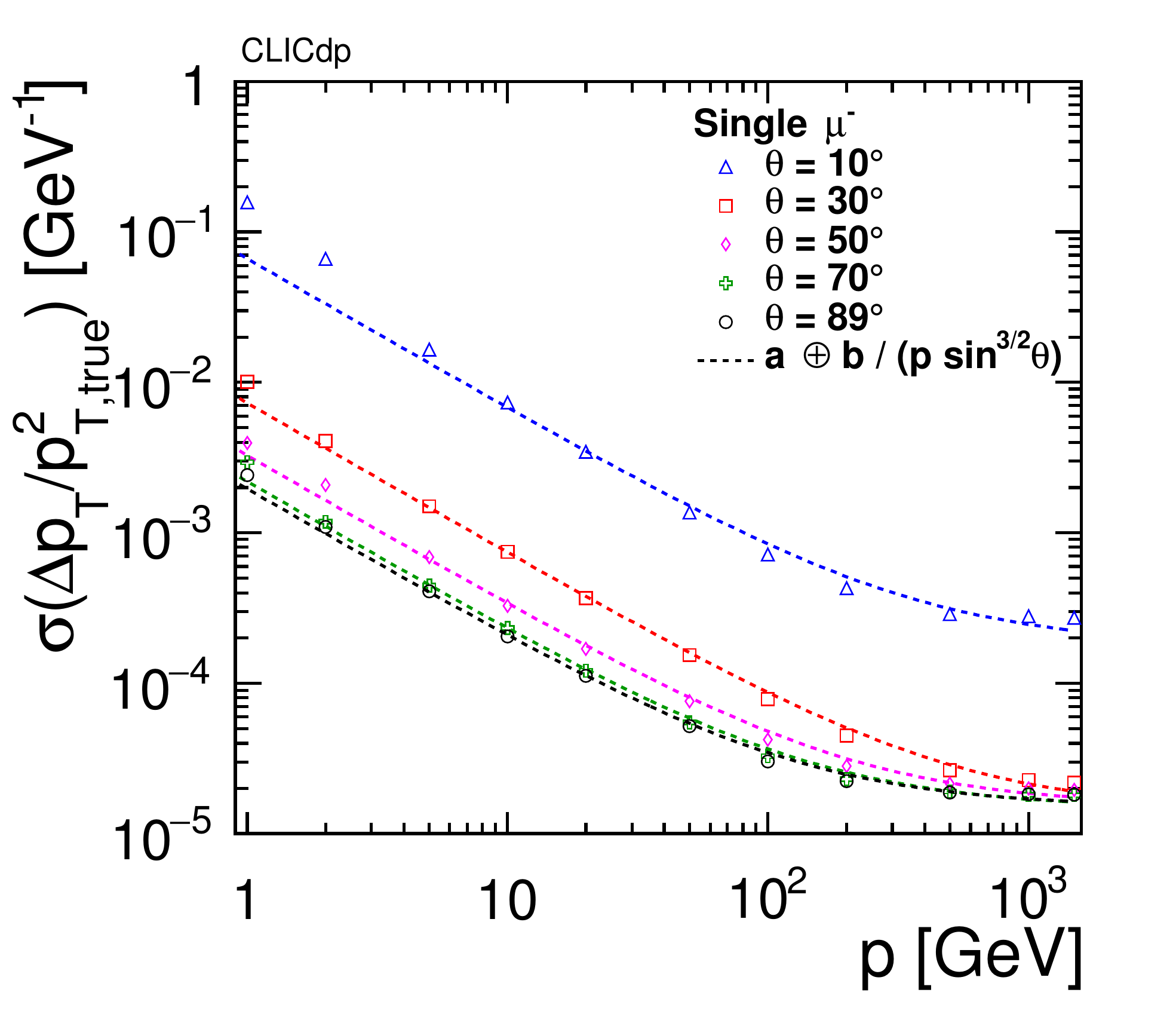}%
    \phantomsubcaption\label{fig:mom_res_p}
  \end{subfigure}
  \vspace{-2mm}
  \caption{Transverse momentum resolution as a function of the polar angle for muons with momenta of
    \SIlist{1;10;100}{GeV}~\subref{fig:mom_res_theta} and as a function of the momentum for muons at polar angles
    $\theta=$10\degrees, 30\degrees, 50\degrees, 70\degrees, 89\degrees~\subref{fig:mom_res_p}.  The lines represent the
    fit of each curve with the parametrization $a \oplus b/(p\cdot\sin^{3/2}\theta)$.}
  \label{fig:mom_res}
\end{figure}

The \pT{} resolution $\sigma(\Delta \pT/\pT^2)$ for single muons is determined from a single Gaussian fit of the distribution $(\pTmc-\pTrec)/\pTmc^2$ and is shown in \cref{fig:mom_res} as a function of the polar angle $\theta$ and of the momentum.
Each data point corresponds to 10\,000 muons simulated at a fixed energy and polar angle. In the barrel region, the \pT{} resolution reaches \SI{3e-5}{\per\gev}
for muons with a momentum of \SI{100}{GeV}\@. The dashed lines correspond to the fit of the data points according to the parametrization:

\begin{equation}
\sigma(\Delta \pT/\pT^{2}) = a \oplus \frac{b}{p \sin^{3/2}\theta}
\label{eq:momRes}
\end{equation}
where parameter $a$ represents the contribution from the curvature measurement and parameter $b$ is the multiple-scattering contribution. The values of these parameters for the different fitted curves are summarised in \cref{tab:parameters}.
Requirements on the momentum resolution follow from dedicated studies of BSM scenarios. As an example, the determination of the smuon and neutralino masses from the muon momentum distribution in the process $\epem\rightarrow\smusmu{}\rightarrow\mpmm\neutralino{1}\neutralino{1}$ requires a momentum resolution of \SI{2e-5}{\per\gev} in order for the high momentum part of the spectrum not to be distorted~\cite[Section 2.2.1]{cdrvol2}.
This is achieved for central tracks with momenta larger than \SI{500}{GeV}. For low momentum tracks, the values deviate from the parametrization due to the larger amount of material traversed.

\begin{table}[bt]
  \centering
  \caption{Parameter values from the fit of \cref{eq:momRes} for \cref{fig:mom_res_p}}
  \begin{tabular}{ccc}
    \toprule
    Angle    & Parameter $a$ [\si{\per\gev}] & Parameter $b$ \\\midrule
    \ang{10} & \num{1.8e-4}                 & 0.0048        \\
    \ang{30} & \num{1.4e-5}                 & 0.0026        \\
    \ang{50} & \num{1.5e-5}                 & 0.0022        \\
    \ang{70} & \num{1.5e-5}                 & 0.0020        \\
    \ang{89} & \num{1.5e-5}                 & 0.0019        \\
    \bottomrule{}
  \end{tabular}
  \label{tab:parameters}
\end{table}

\begin{figure}[tbp]
  \renewcommand{\thesubfigure}{(\lr{subfigure})}
  \centering
  \begin{subfigure}{.5\textwidth}
    \centering
    \includegraphics[width=\linewidth]{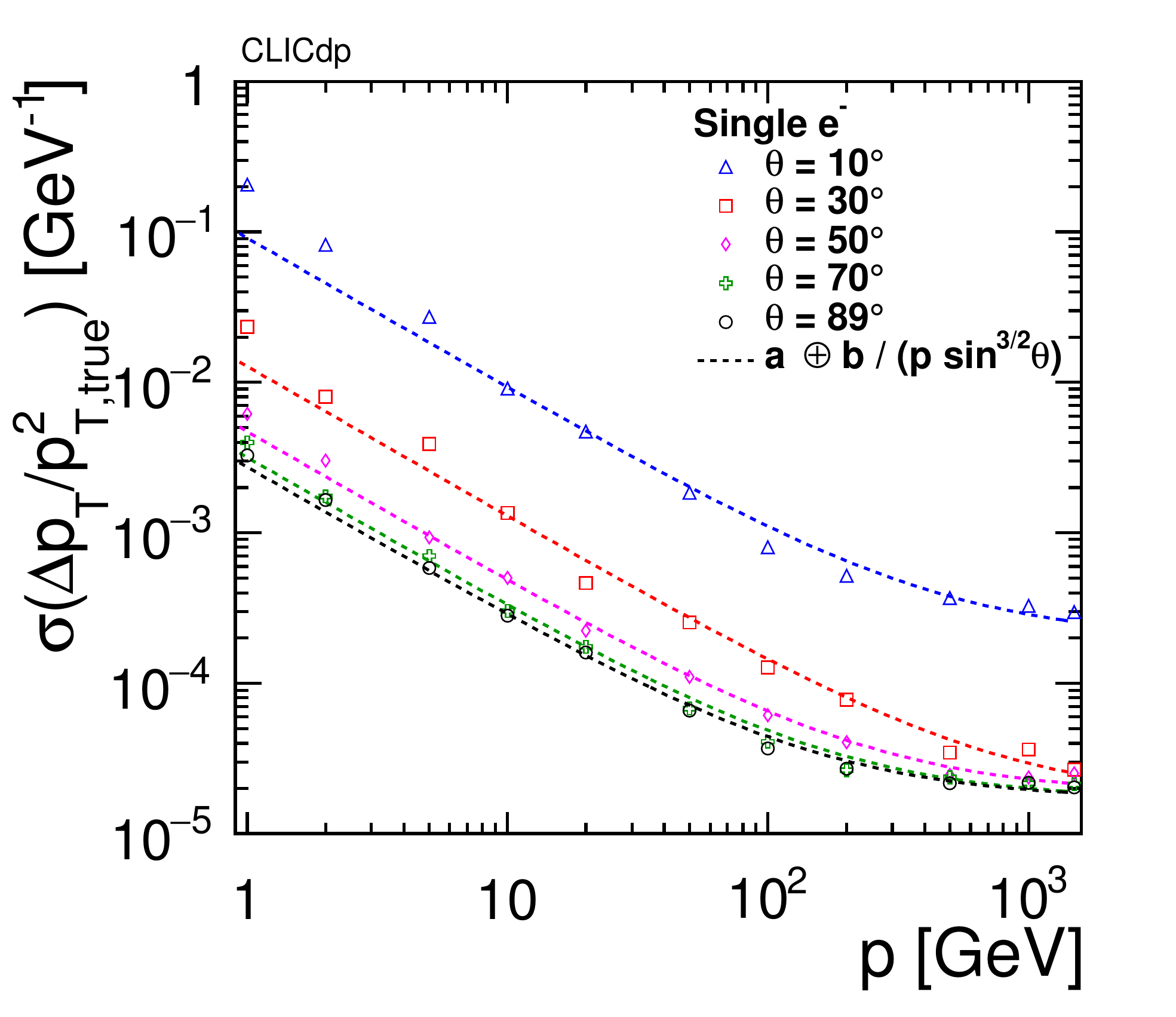}%
    \phantomsubcaption\label{fig:mom_res_p_electrons}
  \end{subfigure}%
  \begin{subfigure}{.5\textwidth}
    \centering
    \includegraphics[width=\linewidth]{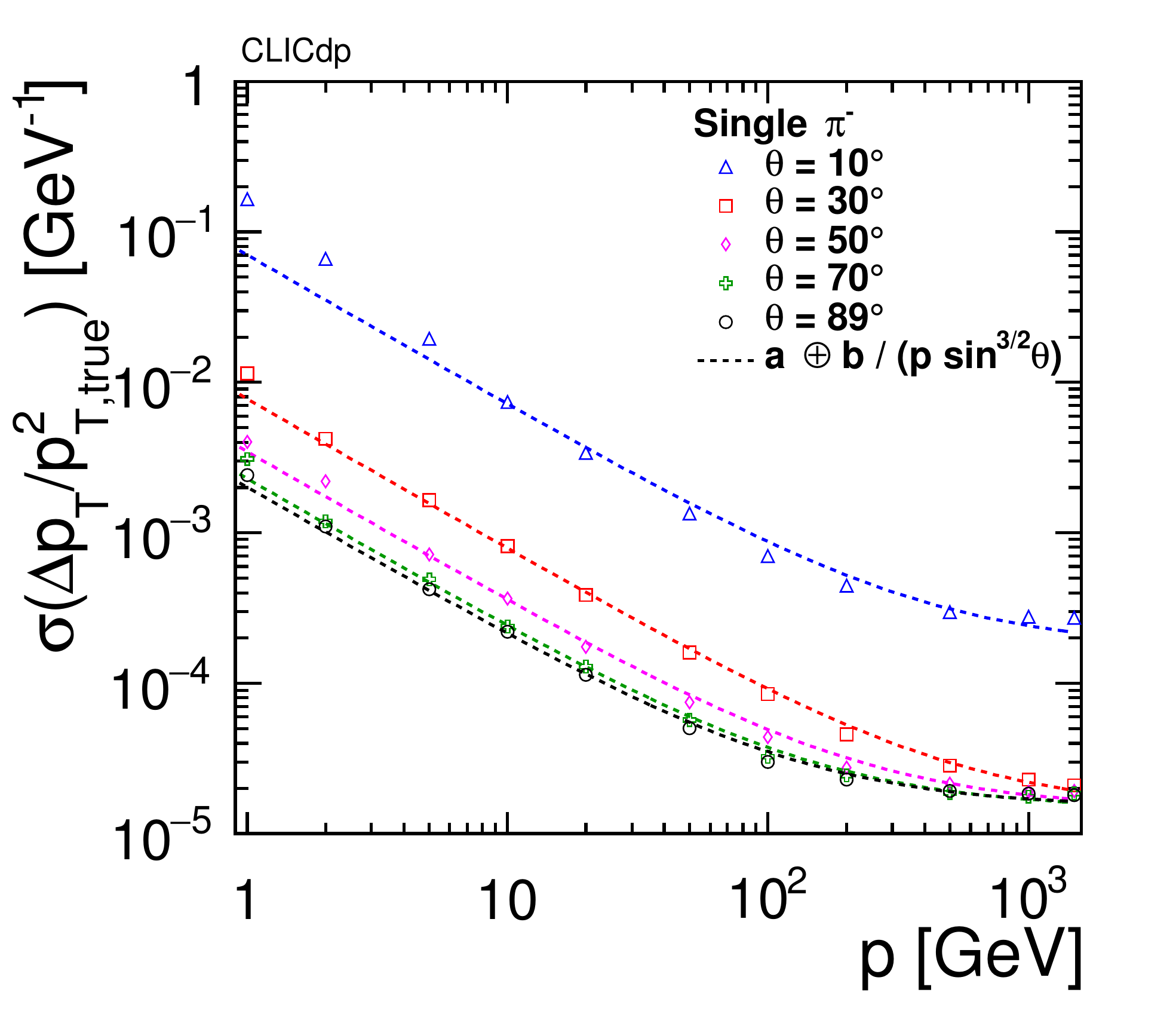}%
    \phantomsubcaption\label{fig:mom_res_p_pions}
  \end{subfigure}
  \vspace{-2mm}
  \caption{Transverse momentum resolution as a function of the momentum for electrons~\subref{fig:mom_res_p_electrons}
    and pions~\subref{fig:mom_res_p_pions} at polar angles $\theta=$10\degrees, 30\degrees, 50\degrees, 70\degrees,
    89\degrees.  The lines represent the fit of each curve with the parametrization
    $a \oplus b/(p\cdot\sin^{3/2}\theta)$.}\label{fig:mom_res_particleSpecies}
\end{figure}

Similarly, the momentum resolution for isolated electron and pion tracks was studied and is shown in \cref{fig:mom_res_particleSpecies}. Performances are comparable with those for isolated muon tracks.
 
\paragraph{Tracking Efficiency}

\begin{figure}[tbp]
  \centering
  \includegraphics[width=0.5\linewidth]{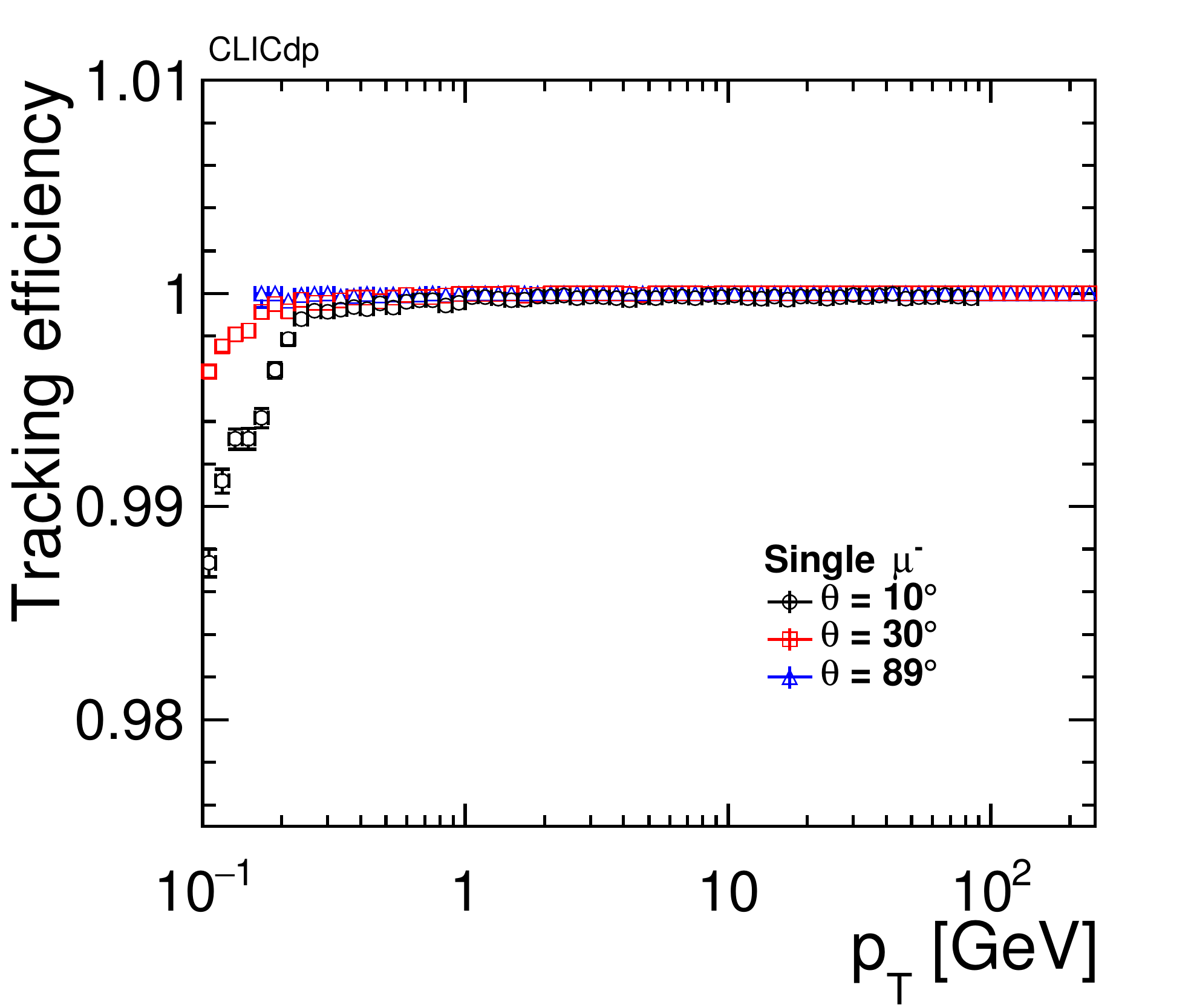}%
  \caption{Tracking efficiency as a function of \pT{} for muons at polar angles $\theta=$10\degrees, 30\degrees, 89\degrees.}%
  \label{fig:muons_eff_vs_pt}
\end{figure}

\begin{figure}[tbp]
  \renewcommand{\thesubfigure}{(\lr{subfigure})}
  \centering
  \begin{subfigure}{.5\textwidth}
    \centering
    \includegraphics[width=\linewidth]{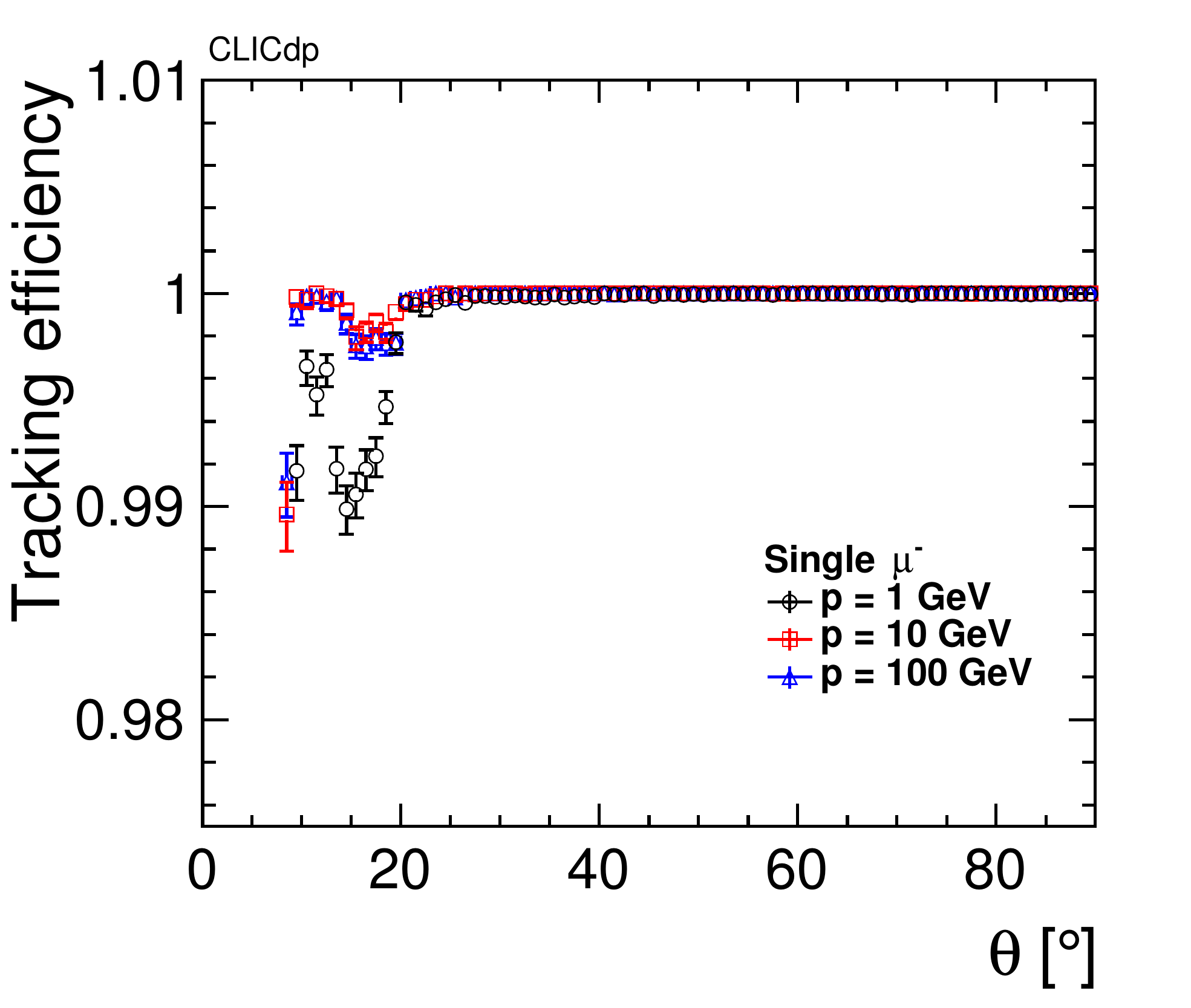}%
    \phantomsubcaption\label{fig:muons_eff_theta}
  \end{subfigure}%
  \begin{subfigure}{.5\textwidth}
    \centering
    \includegraphics[width=\linewidth]{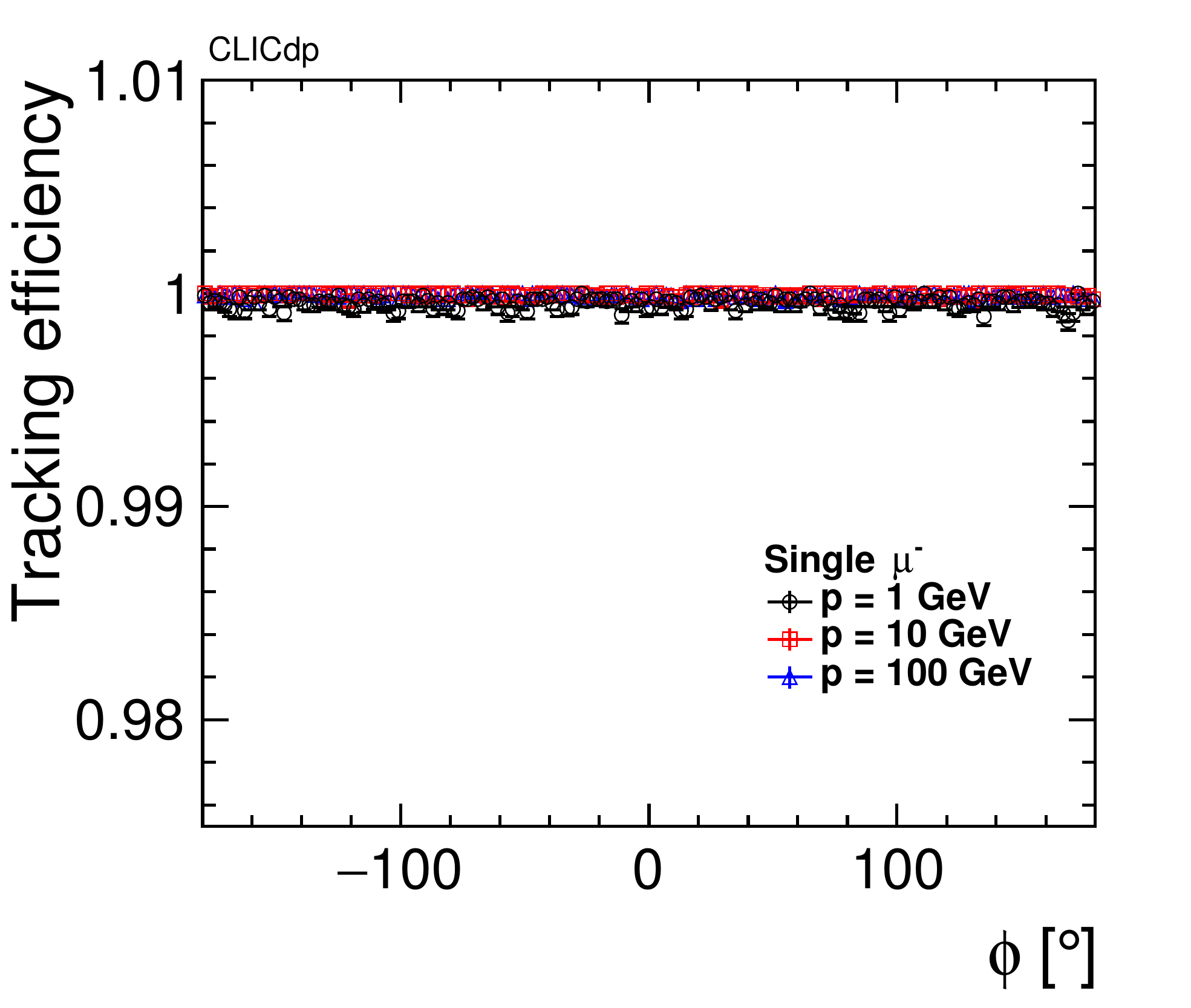}%
    \phantomsubcaption\label{fig:muons_eff_phi}
  \end{subfigure}
  \vspace{-2mm}
  \caption{Tracking efficiency as a function of the polar~\subref{fig:muons_eff_theta} and the
    azimuthal~\subref{fig:muons_eff_phi} angle for muons with momenta of \SIlist{1;10;100}{GeV}\@.}
  \label{fig:muons_eff_vs_angle}
\end{figure}

Tracking efficiency is defined as the fraction of reconstructable Monte Carlo particles that have been reconstructed. A particle is considered reconstructable if it is stable at generator level ($\mathrm{genStatus}=1$),
if $\pT > \SI{100}{MeV}$, $|\cos\theta| < 0.99$ and if it has at least 4 unique hits (i.e.\ hits which do not occur on the same subdetector layer).
The efficiency for isolated muon tracks has been computed by reconstructing 2 million muons simulated at polar angles $\theta=$10\degrees, 30\degrees, 89\degrees{} and with a descending power-law energy distribution defined between \SI{100}{MeV} and \SI{500}{GeV}\@.
It is shown in \cref{fig:muons_eff_vs_pt}  as a function of \pT. The tracking efficiency is constant at 100\% for all polar angles, with a maximum drop of roughly 1\% for $\pT < \SI{200}{MeV}$.
\cref{fig:muons_eff_vs_angle} shows the muon efficiency as a function of the polar angle (left) and the azimuthal angle (right).
For all energies, fully efficient performances are obtained at any $\phi$ and $\theta$, except in the region $10\degrees < \theta < 20\degrees$, where the geometrical layout of the tracking system (see \cref{fig:tracker_angles} in \cref{sec:appendix}) contributes to an efficiency loss of at most 1\% at 10\degrees and 15\degrees. It has been demonstrated that this efficiency loss is fully recovered when performing the Kalman filter fit in the inverted direction by starting from the outermost hits. A study aimed at upgrading the fitting procedure is currently ongoing.

\begin{figure}[tbp]
  \renewcommand{\thesubfigure}{(\lr{subfigure})}
  \centering
  \begin{subfigure}{.5\textwidth}
    \centering
    \includegraphics[width=\linewidth]{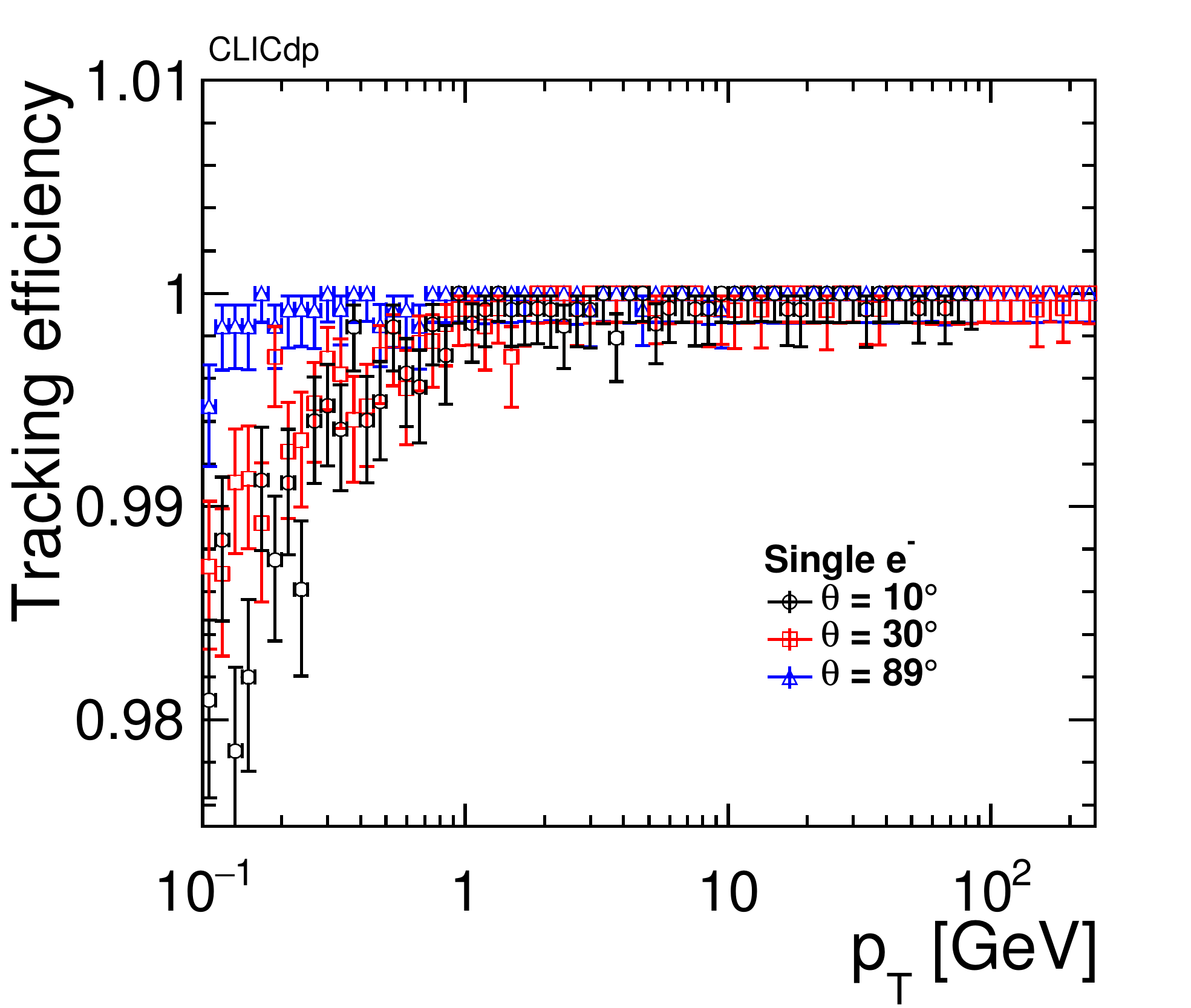}%
    \phantomsubcaption\label{fig:electrons_eff_pt}
  \end{subfigure}%
  \begin{subfigure}{.5\textwidth}
    \centering
    \includegraphics[width=\linewidth]{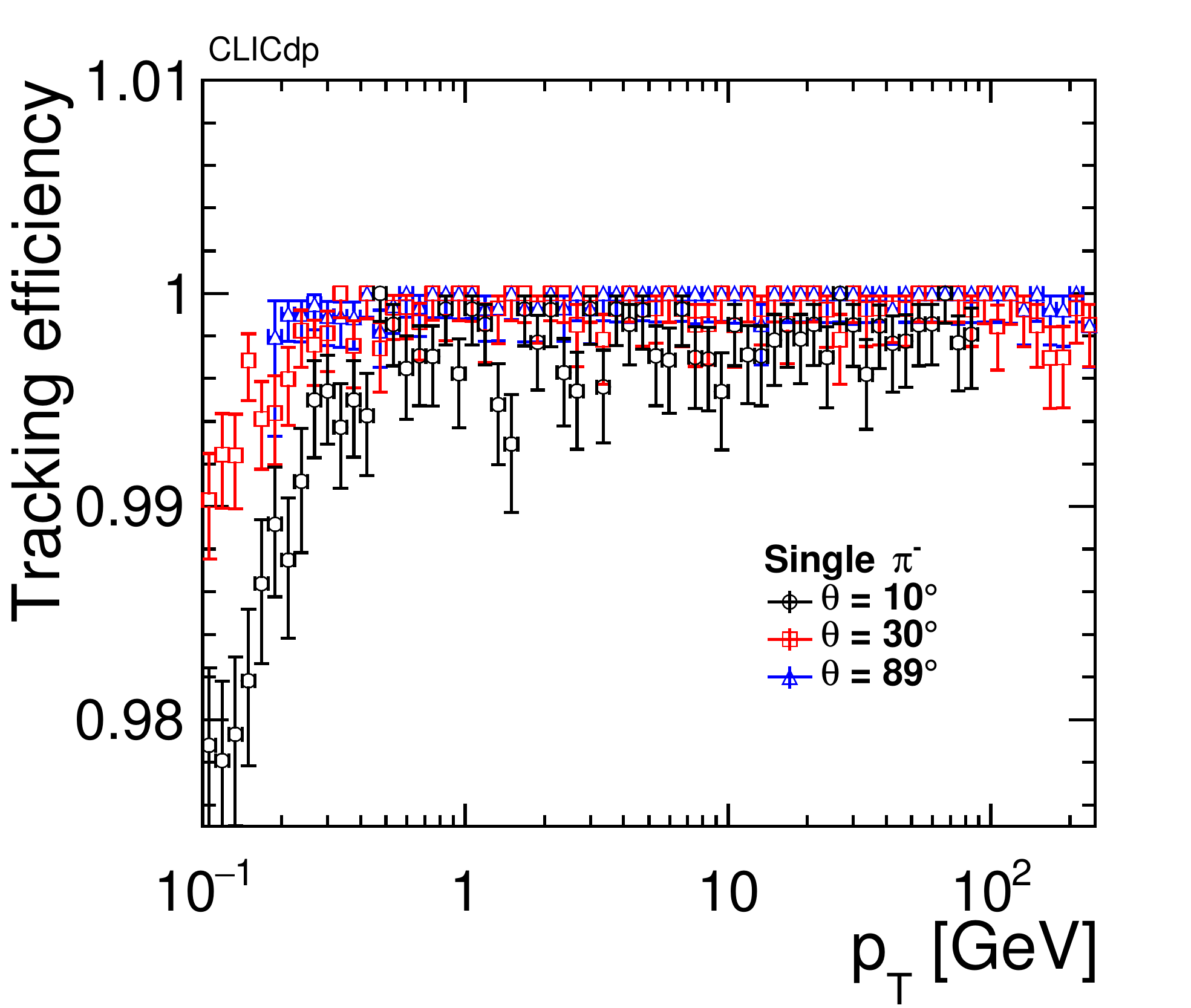}%
    \phantomsubcaption\label{fig:pions_eff_pt}
  \end{subfigure}
  \vspace{-2mm}
  \caption{Tracking efficiency as a function of \pT{} for electrons \subref{fig:electrons_eff_pt} and
    pions~\subref{fig:pions_eff_pt} at polar angles $\theta=$
    10\degrees, 30\degrees, 89\degrees.}\label{fig:eff_pt_particleSpecies}
\end{figure}

Similarly, 100\,000 isolated electrons and pions simulated at polar angles $\theta=$ 10\degrees, 30\degrees, 89\degrees{} and with a descending power-law energy distribution defined between \SI{100}{MeV} and \SI{500}{GeV} have been simulated and reconstructed.
The tracking efficiency is shown in \cref{fig:eff_pt_particleSpecies} as a function of the transverse momentum.
For both particle species, the reconstruction of central tracks is fully efficient, while for lower-energy electrons and pions the efficiency loss appears at higher \pT{} than for muons, and it amounts to a maximum of 2\% within statistical uncertainties.

\begin{figure}[tbp]
  \centering
  \includegraphics[width=0.5\linewidth]{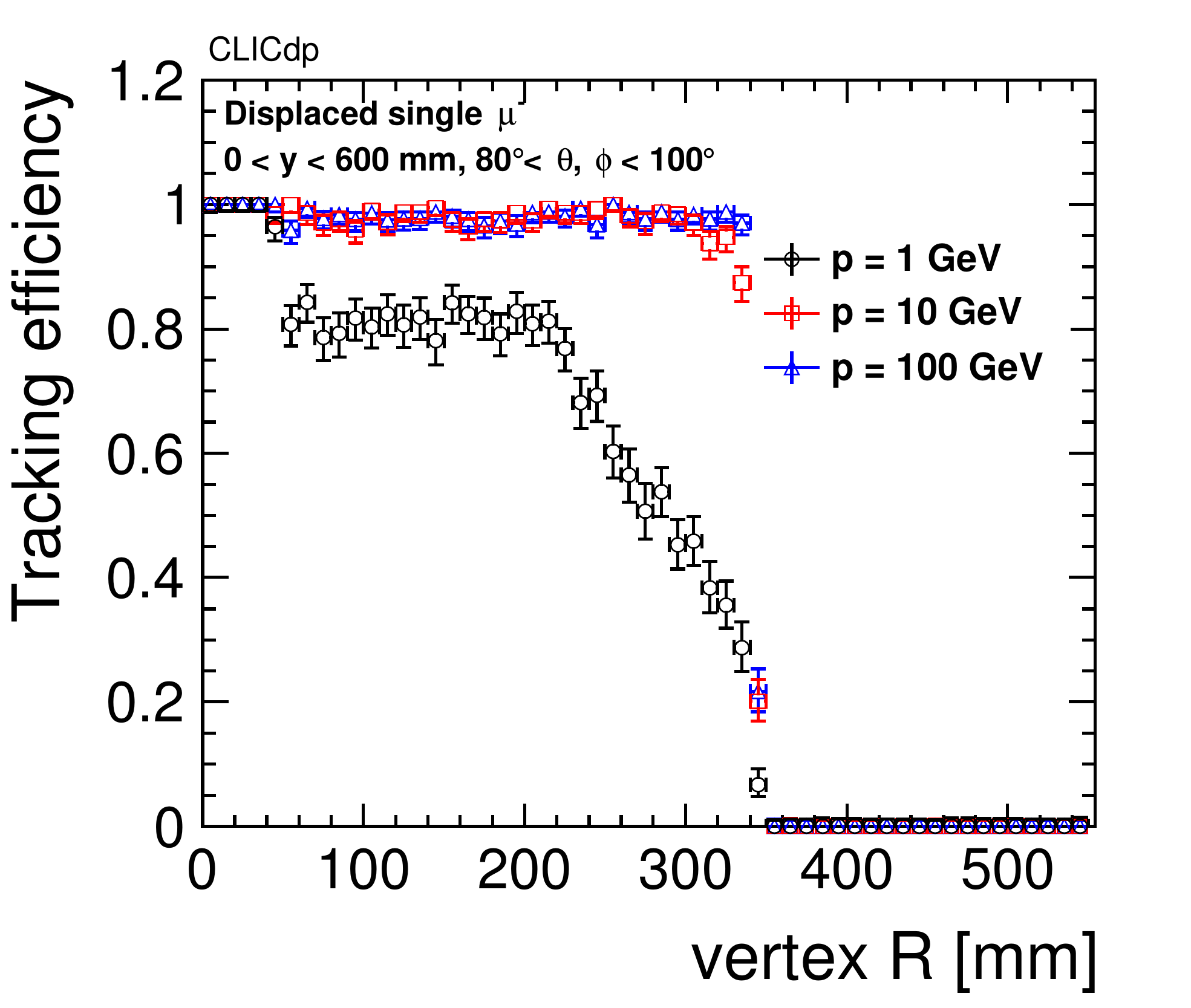}
  \caption{Tracking efficiency as a function of the production vertex radius for muons with momenta of \SIlist{1;10;100}{GeV}, uniformly generated in a range 0 < y < \SI{600}{mm} and \mbox{$80\degrees < \theta,\phi < 100\degrees$}.}
   \label{fig:displaced_muons}
\end{figure}

To probe the tracking performances for displaced tracks, 10\,000 single muons have been simulated, 
requiring their production vertex to be within $\SI{0}{cm} < y < \SI{60}{cm}$ and their angular distribution in a cone of 10\degrees{} width around the y axis, i.e. $80\degrees{} < \theta, \phi < 100\degrees$. This is done for simplicity, such that particles are produced in the barrel and they traverse roughly the same amount of material.
The efficiency as a function of the production vertex radius $R = \sqrt{\smash[b]{x^2+y^2}}$, is shown in \cref{fig:displaced_muons} for muons with momenta of \SIlist{1;10;100}{GeV}\@.
For \SI{1}{GeV} muons that are produced with \mbox{$R \ge \SI{60}{mm}$} outside the vertex detector the efficiency drops by 20\%, due to the fact that by losing energy while traversing the detector layers,
those particles have not enough left-over momentum to reach the required minimum number of hits. 
For higher-energy muons, instead, the efficiency is above 95\% at any radius.
Regardless of the energy, an abrupt fall-off is observed for all tracks with a production radius of \SI{350}{mm} or more.
This is an effect of the reconstruction cuts, since for displaced tracks a minimum number of 5 hits is required to reconstruct the track, 
while only 4 sensitive layers are traversed by tracks starting beyond $R=\SI{350}{mm}$.

\paragraph{Particle Reconstruction and Identification}

Particles are reconstructed and identified using the PandoraPFA C++ Software Development Kit~\cite{Marshall:2015rfaPandoraSDK}. 
These particle flow reconstruction algorithms have been studied extensively in full \geant{} simulations of the CLIC\_ILD detector concept~\cite{Marshall:2012ryPandoraPFA}. 
Particle flow aims to reconstruct each visible particle in the event using information from all subdetectors. The high granularity of calorimeters is essential in achieving the desired precision measurements. 
Electrons are identified using clusters largely contained within ECAL and matched with a track. Muons are determined from a track and a matched cluster compatible with a minimum ionising particle signature in ECAL and HCAL, 
plus corresponding hits in the muon system. A hadronic cluster in ECAL and HCAL matched to a track is used in reconstructing charged hadrons. Hadronic clusters without a corresponding track give rise to neutrons,
 and photons are reconstructed from an electromagnetic cluster in ECAL\@. In jets typically 60\% of the energy originates from charged hadrons and 30\% from photons. The remaining 10\% of the jet energy are mainly carried by neutral hadrons.

\begin{figure}[tpb]
  \renewcommand{\thesubfigure}{(\lr{subfigure})}
  \centering
  \begin{subfigure}{.5\textwidth}
    \centering
    \includegraphics[width=\linewidth]{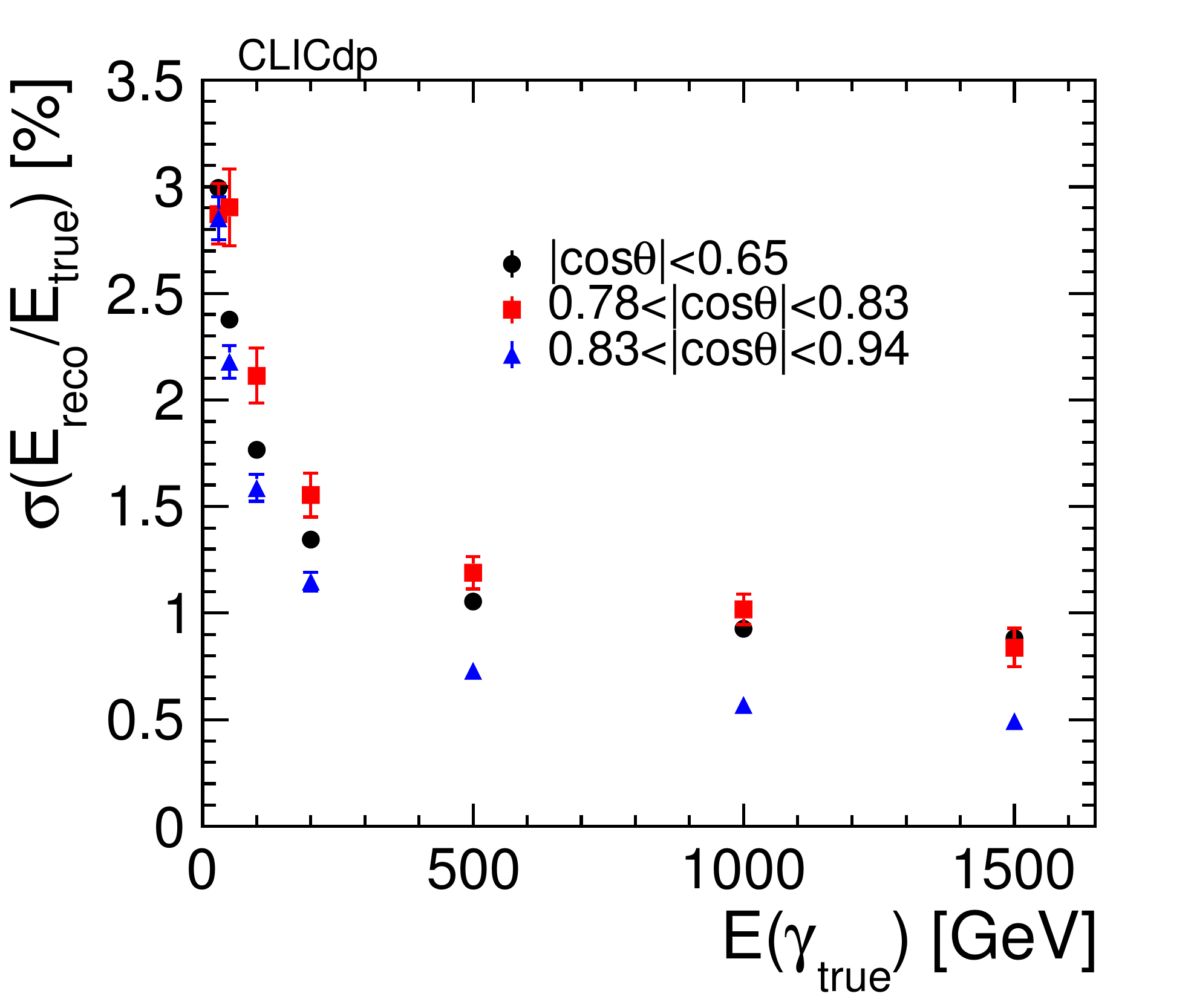}%
    \phantomsubcaption\label{fig:photonResolutionVsEnergy}
  \end{subfigure}%
  \begin{subfigure}{.5\textwidth}
    \centering
    \includegraphics[width=\linewidth]{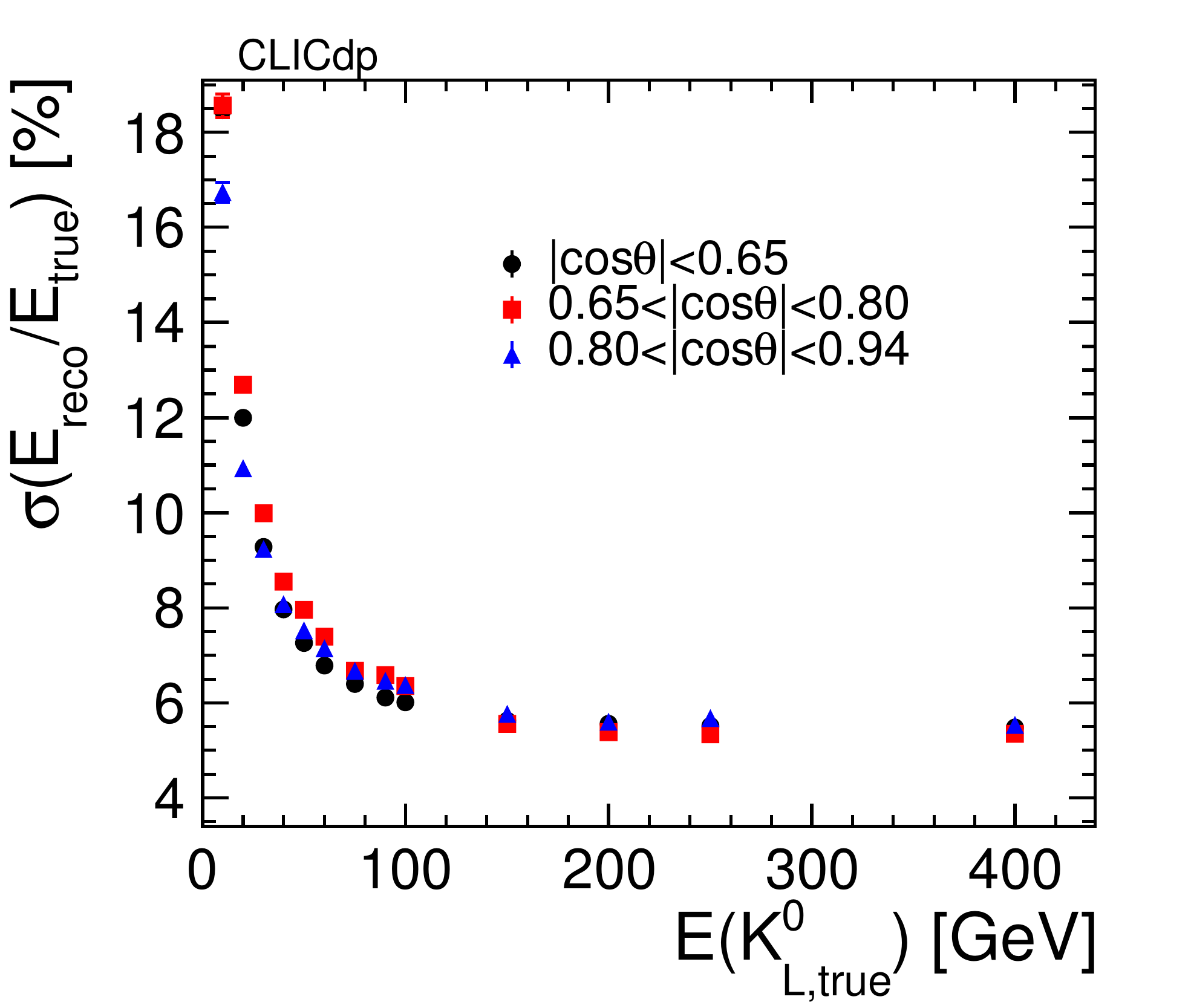}%
    \phantomsubcaption\label{fig:K0LResolutionVsEnergy}
  \end{subfigure}
  \vspace{-2mm}
  \caption{Photon energy resolution~\subref{fig:photonResolutionVsEnergy} and neutral hadron resolutions of
    \PKzL's~\subref{fig:K0LResolutionVsEnergy} as a function of the energy. Results are shown for the barrel region,
    transition region and endcap.}
\end{figure}

The performance of the Pandora reconstruction algorithms is studied in single particle events at several energies, generated as flat distributions in $\cos\theta$. The ECAL energy resolution is studied using single photon events. At each energy point in three different regions (barrel, endcap, and transition region) the photon energy response distribution is fitted with a Gaussian. The $\sigma$ of the Gaussian is a measure for the energy resolution in ECAL\@. At very large photon energies the photon energy resolution is affected by energy leakage into HCAL\@. The energy dependence of the photon energy resolution of CLICdet is shown in \cref{fig:photonResolutionVsEnergy}, for the three detector regions. The stochastic term is $15\%/\sqrt{E}$, determined from a two parameter fit within the energy range between \SI{5}{GeV} and \SI{200}{GeV}\@.

For hadrons, the HCAL hits are reweighted using the software compensation technique implemented within PandoraPFA, and developed by the CALICE (Calorimeter for Linear Collider Experiment) collaboration~\cite{CALICE_sc_2012, Tran:2017tgrSoftwareCompensation}. In the non-compensating calorimeters of CLICdet the detector response for electromagnetic subshowers is typically larger than for hadronic showers. On average the electromagnetic component of the shower has larger hit energy densities. The weights depend on the hit energy density and the unweighted energy of the calorimeter cluster, where hits with larger hit energy densities receive smaller weights. In a dedicated calibration procedure within PandoraPFA, software compensation weights are determined using single neutron and \PKzL{} events over a wide range of energy points. At each energy point equal statistics is required, i.e.\ using the same number of events for neutrons and \PKzL{}. Only events with one cluster fully contained within ECAL plus HCAL are used in this calibration. Software compensation improves the energy resolution of hadronic clusters. 
The resulting energy resolution of neutral hadrons is shown for \PKzL as a function of the energy in \cref{fig:K0LResolutionVsEnergy}.

\begin{figure}[tbp]
  \renewcommand{\thesubfigure}{(\lr{subfigure})}
  \centering
  \begin{subfigure}{.5\textwidth}
    \centering
    \includegraphics[width=\linewidth]{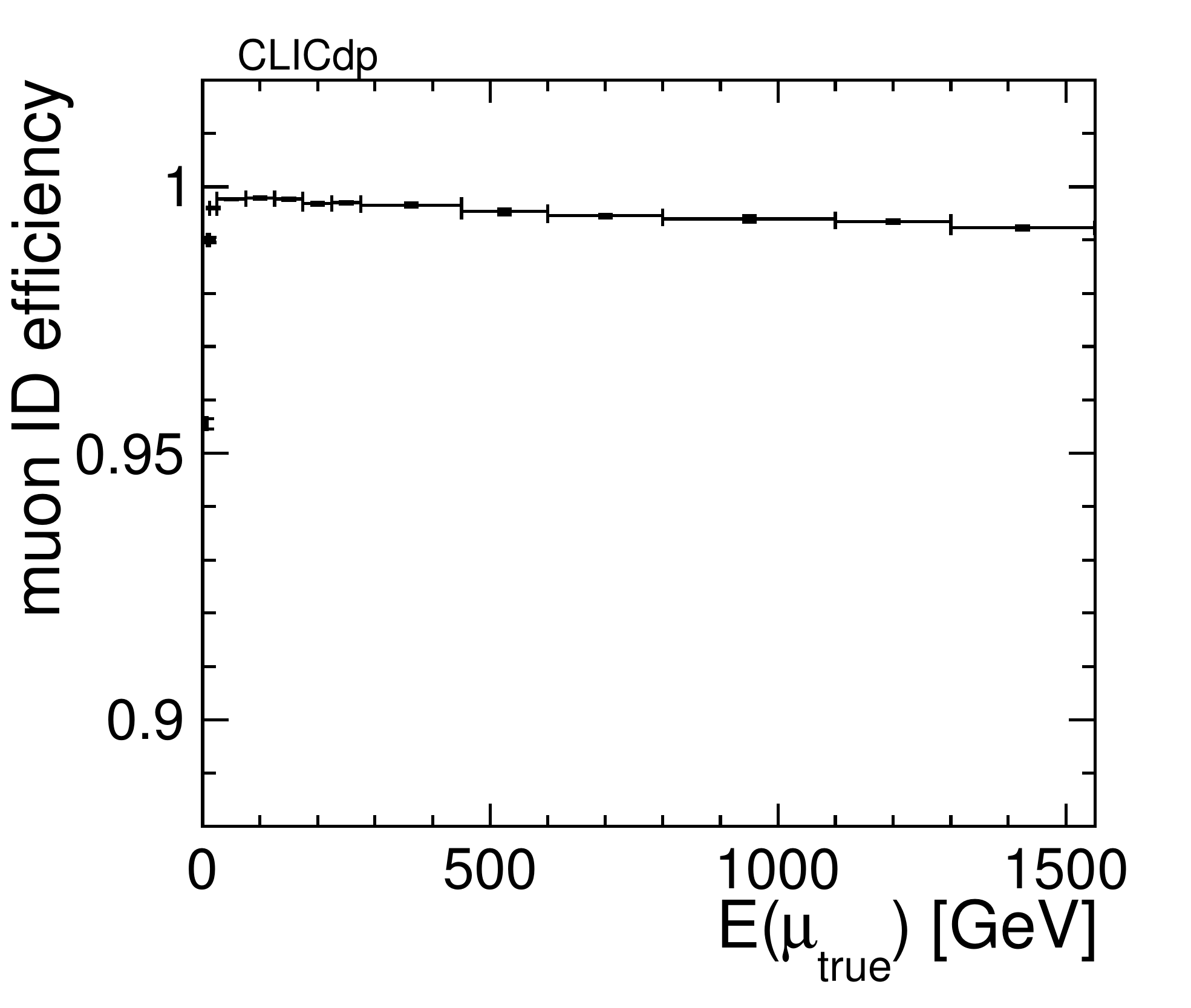}%
    \phantomsubcaption\label{fig:particleGun_muIDEffVsE}
  \end{subfigure}%
  \begin{subfigure}{.5\textwidth}
    \centering
    \includegraphics[width=\linewidth]{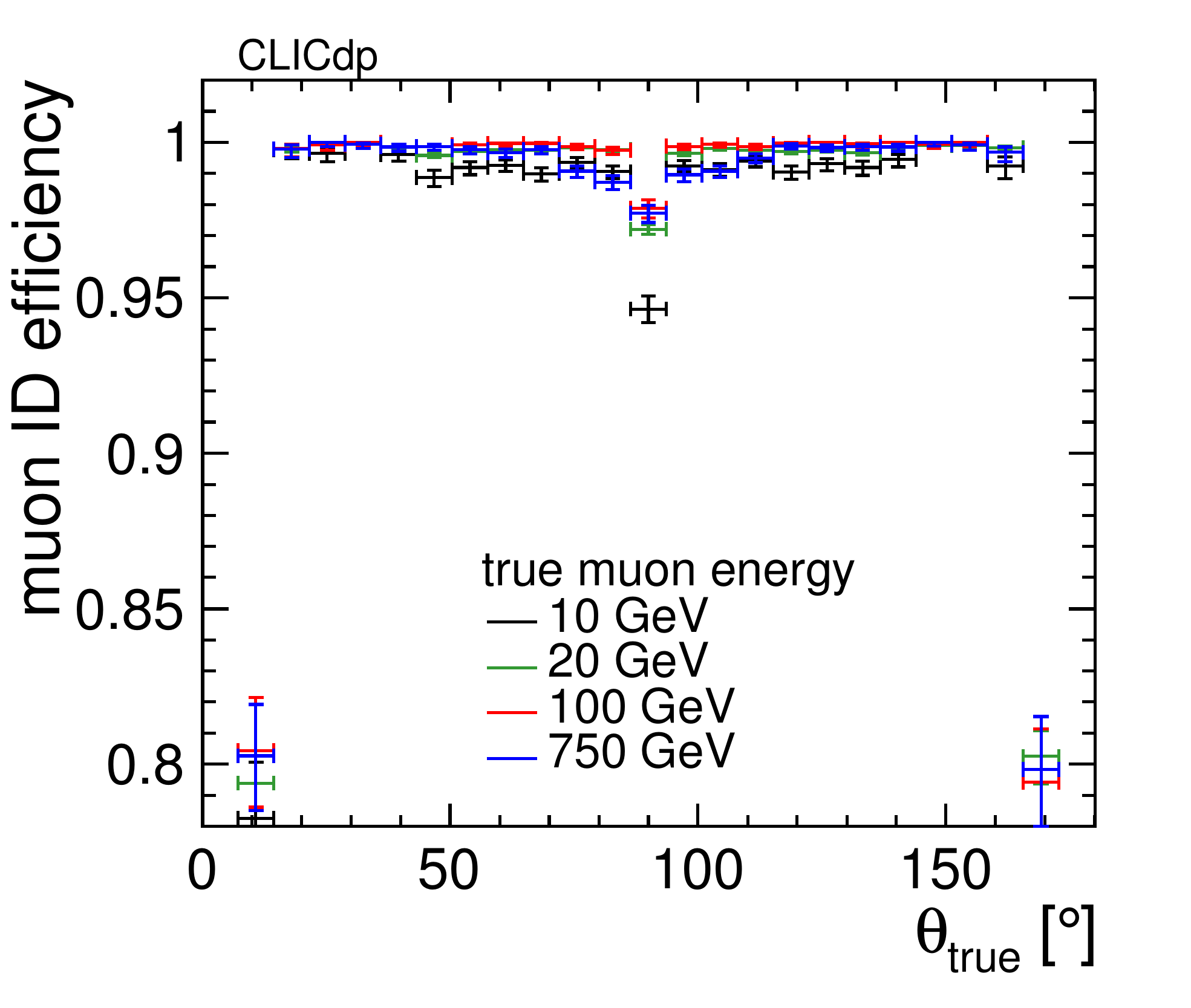}%
    \phantomsubcaption\label{fig:particleGun_muIDEffVsTheta}
  \end{subfigure}
  \vspace{-2mm}
  \caption{Particle identification efficiency for muons as a function of the energy~\subref{fig:particleGun_muIDEffVsE} and as a function of the polar angle $\theta$ for four different energies~\subref{fig:particleGun_muIDEffVsTheta}.}
\end{figure}

\begin{figure}[tbp]
  \renewcommand{\thesubfigure}{(\lr{subfigure})}
  \centering
  \begin{subfigure}{.5\textwidth}
    \centering
    \includegraphics[width=\linewidth]{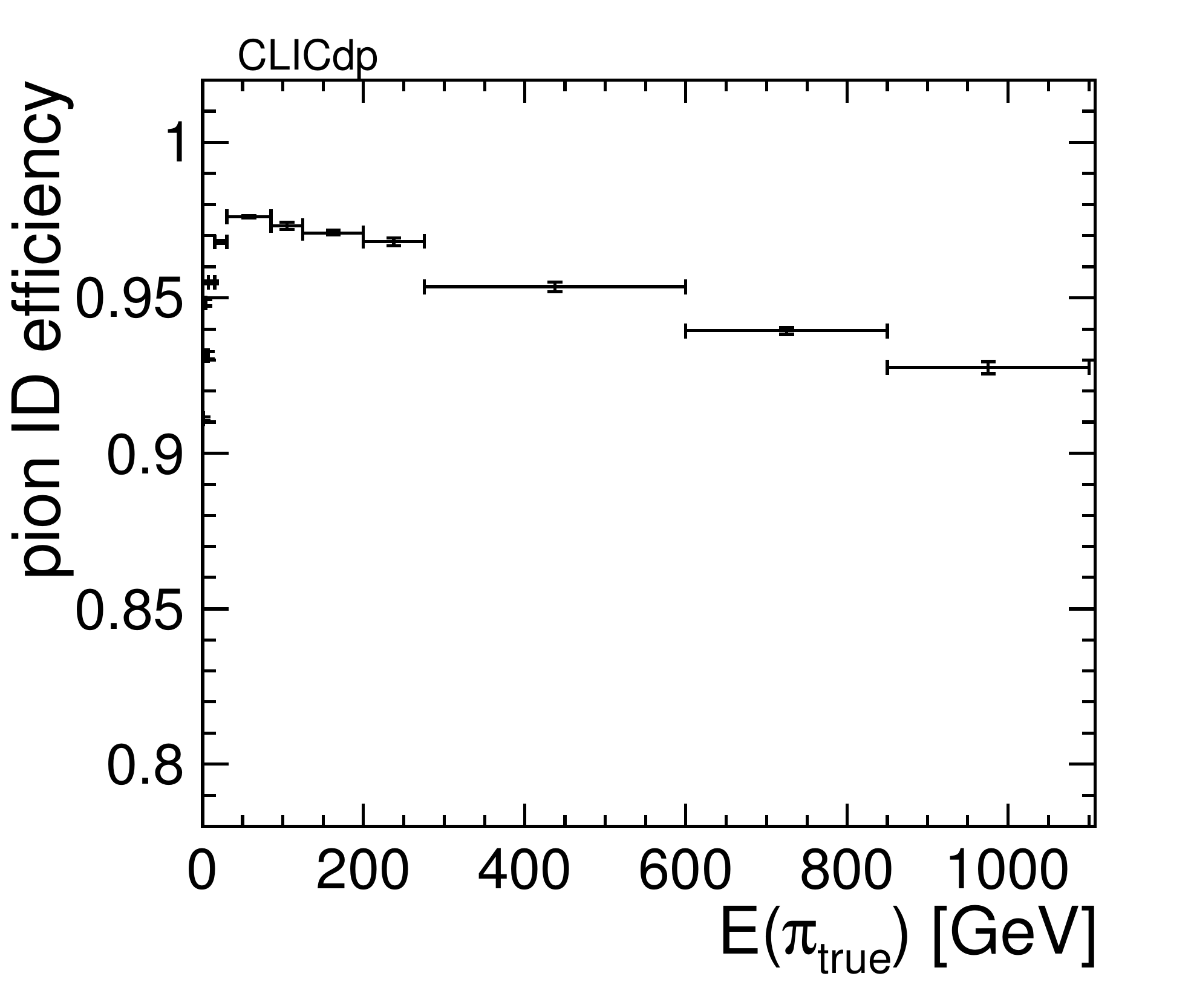}%
    \phantomsubcaption\label{fig:particleGun_piIDEffVsE}
  \end{subfigure}%
  \begin{subfigure}{.5\textwidth}
    \centering
    \includegraphics[width=\linewidth]{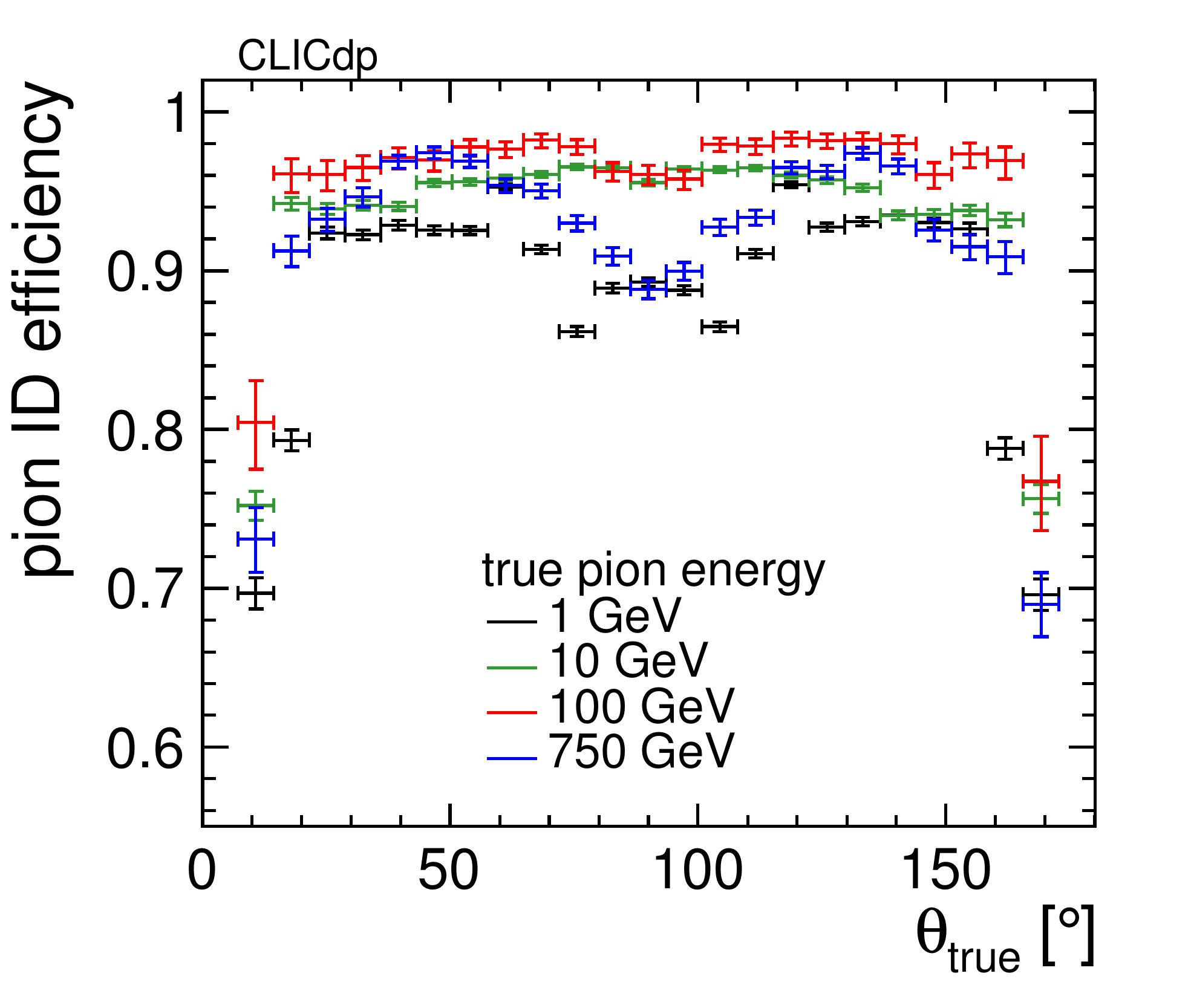}%
    \phantomsubcaption\label{fig:particleGun_piIDEffVsTheta}
  \end{subfigure}
  \vspace{-2mm}
  \caption{Particle identification efficiency for pions as a function of the energy~\subref{fig:particleGun_piIDEffVsE} and as a function of the polar angle $\theta$ for four different energies~\subref{fig:particleGun_piIDEffVsTheta}.}
\end{figure}

\begin{figure}[tbp]
  \renewcommand{\thesubfigure}{(\lr{subfigure})}
  \centering
  \begin{subfigure}{.5\textwidth}
    \centering
    \includegraphics[width=\linewidth]{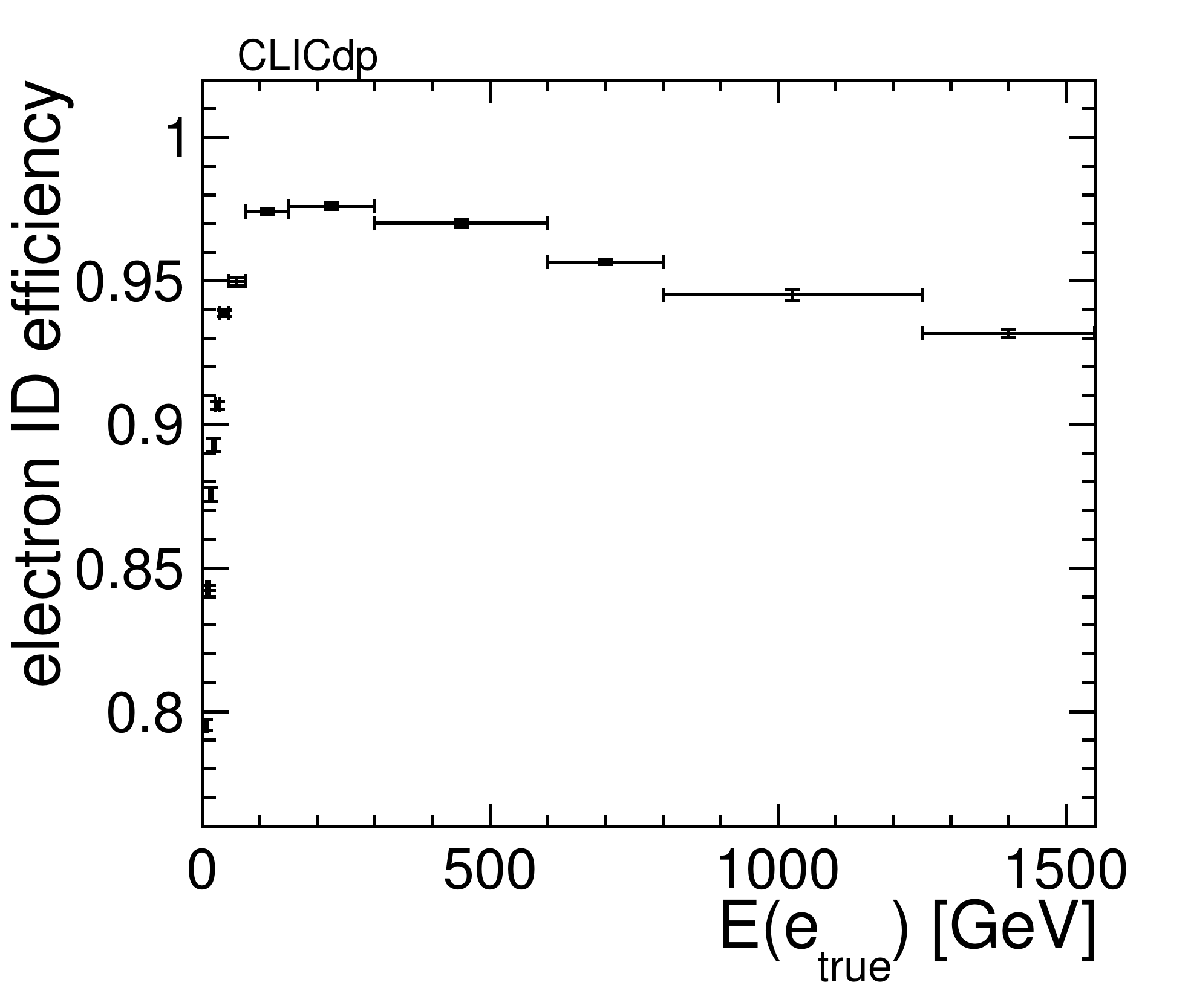}%
    \phantomsubcaption\label{fig:particleGun_elIDEffVsE}
  \end{subfigure}%
  \begin{subfigure}{.5\textwidth}
    \centering
    \includegraphics[width=\linewidth]{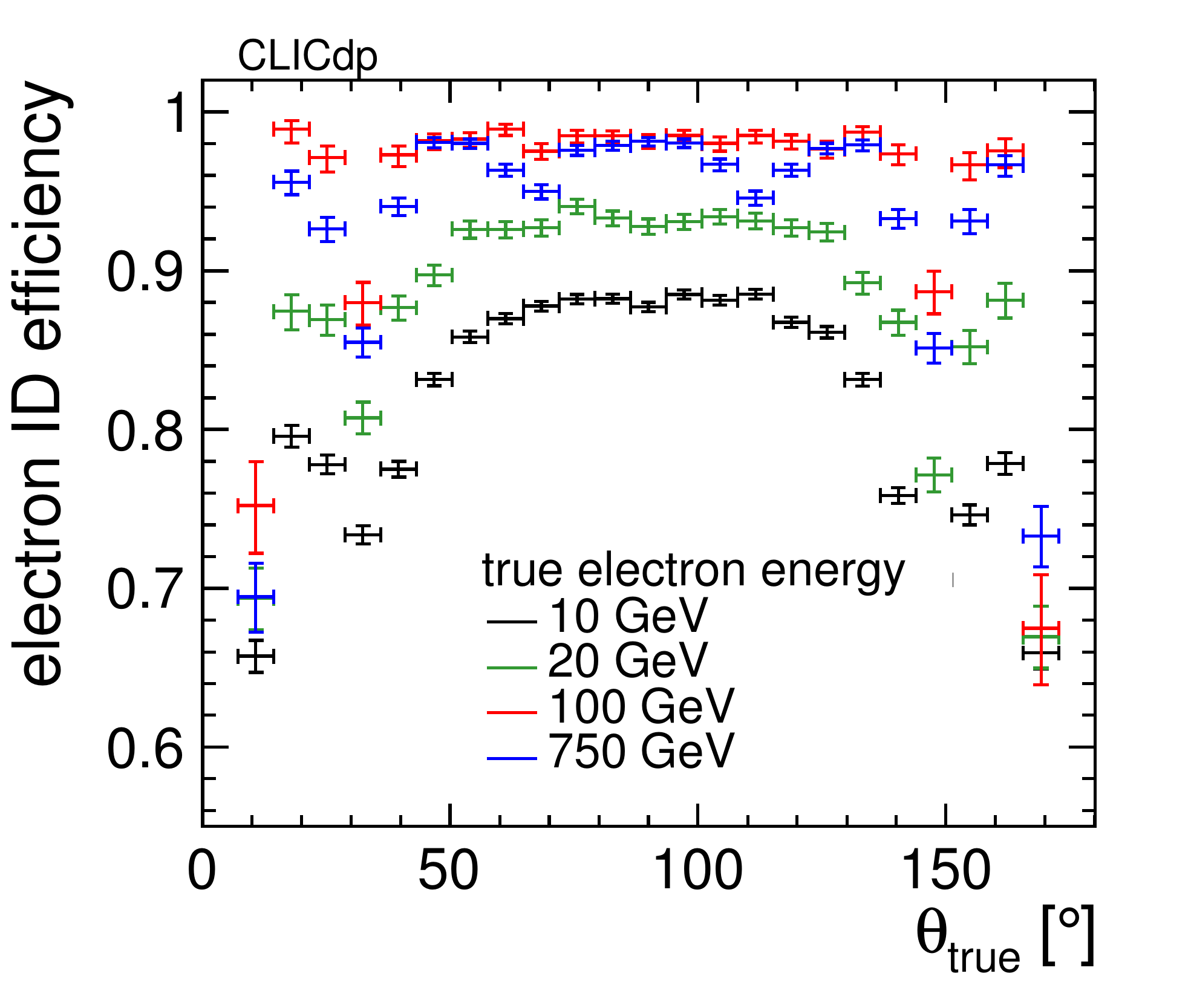}%
    \phantomsubcaption\label{fig:particleGun_elIDEffVsTheta}
  \end{subfigure}
  \vspace{-2mm}
  \caption{Particle identification efficiency for electrons as a function of the energy~\subref{fig:particleGun_elIDEffVsE} and as a function of the polar angle $\theta$ for four different energies~\subref{fig:particleGun_elIDEffVsTheta}.}
\end{figure}

The particle identification efficiency of Pandora particle flow algorithms is studied in single particle events separately for muons, electrons, photons and pions. The events are produced as flat distributions in $\cos\theta$. 
The reconstructed particle is required to be of the same type as the generated particle, and spatially matched within 1\degrees{} around it. The identification efficiency is studied as a function of the energy and polar angle $\theta$. 
To avoid any bias from lower efficiencies in the forward tracking, the energy dependence is studied for events where the true particle direction is restricted to $|\cos\theta|<0.95$.
 Identification for muons are illustrated in \cref{fig:particleGun_muIDEffVsE} as a function of the energy and in \cref{fig:particleGun_muIDEffVsTheta} as a function of the polar angle $\theta$.
The efficiency reaches a plateau beyond 99\% from energies of \SI{10}{GeV} up to \SI{1.5}{TeV}, flat as a function of the polar angle.
The drop at 90\degrees{} appears to be an artifact of the reconstruction, which is currently being fixed.
The efficiency of pion identification is beyond 90\% already from \SI{1}{GeV} on, about 98\% at \SI{100}{GeV}, but drops below 95\% at very high energies (\cref{fig:particleGun_piIDEffVsE}),
particularly due to inefficiencies around polar angles of 90\degrees{} (\cref{fig:particleGun_piIDEffVsTheta}).
The electron identification efficiency as a function of the energy is shown in \cref{fig:particleGun_elIDEffVsE}.
At \SI{10}{GeV}, the efficiency is around 85\% in the barrel and around 80\% in the endcaps, as can be seen in \cref{fig:particleGun_elIDEffVsTheta}.
For energies of about \SI{30}{GeV} and higher, the efficiencies reach 90\% both in the endcap and barrel.
The lack of electron identification efficiency at low energies can be attributed to incomplete recovery of Bremsstrahlung photon energy.
 In the transition region between barrel and endcaps the efficiency is 5\%--10\% lower than in the endcaps.

While for muons and pions the reconstructed energy is typically well within 10\% of the true energy, for electrons the reconstructed energy has, due to Bremsstrahlung, a long tail towards lower values compared to the true energy.
Work is ongoing to develop a Bremsstrahlung recovery algorithm, using close by photons to dress the electron momentum by summing their four momenta.
This procedure shows a good performance for electron energies below \SI{250}{GeV}\@. The energy response is recovered and the width of the energy response is within 10\%.
However, at higher electron energies sizeable tails are introduced. Thus at large electron energies the recovery algorithm needs to be improved.

\begin{figure}[tbp]
  \renewcommand{\thesubfigure}{(\lr{subfigure})}
  \centering
  \begin{subfigure}{.5\textwidth}
    \centering
    \includegraphics[width=\linewidth]{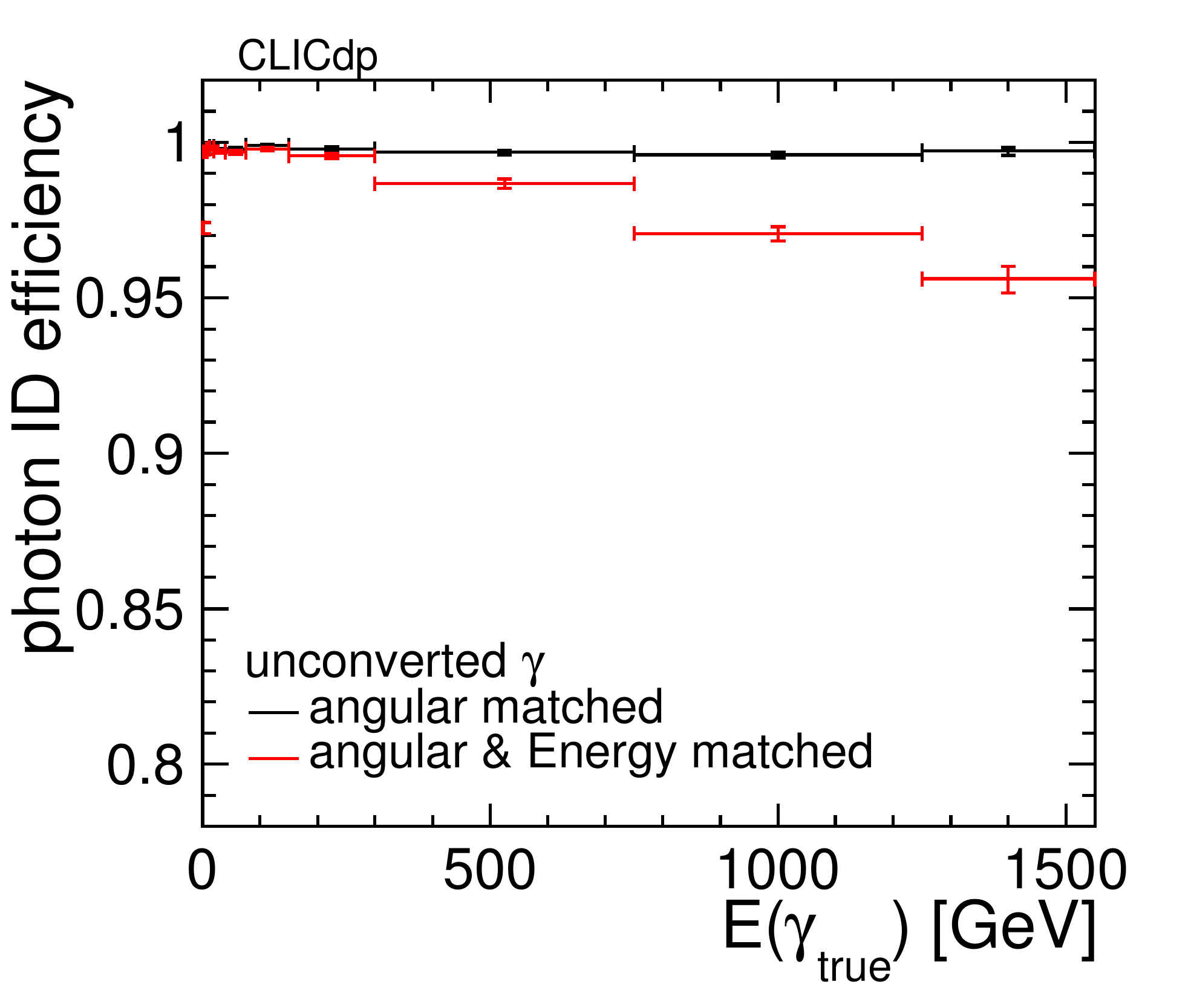}%
    \phantomsubcaption\label{fig:particleGun_phUnconvIDEffVsE}
  \end{subfigure}%
  \begin{subfigure}{.5\textwidth}
    \centering
    \includegraphics[width=\linewidth]{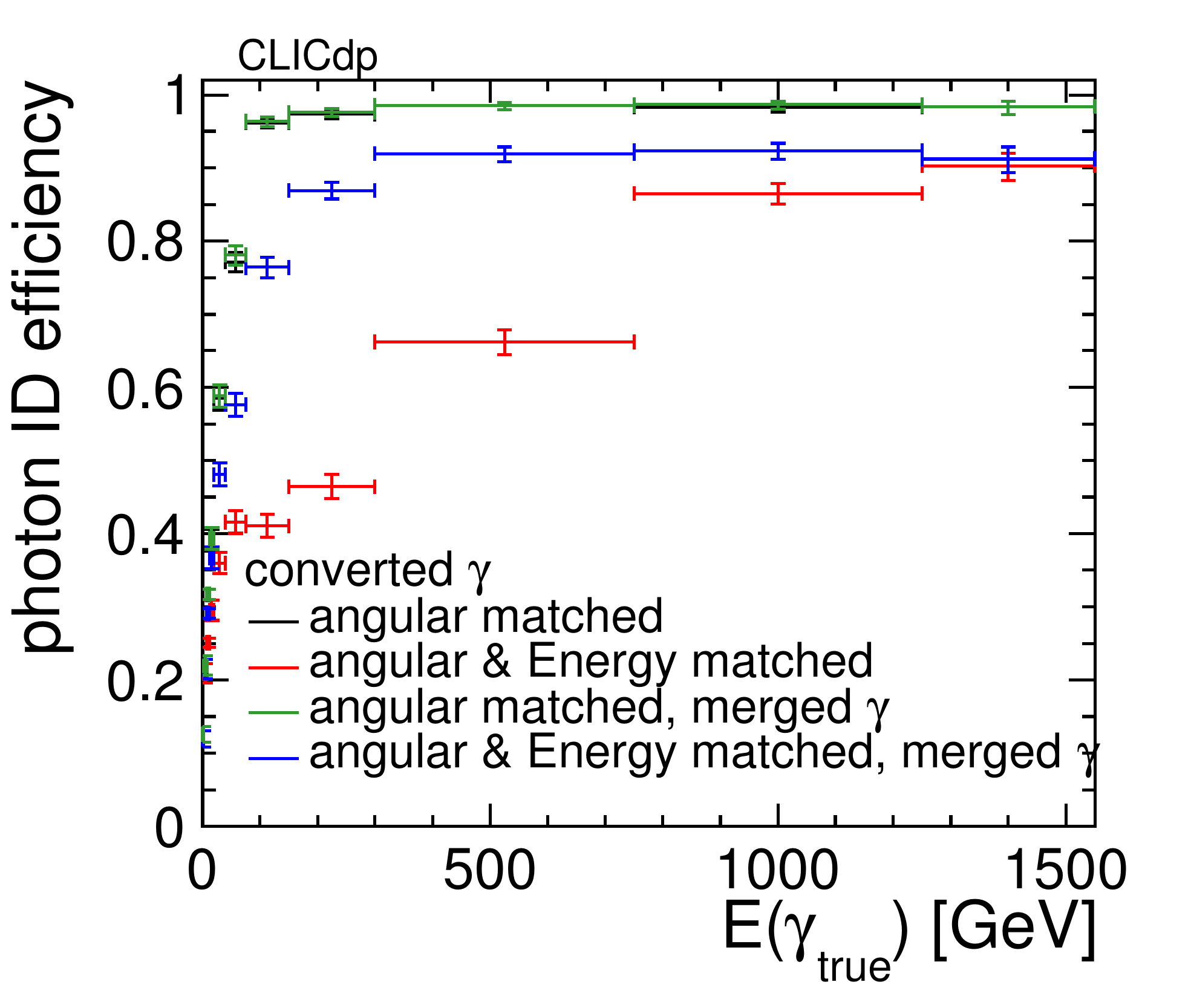}%
    \phantomsubcaption\label{fig:particleGun_phConvIDEffVsTheta}
  \end{subfigure}
  \vspace{-2mm}
  \caption{Particle identification efficiency for unconverted~\subref{fig:particleGun_phUnconvIDEffVsE} and converted photons~\subref{fig:particleGun_phConvIDEffVsTheta} as a function of the energy. In the case of unconverted photons, the efficiency is shown when requiring either angular, or angular and energy, matching. 
In the case of converted photons both criteria are, additionally, applied after merging leading photon candidates (i.e.\ their electromagnetic clusters) into one new photon candidate.}
   \label{fig:photon_id_eff}
\end{figure}

For single photons, the signatures for unconverted and converted photons are considered separately. 
The fraction of converted photons is around 11\% overall, for all energy points. 
This fraction increases from around 7\% at 90\degrees{} to around 20\% for very forward polar angles.
The particle identification efficiency for unconverted and converted photons is shown in \cref{fig:photon_id_eff}.
For unconverted photons the identification efficiency is beyond 99\% for all energies if only angular matching is required. 
An additional requirement on the reconstructed energy (must be within 5~$\sigma_{E}$, based on the two parameter fit on the resolution curves shown in \cref{fig:photonResolutionVsEnergy})
leads to a slight reduction of the efficiency, in particular at the highest photon energies, as shown in \cref{fig:particleGun_phUnconvIDEffVsE}:
At such high photon energies, leakage into HCAL leads to larger tails in the photon energy response distribution. 

In addition the photon cluster might be split into two clusters. In a separate study it has been found that merging of these two clusters  increases the angular and energy matched ID efficiency from 95\% to 97\% for \SI{1.5}{TeV} photons.

For converted photons, requiring only angular matching results in a high efficiency at high energies, but is only about 75\% at \SI{50}{GeV}\@.
Adding the energy matching criterion to the leading photon in the event leads to a strongly reduced efficiency, as expected (see the red curve in \cref{fig:particleGun_phConvIDEffVsTheta}).
 For high energies both electrons from the converted photon are so collimated that a single cluster is reconstructed, 
thus efficiencies beyond 80\% are reached even when requiring energy matching. 
In many conversion events PandoraPFA reconstructs two photons in its default configuration. 
Merging both reconstructed candidate clusters, if they are within a distance of 2\degrees, and applying the identification criteria on the merged candidate,
 significantly improves the efficiency for the angular and energy matched case (see the blue curve in \cref{fig:particleGun_phConvIDEffVsTheta}).

 Around 60\% of all conversions occur before reaching the last 4 layers of the tracker. 
The tracking algorithm requires at least four hits in the tracker. Work has started on a CLIC specific conversion algorithm in Pandora which should improve identification of converted photons particularly at low energies.  

\subsubsection{Performances for Complex Events}

\paragraph{Tracking Efficiency}

The tracking performances for particles in jets have been studied in samples of different energy and event type. In the following, results will be presented for 10\,000 \ttbar{} and \bb{} events at \SI{3}{TeV} centre-of-mass energy and for 10\,000 light flavour \PZgstarToqq events of \SI{380}{GeV} and \SI{500}{GeV} centre-of-mass energy.

Tracking efficiency is defined as the fraction of reconstructable\footnote{The definition of reconstructable particle is the same as given for single particle efficiency in \cref{sec:single_particle}.} Monte Carlo particles which have been reconstructed as \emph{pure} tracks. A track is considered pure if at least 75\% of its hits belong to the same Monte Carlo particle. The fraction of reconstructed tracks that are not pure defines the fake rate.

\begin{figure}[tbp]
  \centering
  \includegraphics[width=0.49\linewidth]{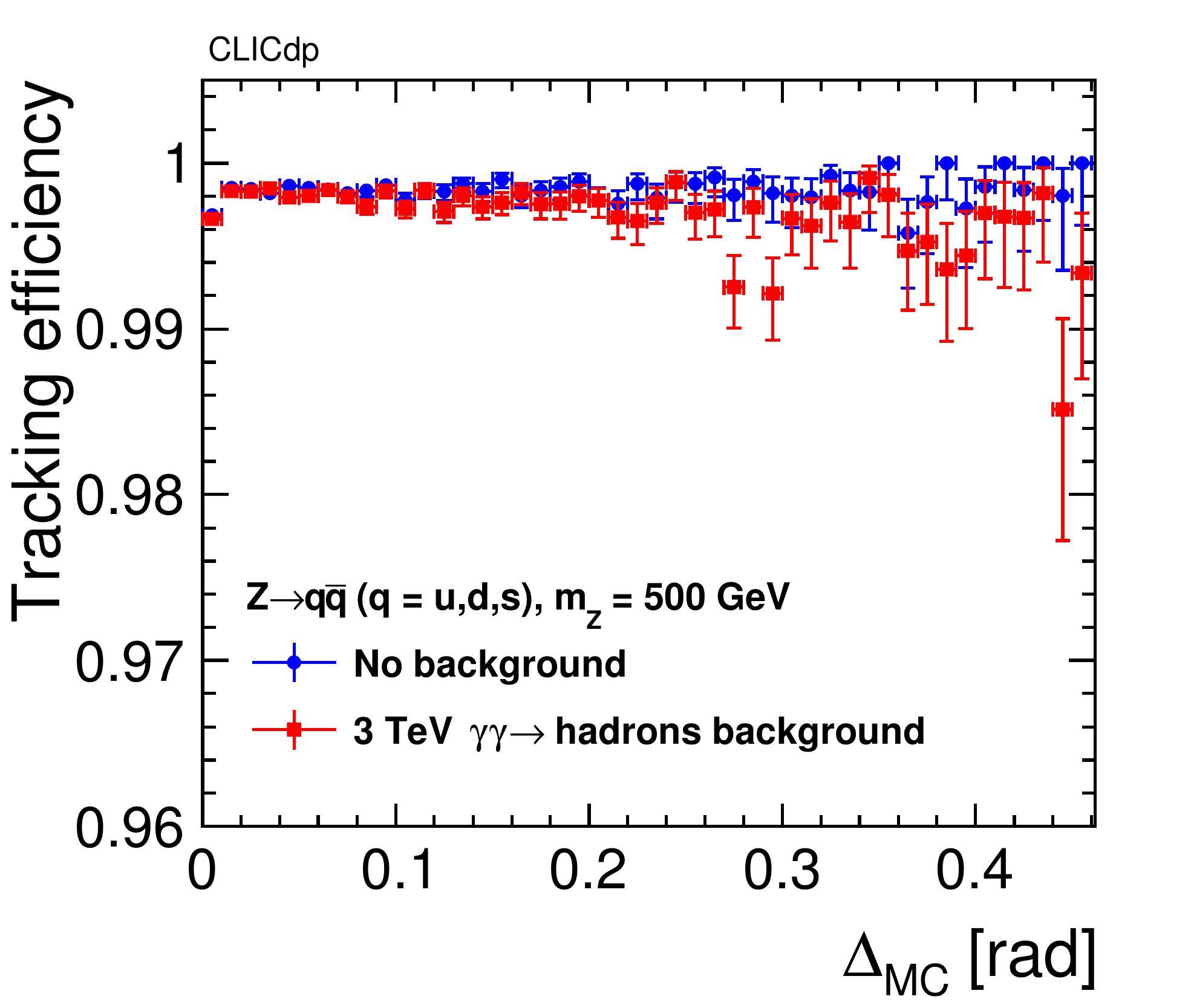}%
  \caption{Tracking efficiency as a function of the particle proximity \deltamc{} for \SI{500}{GeV} light flavour \PZgstarToqq events, with and without \SI{3}{TeV} \gghadron{} background overlay.}\label{fig:Zuds500GeV_eff_dist}
\end{figure}

In jet events, the vicinity of other particles may affect the performance of the pattern recognition in assigning the right hits to the proper track. Therefore, the tracking efficiency in \PZgstar events decaying to light quarks at \SI{500}{GeV} has been monitored as a function of the particle proximity \deltamc. This is defined as the smallest distance between the Monte Carlo particle associated to the track and any other Monte Carlo particle, $\deltamc = \sqrt{\smash[b]{{(\Delta\eta)}^{2}+{(\Delta\phi)}^{2}}}$, where $\eta$ is the pseudorapidity. The efficiency is shown in \cref{fig:Zuds500GeV_eff_dist}, in which the following cuts are applied: $10\degrees{} < \theta < 170\degrees{}$, $\pT > \SI{1}{GeV}$ and production radius smaller than \SI{50}{mm}.

The efficiency has been estimated for events with and without overlay of 30 \ac{BX} of \gghadron{} background for the \SI{3}{TeV} CLIC machine, with negligible differences observed between the two cases. A cut on the particle proximity of $\deltamc > \SI{0.02}{rad}$ has been applied in all following tracking efficiency results.

\begin{figure}[tbp]
  \renewcommand{\thesubfigure}{(\lr{subfigure})}
  \centering
  \begin{subfigure}{.5\textwidth}
    \centering
    \includegraphics[width=\linewidth]{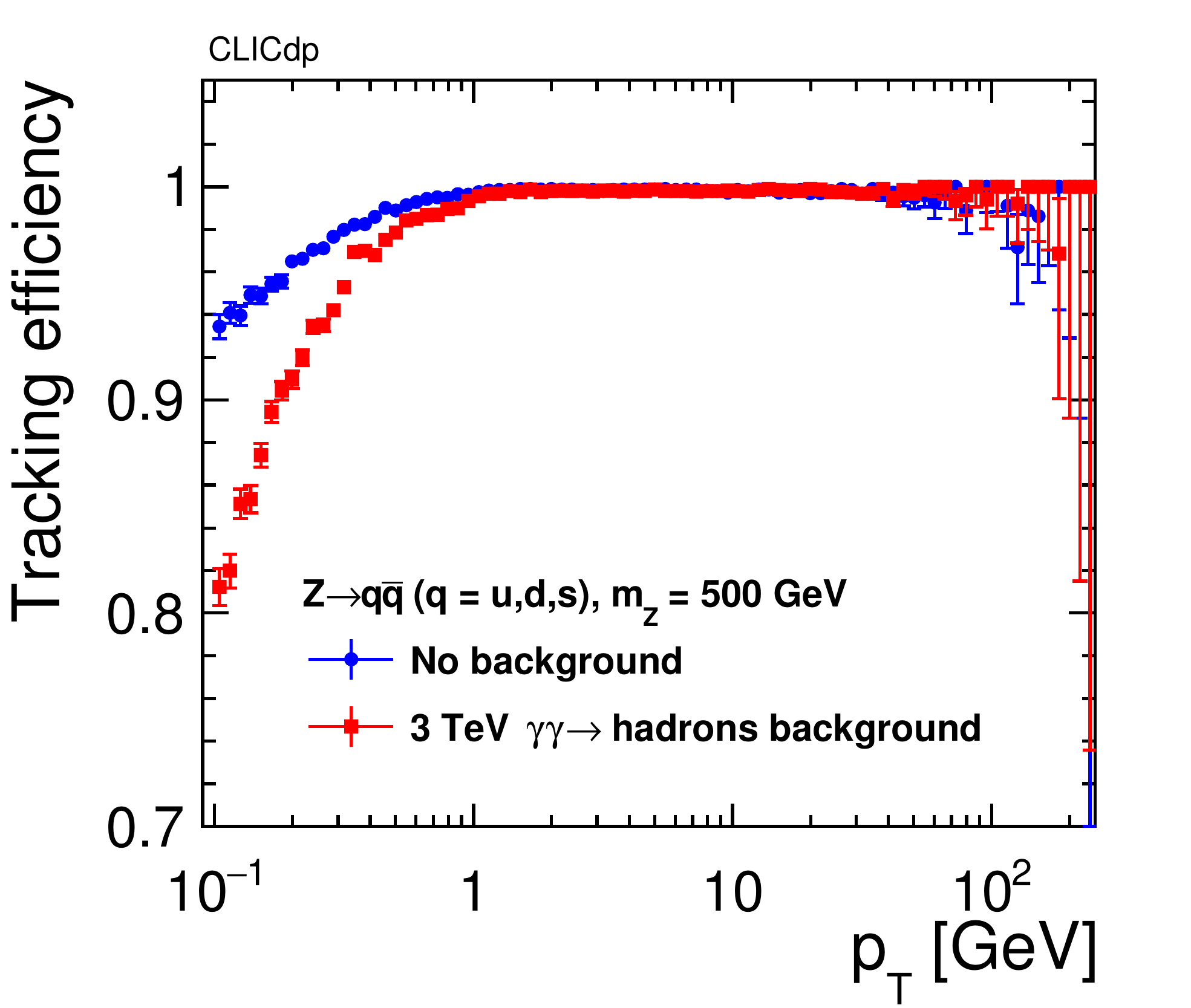}%
    \phantomsubcaption\label{fig:Zuds500GeV_eff_pt}
  \end{subfigure}%
  \begin{subfigure}{.5\textwidth}
    \centering
    \includegraphics[width=\linewidth]{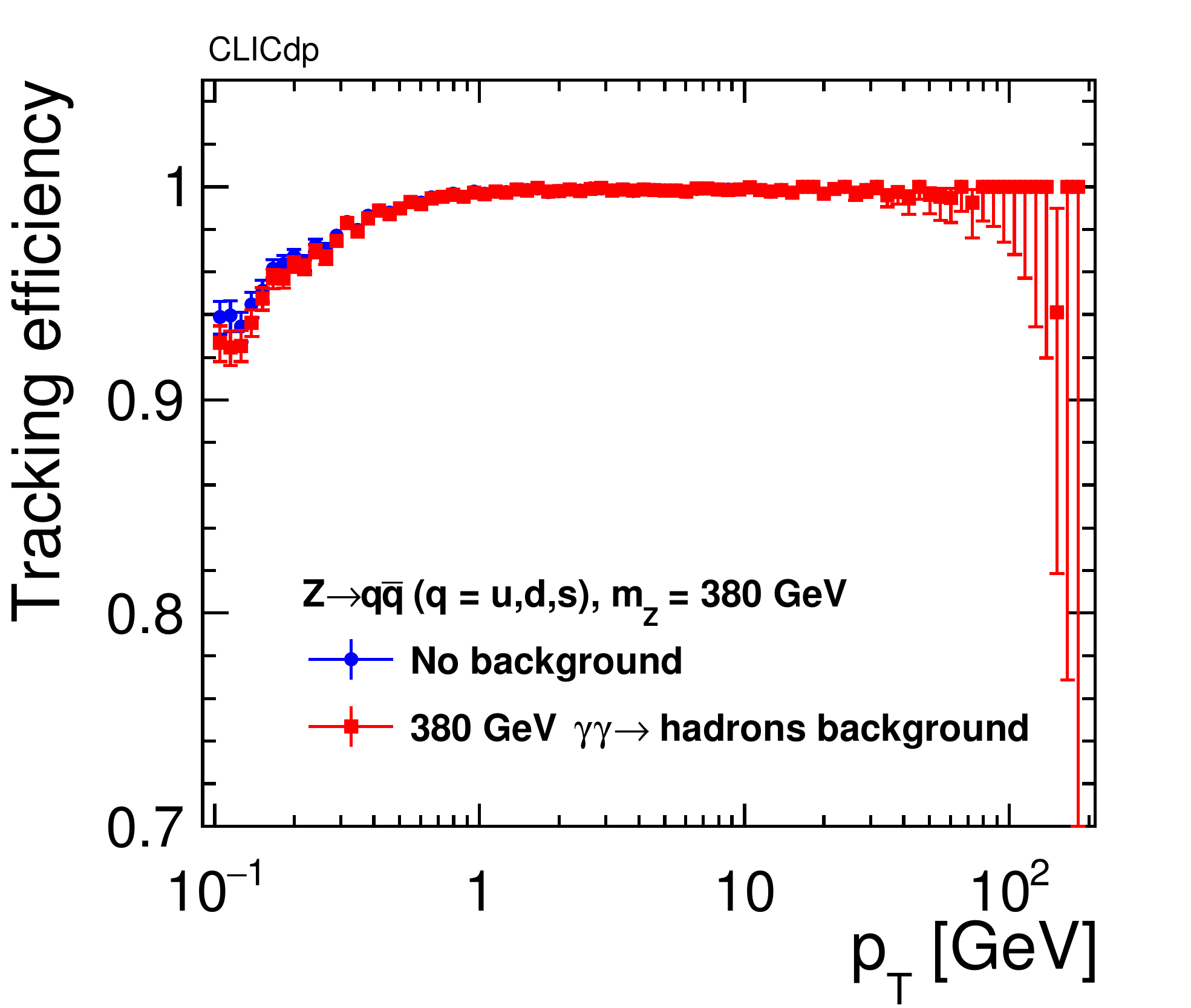}%
    \phantomsubcaption\label{fig:Zuds380GeV_eff_pt}
  \end{subfigure}
  \vspace{-2mm}
  \caption{Tracking efficiency in light flavour \PZgstarToqq events as a function of \pT{} at \SI{500}{GeV}, with and without \SI{3}{TeV} \gghadron{} background overlay~\subref{fig:Zuds500GeV_eff_pt}, and at \SI{380}{GeV}, with and without \SI{380}{GeV} \gghadron{} background overlay~\subref{fig:Zuds380GeV_eff_pt}.}\label{fig:Zuds500_vs_380}
\end{figure}

\cref{fig:Zuds500GeV_eff_pt} shows the tracking efficiency in \SI{500}{GeV} light flavour \PZgstarToqq events, with and without overlay of \SI{30}{BX} of \gghadron{} background for the \SI{3}{TeV} CLIC machine, as a function of the transverse momentum. The following cuts are applied for each particle in this plot: $10\degrees{} < \theta < 170\degrees{}$, $\deltamc > \SI{0.02}{rad}$, and a production radius smaller than \SI{50}{mm}.
The effect of the \gghadron{} background is visible mostly in the low-$\pT$ region, since background particles have mostly $\pT < \SI{5}{GeV}$, where it results at most in a 10\% efficiency loss at \SI{100}{MeV} with respect to the case without background. In the region at intermediate and high $\pT$,
the impact from background is negligible,
and the efficiency reaches 100\%.
For comparison, \SI{30}{BX} of \gghadron{} background for the \SI{380}{GeV} CLIC machine have been overlaid to a \SI{380}{GeV} light flavour \PZgstarToqq sample. The resulting efficiency as a function of the transverse momentum is shown in \cref{fig:Zuds380GeV_eff_pt}. The effect of \SI{380}{GeV} \gghadron{} background is negligible. Therefore, performances are shown hereafter only for the overlay of \SI{3}{TeV} \gghadron{} background.

\begin{figure}[tbp]
  \renewcommand{\thesubfigure}{(\lr{subfigure})}
  \centering
  \begin{subfigure}{.5\textwidth}
    \centering
    \includegraphics[width=\linewidth]{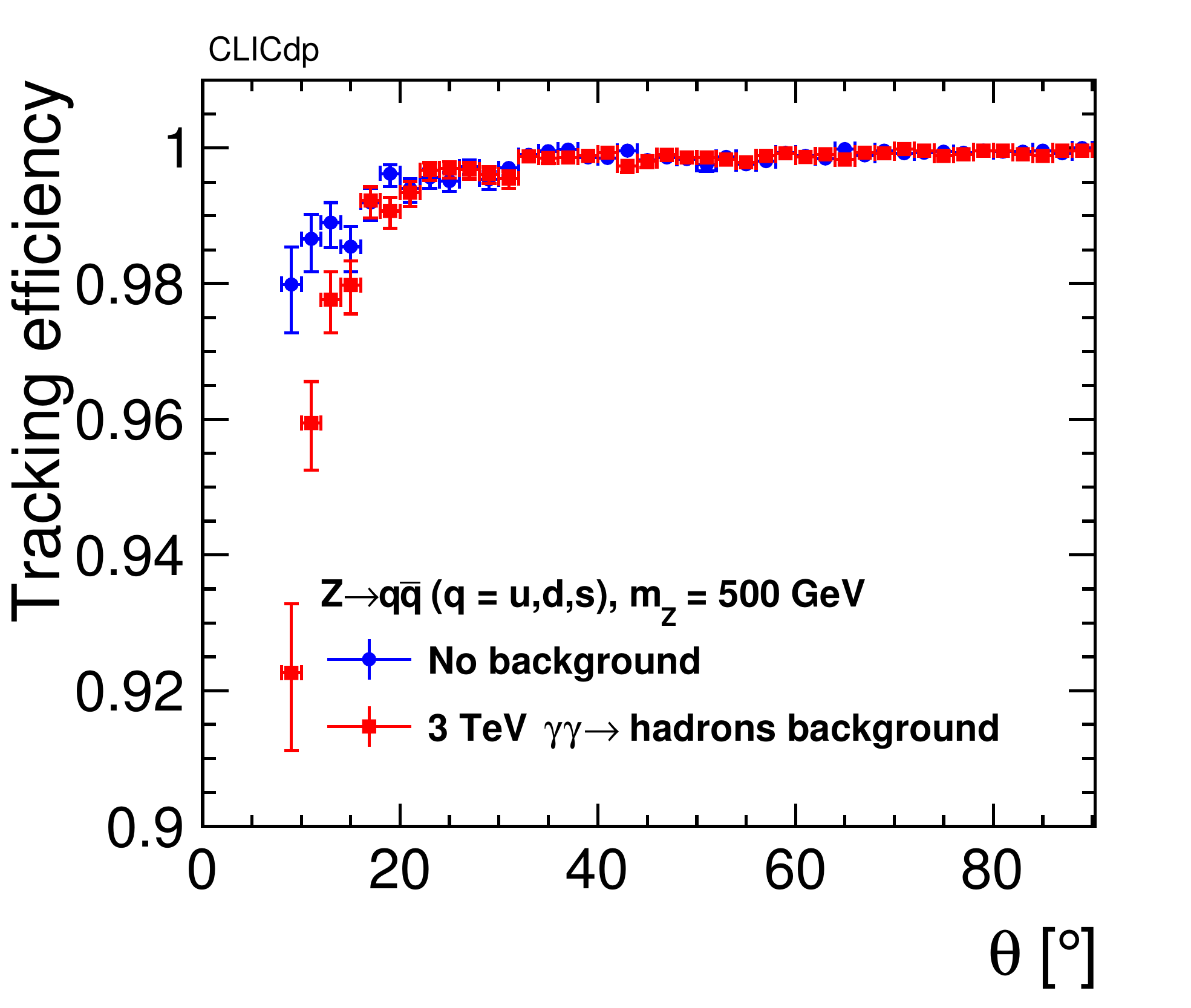}%
    \phantomsubcaption\label{fig:Zuds500GeV_eff_theta}
  \end{subfigure}%
  \begin{subfigure}{.5\textwidth}
    \centering
    \includegraphics[width=\linewidth]{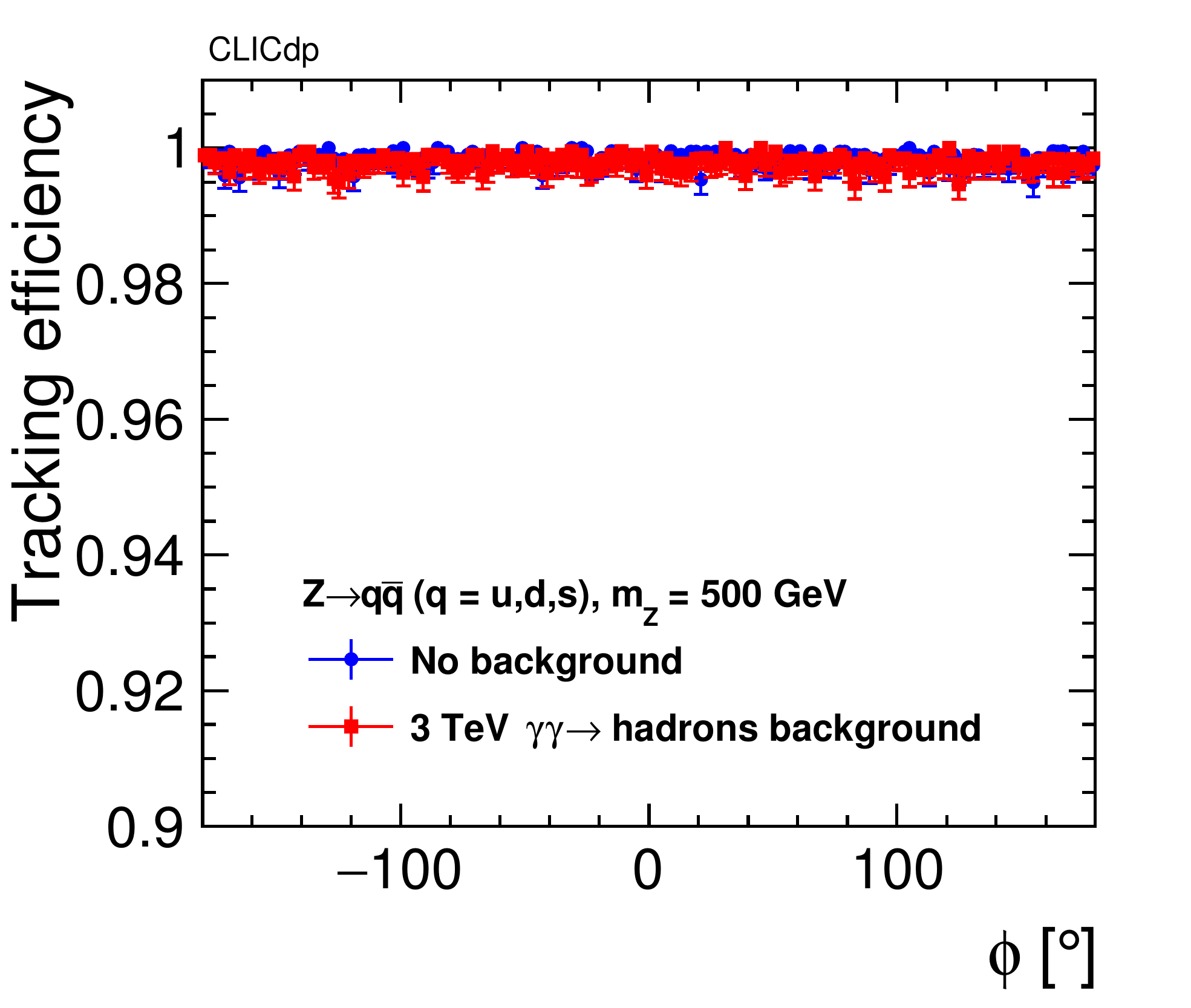}%
    \phantomsubcaption\label{fig:Zuds500GeV_eff_phi}
  \end{subfigure}
  \vspace{-2mm}
  \caption{Tracking efficiency as a function of the polar~\subref{fig:Zuds500GeV_eff_theta} and the
    azimuthal~\subref{fig:Zuds500GeV_eff_phi} angle for \SI{500}{GeV} light flavour \PZgstarToqq events, with and without \SI{3}{TeV} \gghadron{} background overlay.}\label{fig:Zuds500GeV_eff_angle}
\end{figure}

For \SI{500}{GeV} light flavour \PZgstarToqq events, the tracking efficiency is shown in \cref{fig:Zuds500GeV_eff_angle} as a function of the polar (left) and the azimuthal angle (right). The following cuts are applied in these plots: $\pT > \SI{1}{GeV}$, $\deltamc > \SI{0.02}{rad}$, and a production radius smaller than \SI{50}{mm}, with an additional cut in polar angle ($10\degrees < \theta < 170\degrees$) for \cref{fig:Zuds500GeV_eff_phi}.
Fully efficient performances are observed at all $\phi$ angles and in the whole $\theta$ range, with the exception of the region $10\degrees < \theta < 20\degrees$, where a maximum efficiency loss of 8\% occurs with and a loss of 2\% occurs without background overlay. 

\begin{figure}[tbp]
  \centering
  \includegraphics[width=0.523\linewidth]{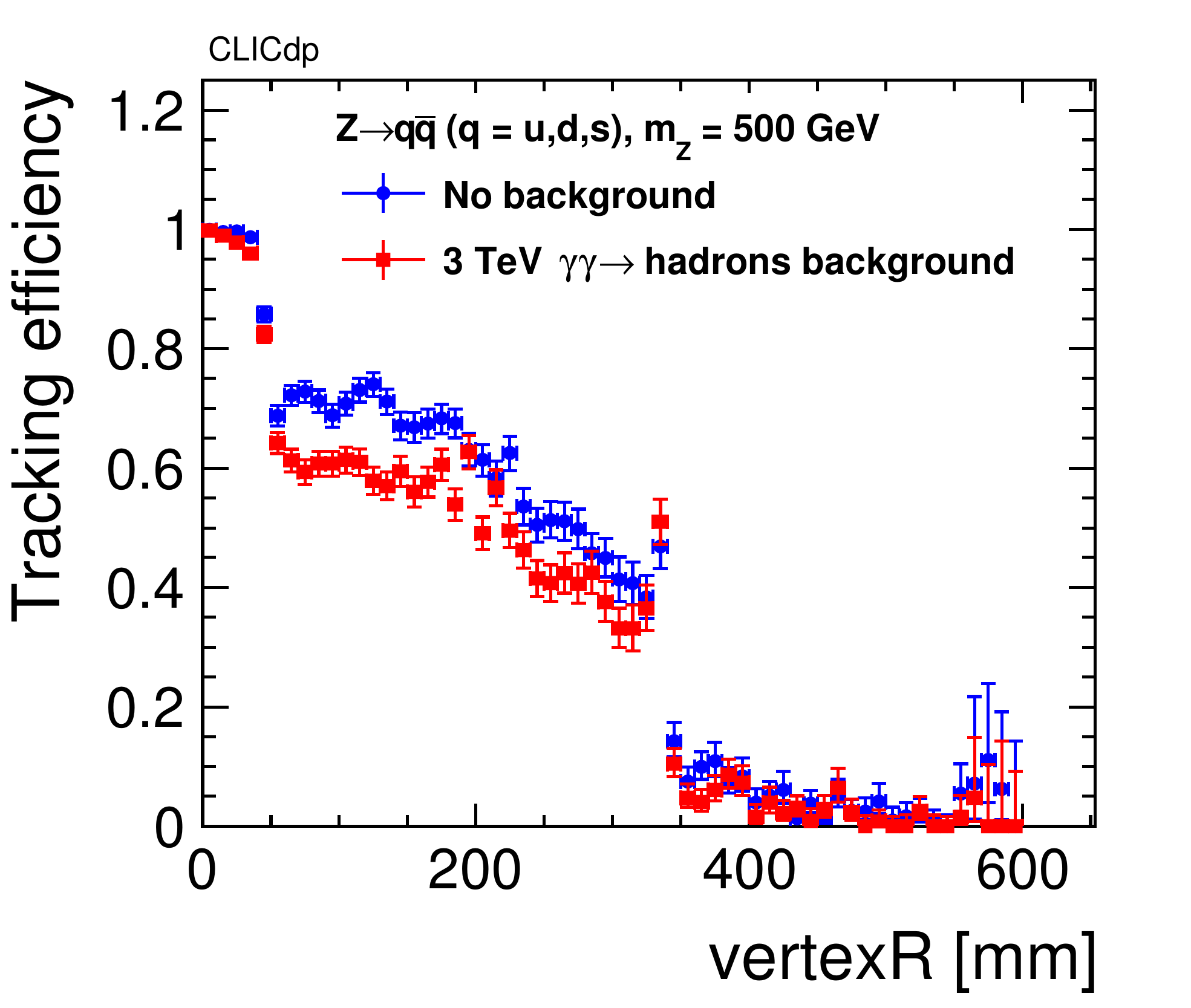}
\caption{Tracking efficiency as a function of the production vertex radius for \SI{500}{GeV} light flavour \PZgstarToqq events, with and without \SI{3}{TeV} \gghadron{} background overlay.}
\label{fig:Zuds500GeV_eff_vertexR}
\end{figure}

Finally, for the same events, the tracking efficiency as a function of the production vertex radius is shown in \cref{fig:Zuds500GeV_eff_vertexR}. In this plot, the following cuts are applied: $\pT > \SI{1}{GeV}$, $10\degrees{} < \theta < 170\degrees{}$ and $\deltamc > \SI{0.02}{rad}$. The trend reflects the same behaviour observed for single displaced low-momentum muons in \cref{fig:displaced_muons}, since the low-energy component of the particle spectrum dominates. Selecting $\pT > \SI{10}{GeV}$ allows one to remove the efficiency drop at \mbox{$R = \SI{50}{mm}$}.  The effect of background amounts to at most a 1\% efficiency loss for particles produced outside the vertex detector with $R \ge \SI{60}{mm}$.

\begin{figure}[tbp]
  \renewcommand{\thesubfigure}{(\lr{subfigure})}
  \begin{subfigure}{.5\textwidth}
  \centering
    \centering
    \includegraphics[width=\linewidth]{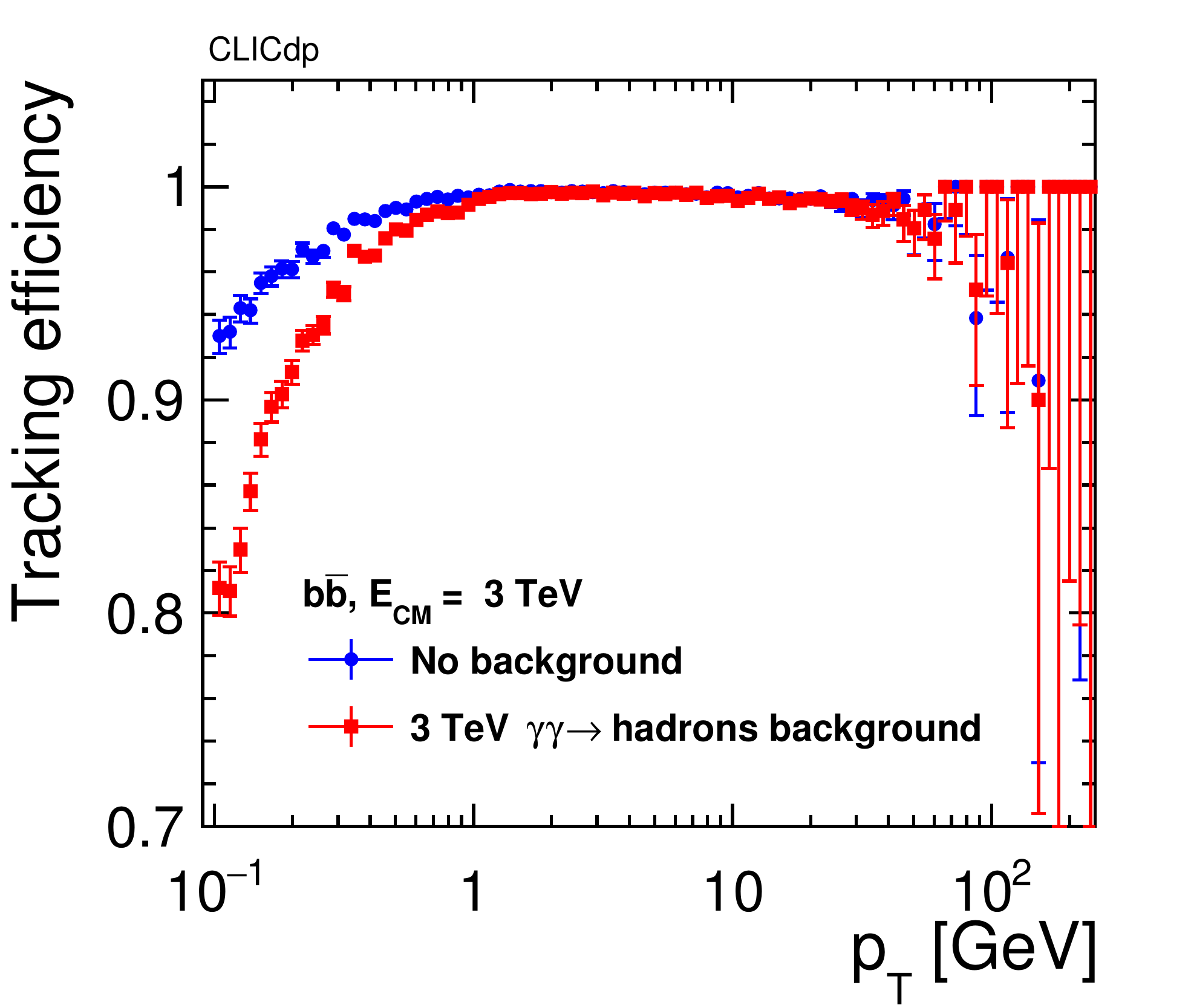}%
    \phantomsubcaption\label{fig:bbbar3TeV_eff_pt}
  \end{subfigure}%
  \begin{subfigure}{.5\textwidth}
    \centering
    \includegraphics[width=\linewidth]{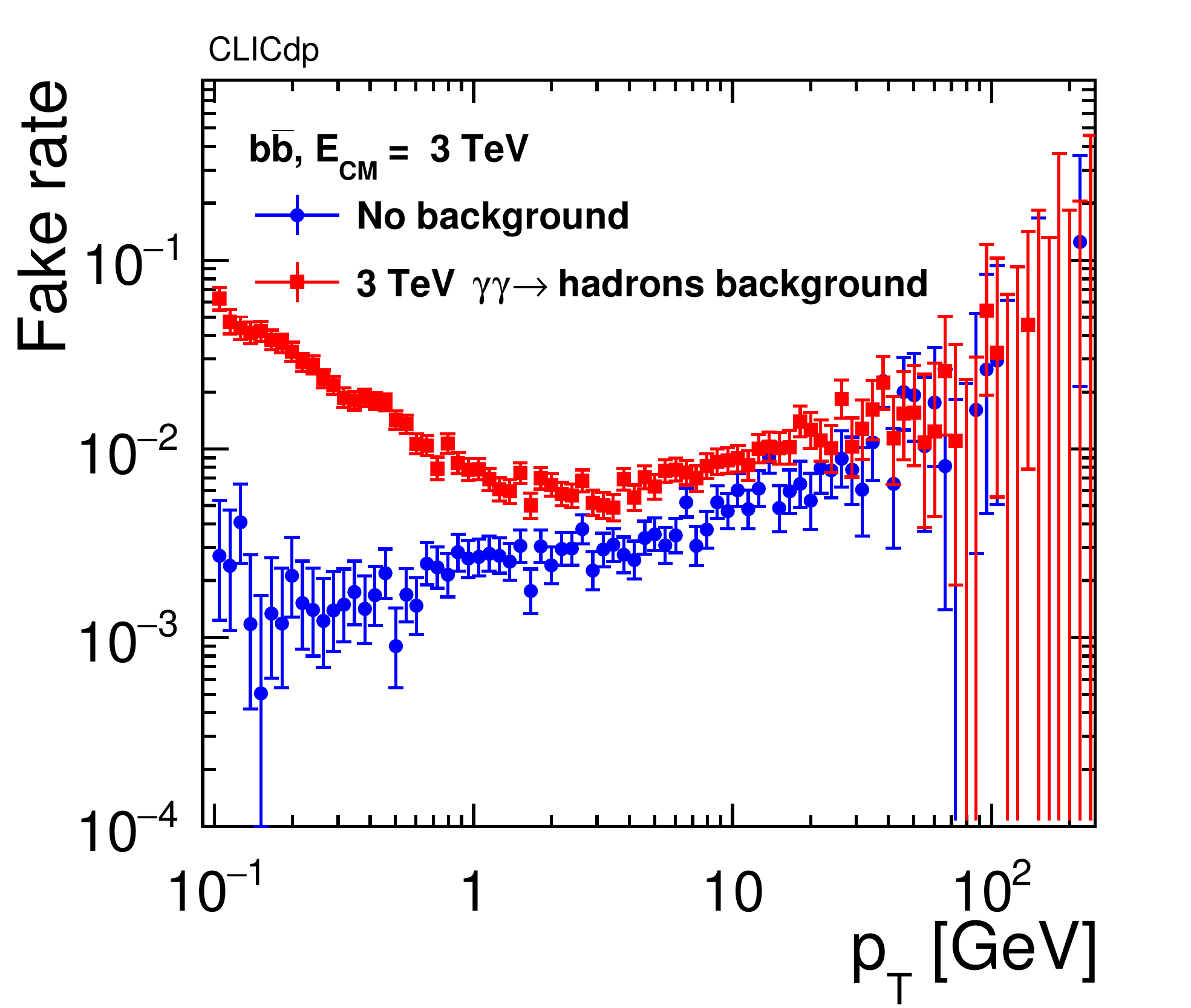}%
    \phantomsubcaption\label{fig:bbbar3TeV_fake_pt}
  \end{subfigure}
  \vspace{-2mm}
  \caption{Tracking efficiency~\subref{fig:bbbar3TeV_eff_pt} and fake rate~\subref{fig:bbbar3TeV_fake_pt} as a function of \pT{} for \bb{} events at \SI{3}{TeV}, with and without \SI{3}{TeV} \gghadron{} background overlay.}
  \label{fig:bb_pt}
\end{figure}

\begin{figure}[tbp]
  \renewcommand{\thesubfigure}{(\lr{subfigure})}
  \centering
  \begin{subfigure}{.5\textwidth}
    \centering
    \includegraphics[width=\linewidth]{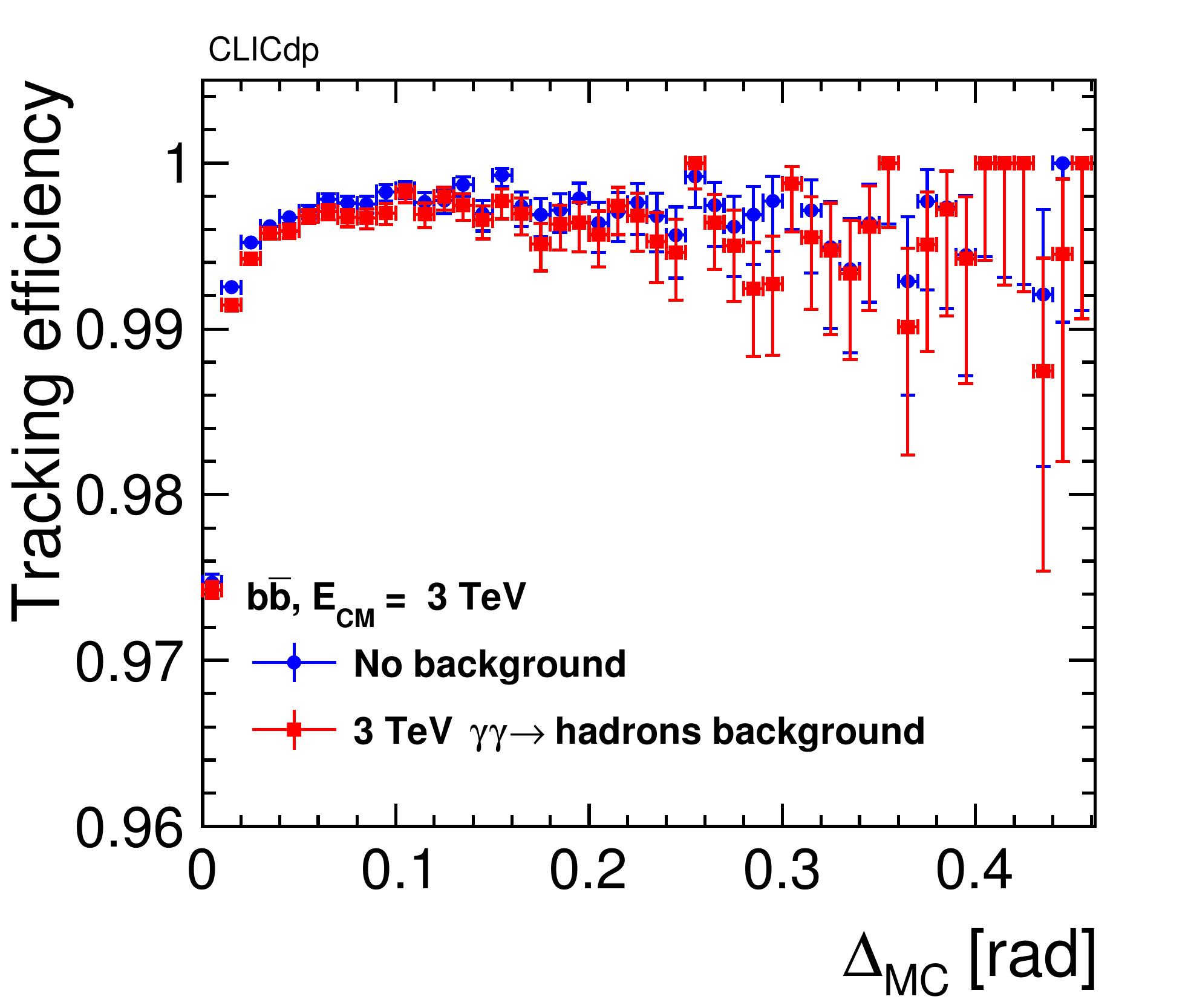}%
    \phantomsubcaption\label{fig:bbbar3TeV_eff_dist}
  \end{subfigure}%
  \begin{subfigure}{.5\textwidth}
    \centering
    \includegraphics[width=\linewidth]{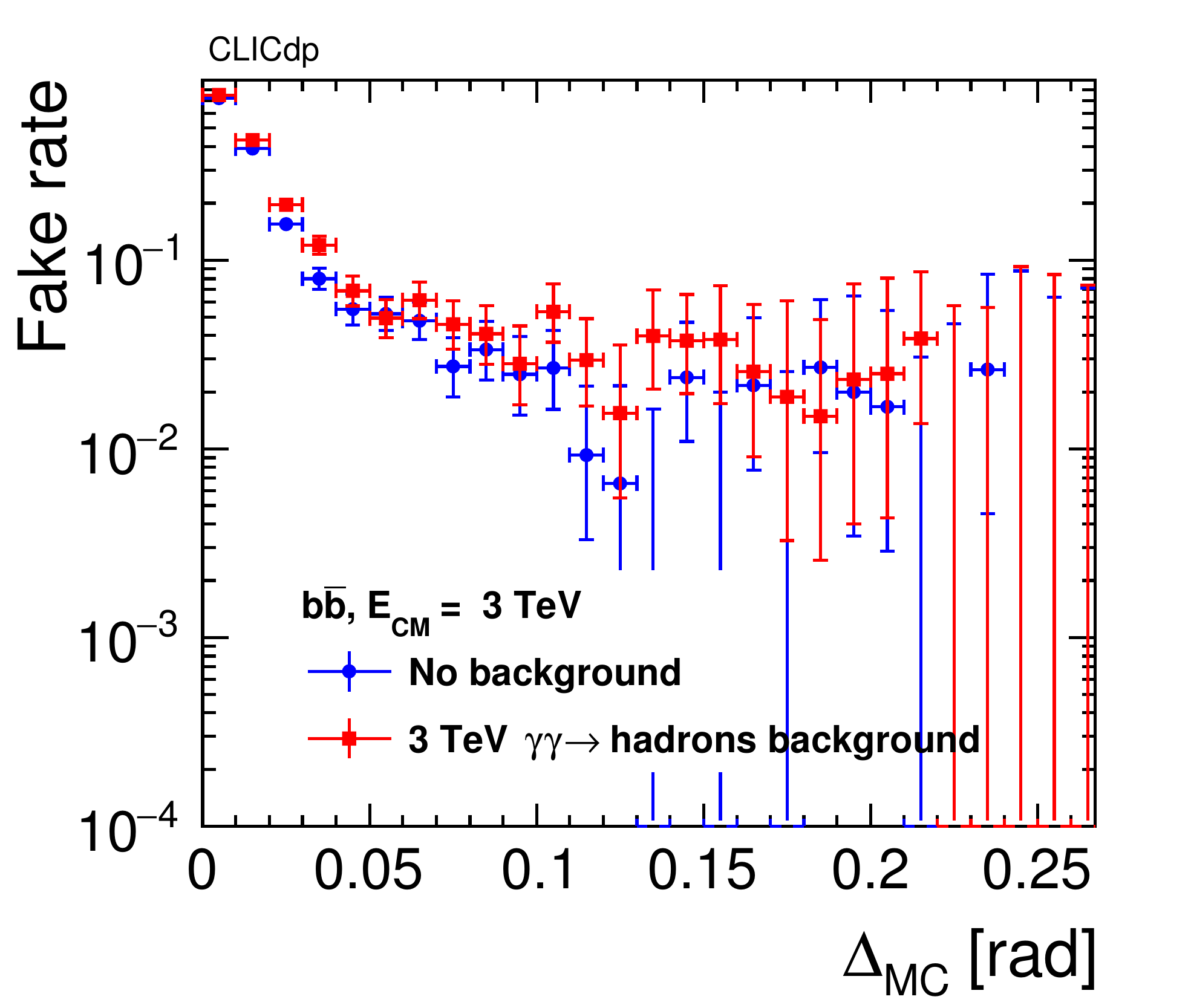}%
    \phantomsubcaption\label{fig:bbbar3TeV_fake_dist}
  \end{subfigure}
  \vspace{-2mm}
  \caption{Tracking efficiency~\subref{fig:bbbar3TeV_eff_dist} and fake rate~\subref{fig:bbbar3TeV_fake_dist} as a
    function of the particle proximity \deltamc{} for \bb{} events at \SI{3}{TeV}, with and without
    \SI{3}{TeV} \gghadron{} background overlay.}\label{fig:bb_dist}
\end{figure}

\begin{figure}[tbp]
  \renewcommand{\thesubfigure}{(\lr{subfigure})}
  \centering
  \begin{subfigure}{.5\textwidth}
    \centering
    \includegraphics[width=\linewidth]{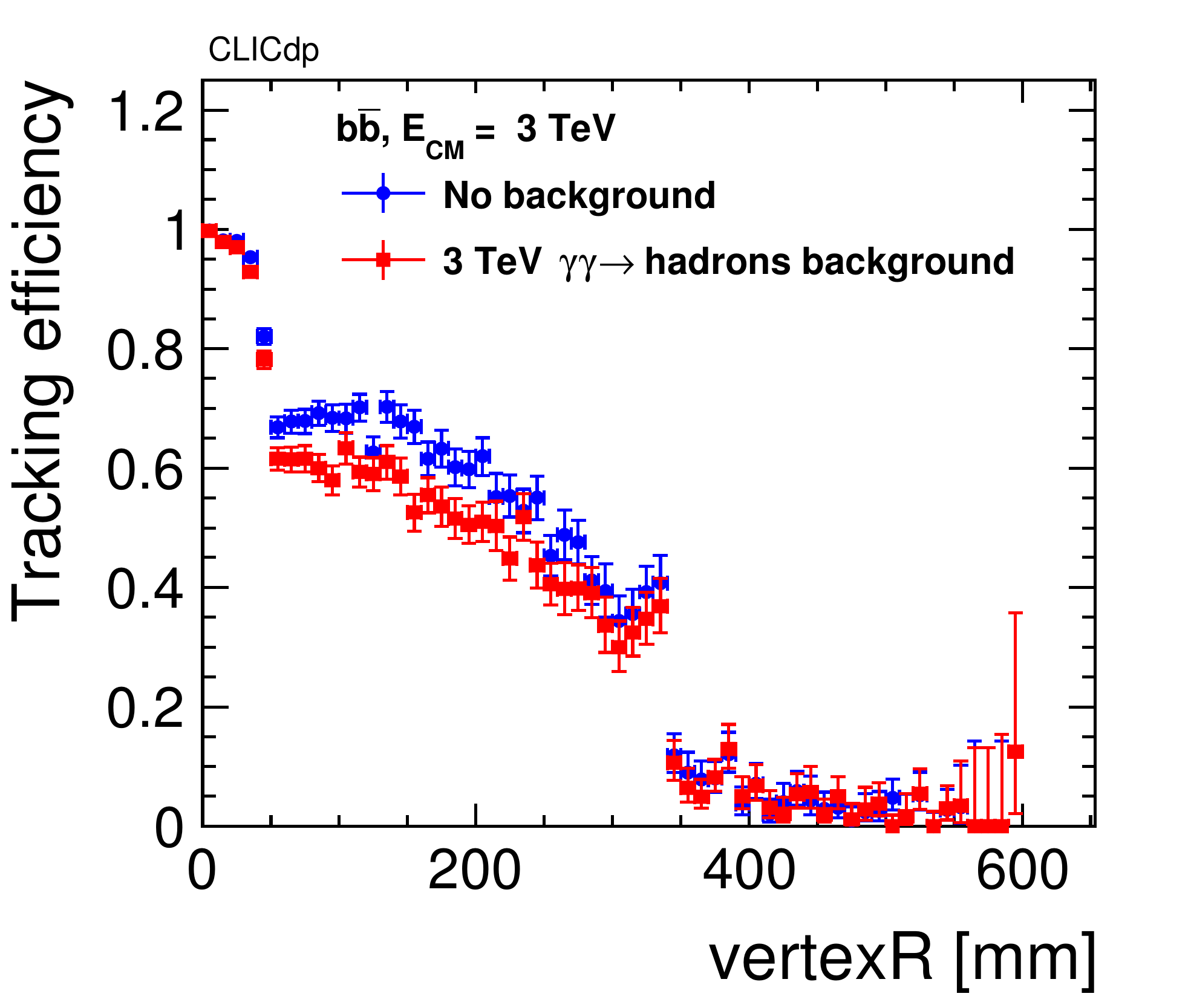}%
    \phantomsubcaption\label{fig:bbbar3TeV_eff_vertexR}
  \end{subfigure}%
  \begin{subfigure}{.5\textwidth}
    \centering
    \includegraphics[width=\linewidth]{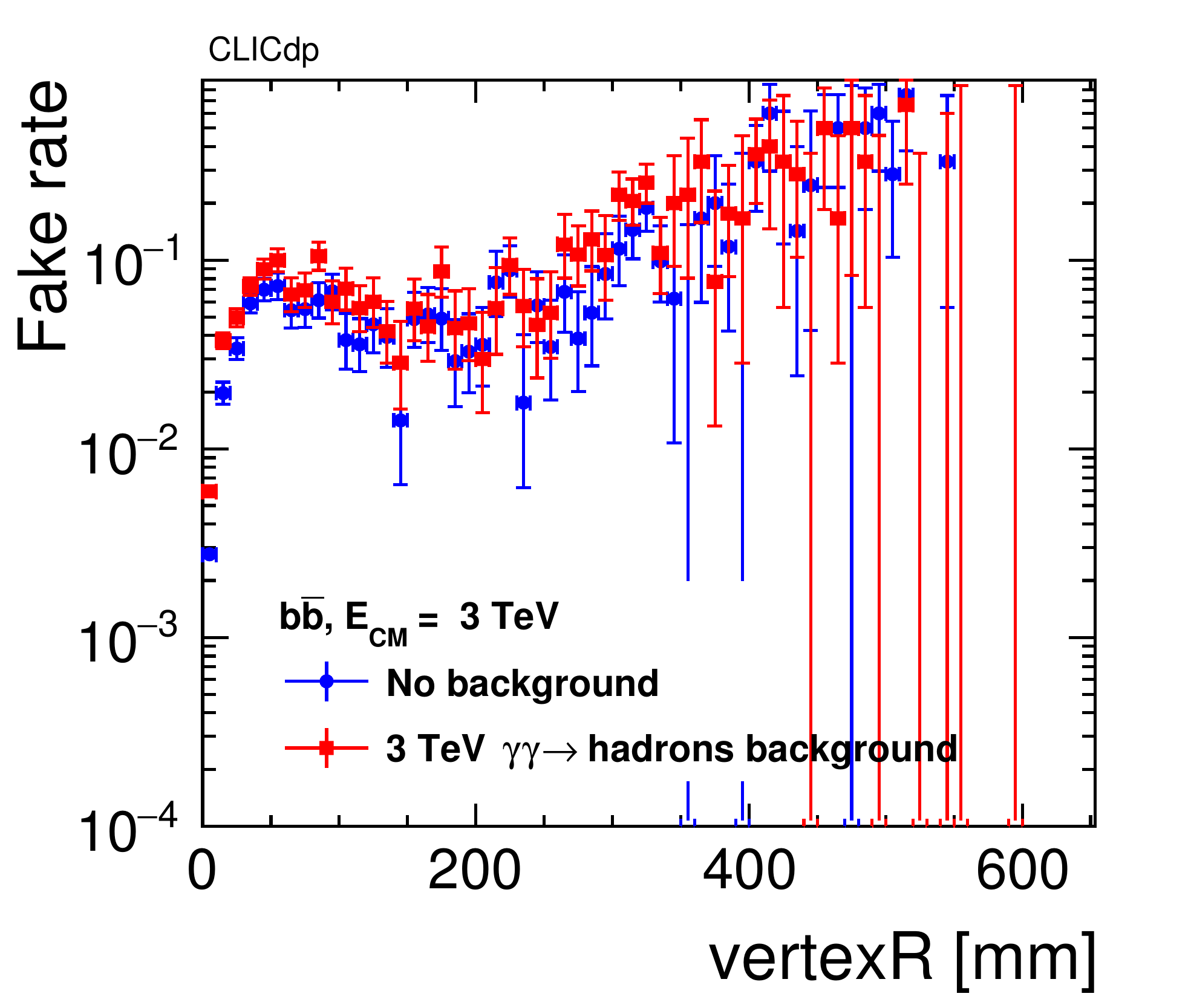}%
    \phantomsubcaption\label{fig:bbbar3TeV_fake_vertexR}
  \end{subfigure}
  \vspace{-2mm}
  \caption{Tracking efficiency~\subref{fig:bbbar3TeV_eff_vertexR} and fake rate~\subref{fig:bbbar3TeV_fake_vertexR} as a
    function of the production vertex radius for \bb{} events at \SI{3}{TeV}, with and without \SI{3}{TeV} \gghadron{}
    background overlay.}\label{fig:bb_vertexR}
\end{figure}

More complex events allow one to complete the performance study with the evaluation of fake rate. 
\crefrange{fig:bb_pt}{fig:bb_vertexR} show the efficiency (left) and fake rate (right) for \bb{} events at \SI{3}{TeV} as a function of the transverse momentum, particle proximity and production vertex radius. The corresponding cuts used for each plot are the same as for Z-boson events. Results are presented for the cases with and without \SI{3}{TeV} \gghadron{} background overlay.
The trend of tracking efficiency as a function of all observables does not show a deviation from that of \SI{500}{GeV} light flavour \PZgstarToqq events, proving that the performance is overall independent of the physics process and centre-of-mass energy.
The fake rate increases as a function of the transverse momentum (\cref{fig:bbbar3TeV_fake_pt}) and with decreasing particle proximity (\cref{fig:bbbar3TeV_fake_dist}), due to the increased confusion in distinguishing, respectively, too straight tracks and tracks in more dense environment. Moreover, the fake rate increases with the production vertex radius (\cref{fig:bbbar3TeV_fake_vertexR}), suffering from the lower number of measurements (traversed layers) available. The effect of the \gghadron{} background is particularly large for \mbox{$\pT < \SI{1}{GeV}$}, in which case the fake rate increases by roughly one order of magnitude down to \SI{100}{MeV}, where it reaches a maximum of 6\%.

\begin{figure}[tbp]
  \renewcommand{\thesubfigure}{(\lr{subfigure})}
  \centering
  \begin{subfigure}{.5\textwidth}
    \centering
    \includegraphics[width=\linewidth]{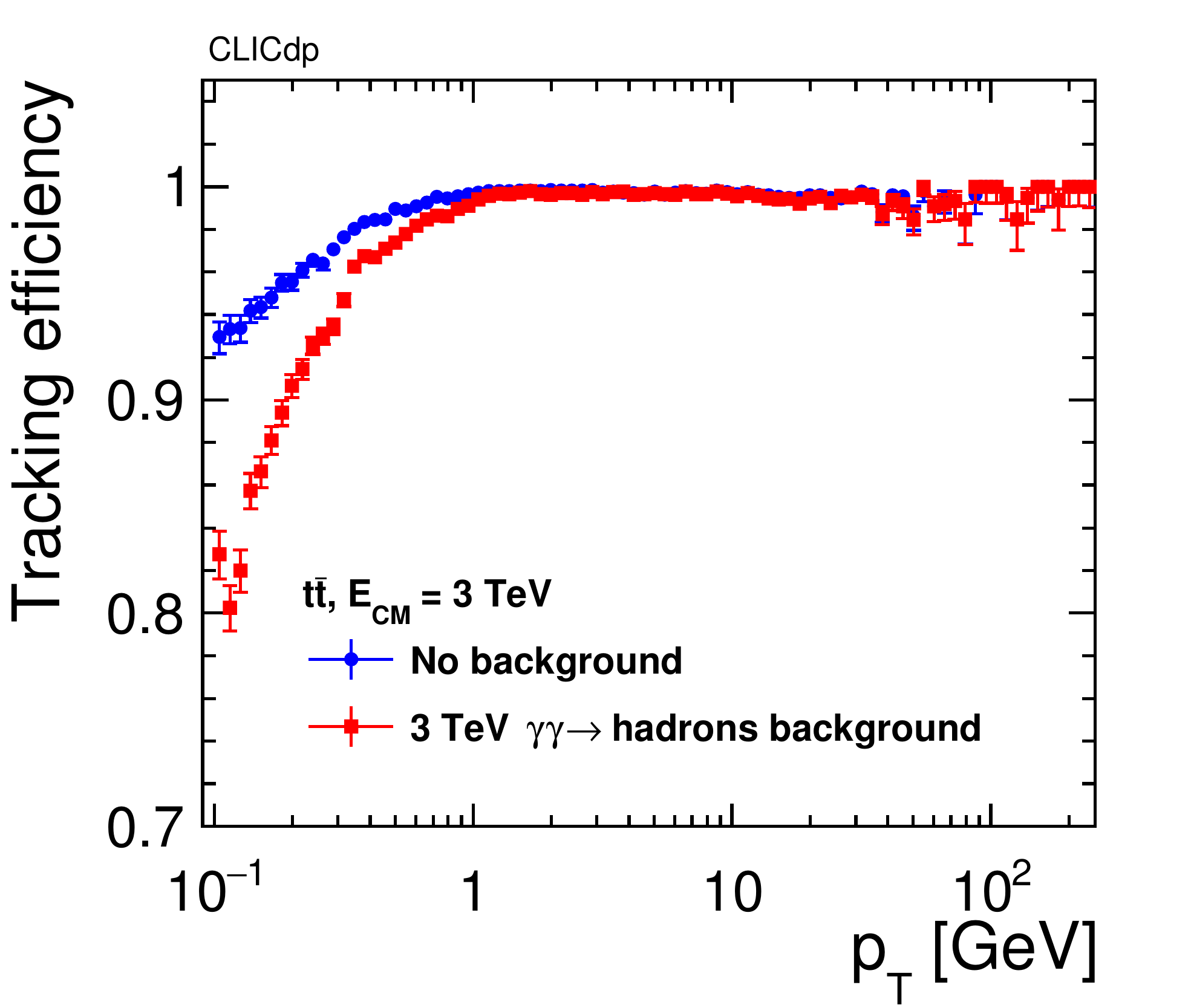}%
    \phantomsubcaption\label{fig:ttbar3TeV_eff_pt}
  \end{subfigure}%
  \begin{subfigure}{.5\textwidth}
    \centering
    \includegraphics[width=\linewidth]{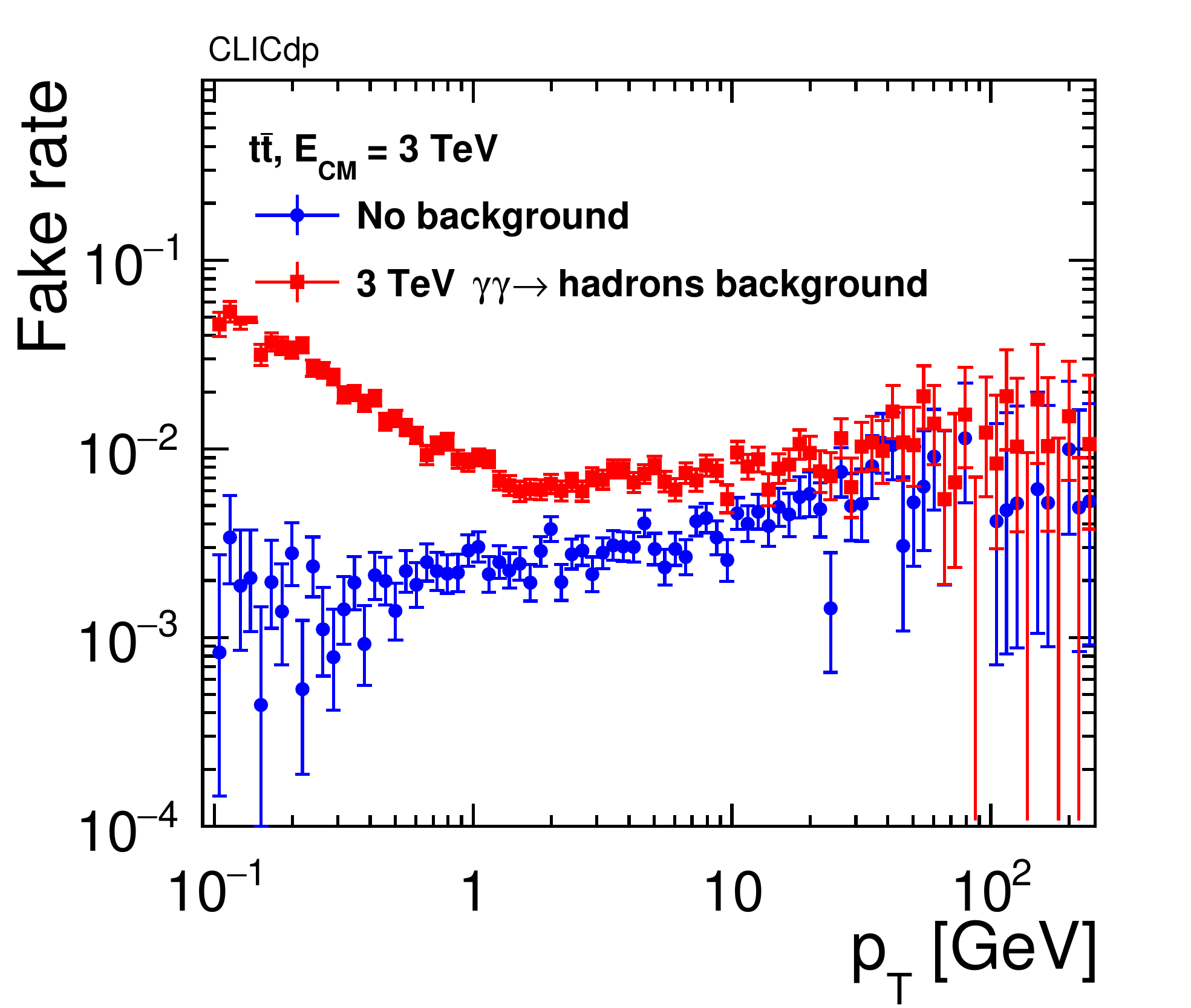}%
    \phantomsubcaption\label{fig:ttbar3TeV_fakes_pt}
  \end{subfigure}
  \vspace{-3mm}
  \caption{Tracking efficiency~\subref{fig:ttbar3TeV_eff_pt} and fake rate~\subref{fig:ttbar3TeV_fakes_pt} as a function
    of \pT{} for \ttbar{} events at \SI{3}{TeV}, with and without \SI{3}{TeV} \gghadron{} background overlay.}\label{fig:tt_pt}
\end{figure}

\begin{figure}[tbp]
  \renewcommand{\thesubfigure}{(\lr{subfigure})}
  \centering
  \begin{subfigure}{.5\textwidth}
    \centering
    \includegraphics[width=\linewidth]{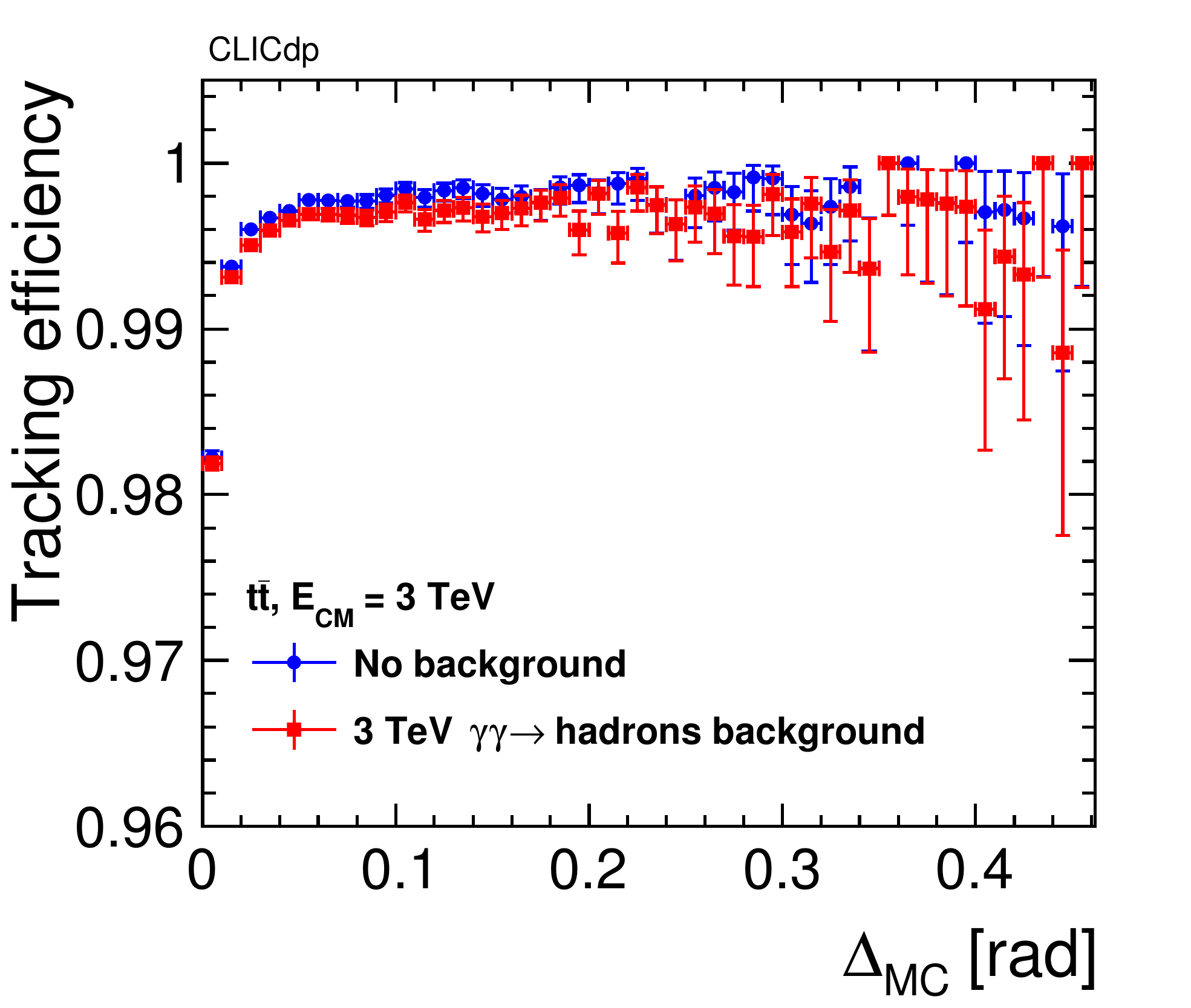}%
    \phantomsubcaption\label{fig:ttbar3TeV_eff_dist}
  \end{subfigure}%
  \begin{subfigure}{.5\textwidth}
    \centering
    \includegraphics[width=\linewidth]{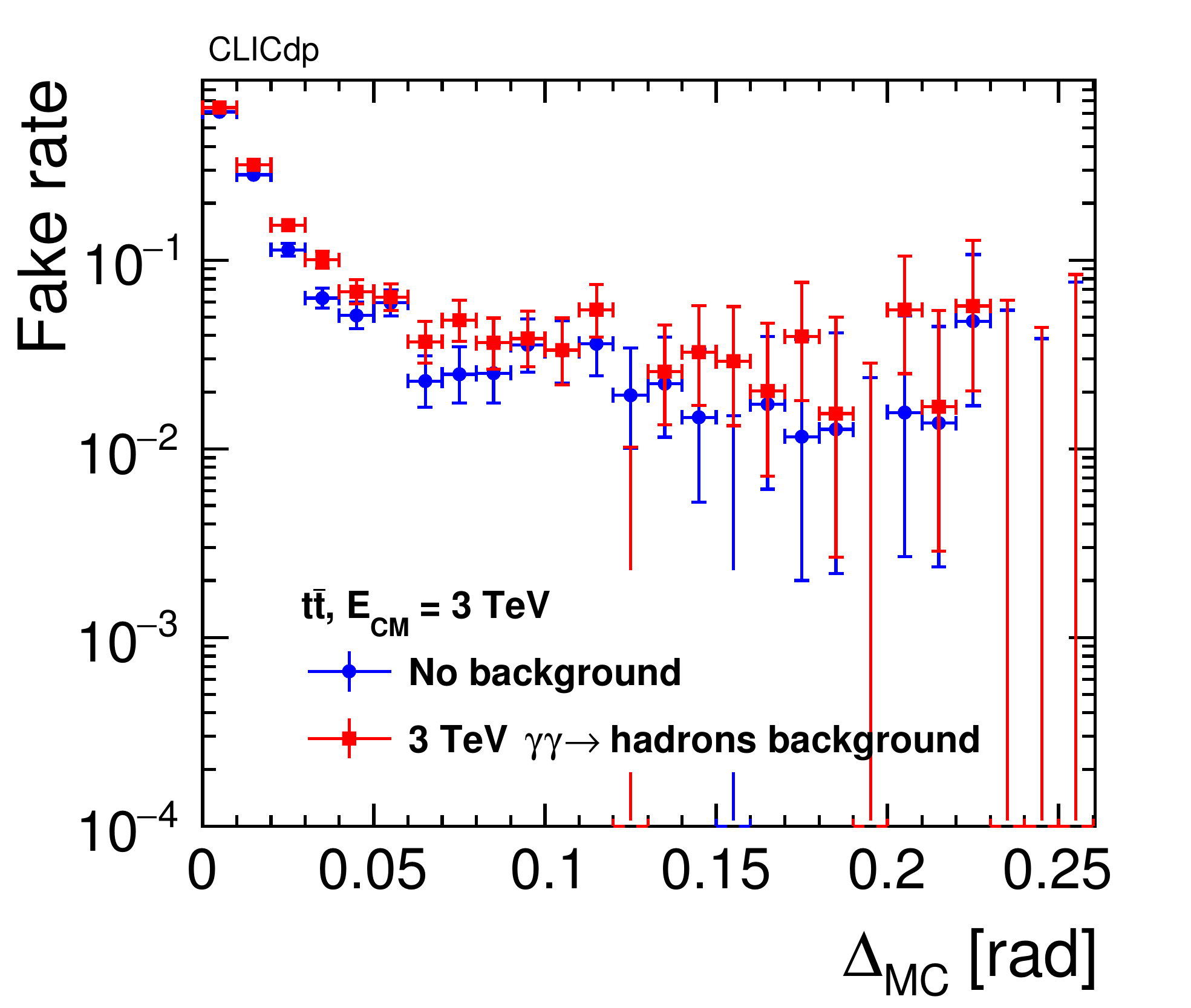}%
    \phantomsubcaption\label{fig:ttbar3TeV_fakes_dist}
  \end{subfigure}
  \vspace{-3mm}
  \caption{Tracking efficiency~\subref{fig:ttbar3TeV_eff_dist} and fake rate~\subref{fig:ttbar3TeV_fakes_dist} as a
    function of the particle proximity \deltamc{} for \ttbar{} events at \SI{3}{TeV}, with and
    without \SI{3}{TeV} \gghadron{} background overlay.}\label{fig:tt_dist}
\end{figure}

\begin{figure}[tbp]
  \renewcommand{\thesubfigure}{(\lr{subfigure})}
  \centering
  \begin{subfigure}{.5\textwidth}
    \centering
    \includegraphics[width=\linewidth]{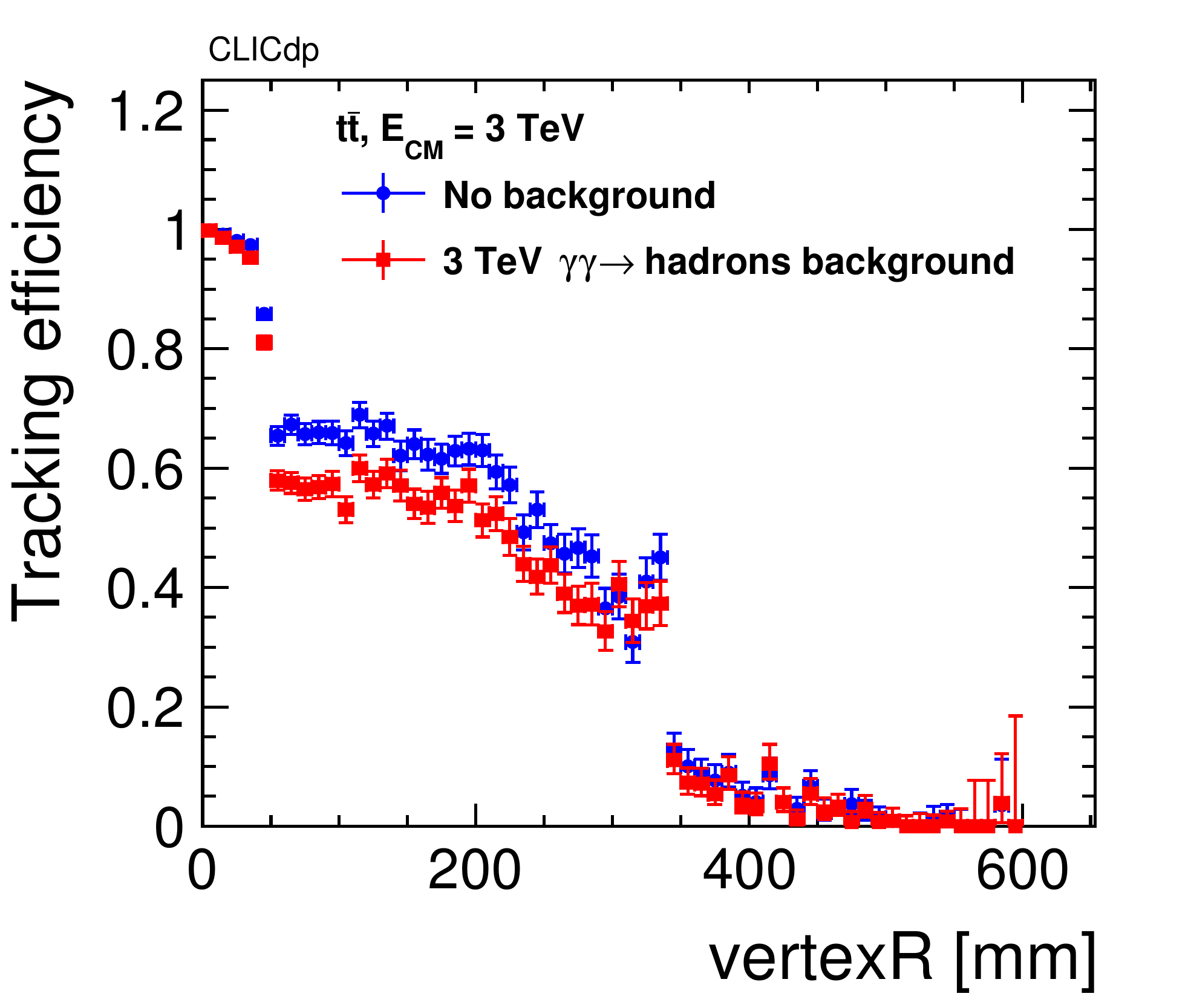}%
    \phantomsubcaption\label{fig:ttbar3TeV_eff_vertexR}
  \end{subfigure}%
  \begin{subfigure}{.5\textwidth}
    \centering
    \includegraphics[width=\linewidth]{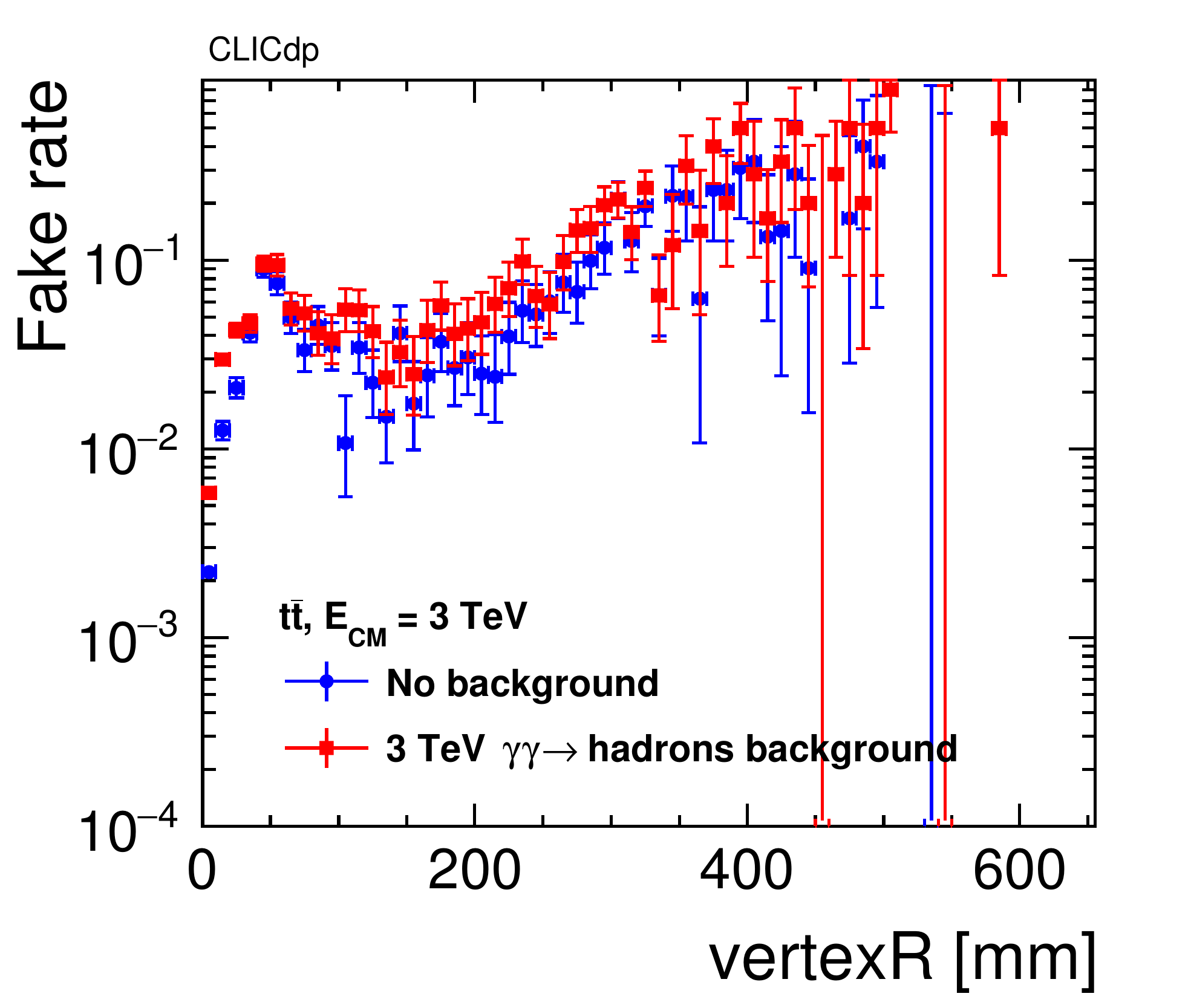}%
    \phantomsubcaption\label{fig:ttbar3TeV_fakes_vertexR}
  \end{subfigure}
  \vspace{-3mm}
  \caption{Tracking efficiency~\subref{fig:ttbar3TeV_eff_vertexR} and fake rate~\subref{fig:ttbar3TeV_fakes_vertexR} as
    a function of the production vertex radius for \ttbar{} events at \SI{3}{TeV}, with and without \SI{3}{TeV}
    \gghadron{} background overlay.}\label{fig:tt_vertexR}
\end{figure}

For comparison, the same study has been performed for \ttbar{} events at \SI{3}{TeV} centre-of-mass energy.
Efficiencies and fake rates are shown in \crefrange{fig:tt_pt}{fig:tt_vertexR} as a function of the transverse momentum, particle proximity and production vertex radius, with and without \SI{3}{TeV} \mbox{\gghadron} background overlay.

The remarkable effect of \gghadron{} background on the fake rate for tracks with \pT < \SI{1}{GeV} is similar in \ttbar{} and \bb{} events. Effort is ongoing to improve the tracking efficiency and reduce the fake rate for prompt and displaced tracks at low transverse momenta.

\clearpage

\paragraph{Lepton Identification}

\begin{figure}[tp]
  \renewcommand{\thesubfigure}{(\lr{subfigure})}
  \centering
  \begin{subfigure}{.5\textwidth}
    \centering
    \includegraphics[width=\linewidth]{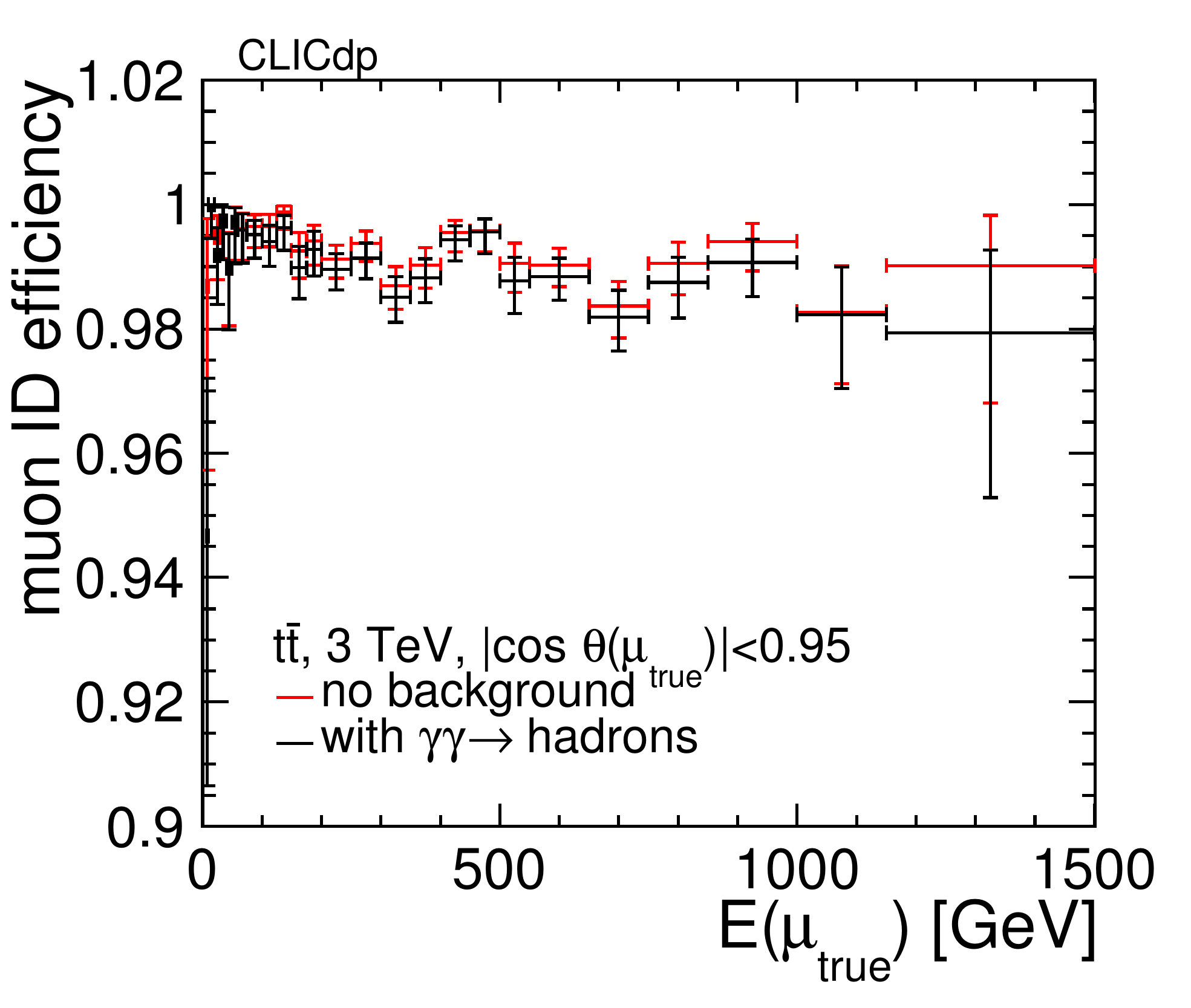}%
    \phantomsubcaption\label{fig:ttbar3TeV_muonIDEffVsE}
  \end{subfigure}%
  \begin{subfigure}{.5\textwidth}
    \centering
    \includegraphics[width=\linewidth]{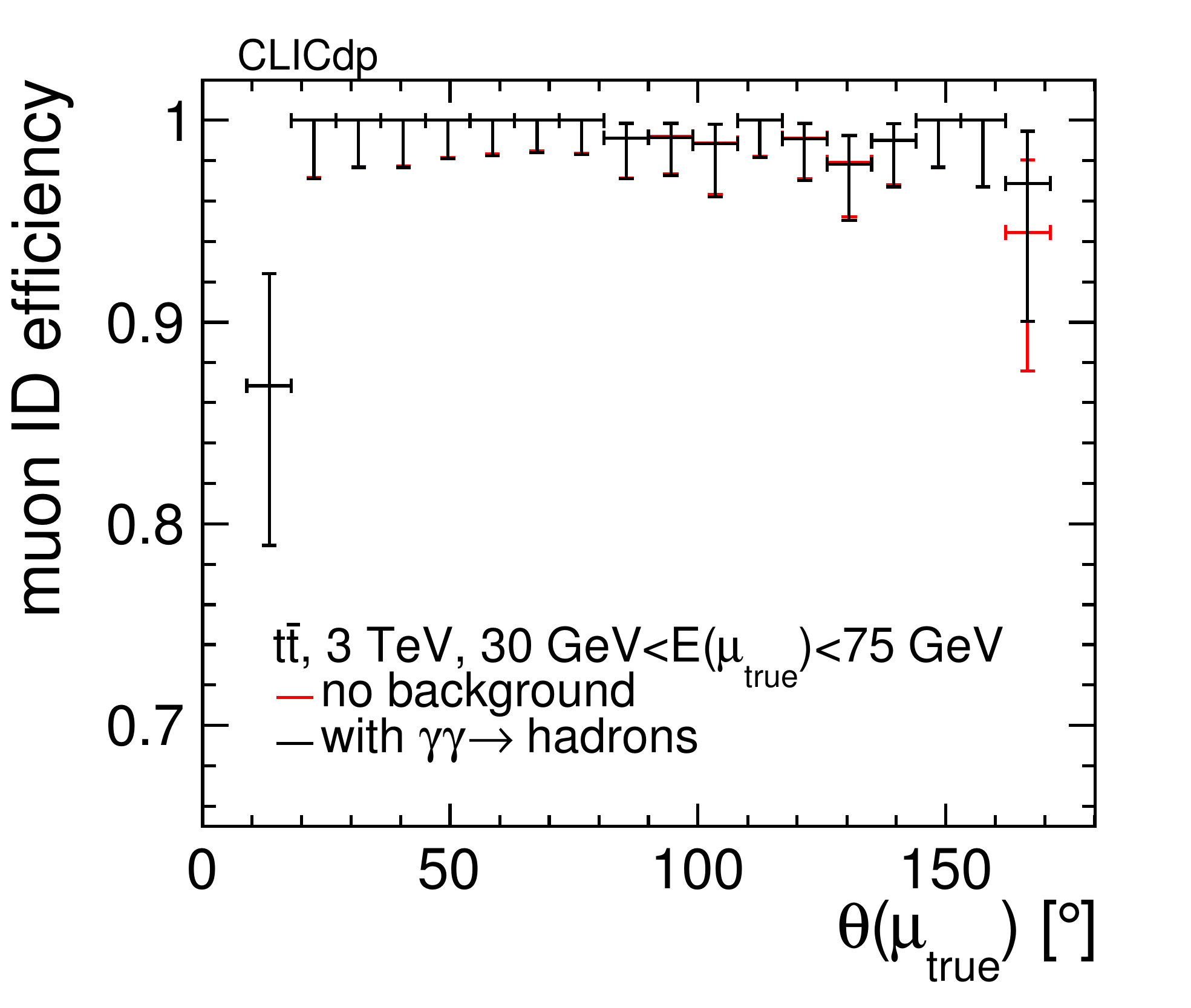}%
    \phantomsubcaption\label{fig:ttbar3TeV_muonIDEffVsTheta}
  \end{subfigure}
  \vspace{-2mm}
  \caption{Muon identification efficiency in \ttbar{} events at \SI{3}{TeV}, without and with \gghadron{} background
    overlay as a function of the energy for
    $|\cos \theta_{\PGm^{\mathrm{true}}}|<0.95$~\subref{fig:ttbar3TeV_muonIDEffVsE} and as a function of the polar
    angle $\theta$ in events with
    $\SI{30}{GeV}<E_{\PGm^{\mathrm{true}}}<\SI{75}{GeV}$~\subref{fig:ttbar3TeV_muonIDEffVsTheta}.}
\end{figure}

Lepton identification efficiencies for muons and electrons are studied in more complex \ttbar{} sample at \SI{3}{TeV}\@.
Investigating the identification efficiency as a function of the lepton energy, an additional restriction of $|\cos\theta_{\mathrm{lep}}|<0.95$ is imposed to ensure the presence of a well-reconstructed track.
 In this study direct leptons from W decays are considered. Reconstructed leptons are required to be spatially matched within an angle of 1\degrees{} around the ``true'' lepton momentum.
The impact of beam background is evaluated by overlaying \SI{30}{BX} of \gghadron{} events. Muons are identified with more than 98\% efficiency for all energies and polar angles,
as shown in \cref{fig:ttbar3TeV_muonIDEffVsE,fig:ttbar3TeV_muonIDEffVsTheta}.
The impact of beam background on muon identification is negligible.

\begin{figure}[tp]
  \renewcommand{\thesubfigure}{(\lr{subfigure})}
  \centering
  \begin{subfigure}{.5\textwidth}
    \centering
    \includegraphics[width=\linewidth]{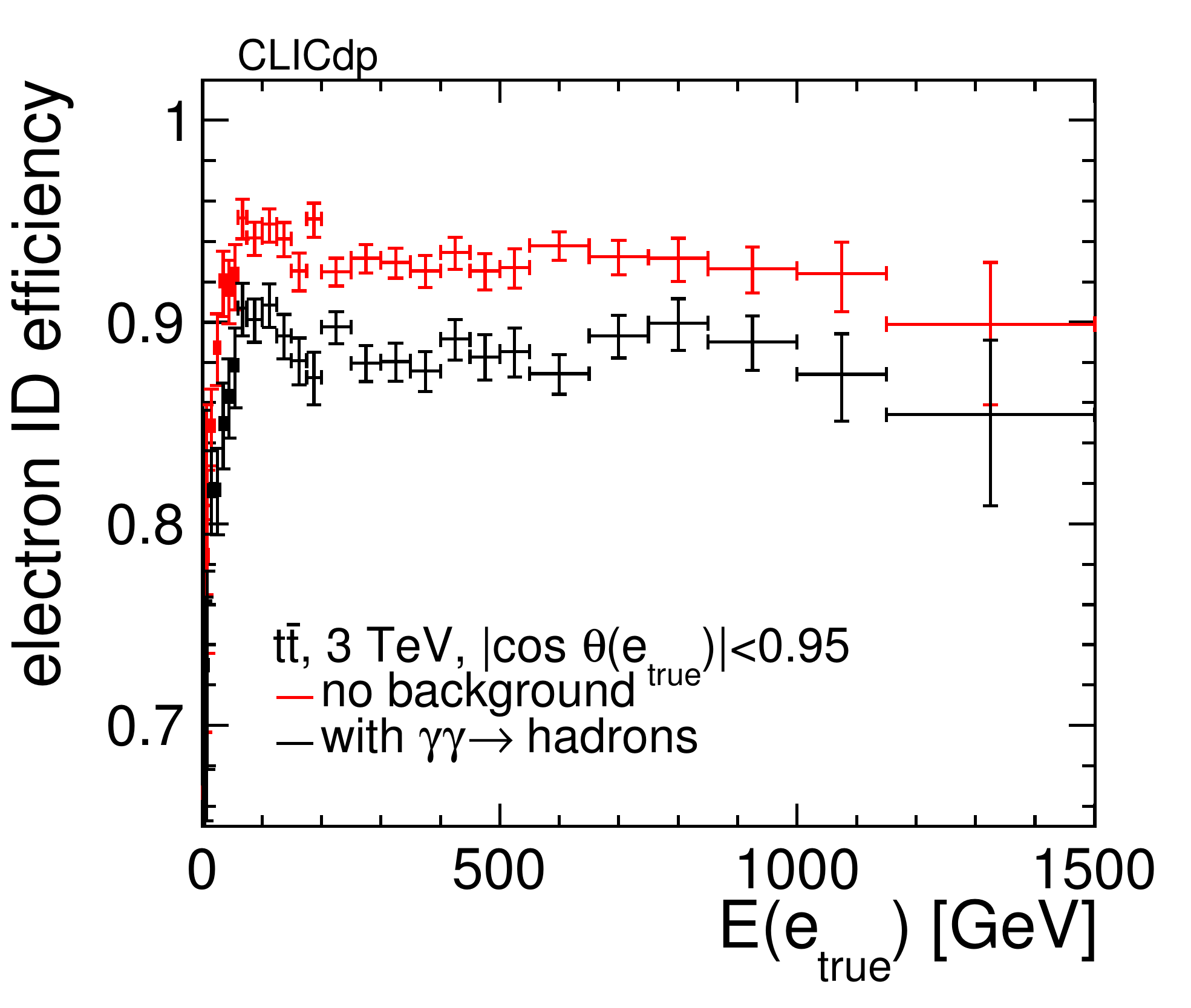}%
    \phantomsubcaption\label{fig:ttbar3TeV_electronIDEffVsE}
  \end{subfigure}%
  \begin{subfigure}{.5\textwidth}
    \centering
    \includegraphics[width=\linewidth]{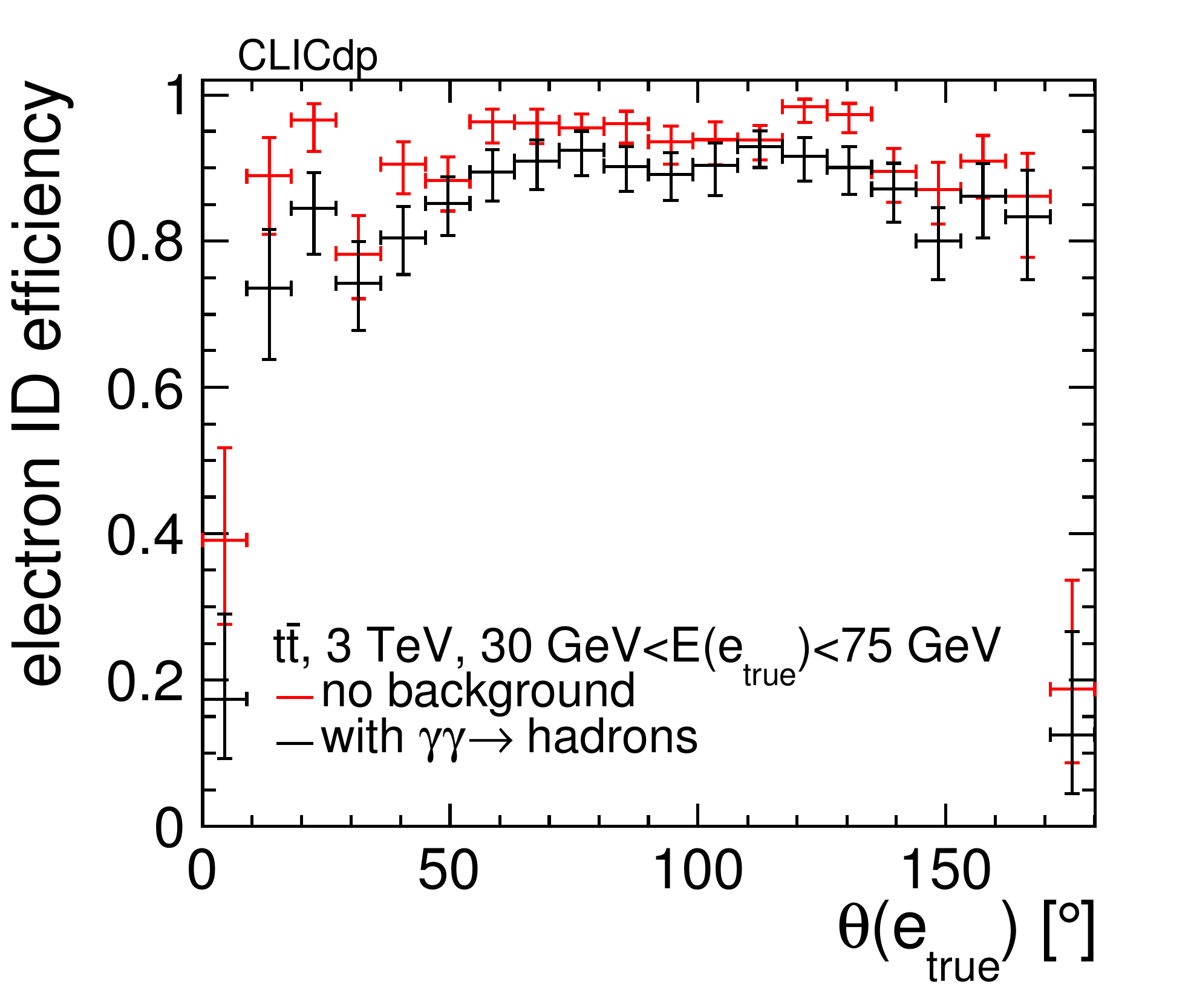}%
    \phantomsubcaption\label{fig:ttbar3TeV_electronIDEffVsTheta}
  \end{subfigure}
  \vspace{-2mm}
  \caption{Electron identification efficiency in \ttbar{} events at \SI{3}{TeV}, without and with \gghadron{} background
    overlay as a function of the energy for
    $|\cos\theta_{\Pe^{\mathrm{true}}}|<0.95$~\subref{fig:ttbar3TeV_electronIDEffVsE} and as a function of the polar
    angle $\theta$ in events with $\SI{30}{GeV}<E_{\Pe^{\mathrm{true}}}<\SI{75}{GeV}$~\subref{fig:ttbar3TeV_electronIDEffVsTheta}.}\label{el_id}
\end{figure}

Electrons are correctly identified in 90\% to 95\% of all cases at energies of \SI{20}{GeV} and higher, as illustrated in \cref{fig:ttbar3TeV_electronIDEffVsE}.
The identification efficiency is lower in the endcaps by about 5\%, see \cref{fig:ttbar3TeV_electronIDEffVsTheta}.
In the presence of beam background the identification efficiency decreases by about 5\% for all energies, both in barrel and endcaps. 

\subsubsection{Jet Energy Resolution}
\label{sec:jets_E}


The accurate jet energy resolution obtained using highly granular calorimeters and Particle Flow algorithms,  allows differentiating between different decays, e.g.\ between jets originating from W and Z boson decays.
The jet performance in CLICdet is studied in di-jet events using \PZgstar decays into light quarks (u, d, s) at several centre-of-mass energies. Tracks are reconstructed using either the ConformalTracking or TruthTracking.
The Pandora particle flow algorithms~\cite{Marshall:2015rfaPandoraSDK,Thomson:2009rp,Marshall:2012ryPandoraPFA} are used to reconstruct each particle, combining information from tracks, calorimeter clusters and hits in the muon system. 
Software compensation is applied to clusters of reconstructed hadrons to improve their energy measurement, using local energy density information provided by the high granularity of the calorimeter system~\cite{Tran:2017tgrSoftwareCompensation}.
The jet energy resolution is determined using the energy sum of all reconstructed particles $E_{\mathrm{tot}}^{\mathrm{PFOs}}$ compared to the sum of all stable visible particles on MC truth level~\cite{Buttar:2008jxParticleLevel} $E_{\mathrm{true}}$.  Since the vast majority of \PZgstarToqq events is reconstructed in a di-jet signature, this procedure effectively measures the energy resolution of jets with an energy of half the centre-of-mass energy $E_{\mathrm{cm}}$, assuming that all particles are clustered into two jets. \rmsninety{} is used as a measure for the jet energy resolution. \rmsninety{} is defined as the RMS in the smallest range of the reconstructed energy containing 90\% of the events~\cite{Marshall:2012ryPandoraPFA}. This measure is a good representation for the resolution of the bulk of events, while it is relatively insensitive to the presence of tails. The relative energy resolution for a jet energy of $E_{j}=1/2\cdot E_{\mathrm{cm}}$ is then calculated as $\Delta E_{j}/E_{j}=\sqrt{2}\cdot \rmsninety (E_{\mathrm{tot}}^{\mathrm{PFOs}}/E_{\mathrm{true}})/\mathrm{Mean}_{90}(E_{\mathrm{tot}}^{\mathrm{PFOs}}/E_{\mathrm{true}})$.

\begin{figure}[tbp] \centering
  \begin{subfigure}{\subfigwidth}
    \includegraphics[width=1.0\textwidth]{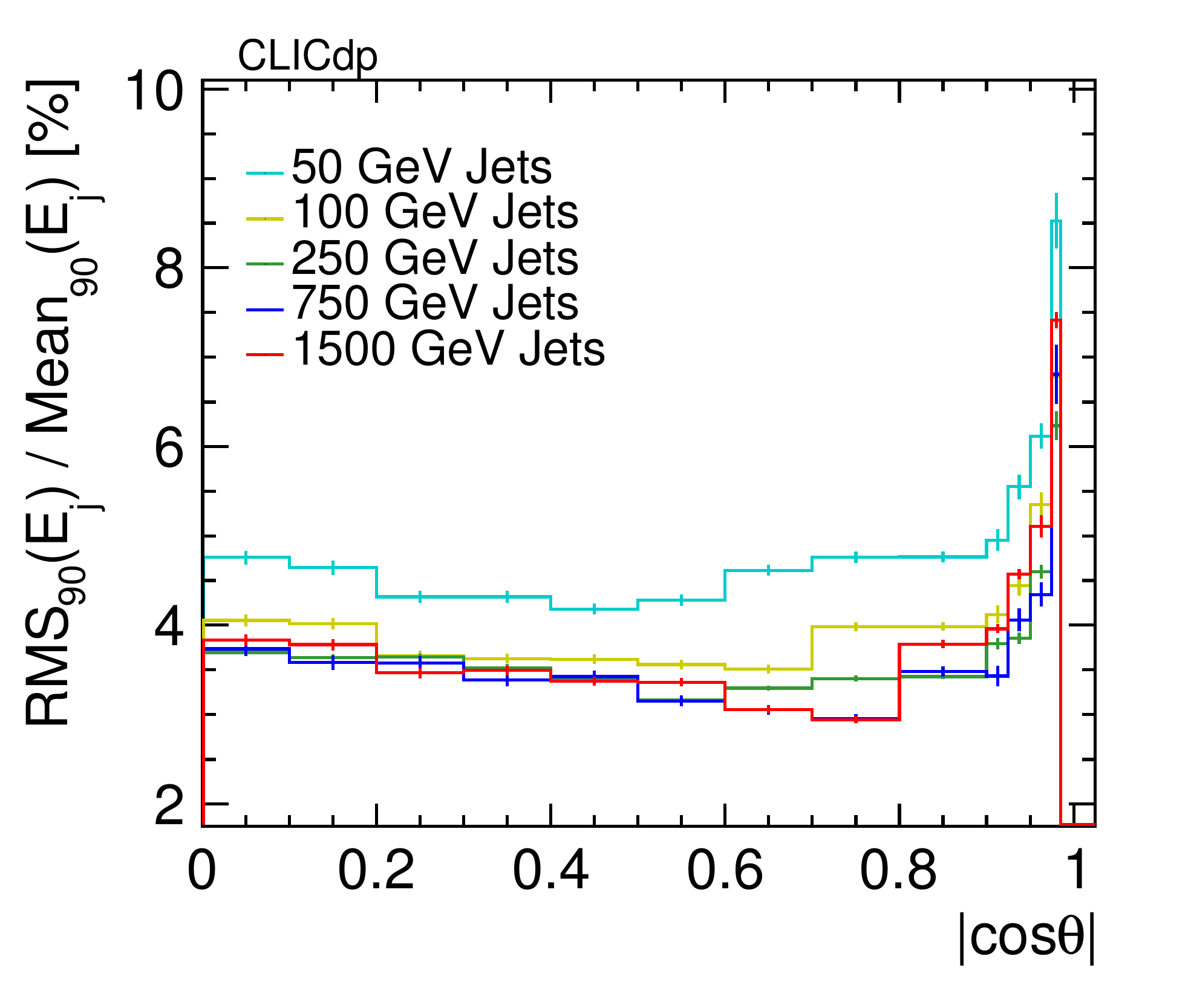}%
  \end{subfigure}
  \begin{subfigure}{\subfigwidth}
    \includegraphics[width=1.0\textwidth]{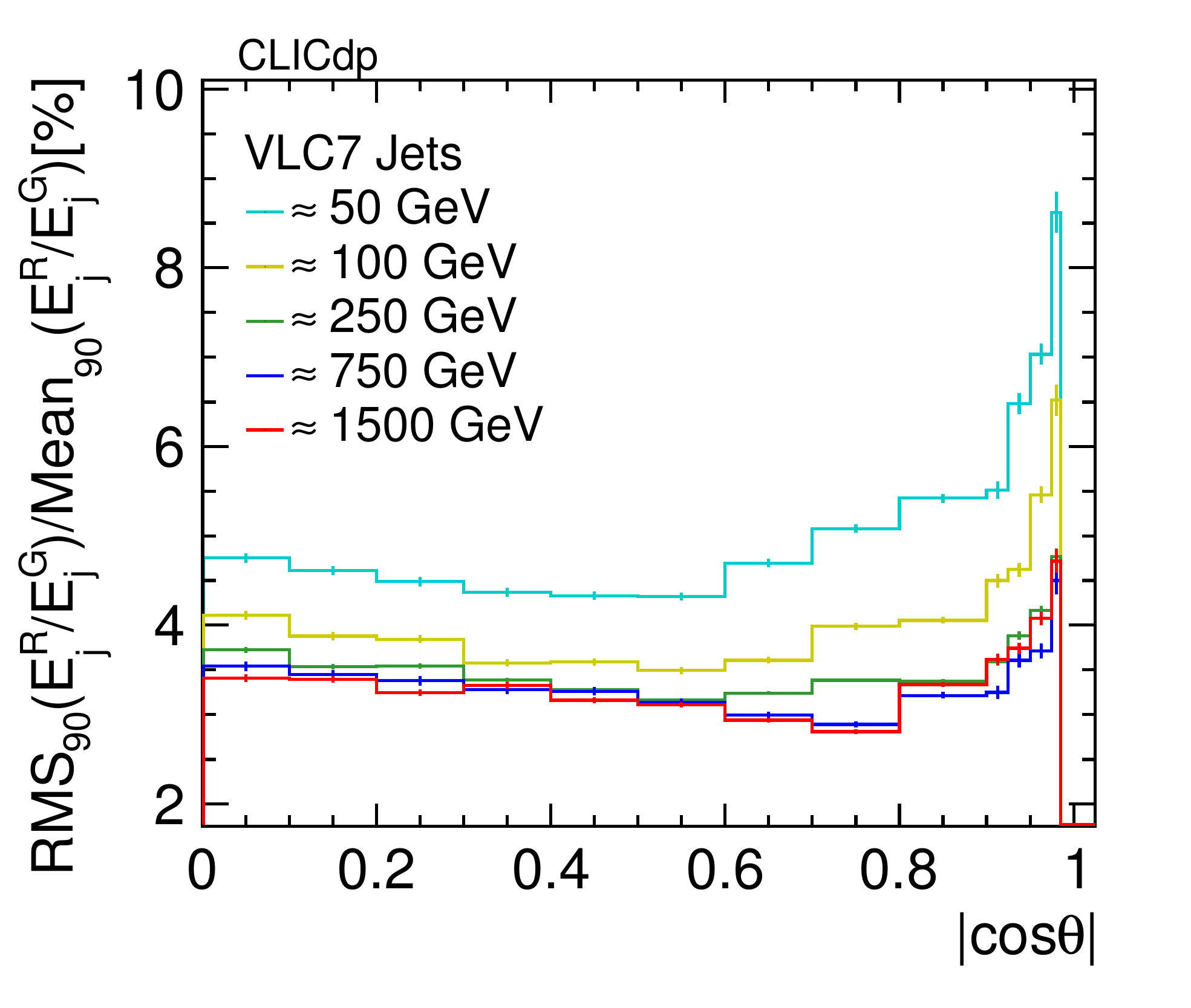}%
  \end{subfigure}
  \caption{Jet energy resolution distributions for various jet energies as a function of the $|\cos\theta|$ of the quark
    using two methods. The first method compares the total reconstructed energy with the energy sum from all visible
    particles on MC truth (left). The second method compares the jet energy of reconstructed jets and matched MC truth
    particle jets, using the VLC algorithm with an $R=0.7$ (VLC7, right)}\label{fig:JER_totE_vs_jetE}
\end{figure}

In a second method the response of particle-level jets (clustering stable visible particles, $\mathrm{j_{G}}$) is compared to reconstructed jets at detector level (clustering PandoraPFOs, $\mathrm{j_{R}}$), 
using 
the VLC algorithm~\cite{Boronat:2016tgdVLC} as implemented in the FastJet\footnote{FastJet version 3.2.1.\ and FastJet Contrib version 1.025. The difference in the beam-distance calculation $d_{i\mathrm{B}}=E_{i}^{2\beta}\left(\frac{p_{\mathrm{T},i}}{X}\right)^{2\gamma}$, where $X$ is $E_{i}$ instead of $p_{i}$ (corrected in FastJet Contrib from 1.040), does not affect the results presented in this note.} library~\cite{Cacciari:2011maFastJet} in exclusive mode to force the event into two jets.
The VLC algorithm combines a Durham-like inter-particle distance based on energy and polar angle with a beam distance. The algorithm applies a sequential recombination procedure, similar to those present in hadron collider algorithms, providing a robust performance at \epem colliders with non-negligible background. For these studies, the VLC parameter values are $\gamma=\beta=1.0$ and $R=0.7$.
The two reconstructed jets are required to be matched to each of the particle level jets within an angle of \ang{10}. Studies using both methods in di-jet events lead to equivalent values of the jet energy resolution for most of the range as shown in \cref{fig:JER_totE_vs_jetE} for several jet energies as a function of the quark $|\cos\theta|$. For low energy jets at \SI{50}{GeV},  the jet energy resolution values are around 4.5--5.5\% for barrel $(|\cos\theta|<0.7$) and endcap jets $(0.80<|\cos\theta|<0.925$). For jets beyond \SI{150}{GeV}, the jet energy resolutions are between 3--4.0\% over most of the angular range. For forward jets with $|\cos\theta|$ between 0.925 and 0.975 the jet energy resolutions increase by typically 0.5--2.0\% points. For very forward jets with $0.975<|\cos\theta|<0.985$, the jet can be partly outside of the tracker volume.

\begin{figure}[tbp]
  \centering
  \begin{subfigure}{\subfigwidth}
    \includegraphics[width=1.0\textwidth]{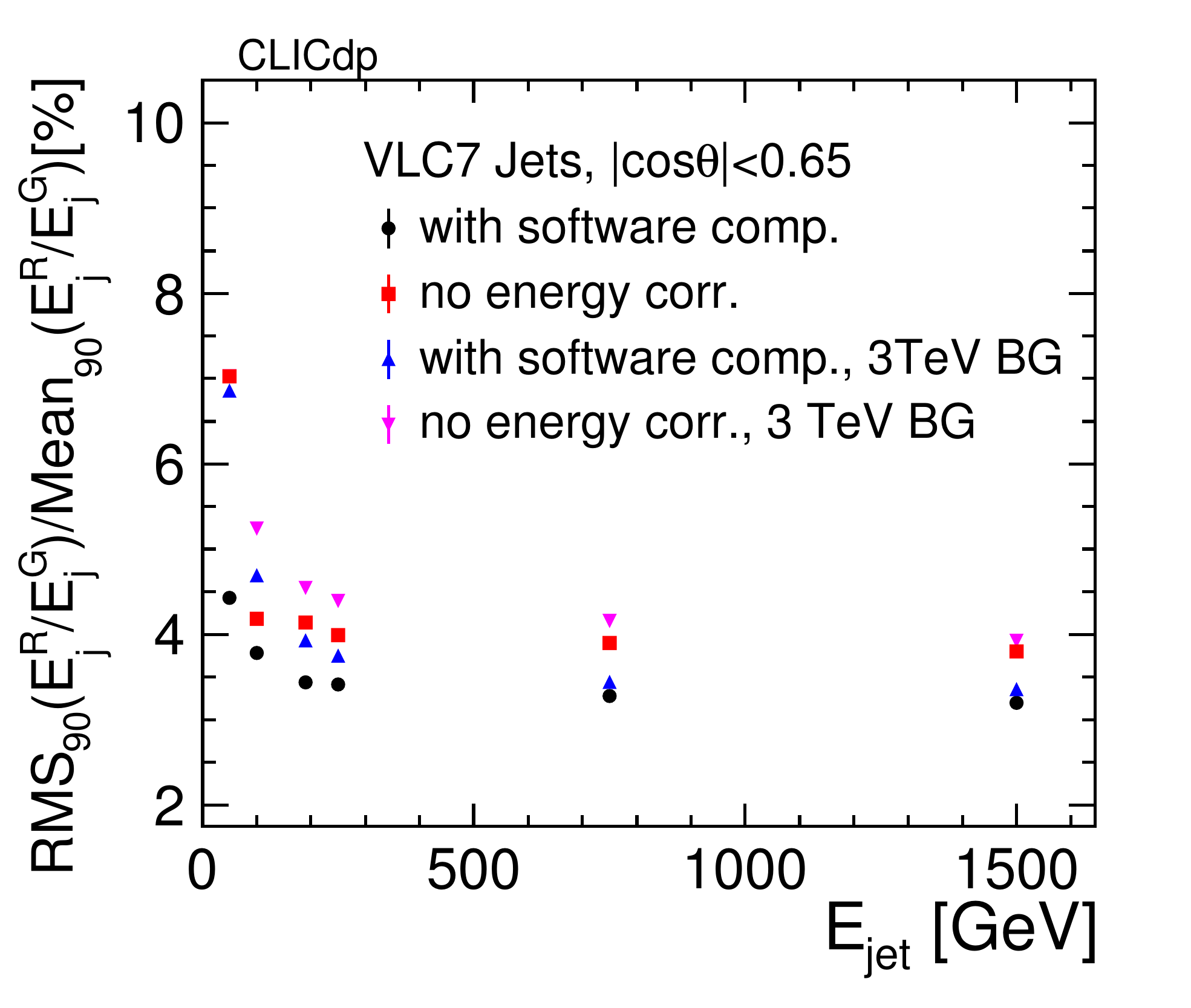}%
  \end{subfigure}
  \begin{subfigure}{\subfigwidth}
    \includegraphics[width=1.0\textwidth]{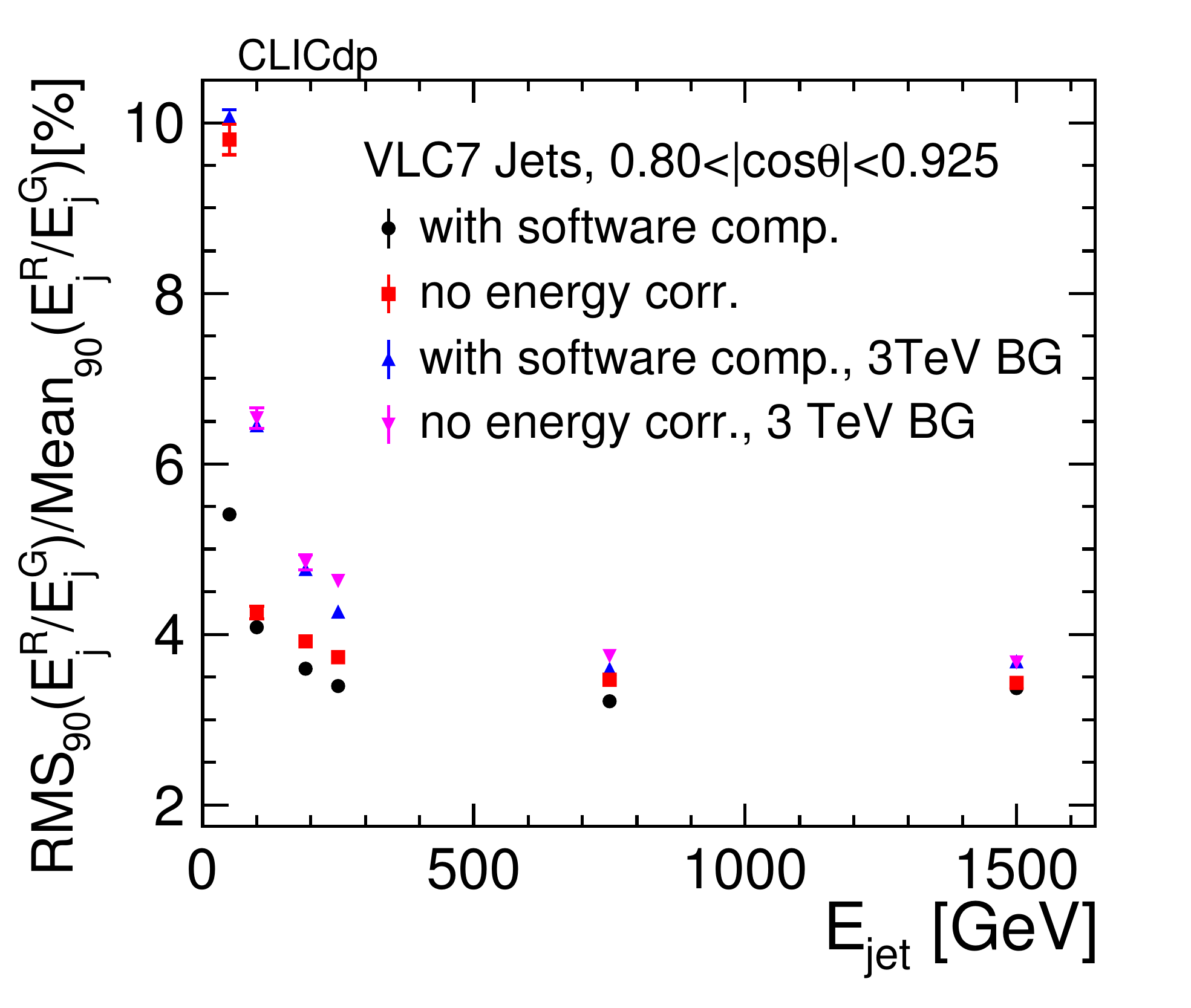}%
  \end{subfigure}
  \caption{Jet energy resolution for central light flavour jets with $|\cos\theta|<0.65$ (left) and endcap jets with
    \mbox{$0.8<|\cos\theta|<0.925$ (right)} in \PZgstarToqq events as a function of the jet energy with and without \SI{3}{TeV} beam-induced
    backgrounds from \gghadrons. PFO reconstruction without energy correction is compared to PFO reconstruction applying
    software compensation.}\label{fig:JER_vs_E_SC_vs_noSC}
\end{figure}

\Cref{fig:JER_vs_E_SC_vs_noSC} shows that applying software compensation improves the energy resolution of jets significantly for most jet energies, for jets with $|\cos\theta|<0.65$ by about 5--15\%, for endcap jets with $0.8<|\cos\theta|<0.925$ by 5--10\%. In events including beam-induced backgrounds from \gghad{} from the \SI{3}{TeV} collider the improvement is on a similar level for both detector regions.

\begin{figure}[bp]
  \centering
  \begin{subfigure}{.5\textwidth}
    \centering
    \includegraphics[width=1.0\textwidth]{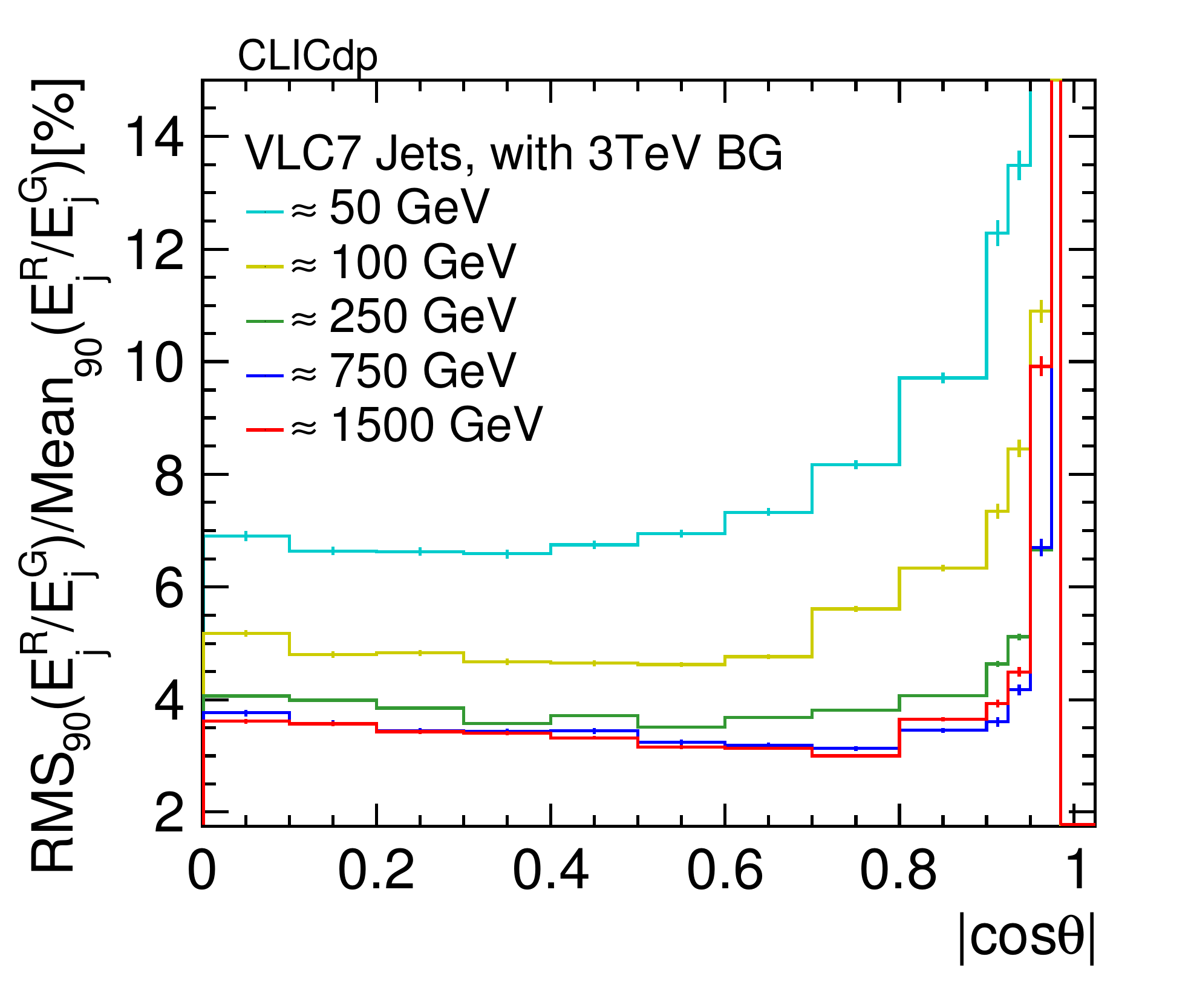}%
  \end{subfigure}%
  \begin{subfigure}{.5\textwidth}
    \includegraphics[width=1.0\textwidth]{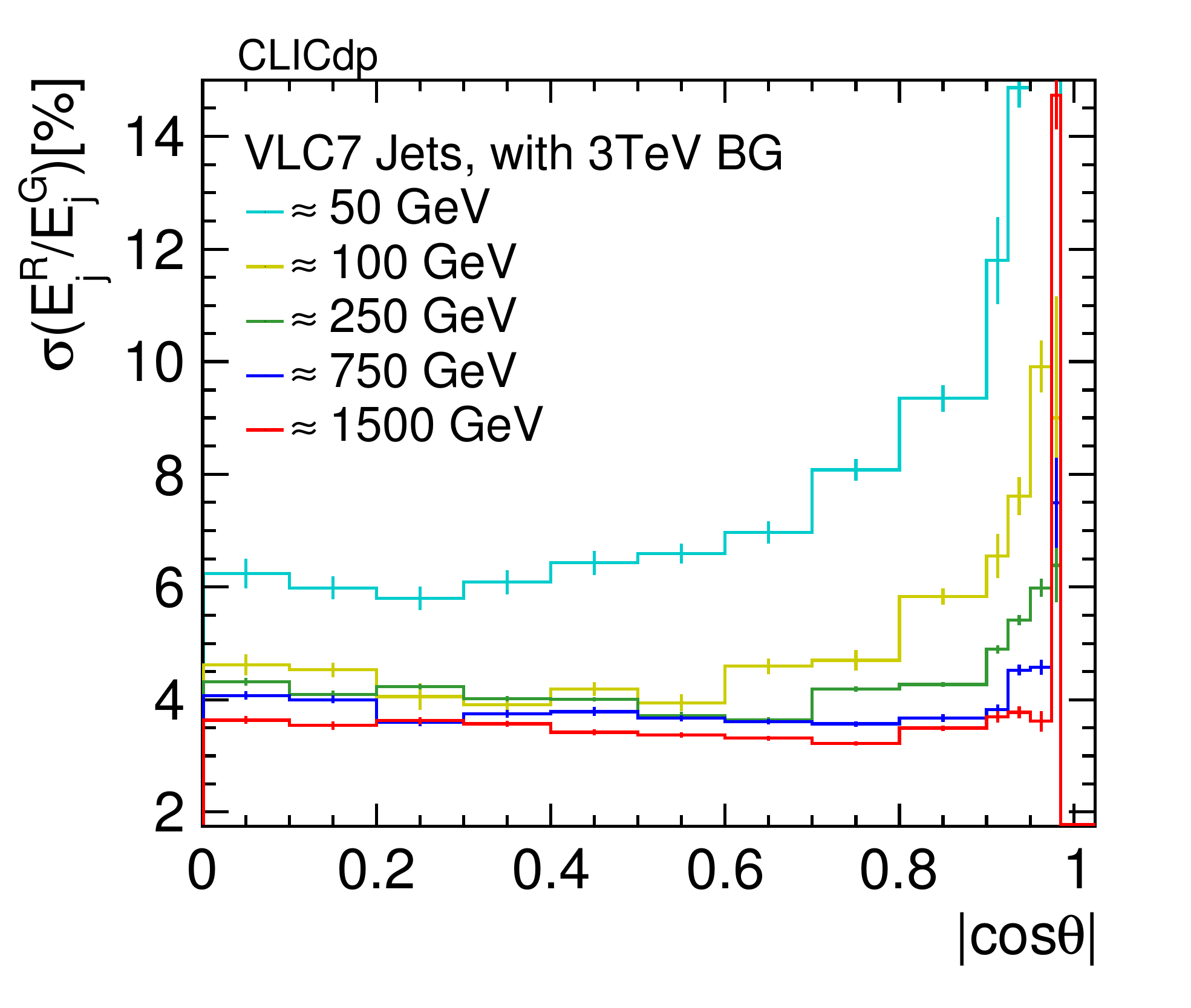}%
  \end{subfigure}
  \caption{Jet energy resolution for various jet energies as a function of the $|\cos\theta|$ of the quark with 3~TeV
    \gghadron{} background overlaid on the physics di-jet event. In the first method \rmsninety is used as measure of
    the jet energy resolution (left), the standard deviation $\sigma$ of the Gaussian core of the double-sided Crystal
    Ball fit quantifies the jet energy resolution in the second method (right). Tight PFO selection cuts are used.}
  \label{fig:jet_response_jets_wBG}
\end{figure}

As alternative the response distribution is fitted with a double sided Crystal ball function~\cite{Oreglia:1980cs}, using the Minuit2 library~\cite{James:1994vla} as implemented in \ROOT~6.08.00~\cite{Antcheva:2009zz}. The procedure starts by fitting a Gaussian over the full range, iteratively changing the fit range until the standard deviation $\sigma$ of the fit stabilises within 5\%. The range of the $\sigma$ parameter of the Crystal Ball fit is restricted to be within a factor of 2 around the width of the Gaussian fit. Non-Gaussian tails are particularly significant in simulated data that include \gghadron{} backgrounds. \cref{fig:jet_response_jets_wBG} compares the resolutions obtained with \rmsninety{} and the one from the Crystal Ball $\sigma$ for different jet energies in events with \SI{3}{TeV} \gghadron{} beam-induced background. In events where \SI{3}{TeV} beam-induced backgrounds from \gghadrons are overlaid on the physics event, \textit{tight}~\cite{Marshall:2012ryPandoraPFA} selection cuts are applied to the PandoraPFOs prior to jet clustering. When \SI{380}{GeV} \gghadron{} backgrounds are used, the \textit{low energy loose}~\cite{Brondolin:2641311} selection cuts are applied. These beam-induced backgrounds represent 30 bunch crossings. The mean of the jet energy response distribution $E_{\mathrm{recojet}}/E_{\mathrm{genjet}}$ between the reconstructed and the matched MC truth particle jet for events with and without beam-induced backgrounds ranges from 0.98 to 1.01 for jet energies between \SI{50}{GeV} and \SI{1.5}{TeV} for polar angles $|\cos\theta|<0.95$, and 0.94 to 1.03 for forward jets with $0.95<|\cos\theta|<0.975$~\cite{JetPerformanceNote}. For more forward jets with $0.975<|\cos\theta|<1.00$ the peak of the response distribution is still between 0.95 and 1.05 (depending on the jet energy), tails to lower energies are substantial and the mean of the distribution is between 0.98 and 0.85, gradually decreasing with jet energy. 
 In general for the jet energy response distributions, the standard deviation $\sigma$ of the Gaussian core of the double-sided Crystal Ball fits are in good agreement with \rmsninety values for almost all jet energies and polar angles, with jet energy resolution values of 3.5--10\% for barrel and endcap jets in the presence of \SI{3}{TeV} backgrounds (\cref{fig:jet_response_jets_wBG}). For \SI{50}{GeV} jets, the $\sigma$ of the fit is considerably lower than the \rmsninety values, where a decrease is observed from 7\% to 6\% in the detector barrel for events with beam-induced backgrounds. For very forward jets, the $\sigma$ of the double-sided Crystal Ball function does not account for the sizeable non-Gaussian tails of the jet energy response and thus the \rmsninety values are considerably larger. Jet energy resolutions are around 3.5--4.5\% for large jet energies beyond \SI{200}{GeV}, using either measure for quantification. In the forward region ($|\cos\theta|>0.925$) the $\sigma$ of the fit is below 6\% for most jet energies. The beam-induced background leads to larger tails in the jet energy response distribution in this detector region, which are reflected in the larger values of the \rmsninety{} measure.
Compared to jet energy resolution values in events without \gghadron{} backgrounds (\cref{fig:JER_totE_vs_jetE}), a degradation of the jet energy resolution is observed for all jet energies. The effect is most pronounced for low energy jets, e.g.\ for \SI{50}{GeV} jets, where the increase is from around 4.5\% to 7.5\%. For high energy jets, the jet energy resolution increase is limited to less than 0.5\% points for most of the $|\cos\theta|$ range. Since hadrons from beam-induced backgrounds tend to be produced more in the  forward direction,  their impact is larger for endcap and forward jets than for barrel jets. For low energy jets the jet energy resolution values are significantly better than those obtained by ATLAS~\cite{ATLASPFAaboud:2017aca} and CMS~\cite{CMSJERKhachatryan:2016kdb}.

\begin{figure}[tbp]
  \centering
  \begin{subfigure}{\subfigwidth}
    \includegraphics[width=1.0\textwidth]{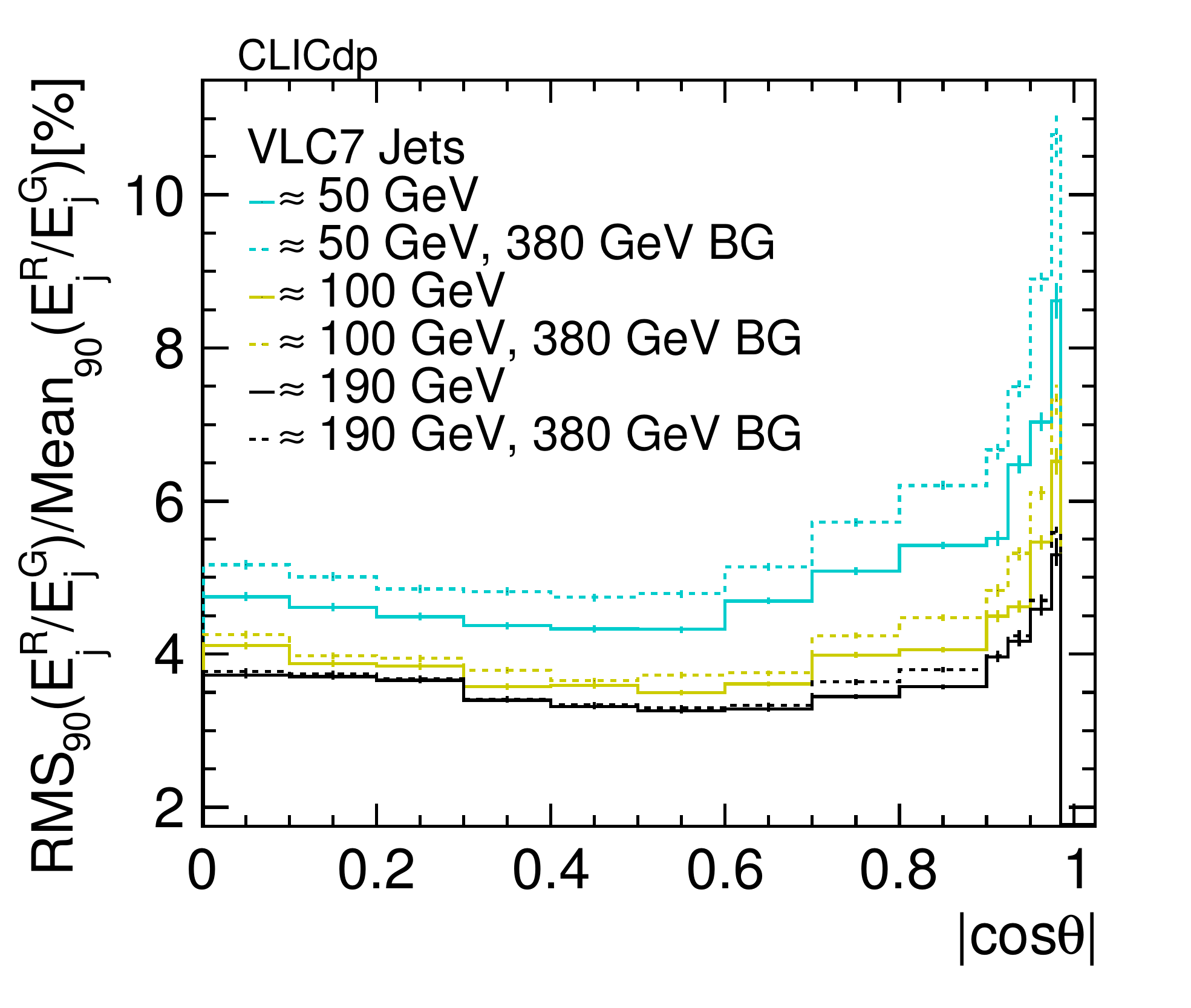}%
  \end{subfigure}
  \caption{Jet energy resolution for various jet energies as a function of the $|\cos\theta|$ of the quark with and
    without \SI{380}{GeV} \gghadron{} background overlay on the physics di-jet event. \rmsninety is used as measure of
    the jet energy resolution. Low energy loose PFO selection cuts are used for events with
    background.}\label{fig:jet_resp_jets_wO_380}
\end{figure}

In events with overlay of \SI{380}{GeV} beam-backgrounds from \gghadrons, \textit{low energy loose} selection cuts~\cite{Brondolin:2641311} are applied on the PandoraPFOs, to reflect the lower levels of beam-induced backgrounds of the \SI{380}{GeV} accelerator relative to the \SI{3}{TeV} accelerator. The impact of beam-induced backgrounds for the \SI{380}{GeV} accelerator on the jet energy resolution is illustrated by \cref{fig:jet_resp_jets_wO_380}. For jet energies above \SI{100}{GeV}, the \SI{380}{GeV} beam-induced background levels lead to almost no increase of the jet energy resolution for barrel and endcap jets. Around 0.5--1.0\% points can be observed for forward jets. Even for \SI{50}{GeV} jets in the barrel, \SI{380}{GeV} beam -induced backgrounds lead only to a mild increase of the jet energy resolution to about 5\%. In the outermost part of the barrel and endcaps the jet energy resolution for \SI{50}{GeV} jets is increased to about 5.5--6.0\%, as well as an absolute increase of 2\% points and more for forward jets.

\begin{figure}[htbp!]
  \centering
  \begin{subfigure}{.5\textwidth}
    \includegraphics[width=1.0\textwidth]{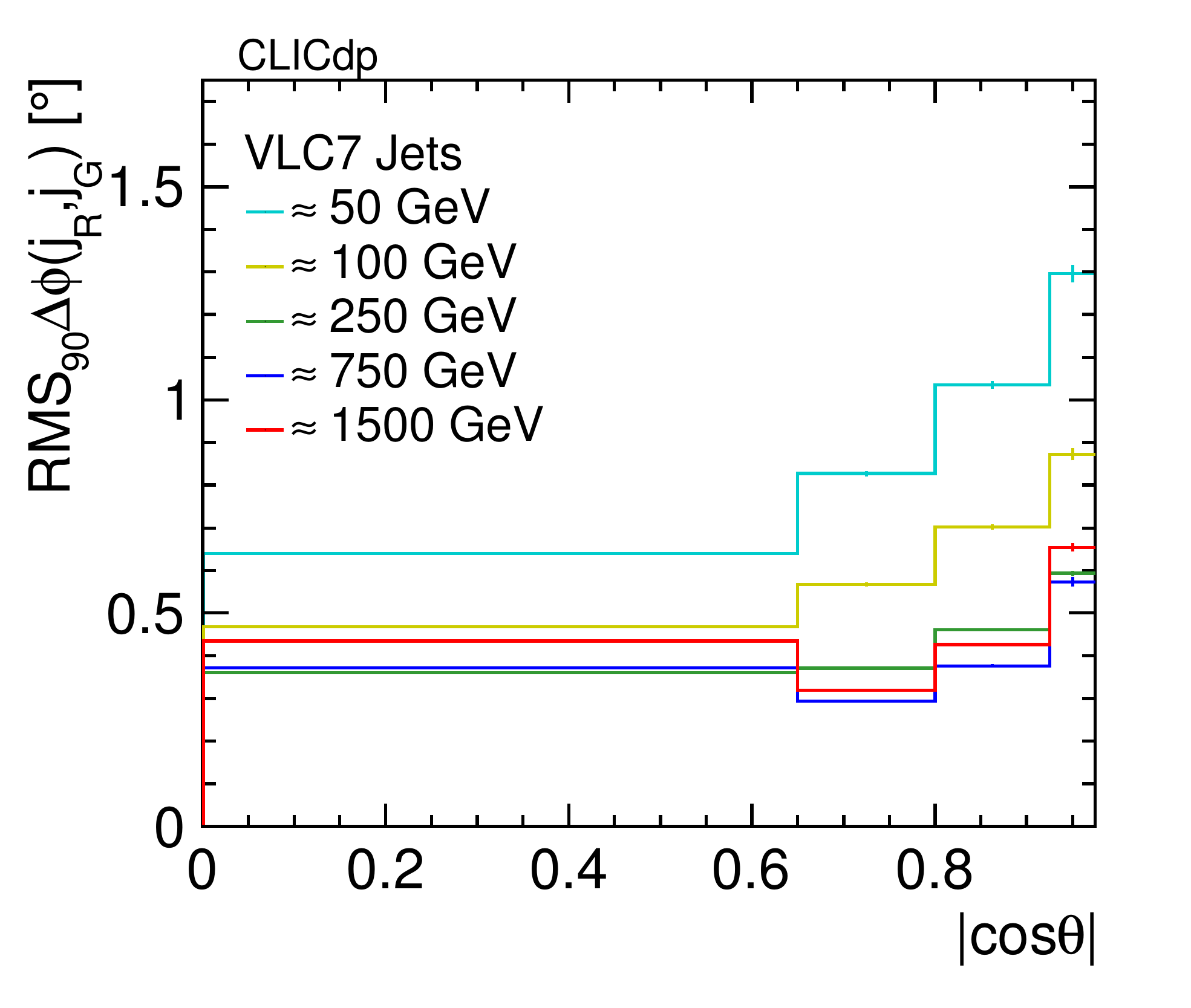}%
  \end{subfigure}%
  \begin{subfigure}{.5\textwidth}
    \centering
    \includegraphics[width=1.0\textwidth]{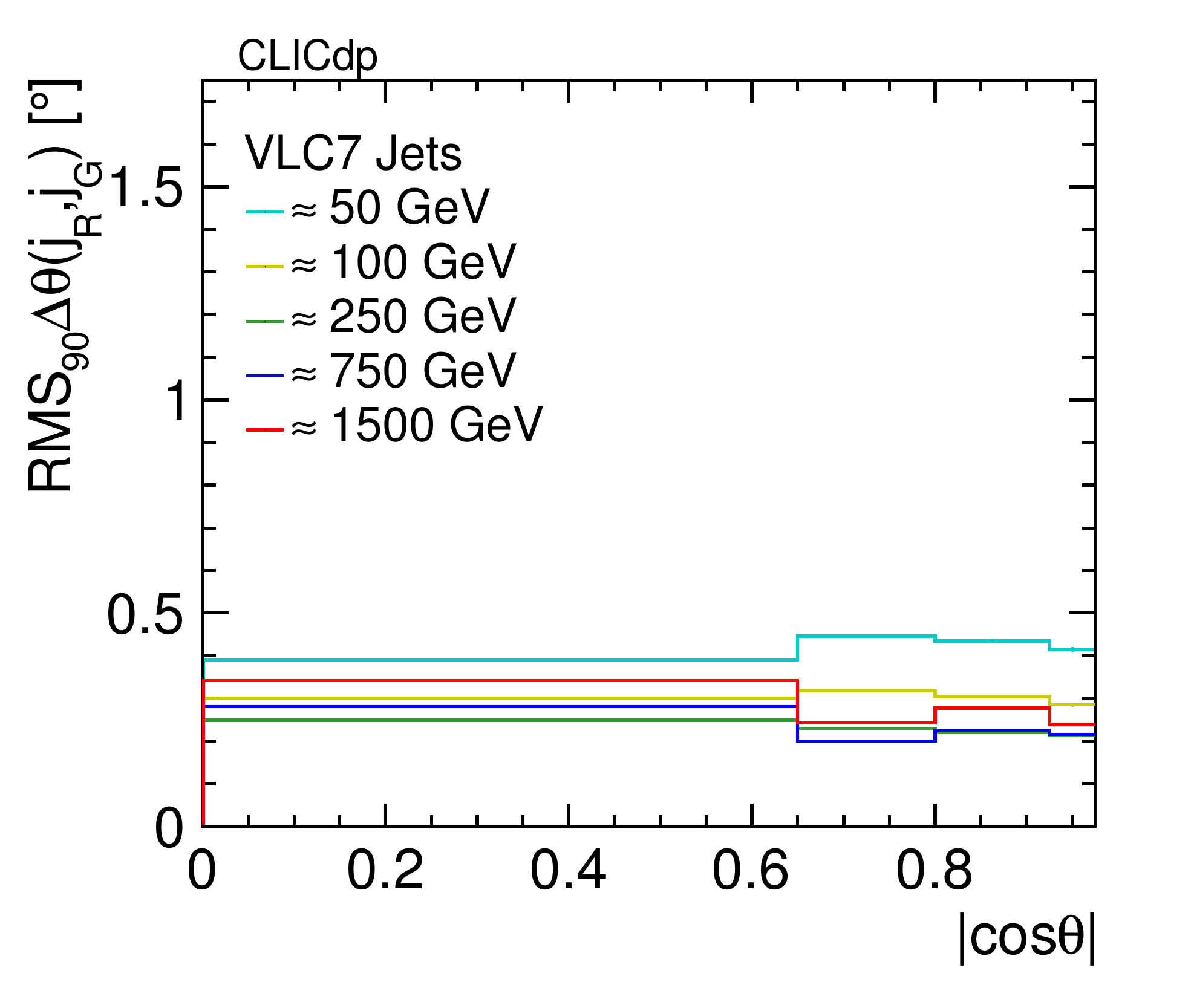}%
  \end{subfigure}
  \caption{Jet $\phi$ (left) and $\theta$ (right) resolutions for several jet energies in four $|\cos\theta|$ bins in events without any simulation of beam-induced background effects.}\label{fig:phi_theta_res_vs_E_noO}
\end{figure}

\begin{figure}[htbp!]
  \centering
  \begin{subfigure}{.5\textwidth}
    \centering
    \includegraphics[width=1.0\textwidth]{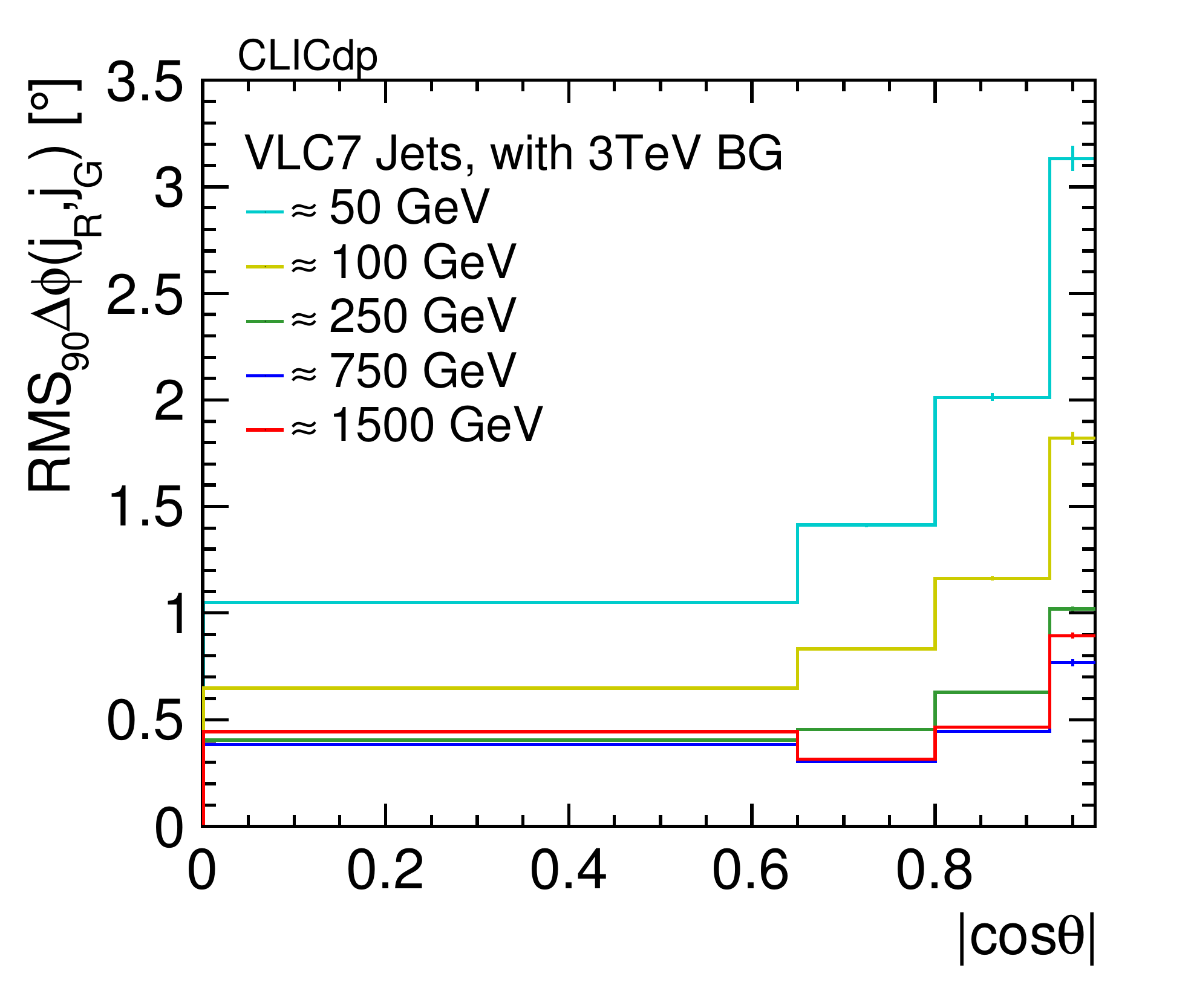}%
  \end{subfigure}%
  \begin{subfigure}{.5\textwidth}
    \centering
    \includegraphics[width=1.0\textwidth]{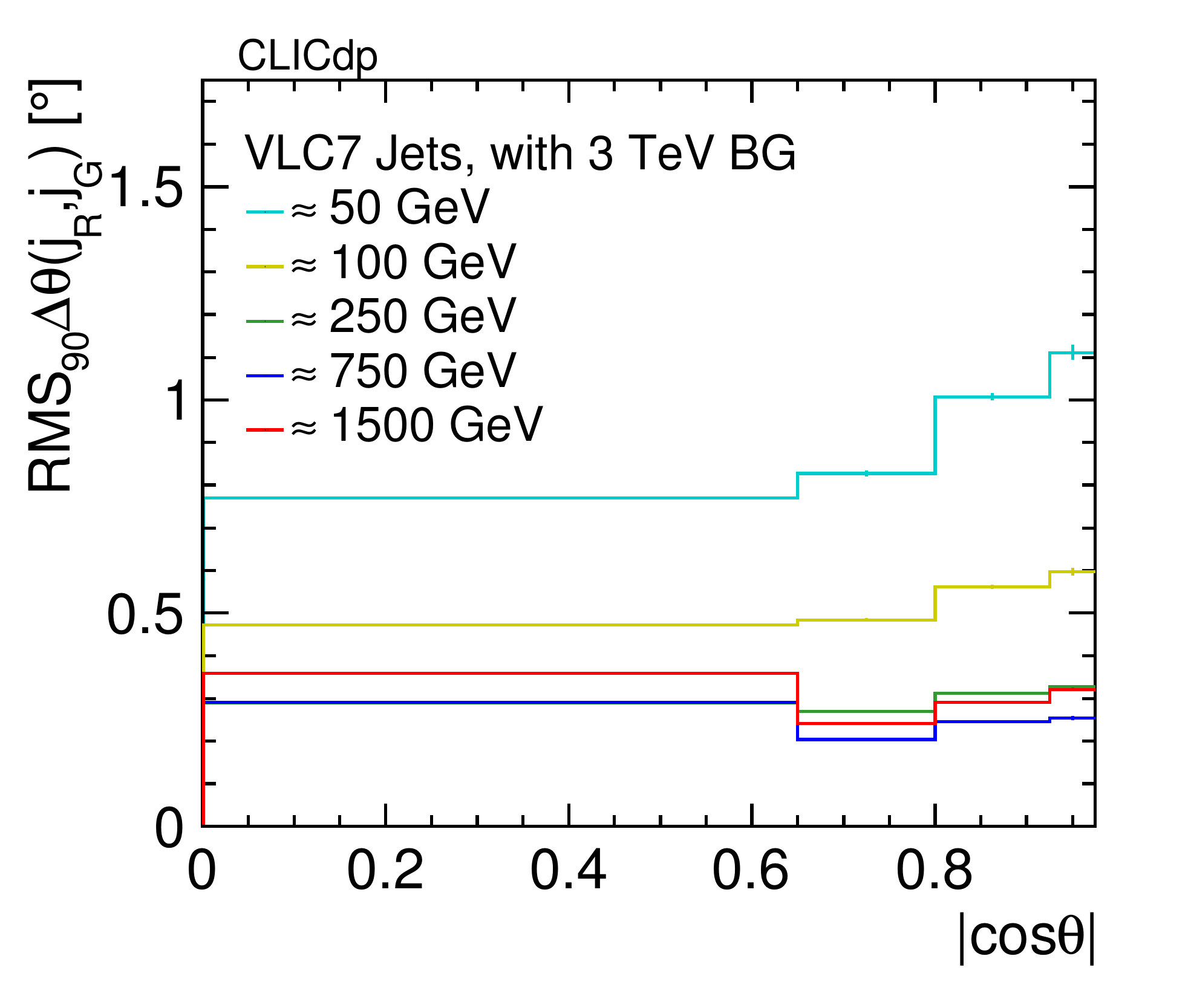}%
  \end{subfigure}
  \caption{Jet $\phi$ (left) and $\theta$ (right) resolutions for several jet energies in four $|\cos\theta|$ bins in events with \SI{3}{TeV} beam-induced backgrounds from \gghadrons, using tight PFO selection cuts.}\label{fig:phi_theta_res_vs_E_wO}
\end{figure}

The jet angular resolutions in azimuth $\phi$ and polar angle $\theta$ are studied as functions of jet energies for different regions in polar angle. For events having significant final state gluon radiation, three jets reflect the event topology better than two jets. Since the jet algorithm is run in exclusive mode with two jets, this can lead to a significant bias in jet angular resolutions. In order to diminish the impact of this bias, events are preselected, where the two particle level MC truth jets are back-to-back in azimuth $\Delta\phi\mathrm{(j1,j2)}>2.8\approx160\degrees$, which vetoes against underlying multi-jet topologies. Each of the reconstructed jets is matched to its closest MC truth particle level jet. The distribution of $\phi$ resolutions are studied as a function of the polar angle, and as a function of the jet energy. The detector is divided into four regions of $|\cos\theta|$: the barrel, the transition region (where the jet energy is reconstructed both in barrel and endcap parts of the detector), the endcap, and the forward region. Resolutions in $\phi$ vary for all energies for almost all regions between 0.3\degrees{} and 1.0\degrees, as \cref{fig:phi_theta_res_vs_E_noO} shows. The jet $\theta$ resolutions values are slightly better with \rmsninety{} values between 0.2\degrees{} and 0.5\degrees{} for all energies in almost all detector regions. Once beam backgrounds from \gghadrons are overlaid, the $\theta$ resolutions increase for \SI{50}{GeV} jets from 0.5\degrees{} to about 1.0\degrees{} (see \cref{fig:phi_theta_res_vs_E_wO}), while for the remaining jet energies the $\theta$ resolutions remain around 0.3\degrees{} to 0.5\degrees. A slight increase in angular resolution for more forward jets can be observed. For jet $\phi$ resolutions in the barrel region and for most jet energies, the values remain at a similar level between 0.4\degrees{} to 0.7\degrees; for more forward jets and for all energies, the $\phi$ resolutions increase relatively by around 25--50\%.

\subsubsection{Missing Transverse Energy Resolution}\label{sec:met}

\begin{figure}[htbp!]
  \renewcommand{\thesubfigure}{(\lr{subfigure})}
  \centering
  \begin{subfigure}{.5\textwidth}
    \centering
    \includegraphics[width=\linewidth]{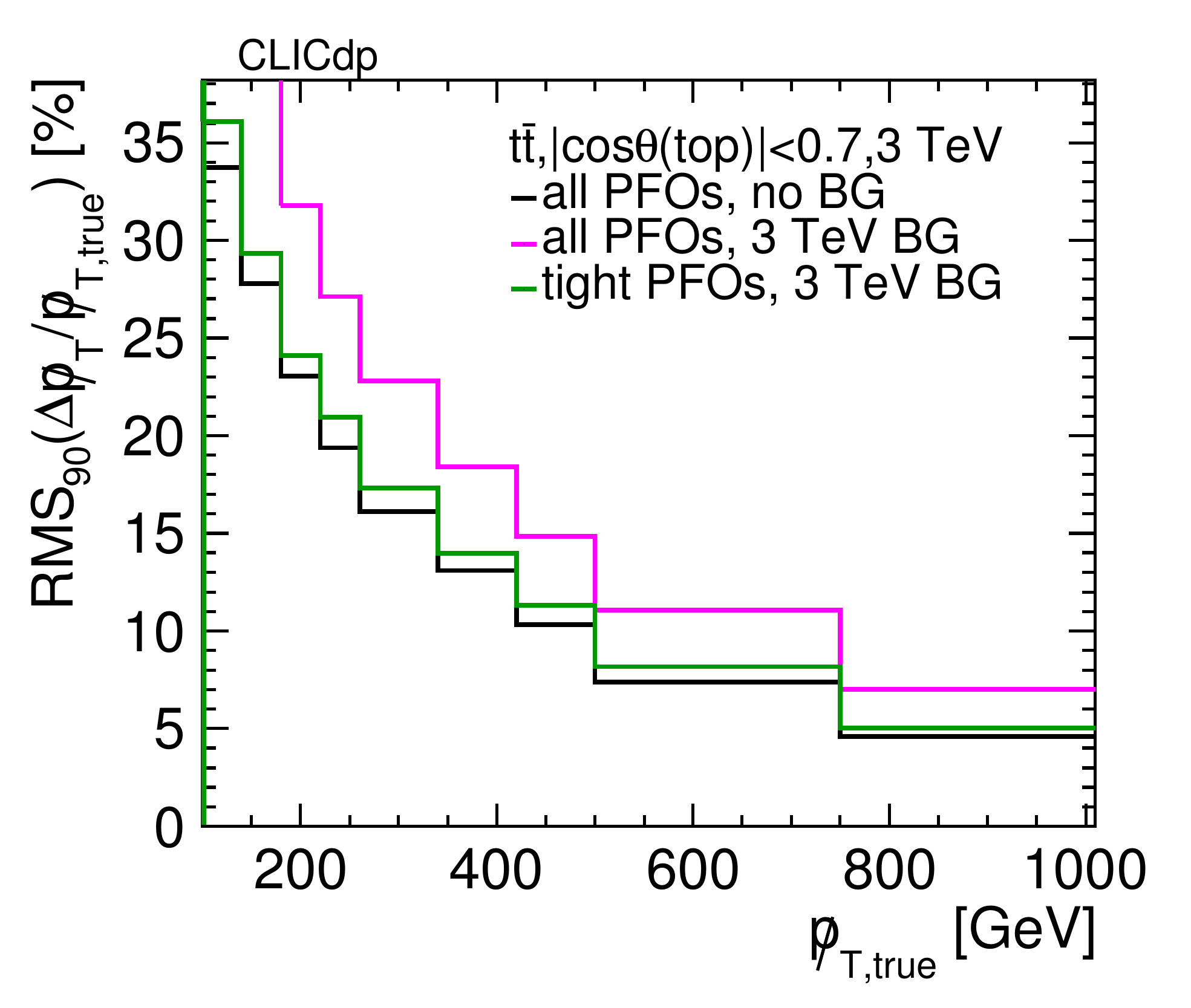}%
    \phantomsubcaption\label{fig:ttbar3TeV_deltaMET_background_impact}
  \end{subfigure}%
  \begin{subfigure}{.5\textwidth}
    \centering
    \includegraphics[width=\linewidth]{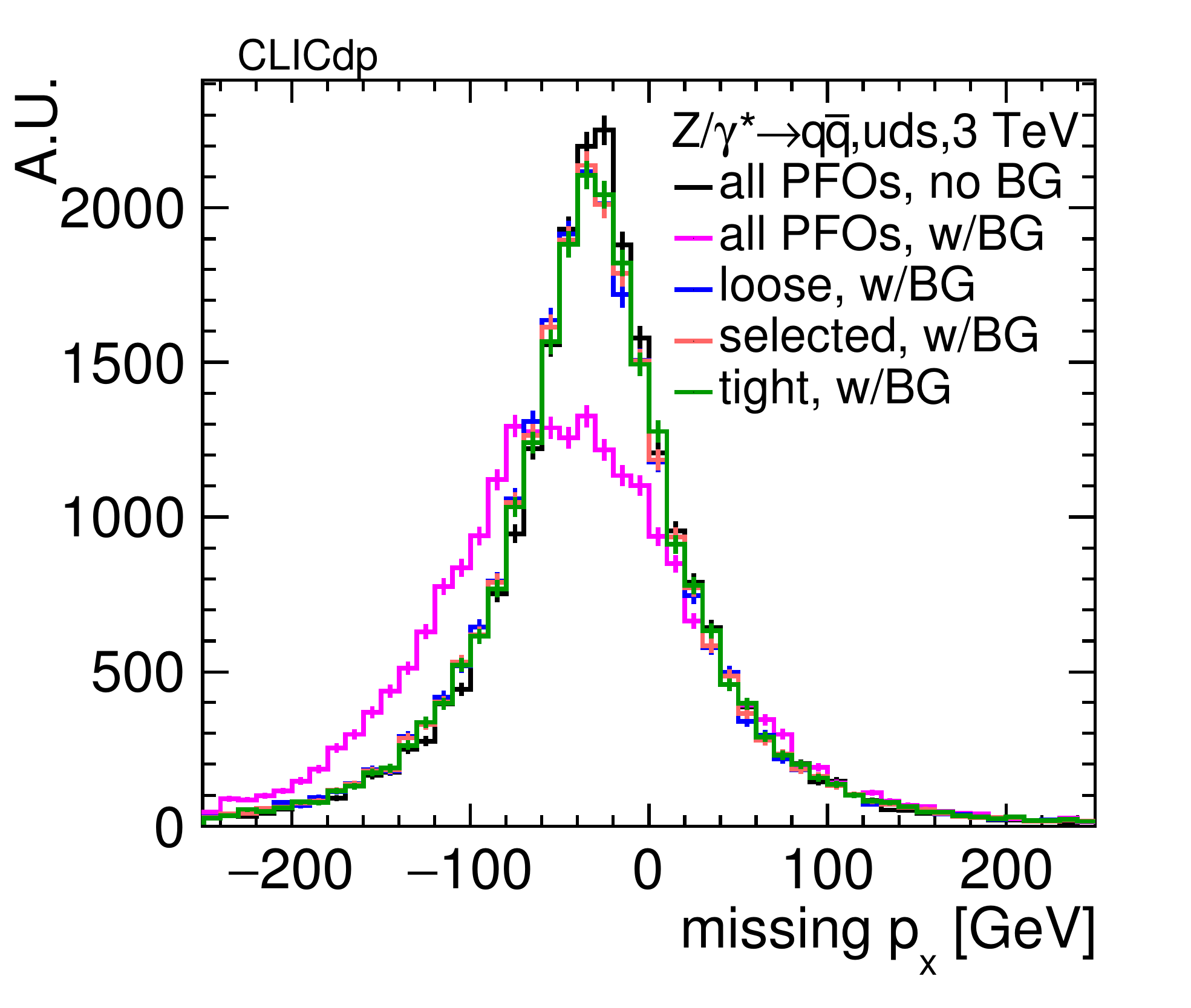}%
    \phantomsubcaption\label{fig:px_miss_Zuds3000_background_impact}
  \end{subfigure}
  \vspace{-2mm}
  \caption{Impact of \SI{3}{TeV} beam-induced \gghadron{} background overlay on the missing momentum distribution using
    different preselections on particle flow objects in semi- and di-leptonic $\ttbar$ events with genuine missing
    momentum~\subref{fig:ttbar3TeV_deltaMET_background_impact} and in light flavour di-jet events with no genuine
    missing momentum~\subref{fig:px_miss_Zuds3000_background_impact}.}
\end{figure}

The impact of beam background is studied at \SI{3}{TeV} using missing transverse momentum in light flavour di-jet events and events from semi- and di-leptonic $\ttbar$ with genuine missing energy due to neutrinos escaping detection in the detector.
The background is simulated using 30 bunch-crossings with on average 3.2 \gghadron{} events per bunch-crossing, overlaid on top of the physics event. 
At \SI{3}{TeV} around \SI{1.6}{TeV}
of additional energy is deposited  in the reconstruction time window on top of the underlying physics events. 
In order to mitigate beam background effects $p_{\mathrm{T}}$ and timing cuts (see \cref{sec:treatment-background}) have been developed for each reconstructed particle type. 
These aim to reduce the additional energy attributed to beam background to the level of about \SI{100}{GeV}\@.  \cref{fig:ttbar3TeV_deltaMET_background_impact}
shows the distribution of the difference between the true and the reconstructed missing transverse momentum originating from neutrinos. The preselection on centrally produced tops avoids a bias due to detector acceptance. 
The \gghadron{} background leads to a considerable deterioration of the missing transverse momentum resolution.
 Applying the $p_{\mathrm{T}}$ cuts and timing cuts on reconstructed particles improves the missing transverse momentum resolution. 
In events without genuine missing momentum the deterioration in the missing momentum resolution of 20\% is reduced to a level of 5\% after the selection cuts (\cref{fig:px_miss_Zuds3000_background_impact}).
The bias in the missing \pT{} comes from the Lorentz boost of the collision with respect to the detector frame of reference due to the crossing angle, there is a momentum bias towards positive x-direction.

\subsubsection{W and Z Mass Separation}\label{WZmassSeparation}

The precise reconstruction of masses of resonances in hadronic channels over wide ranges of energies is a challenging
task. The ability to separate di-jet masses from hadronic decays of W and Z bosons is studied using the Pandora
reconstruction algorithms. The study is carried out using simulated di-boson events, in which only one of the bosons
decays into di-quarks, i.e.\ $\PZ\PZ \to \PGn\PAGn\PQq\PAQq$\ and $\PW\PW\to \Pl\PGn\PQq\PQq$. The boson energies in this study vary from \SI{125}{GeV}, where both bosons are created almost at rest, up to \SI{1}{TeV}, where the bosons are strongly boosted. For each vector boson energy, samples were produced without background (no BG) and with \SI{3}{TeV} beam-induced backgrounds from \gghadrons overlaid, representing 30 bunch crossings (\SI{3}{TeV} BG). For low energy bosons of \SI{125}{GeV} the impact of \SI{380}{GeV} beam-induced backgrounds is also investigated. Events are reconstructed using the VLC algorithm with parameters $R=0.7$, $\gamma=\beta=1$ in exclusive mode, forcing the event into two jets. Prior to jet clustering at particle level, the true charged lepton from the W is removed (together with any associated photons from final state radiation and Bremsstrahlung). At detector level, all reconstructed particles within a cone of $|\cos\alpha|<0.9$ around the true lepton direction are removed prior to jet clustering. This procedure has virtually no impact on particles from the hadronically decaying W. At particle level, visible stable particles (excluding neutrinos) are used as input for the jet clustering. On the reconstruction level for the no BG samples, all Pandora particle flow objects are used as input for the jet clustering, while tight (low energy loose) selected Pandora particle flow objects are used in the samples including \SI{3}{TeV} (\SI{380}{GeV}) \gghadron backgrounds. To ensure that the event is well contained within the detector acceptance, a cut is imposed on the polar angle of both MC truth jets $|\cos\theta|<0.9$.

\begin{figure}[tbp]
  \renewcommand{\thesubfigure}{(\lr{subfigure})}
  \centering
  \begin{subfigure}{.5\textwidth}
    \centering
    \includegraphics[width=\linewidth]{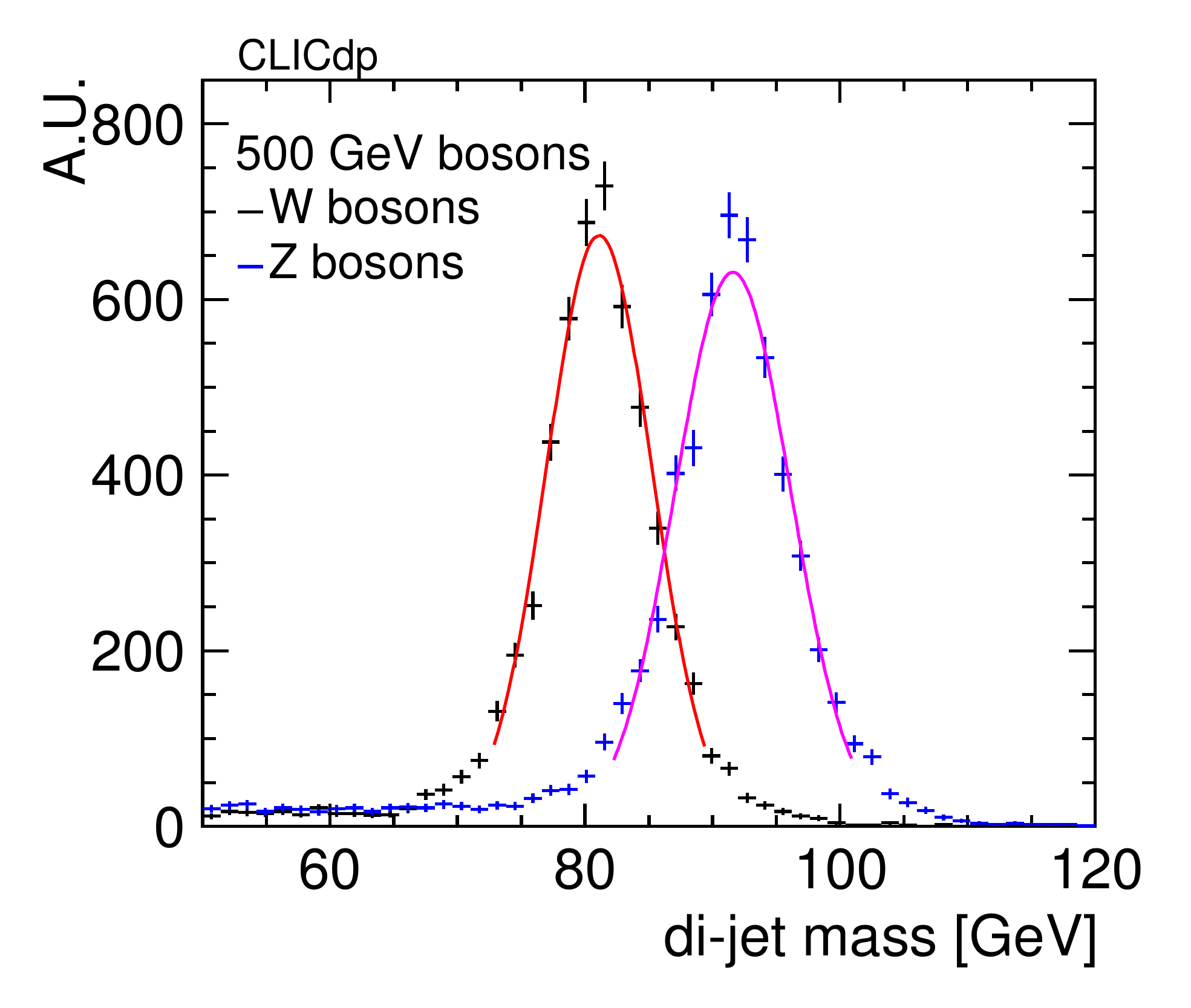}%
    \phantomsubcaption\label{fig:WW_ZZ_500_VLC7}
  \end{subfigure}%
  \begin{subfigure}{.5\textwidth}
    \centering
    \includegraphics[width=\linewidth]{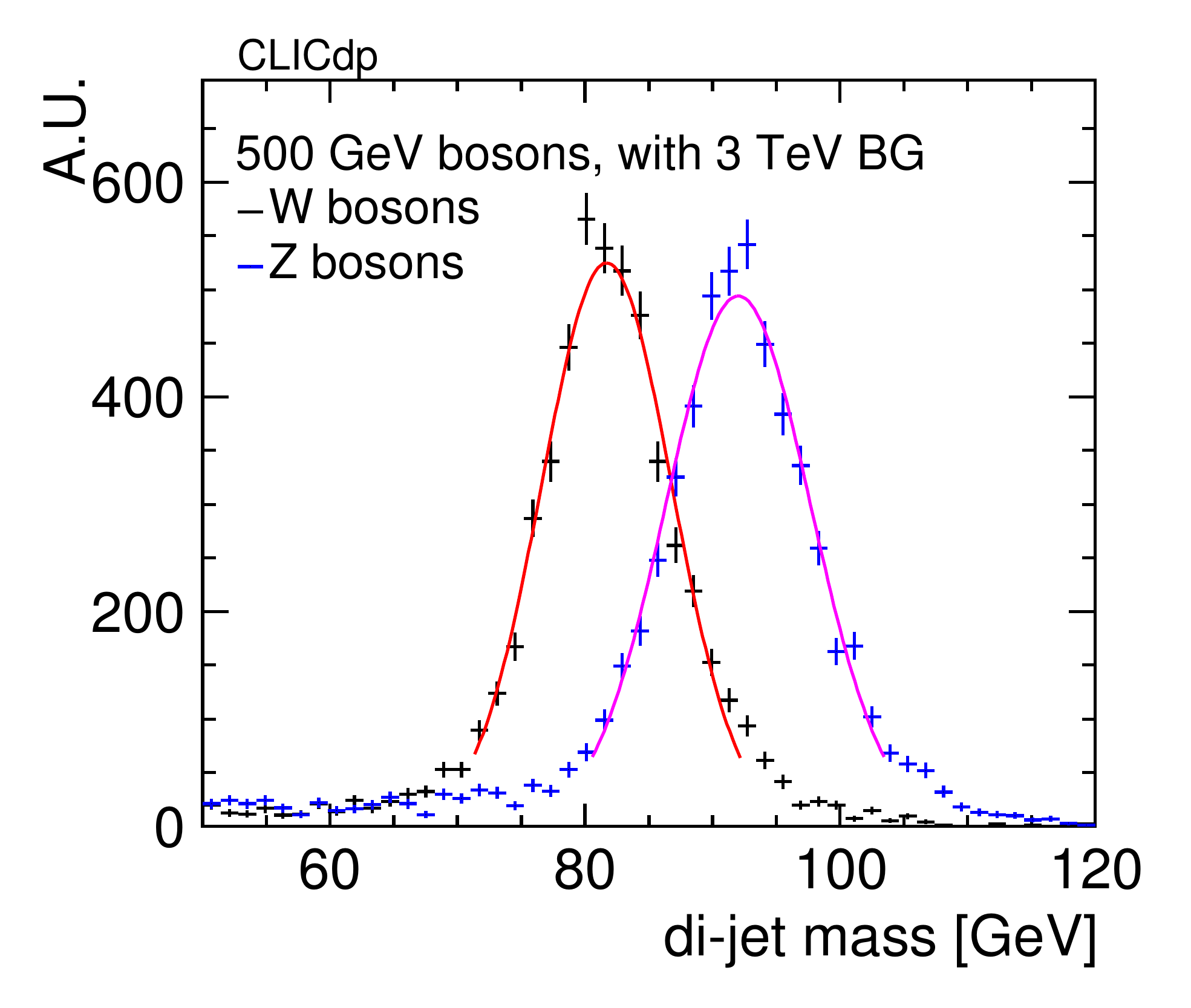}%
    \phantomsubcaption\label{fig:WW_ZZ_500_VLC7_3TeVBG}
  \end{subfigure}
  \vspace{-2mm}
  \caption{Di-jet mass distributions of hadronically decaying W and Z with $E=\SI{500}{GeV}$ in
    $\PW\PW\to\Pl\PGn\PQq\PQq$ and $\PZ\PZ\to \PGn\PAGn\PQq\PAQq$ events, together with Gaussian fits of the di-jet mass
    for events without beam-induced backgrounds~\subref{fig:WW_ZZ_500_VLC7} and overlay of \SI{3}{TeV} beam-induced
    backgrounds from \gghadrons~\subref{fig:WW_ZZ_500_VLC7_3TeVBG}.}\label{fig:WW_ZZ_500_DiJetMass}
\end{figure}

The upper tail and the core of the di-jet mass distribution is described well by a Gaussian function even without this additional preselection criteria for all energies. The di-jet distributions are fitted with a Gaussian, iteratively changing the upper limits of the fit range to 2~$\sigma$ and the lower fit limit to 2~$\sigma$ (1~$\sigma$ without preselection criteria applied to the unclustered energy ratio for \SI{125}{GeV} bosons), until the fitted $\sigma$ stabilises within 5\%. \cref{fig:WW_ZZ_500_DiJetMass} shows the di-jet mass distributions for W and Z bosons with $E=\SI{500}{GeV}$ with the Gaussian fits in events without and with the simulation of \SI{3}{TeV} beam-induced backgrounds from \gghadrons.

Since the di-jet mass distributions are not further calibrated at the moment, the mean of the fit is shifted to the W or Z
mass, fixing the ratio of $\sigma$/mean. The rescaled Gaussian distributions are normalised, such that the integral
of the distributions is 1. The overlap fraction $A_{\mathrm{O}}$ is then defined by
\newcommand{\dx}{\ensuremath{\mathrm{d}x}}
\begin{displaymath}
A_{\mathrm{O}}=\left(\int_{x_{\mathrm{int}}}^{\infty}\mathrm{gauss}_{\mathrm{W}}(x)\dx+\int_{-\infty}^{x_{\mathrm{int}}}\mathrm{gauss}_{\mathrm{Z}}(x)\dx\right)/2,
\end{displaymath}
where $x_{\mathrm{int}}$ is the intersection mass point of the Gaussian fit of the W and Z di-jet mass distributions between the W and Z masses. The efficiency $\epsilon$ of selecting W's or Z's are represented by the integrals of the Gaussian curves up to $x_{\mathrm{int}}$ for W's and from $x_{\mathrm{int}}$ onwards for Z's. The ideal Gaussian separation is evaluated using the quantile function with the normal distribution\footnote{separation calculation using \ROOT~6.08.00: $2\cdot |\mathrm{ROOT::Math::normal\_quantile}(A_{\mathrm{O}},1)|$}. A different approach using the average mass resolution $\sigma_{\mathrm{avg}}=(\sigma_{\mathrm{Z}}+\sigma_{\mathrm{W}})/2$ found the same results for the separation \mbox{$\mathcal{S}=(m_{\mathrm{Z}}-m_{\mathrm{W}})/\sigma_{\mathrm{avg}}$}. 

\begin{table}[tbp]
  \centering
  \caption{Mass resolutions, identification efficiencies, and separation of W and Z peaks for reconstructed W and Z's at
    different energies, with and without overlaid beam-induced backgrounds from \gghadrons. Tight PFO (Low energy loose)
    selection cuts are used for events with \SI{3}{TeV} (\SI{380}{GeV}) background.}\label{tab:WZ_mass_separation_values}
  \begin{tabular}{l c c c c c c c}
    \toprule{}
    Background & $E_{\mathrm{W,Z}}$ & $\sigma_{m(\mathrm{W})}/m(\mathrm{W})$ & $\sigma_{m(\mathrm{Z})}/m(\mathrm{Z})$ & $\epsilon$ & Separation  \\
               & [GeV] & [\%] & [\%] & [\%] &[$\sigma$] \\
    \midrule{}
    \multirow{4}{*}{no BG} & 125  & 5.5 & 5.3 & 88 & 2.3\\
               & 250  & 5.3 & 5.4 & 88 & 2.3\\
               & 500  & 5.1 & 4.9 & 90 & 2.5\\
               & 1000 & 6.6 & 6.2 & 84 & 2.0\\\midrule{}
    \multirow{4}{*}{\SI{3}{TeV} BG} & 125  & 7.8 & 7.1 & 80 & 1.7\\
               & 250  & 6.9 & 6.8 & 82 & 1.8\\
               & 500  & 6.2 & 6.1 & 85 & 2.0\\
               & 1000 & 7.9 & 7.2 & 80 & 1.7\\\midrule{}
    \SI{380}{GeV} BG             & 125  & 6.0 & 5.5 & 87 & 2.2\\
    \bottomrule{}
  \end{tabular}
\end{table}

The di-jet mass resolutions are listed in \cref{tab:WZ_mass_separation_values}, together with identification efficiencies and the separation between W and Z peaks. In events without beam-induced background effects the selection efficiency is between 84\% and 90\%, which corresponds to an overlap fraction of 10--16\%. For very boosted bosons the mass resolution is slightly worse than for bosons at rest or with moderate energies. Once \SI{3}{TeV} beam-induced backgrounds from \gghadrons are taken into account (\SI{3}{TeV} BG), the W and Z selection efficiencies decrease to 80--85\%, which corresponds to overlap fractions of 15--20\%. The degradation is worse for lower boson energies. The peak separation diminishes from 2.0--2.5~$\sigma$ to about 1.7--2.0~$\sigma$ in the presence of beam-induced background levels of the \SI{3}{TeV} accelerator. As alternatives, \emph{loose}, \emph{default} and no selection cuts have been applied to PandoraPFOs prior to jet clustering in simulated events with \SI{3}{TeV} beam-induced backgrounds. These three alternative selection cuts led to wider di-jet mass distributions and a diminished separation power between the two mass peaks. For \SI{380}{GeV} beam-induced background levels there is a mild effect on the separation power, decreasing from $2.3~\sigma$ to $2.2~\sigma$.

\subsubsection{Flavour Tagging}\label{sec:flavTag}

Flavour tagging studies were performed initially in the CDR~\cite[Section 12.3.4]{cdrvol2} for the \clicsid detector model. 
These studies were extended~\cite{Alipour_Roloff_2014} to explore more realistic vertex detector geometries, 
e.g.\ using spirals instead of forward disks, thus allowing for air flow cooling of the vertex detector system.
One of the main findings is related to the material budget: doubling the material per layer in the vertex detector
(as required for the more realistic design adapted for CLICdet) leads to a degradation of the flavour tagging performance. 

\begin{figure}[tbp]
   \renewcommand{\thesubfigure}{(\lr{subfigure})}
   \centering
   \begin{subfigure}{.5\textwidth}
     \centering
     \includegraphics[width=\linewidth]{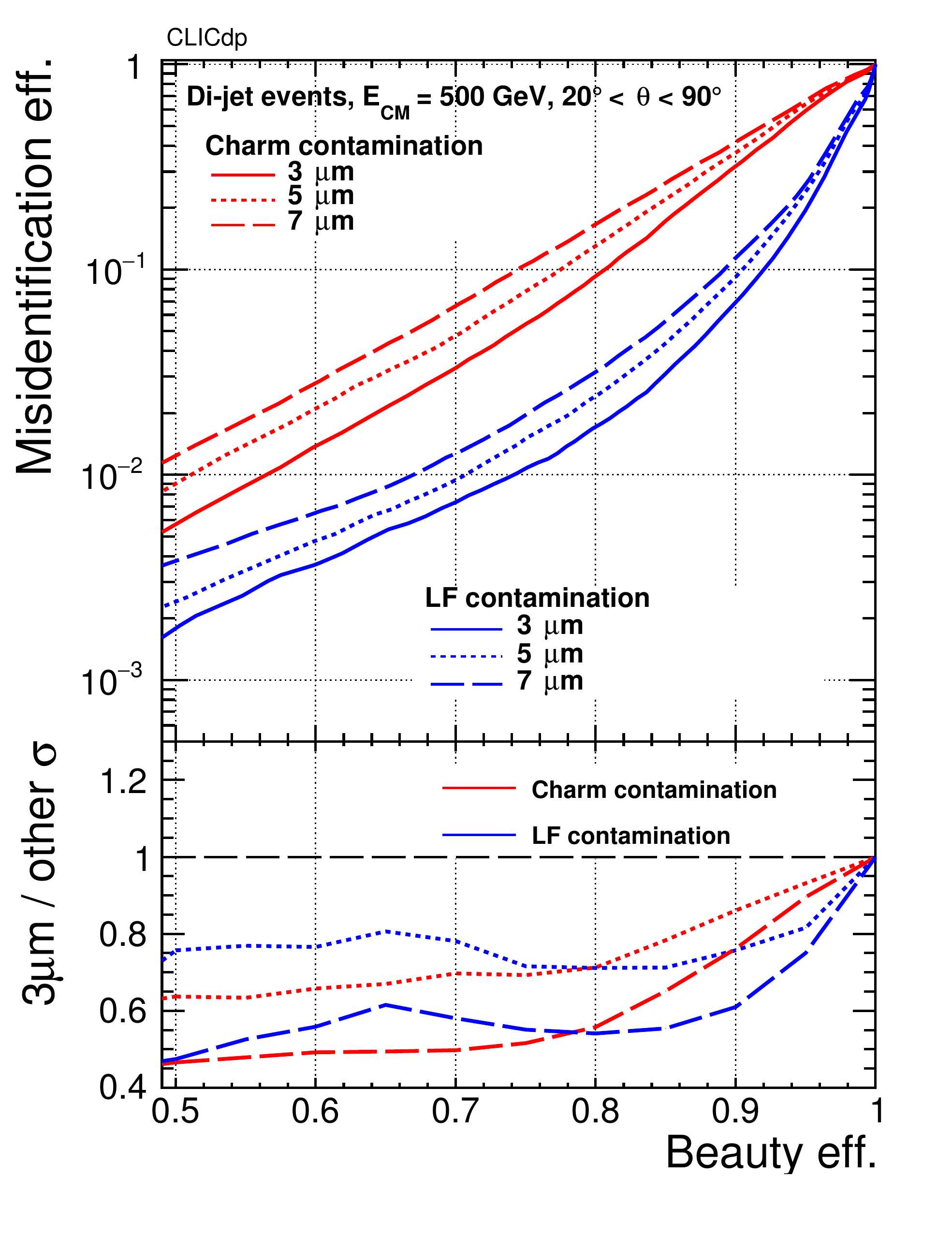}%
     \phantomsubcaption\label{fig:b_mis_eff_spr}
   \end{subfigure}%
   \begin{subfigure}{.5\textwidth}
     \centering
     \includegraphics[width=\linewidth]{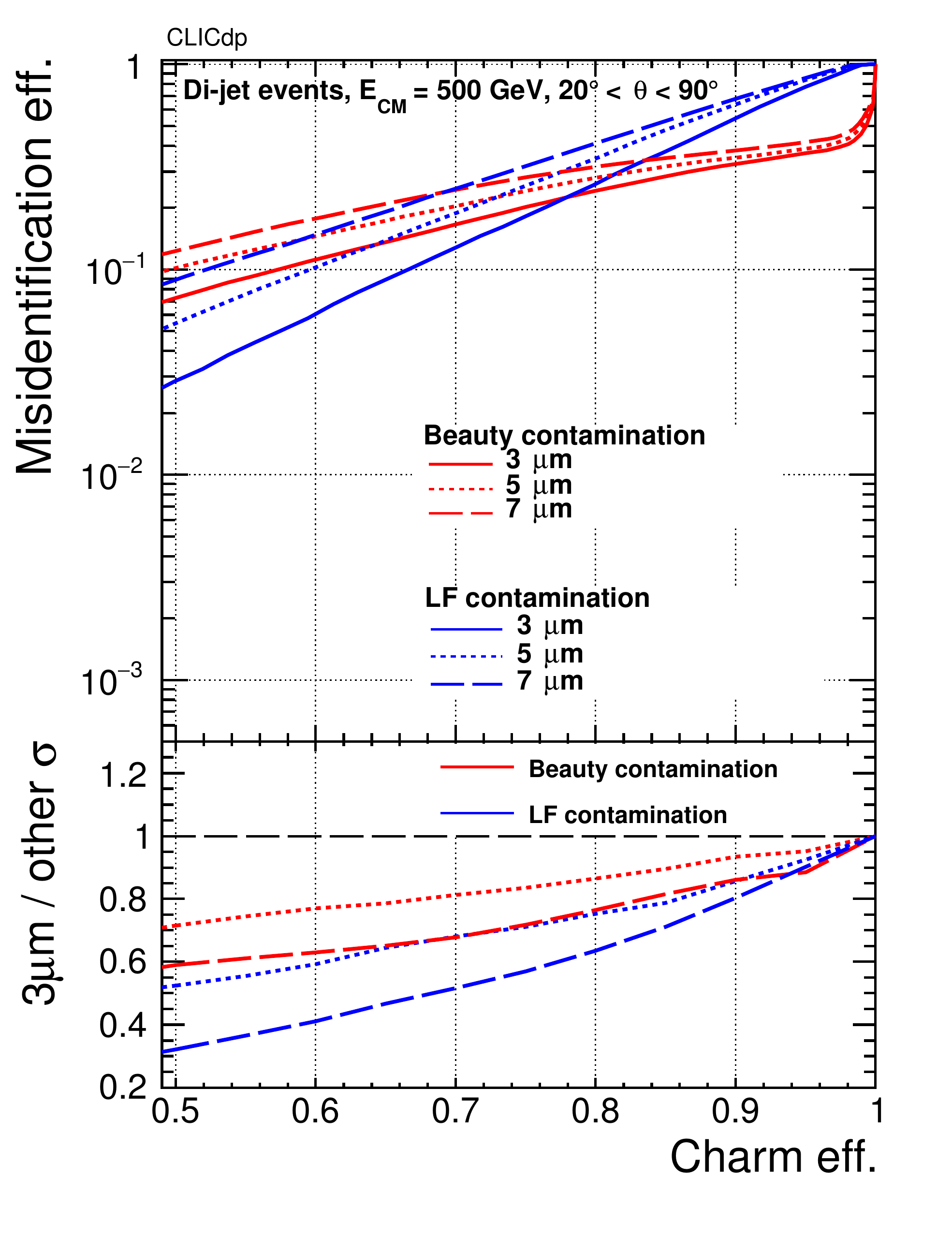}%
     \phantomsubcaption\label{fig:c_mis_eff_spr}
   \end{subfigure}
   \vspace{-5mm}
   \caption{Global performance of beauty tagging~\subref{fig:b_mis_eff_spr} and charm tagging~\subref{fig:c_mis_eff_spr}
     for jets in di-jet events at $\roots = \SI{500}{GeV}$ with a mixture of polar angles between 20\degrees{} and
     90\degrees{}. A comparison of performance obtained with different single point resolutions in the vertex detector
     is presented. On the y-axis, the misidentification probability and the ratio of misidentification probabilities
     with respect to the nominal (\SI{3}{\micron}) single point resolution are given.}\label{fig:beff_ceff_spr}
 \end{figure}

The vertex detector geometry of CLICdet has recently been used for additional studies, using the software chain described in \cref{sec:simreco} and the flavour tagging package LCFIPlus~\cite{Suehara:2015ura}. 
Complementing the above-mentioned exploratory studies, as a first step the dependence of performance on the assumed single point resolution in the vertex detector layers was investigated.
The combined impact on flavour tagging performance, when worsening the single point resolution from the nominal \SI{3}{\micron} to \SI{5}{\micron} or \SI{7}{\micron},
was investigated using di-jet samples at \SI{500}{GeV}\@ and is shown in \cref{fig:beff_ceff_spr}. As expected,
 the results get worse for the less performant vertexing. 

\begin{figure}[tbp]
  \renewcommand{\thesubfigure}{(\lr{subfigure})}
  \centering
  \begin{subfigure}{.5\textwidth}
    \centering
    \includegraphics[width=\linewidth]{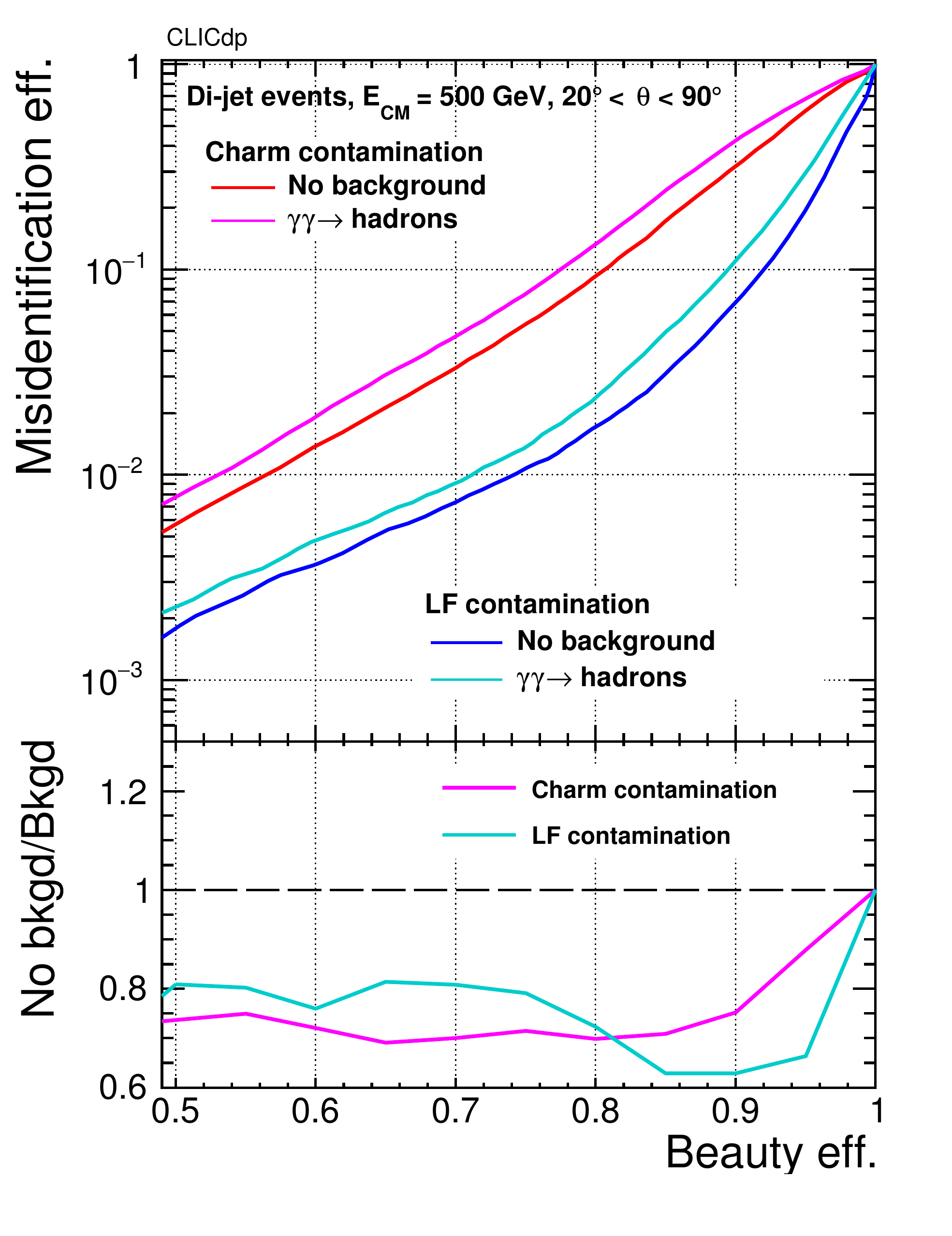}%
    \phantomsubcaption\label{fig:b_mis_eff}
  \end{subfigure}%
  \begin{subfigure}{.5\textwidth}
    \centering
    \includegraphics[width=\linewidth]{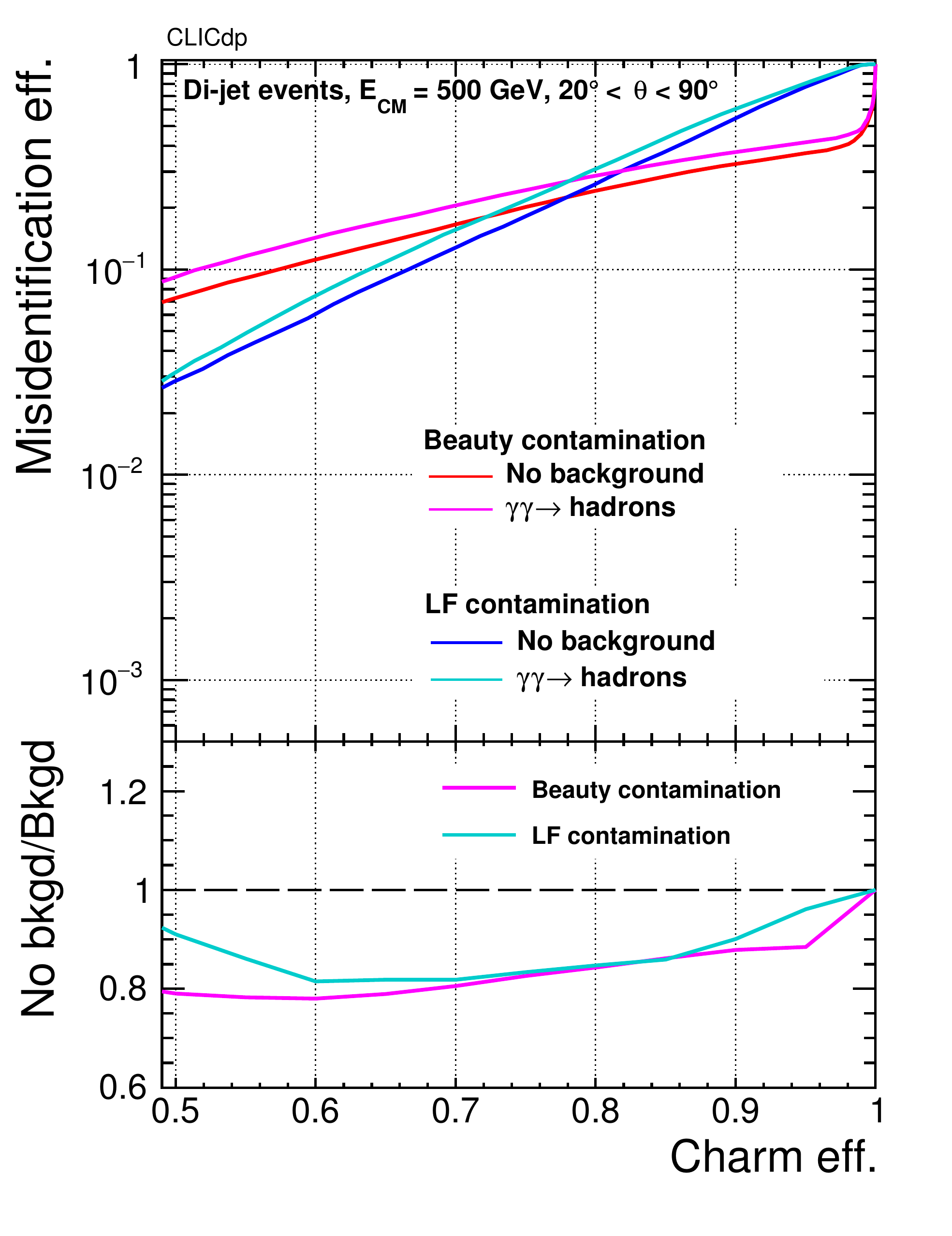}%
    \phantomsubcaption\label{fig:c_mis_eff}
  \end{subfigure}
  \vspace{-5mm}
  \caption{Global performance of beauty tagging~\subref{fig:b_mis_eff} and charm tagging~\subref{fig:c_mis_eff} for jets
    in di-jet events at $\roots = \SI{500}{GeV}$ with a mixture of polar angles between 20\degrees{} and 90\degrees{}. A
    comparison of performance with and without \gghadron{} background is presented. On the y-axis, the misidentification
    probability and the ratio of the misidentification probabilities with and without \gghadron{} background are
    given.}\label{fig:beff_ceff_background}
\end{figure}

The overall performance of flavour tagging at CLICdet has been tested using the nominal parameters (i.e.\ \SI{3}{\micron} position resolution),
and comparing the beauty and charm tagging results without and with overlay of \gghadron{} background corresponding to 30 bunch crossings.
\cref{fig:beff_ceff_background} shows the results obtained.
The beauty misidentification probability (\cref{fig:b_mis_eff}) is assessed separately for charm and light-flavour contamination. At 80\% beauty identification efficiency, the misidentification amounts to 10\% as charm and 1.5\% as light-flavour jets. When \SI{3}{TeV} \gghadron{} background is overlaid, it amounts to 13\% and 2\% for charm and light-flavour respectively.
Similarly, the charm misidentification probability (\cref{fig:c_mis_eff}) is assessed for beauty and light-flavour contamination separately. At 80\% charm identification efficiency, the misidentification amounts to 25\% as beauty as well as light-flavour jets. When \SI{3}{TeV} \gghadron{} background is overlaid, it amounts to 30\% for both types of contamination.



\begin{figure}[htbp]
  \renewcommand{\thesubfigure}{(\lr{subfigure})}
  \centering
  \begin{subfigure}{.5\textwidth}
    \centering
    \includegraphics[width=\linewidth]{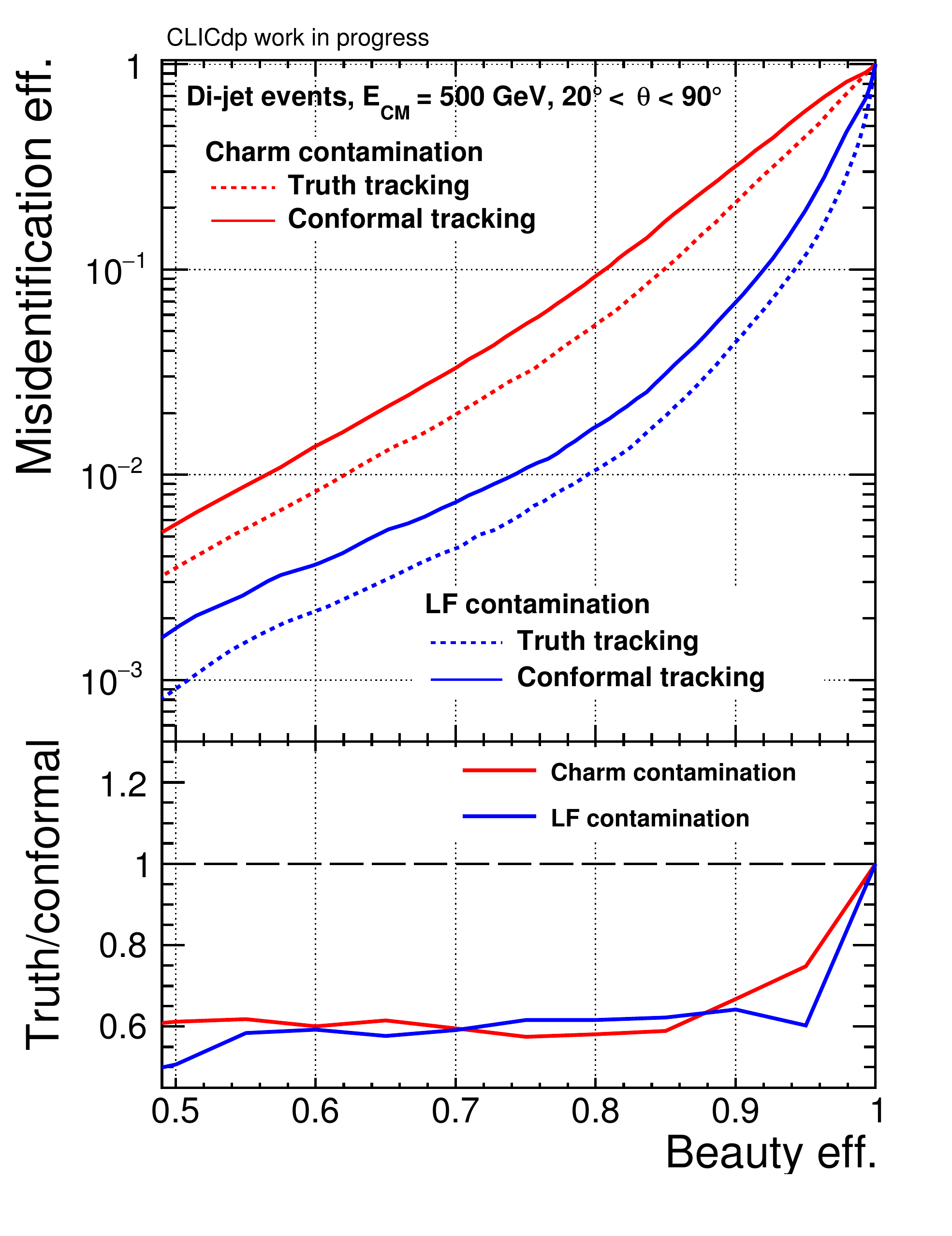}%
    \phantomsubcaption\label{fig:b_eff_TTCT}
  \end{subfigure}%
  \begin{subfigure}{.5\textwidth}
    \centering
    \includegraphics[width=\linewidth]{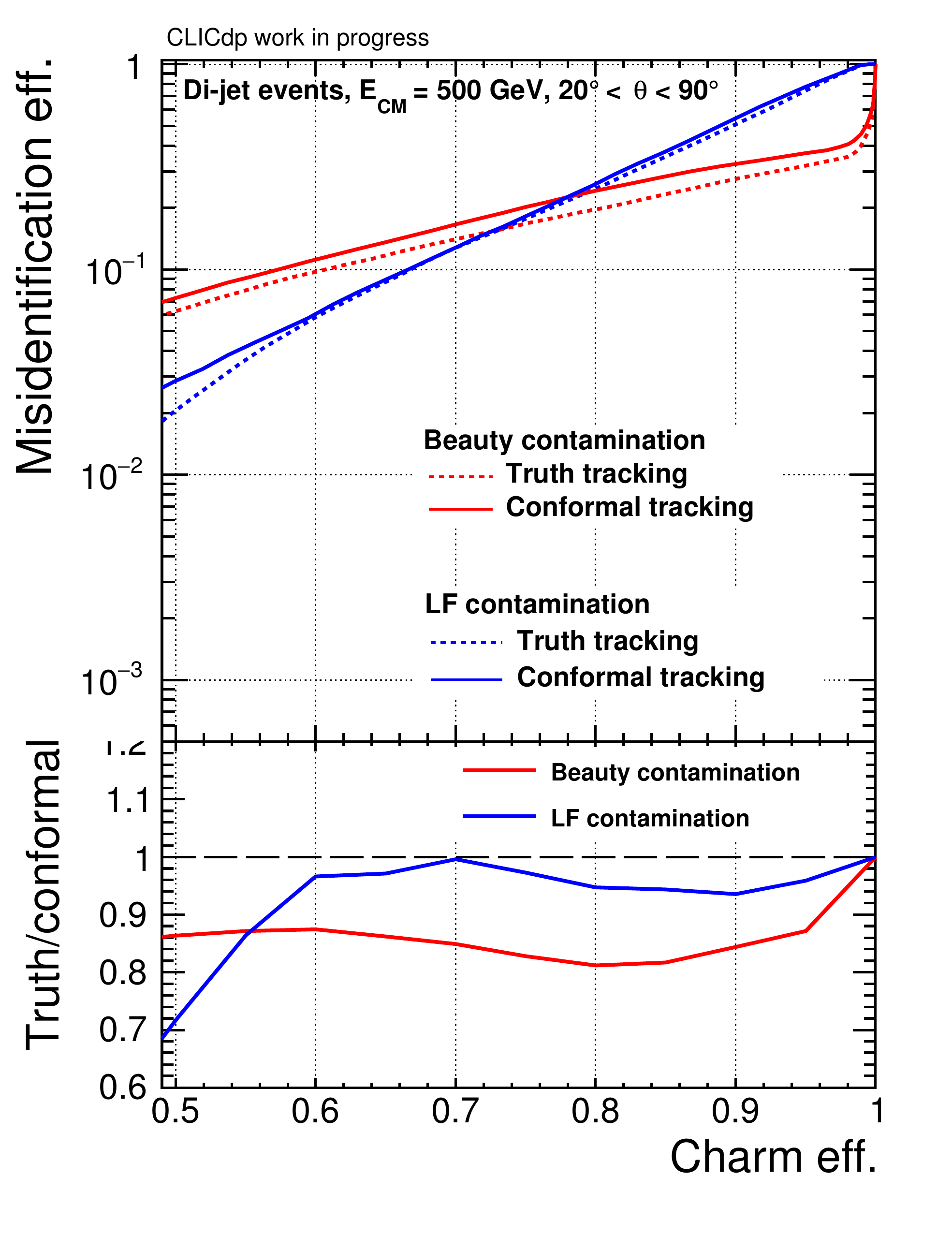}%
    \phantomsubcaption\label{fig:c_eff_TTCT}
  \end{subfigure}
  \vspace{-5mm}
  \caption{Performances of beauty tagging~\subref{fig:b_eff_TTCT} and charm tagging~\subref{fig:c_eff_TTCT} for jets in
    di-jet events at $\roots = \SI{500}{GeV}$ with a mixture of polar angles between 20\degrees{} and 90\degrees{}. A
    comparison of performance using TruthTracking with respect to ConformalTracking is shown. On the y-axis, the
    misidentification probability and the ratio of the misidentification probabilities with and without \gghadron{}
    background are given.}\label{fig:beff_ceff_TTCT}
\end{figure}

In order to estimate the impact of track reconstruction on the flavour tagging, the same study has been performed using the true (Monte Carlo) pattern recognition, dubbed \emph{TruthTracking}. The results, shown in  \cref{fig:beff_ceff_TTCT}, indicate that both beauty and charm tagging can be improved by optimising the pattern recognition.
In particular, in beauty tagging a reduction of misidentification of a b-quark as a c- or light quark by 20\% to 30\% can be expected.


\clearpage
\subsubsection{Performance of Very Forward Calorimetry}\label{sec:VF_perf}

The efficiencies and fake rates for the LumiCal and BeamCal were studied with
mono-energetic electrons of \SIlist{10;25;50;100;190;250;500;1000;1500}{GeV} created in the polar angle range from \SI{9}{mrad} to
\SI{120}{mrad}. The reconstruction was done with incoherent pair background for the CLIC beams of $\roots=\SI{380}{GeV}$ with $\lstar=\SI{6}{m}$ and
$\roots=\SI{3}{TeV}$ with $\lstar=\SI{6}{m}$. Because both the
LumiCal and BeamCal will have to integrate the signal over multiple bunch
crossings, backgrounds equal to \SI{20}{ns} or 40 bunch crossings were overlaid over
each signal electron. Energy deposits for each background event have been
recorded and the standard deviation of the distribution calculated.
For the LumiCal the pulse height threshold for pad selection was chosen
to be 10 standard deviations, and 3 standard deviations for the BeamCal. 
Details of the analysis, using the ``variable energy selection'' procedure for pad selection, 
as well as the methods for tower selection and cluster location are described in~\cite{sailer:bcalreco}.
The reconstruction criteria can be adapted, depending on the requirements of the physics
analysis, whether higher efficiency or lower fake rates are more beneficial.

For the estimate of the efficiencies and fake rates, an electron is considered to be reconstructed if the angle and
energy are matching between the generated electron and the reconstructed cluster. If no matching electron exists a
cluster is considered to be \emph{fake}. The reconstructed cluster has to match the generated electron by
\SI{5}{mrad} in polar angle $\theta$, for the azimuthal angle $\phi$ the conditions\footnote{This approximately corresponds to |$\phi_{\mathrm{rec}}-\phi_{\mathrm{MC}}| = 25\degrees$, but also takes into account the problem of 0\degrees{} being equivalent to 360\degrees.}
\mbox{$|\cos\phi_{\mathrm{rec}}-\cos\phi_{\mathrm{MC}}| < 0.35$} and \mbox{$|\sin\phi_{\mathrm{rec}}-\sin\phi_{\mathrm{MC}}| <
0.35$}
 need to be fulfilled, and the energy needs to be within 50\% of the generated value. The conditions
are aimed at the worse resolutions of the BeamCal due to the large amount of incoherent pairs affecting the angular and
energy resolutions.

\begin{figure}[tbp]
  \begin{subfigure}{\subfigwidth}\centering
    \includegraphics[width=1.0\textwidth]{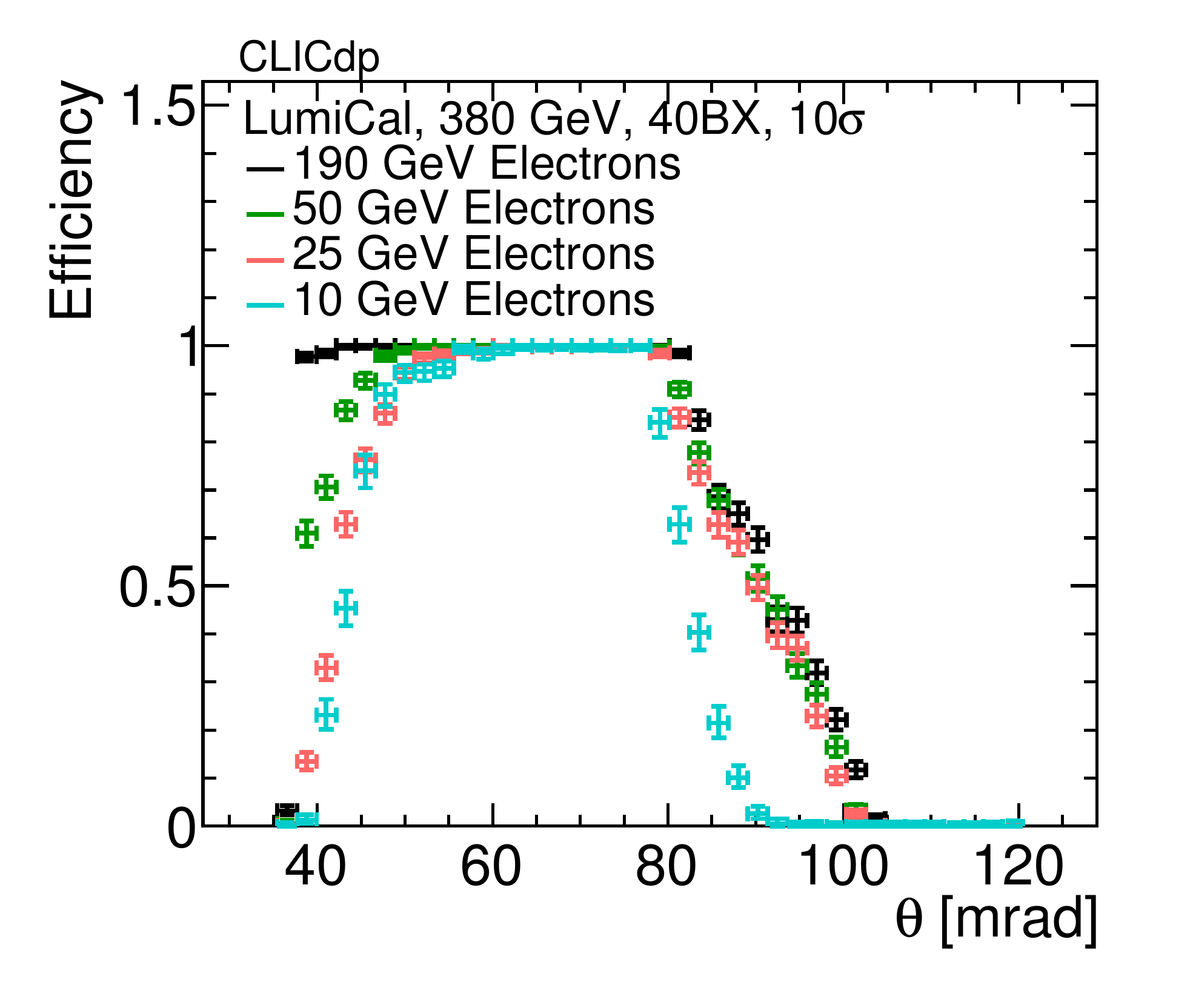}%
  \end{subfigure}
  \begin{subfigure}{\subfigwidth}\centering
    \includegraphics[width=1.0\textwidth]{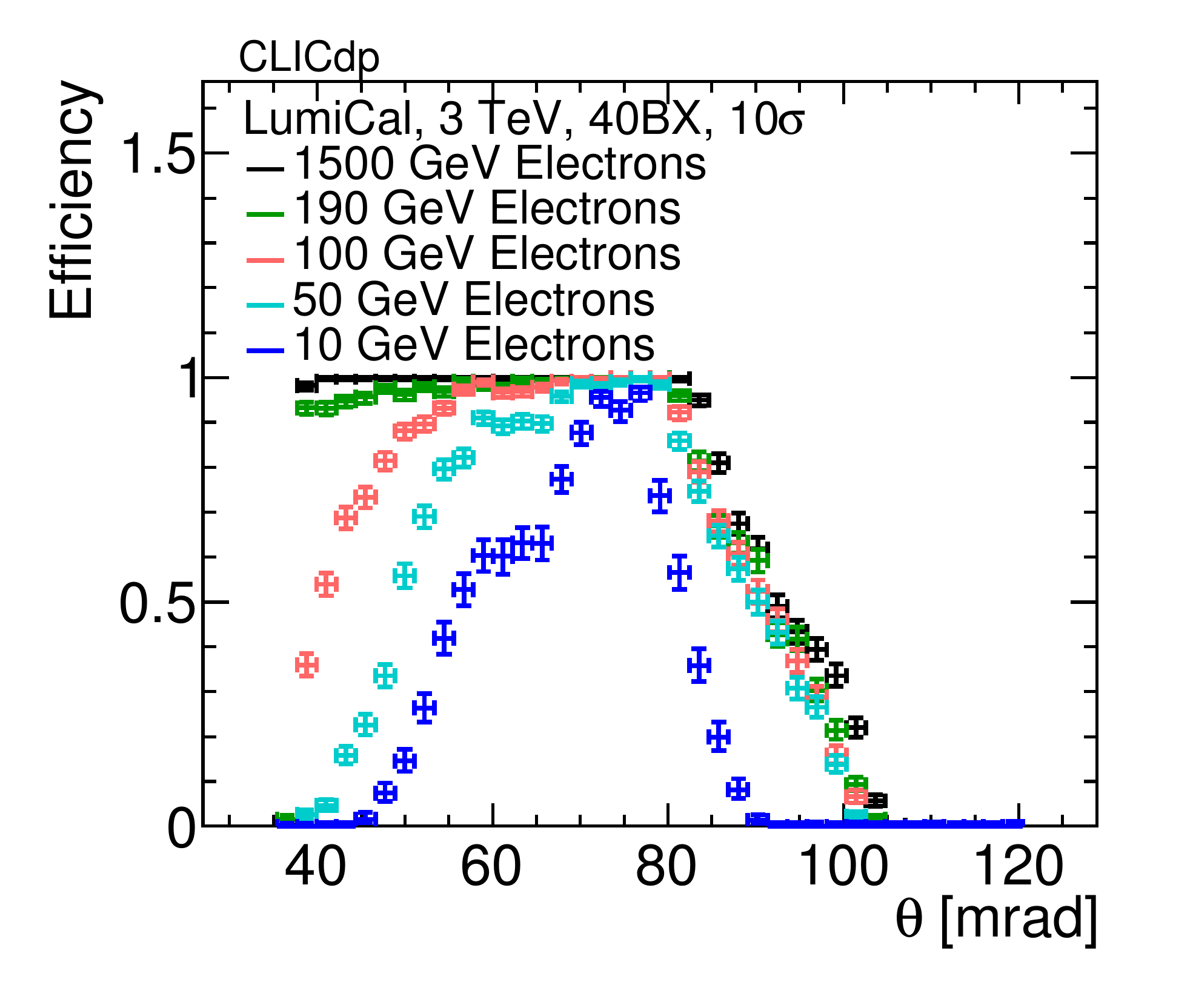}%
  \end{subfigure}
  \vspace{-5mm}
  \caption{Efficiency of electron detection as a function of the polar angle in the LumiCal for \SI{380}{GeV}~(left) and \SI{3}{TeV} (right) collisions.\label{fig:lumiEff}}
\end{figure}

The efficiency for the electron reconstruction in the LumiCal is shown in
\cref{fig:lumiEff}. The \SI{190}{GeV} electrons, the highest energy for the \mbox{$\roots=\SI{380}{GeV}$} case,
are well reconstructed between \SI{40}{mrad} and \SI{85}{mrad}. At the lower edge of
the LumiCal the angular and energy resolutions degrade due to leakage. Starting
at about \SI{85}{mrad} the beam pipe close to the vertex detector starts to shadow the
LumiCal, which causes the electrons to shower and degrades the reconstruction
efficiency. Due to the background and the selection of the pads with energy deposits
significantly above the average backgrounds, the reconstruction
efficiency for lower energy electrons is degraded at the inner edge of the
LumiCal.

For the $\roots=\SI{3}{TeV}$ case electrons between \SI{1.5}{TeV} and \SI{190}{GeV} are reconstructed with
an efficiency above 90\%, for all angles until the LumiCal is shadowed by the beam pipe. For
lower energies the background impacts the reconstruction efficiency more and
more. \SI{10}{GeV} electrons can only be reconstructed with 90\% efficiency in
the narrow angular region from \SI{70}{mrad} to \SI{80}{mrad}.

\begin{figure}[tbp]
  \begin{subfigure}{\subfigwidth}\centering
    \includegraphics[width=1.0\textwidth]{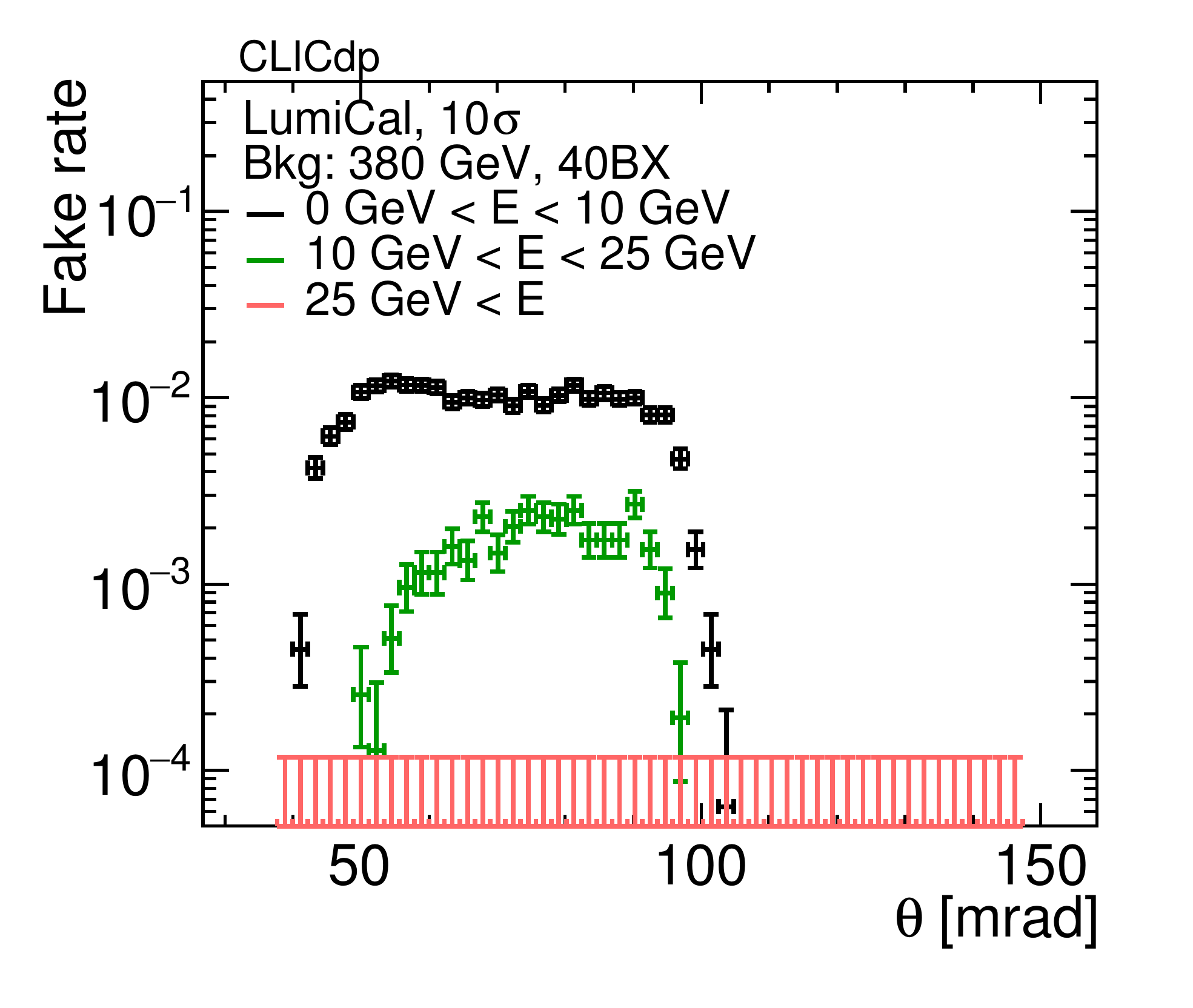}
  \end{subfigure}
  \begin{subfigure}{\subfigwidth}\centering
    \includegraphics[width=1.0\textwidth]{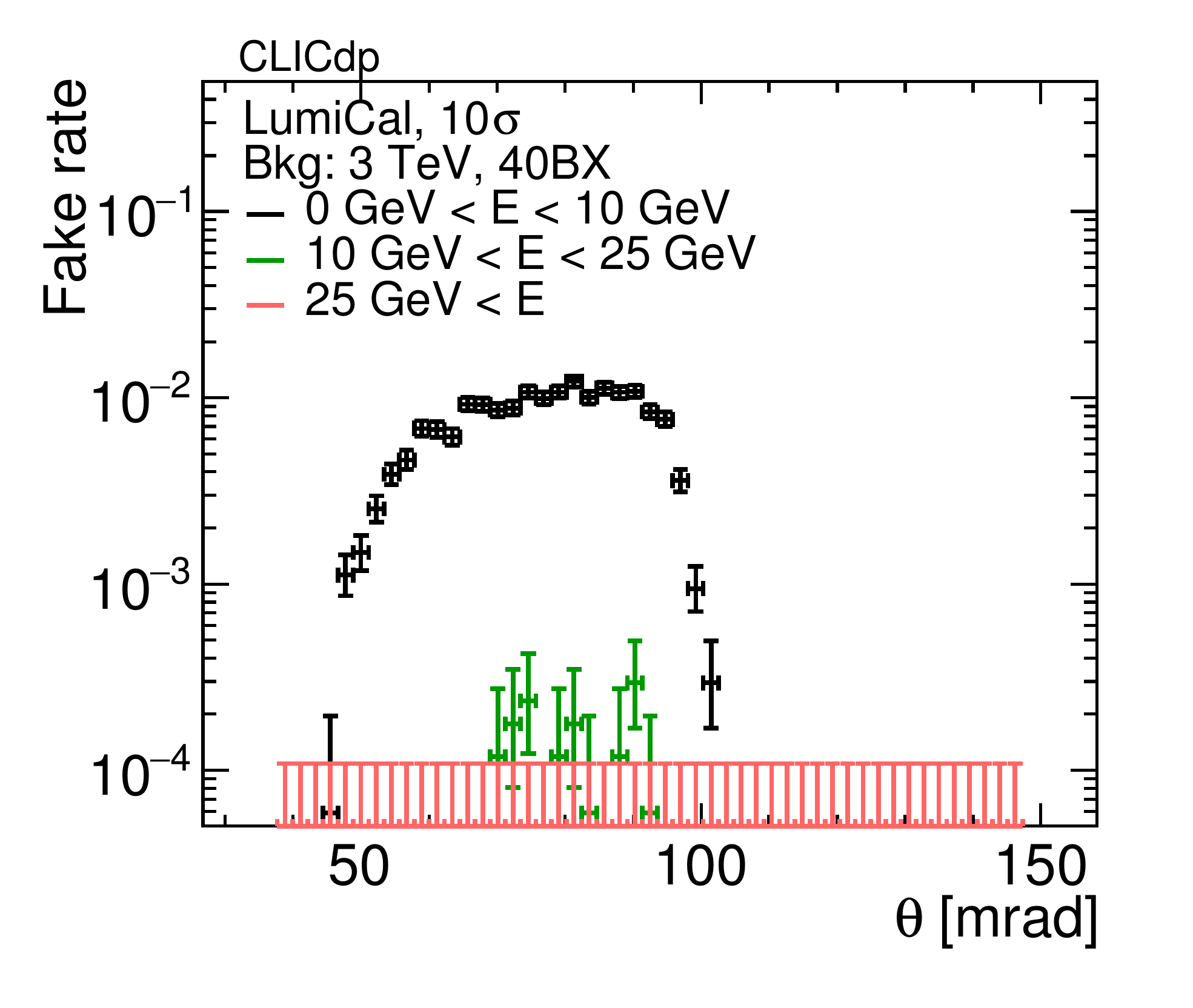}
  \end{subfigure}
  \vspace{-5mm}
  \caption{Fake rate for electron reconstruction as a function of the polar
    angle in the LumiCal at \SI{380}{GeV} (left) and \SI{3}{TeV} (right).\label{fig:lumiFake}}
\end{figure}

The achievable efficiencies to identify signal electrons are closely related to the acceptable fake
rates (i.e.\ a background event is identified as signal electron), which are shown in \cref{fig:lumiFake}.
Both for \SI{380}{GeV} and \SI{3}{TeV} CLIC
the fake rate for energies up to \SI{10}{GeV} is around 1\% between \SI{50}{mrad}
and \SI{80}{mrad}. For a similar angular region, a fake rate of a few permille
is observed for \SI{10}{GeV} to \SI{25}{GeV} clusters, and the probability is below $10^{-4}$ to find
any fake clusters above \SI{25}{GeV}\@. For the absolute luminosity measurement the highest
energy electrons are important, which are already reconstructed with high
efficiency using the procedure described above. For studies that use the LumiCal to complement the electromagnetic
calorimeter coverage further optimisation is possible.

\begin{figure}[tbp]
  \centering
  \begin{subfigure}{\subfigwidth}\centering
    \includegraphics[width=1.0\textwidth]{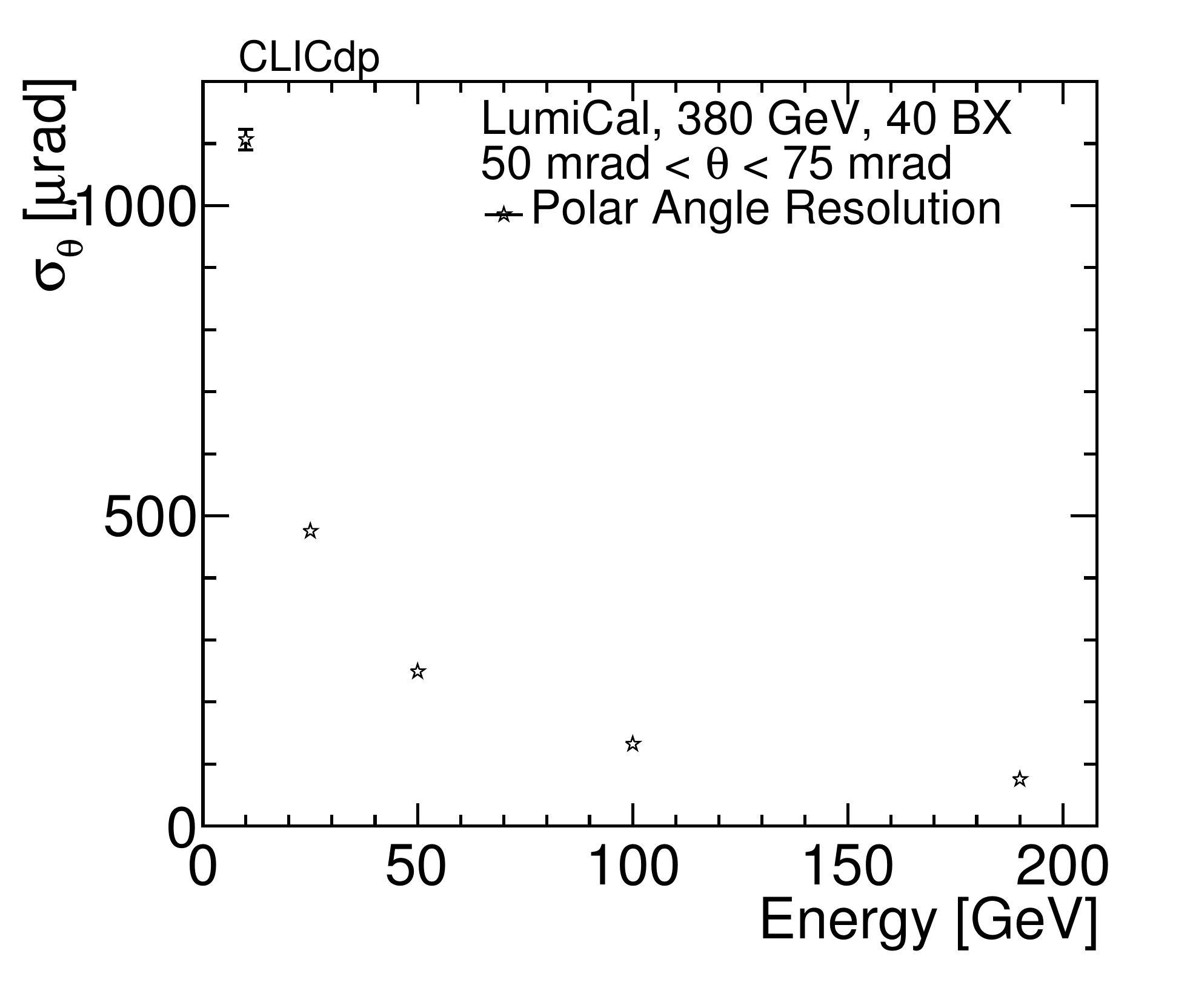}
  \end{subfigure}
  \begin{subfigure}{\subfigwidth}\centering
    \includegraphics[width=1.0\textwidth]{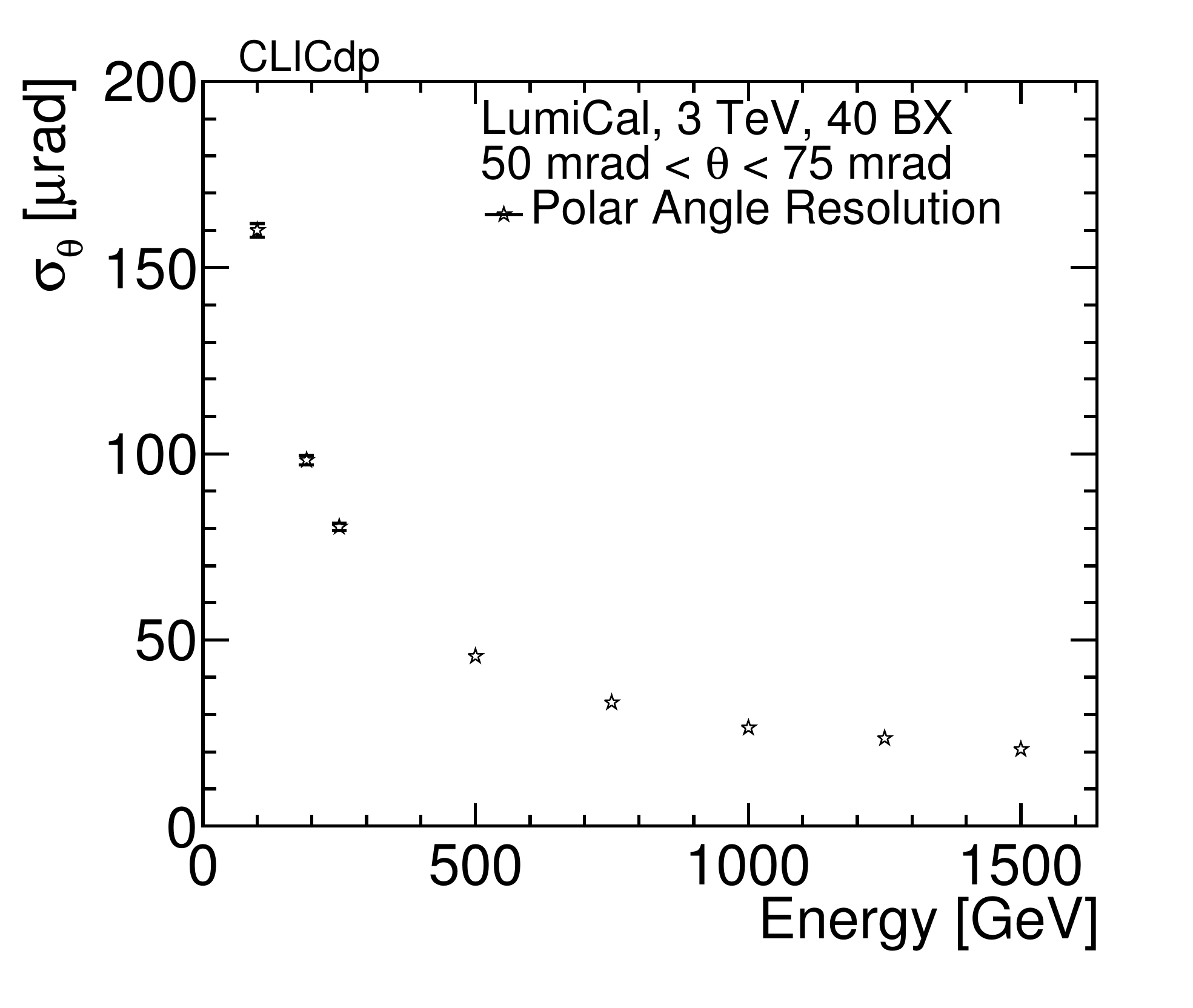}
  \end{subfigure}
  \vspace{-5mm}
  \caption{Polar angle resolution of  reconstructed electrons as a function of
    the electron energy in the LumiCal\label{fig:lumiResT} at \SI{380}{GeV} (left) and \SI{3}{TeV} (right).}
\end{figure}

For an optimal luminosity measurement the polar angle reconstruction plays a
very important role. \Cref{fig:lumiResT} shows the polar angle resolution $\sigma_{\theta}$ for
well reconstructed electrons in the polar angle range between \SI{50}{mrad} and
\SI{75}{mrad}. For \SI{1.5}{TeV} electrons a polar angle resolution of about 20~\microrad{}
is achieved. The resolutions are obtained from Gaussian fits to the distribution of
$\theta_{\mathrm{rec}}-\theta_{\mathrm{MC}}$ for electron energies above \SI{100}{GeV}.
Below this energy the distribution is non-Gaussian and the RMS of the distribution is used instead.

\begin{figure}[tbp]
  \begin{subfigure}{\subfigwidth}\centering
    \includegraphics[width=1.0\textwidth]{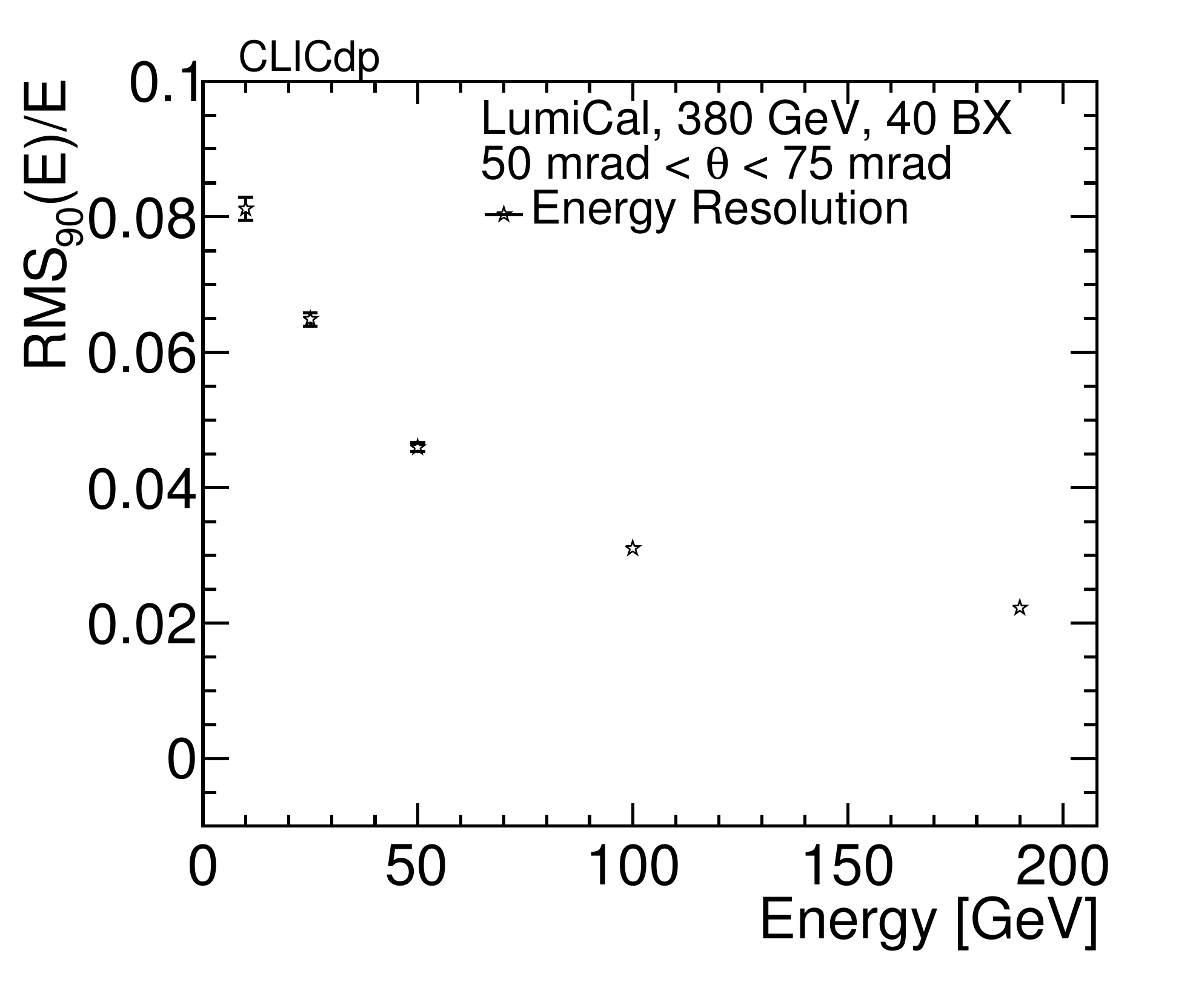}
  \end{subfigure}
  \begin{subfigure}{\subfigwidth}\centering
    \includegraphics[width=1.0\textwidth]{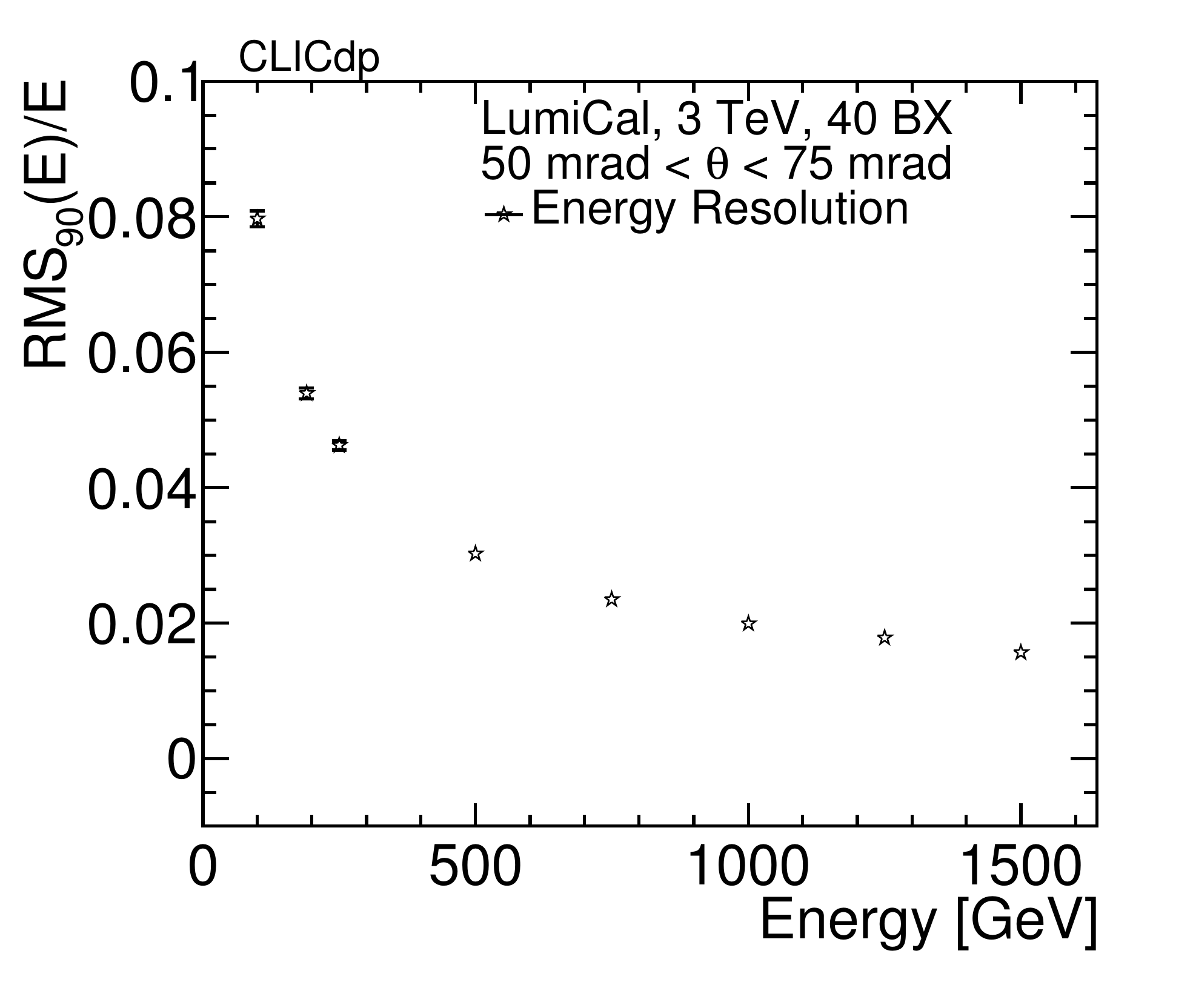}
  \end{subfigure}
  \vspace{-5mm}
  \caption{Energy resolution of reconstructed electrons as a function of the
    electron energy in the LumiCal\label{fig:lumiResE} at \SI{380}{GeV} (left) and \SI{3}{TeV} (right).}
\end{figure}

The energy resolution for the LumiCal is shown in \cref{fig:lumiResE}. The \rmsninety of the distribution
\mbox{$E_{\mathrm{rec}}-E_{\mathrm{MC}}$} is used for all energies, due to the long non-Gaussian tails introduced by the pad
selection criteria.

\begin{figure}[tbp]
  \renewcommand{\thesubfigure}{(\lr{subfigure})}
  \begin{subfigure}{\subfigwidth}\centering
    \includegraphics[width=1.0\textwidth]{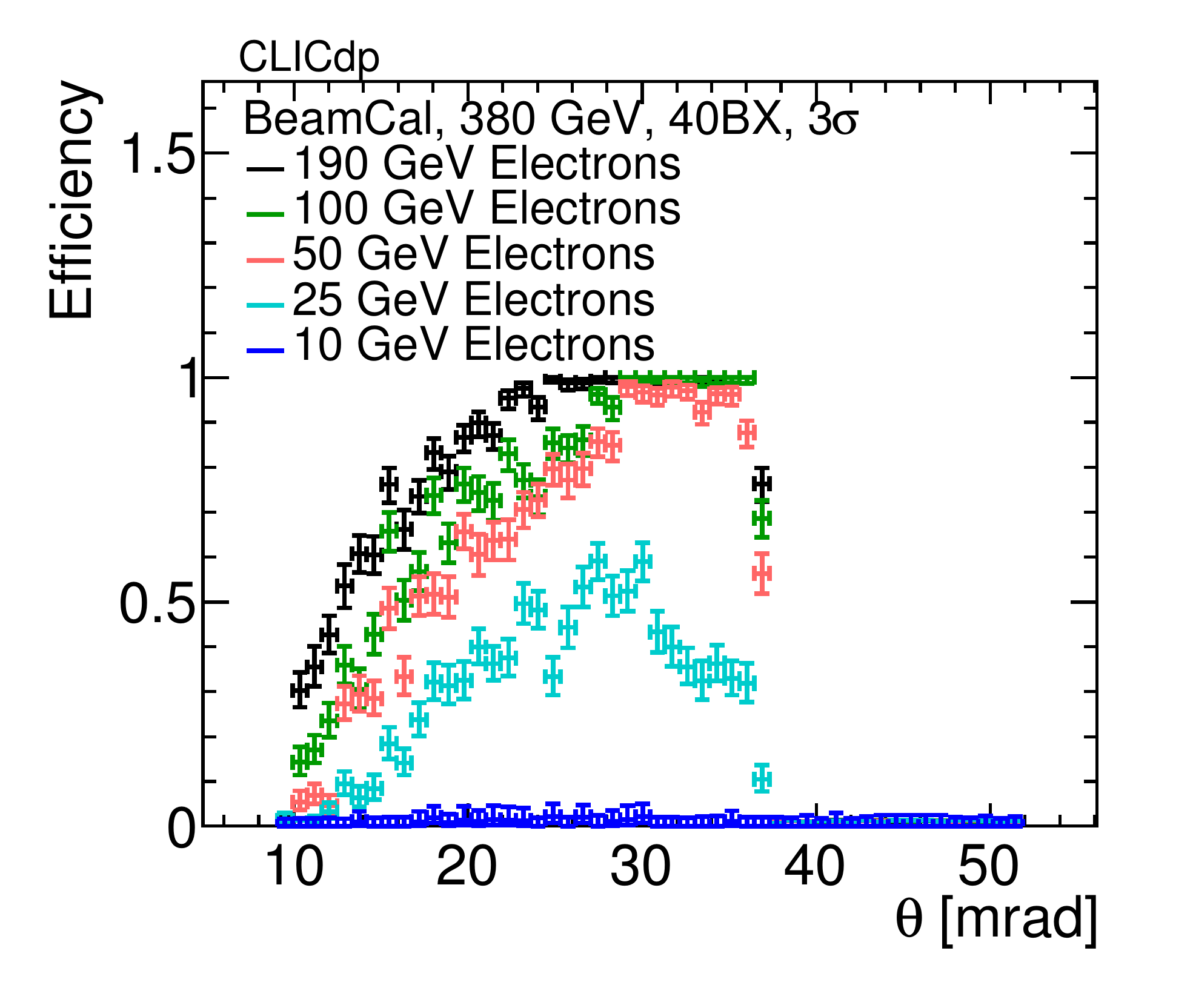}
    \phantomsubcaption\label{fig:beamEff380}
  \end{subfigure}
  \begin{subfigure}{\subfigwidth}\centering
    \includegraphics[width=1.0\textwidth]{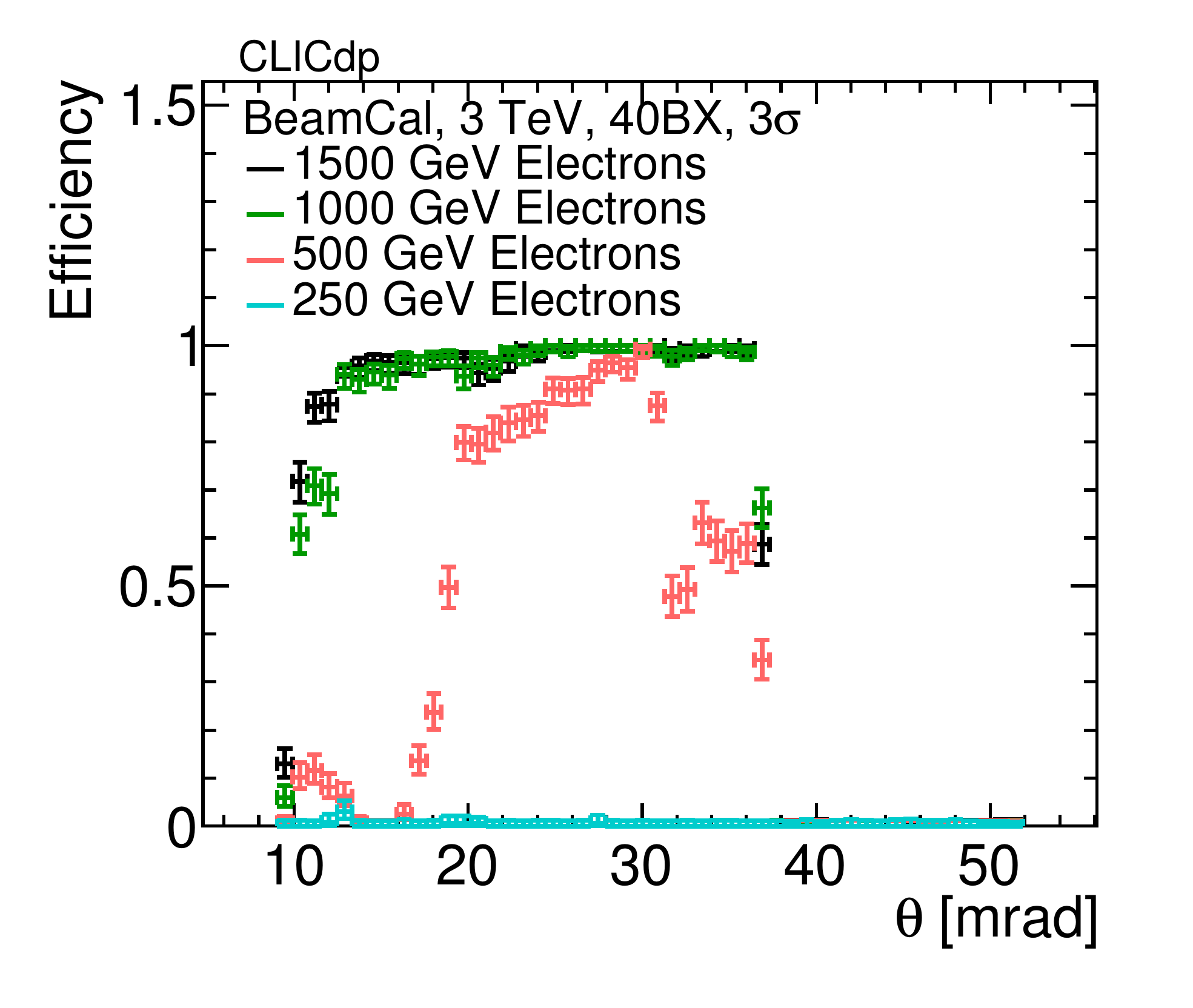}
    \phantomsubcaption\label{fig:beamEff3000}
  \end{subfigure}
  \vspace{-5mm}
  \caption{Reconstruction efficiency for electrons as a function of the polar
    angle in the BeamCal for \SI{380}{GeV} (left) and \SI{3}{TeV} (right).\label{fig:beamEff}}
\end{figure}

\begin{figure}[tbp]
  \renewcommand{\thesubfigure}{(\lr{subfigure})}
  \begin{subfigure}{\subfigwidth}\centering
    \includegraphics[width=1.0\textwidth]{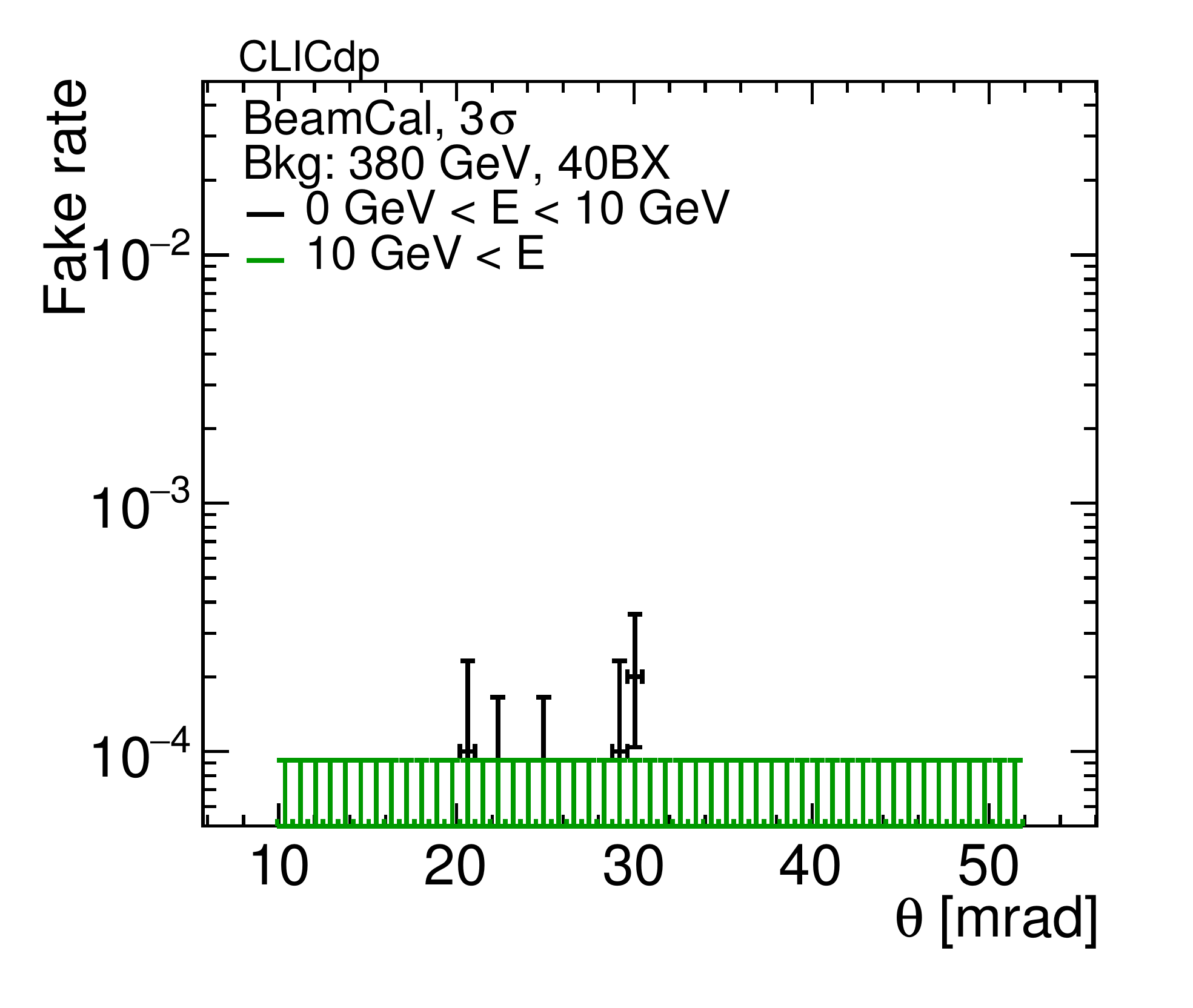}
    \phantomsubcaption\label{fig:beamFake380}
  \end{subfigure}
  \begin{subfigure}{\subfigwidth}\centering
    \includegraphics[width=1.0\textwidth]{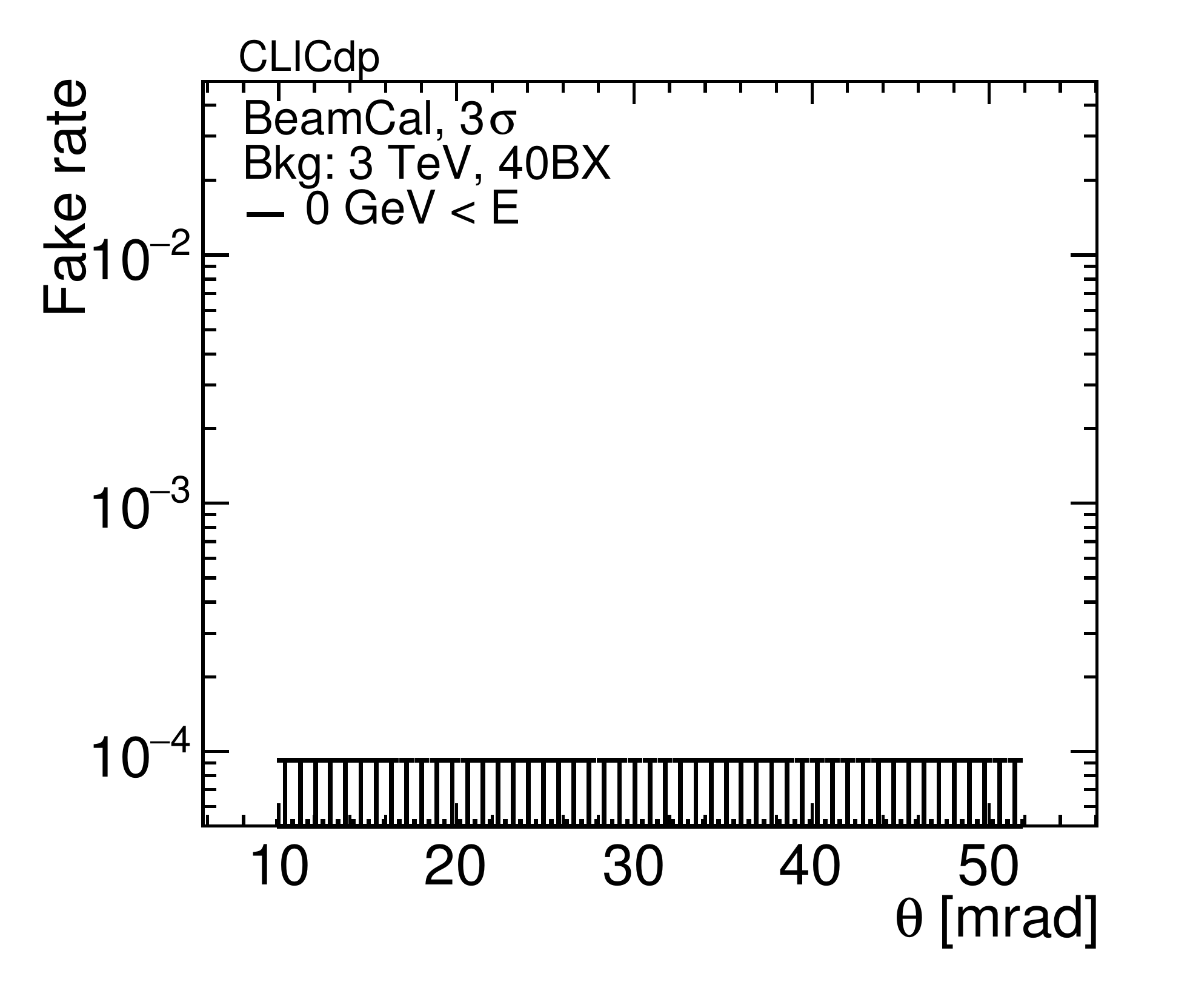}
    \phantomsubcaption\label{fig:beamFake3000}
  \end{subfigure}
  \vspace{-10mm}
  \caption{Fake rate for electron reconstruction as a function of the polar
    angle in the BeamCal at \SI{380}{GeV} (left) and \SI{3}{TeV} (right).\label{fig:beamFake}}
\end{figure}

The reconstruction efficiencies for the BeamCal at \SI{380}{GeV} and \SI{3}{TeV} CLIC are
shown in \cref{fig:beamEff}. At \SI{380}{GeV} even the highest energy electrons are
only reconstructed with 100\% efficiency above \SI{25}{mrad}. At lower radii the
backgrounds are so large that only a fraction of the electrons can be
reconstructed. However, as \cref{fig:beamFake380} shows there are only a few low
energy fake clusters reconstructed so the threshold to select pads can be
reduced, which will increase the reconstruction efficiencies.

\begin{figure}[tbp]
  \renewcommand{\thesubfigure}{(\lr{subfigure})}
  \centering
  \begin{subfigure}{\subfigwidth}\centering
    \includegraphics[width=1.0\textwidth]{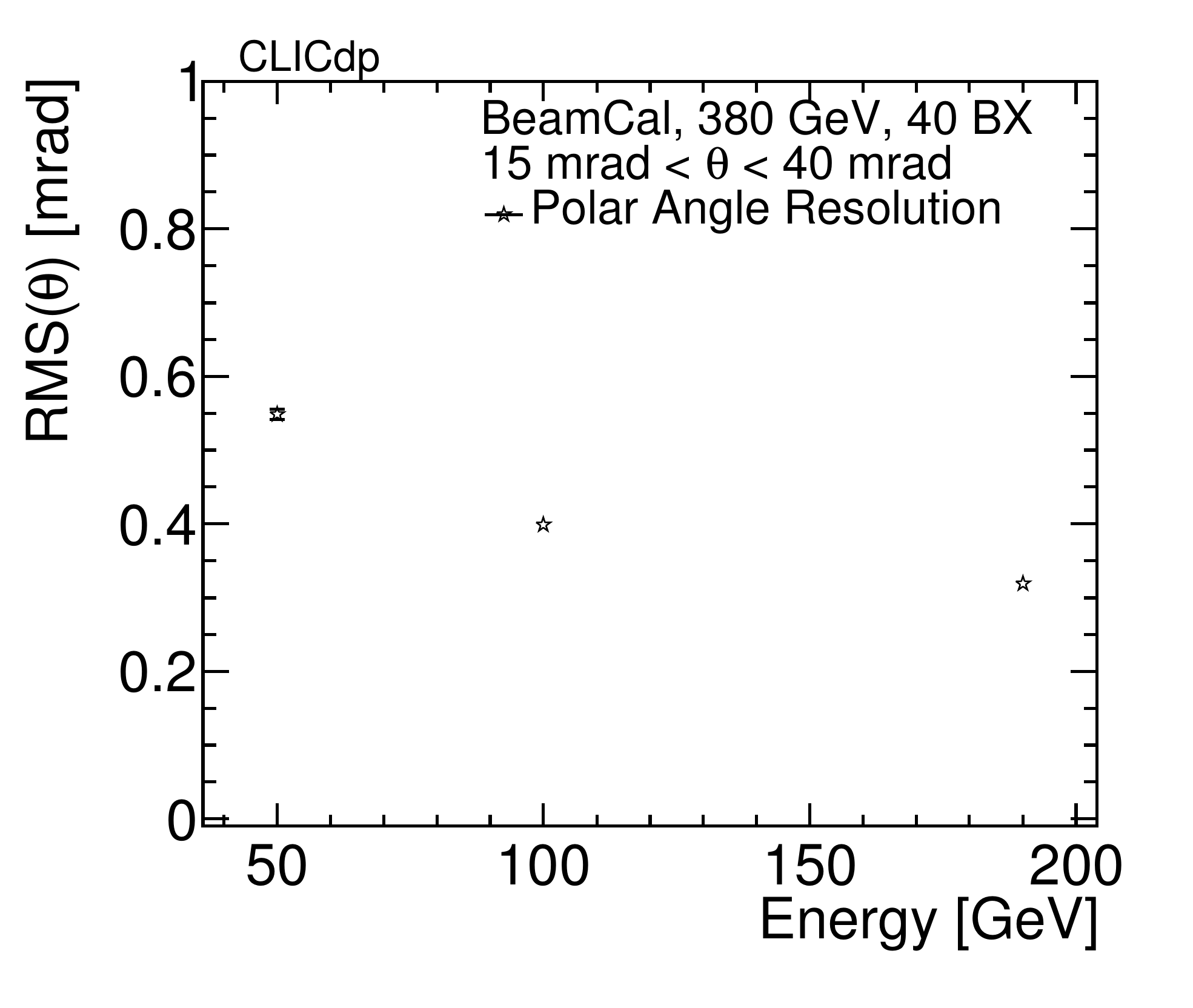}
  \end{subfigure}
  \begin{subfigure}{\subfigwidth}\centering
    \includegraphics[width=1.0\textwidth]{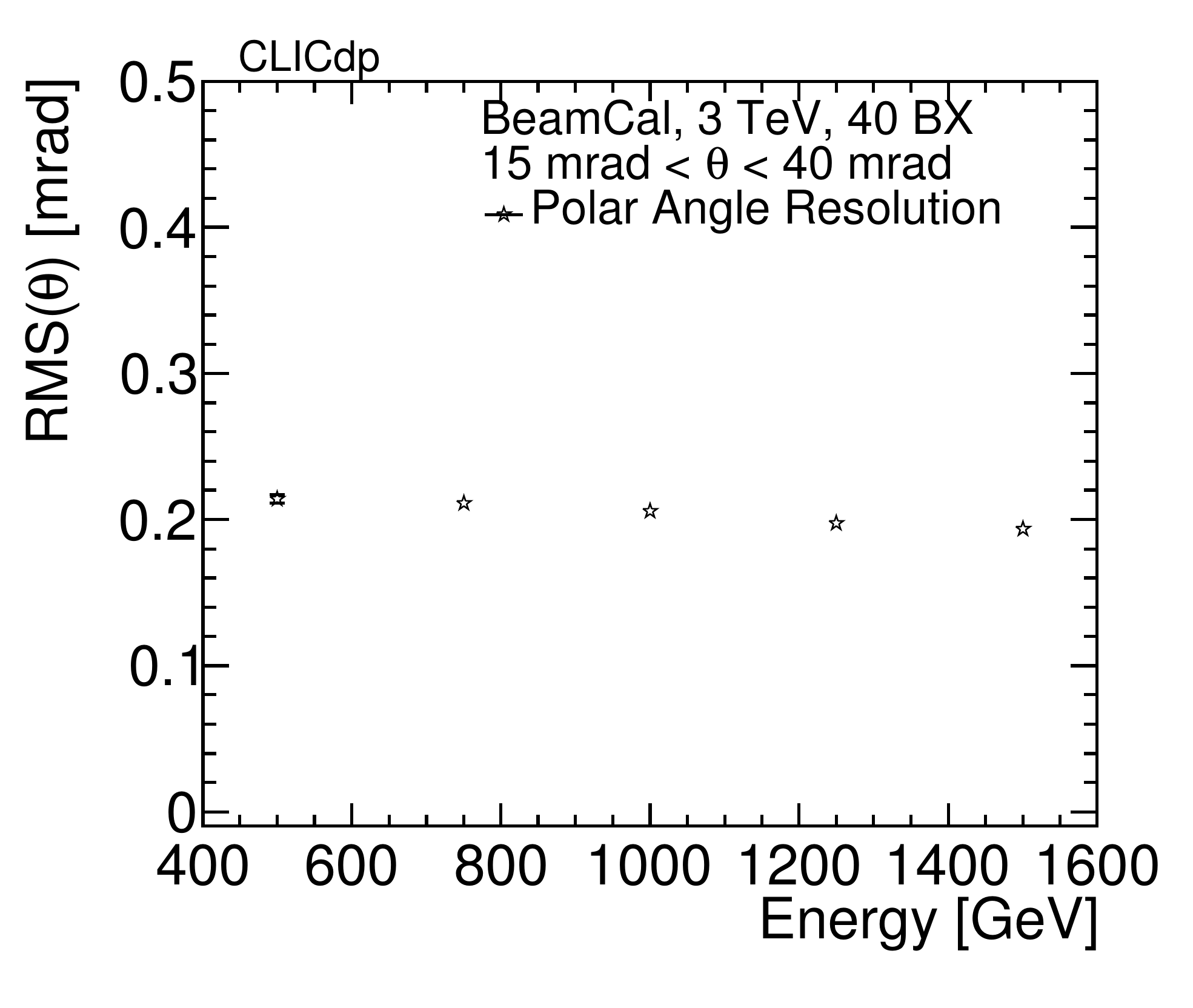}
    \phantomsubcaption\label{fig:beamResT3000}
  \end{subfigure}
  \vspace{-5mm}
  \caption{Polar angle resolution of reconstructed electrons as a function of
    the electron energy in the BeamCal at \SI{380}{GeV} (left) and \SI{3}{TeV} (right).\label{fig:beamResT}}
\end{figure}

\begin{figure}[tbp]
  \renewcommand{\thesubfigure}{(\lr{subfigure})}
  \begin{subfigure}{\subfigwidth}\centering
    \includegraphics[width=1.0\textwidth]{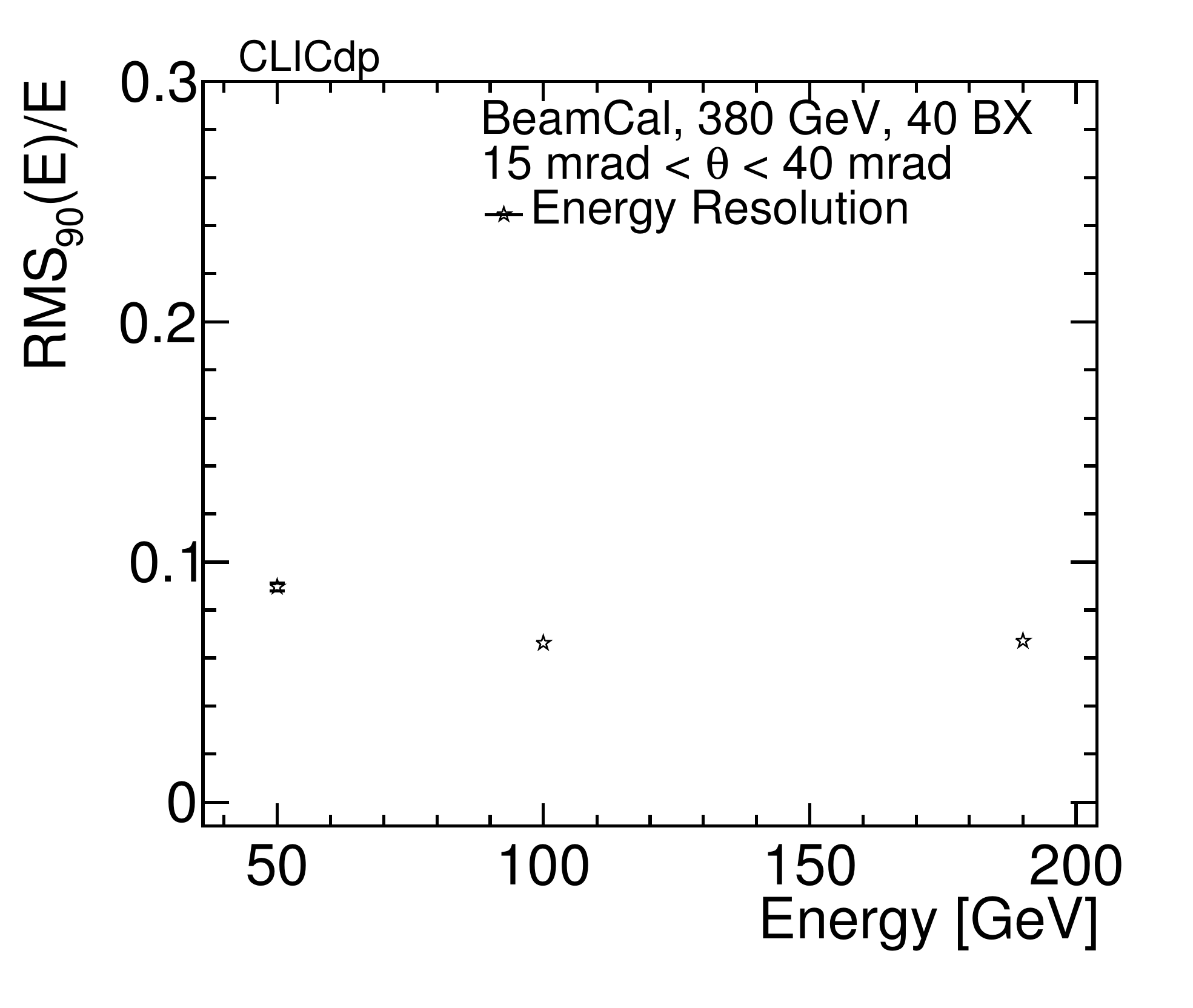}
  \end{subfigure}
  \begin{subfigure}{\subfigwidth}\centering
    \includegraphics[width=1.0\textwidth]{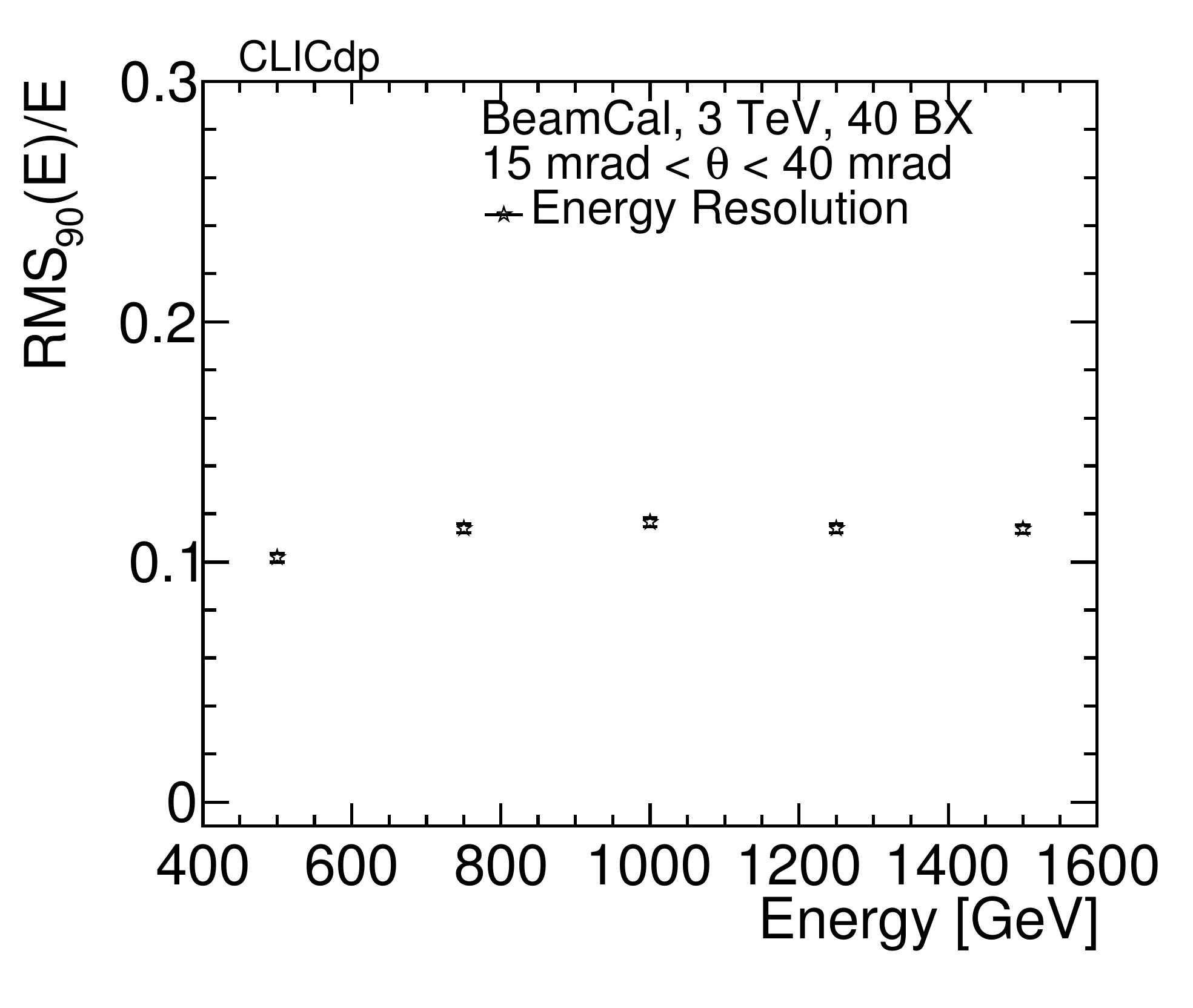}
  \end{subfigure}
  \vspace{-5mm}
  \caption{Energy resolution of reconstructed electrons as a function of the
    electron energy in the BeamCal at \SI{380}{GeV} (left) and \SI{3}{TeV} (right).\label{fig:beamResE}}
\end{figure}

For \SI{3}{TeV}, electrons above \SI{1}{TeV} are reconstructed with high efficiency
starting around \SI{12}{mrad}. Between \SI{12}{mrad} and \SI{22}{mrad} the biggest loss of efficiency
is due to the cutout in the BeamCal sensors due to the incoming beam
pipe. Beyond \SI{22}{mrad} all electrons above \SI{1}{TeV} are reconstructed. \SI{500}{GeV}
electrons can only be reconstructed above \SI{18}{mrad}. The initial 10\% efficiency
at the lower edge of the BeamCal for \SI{500}{GeV} electrons is enhanced by the large
background which pushes the energy deposits from the electron above the
threshold for reconstruction. The fake rate in the BeamCal at \SI{3}{TeV} (shown in
\cref{fig:beamFake3000}) is also below $10^{-4}$. Due to the aggressive rejection
of background energy deposits the polar angle (\cref{fig:beamResT}) and energy
resolutions (\cref{fig:beamResE}) are much
worse than for the LumiCal. For the BeamCal polar angle resolutions, the RMS of the
$\theta_{\mathrm{rec}}-\theta_{\mathrm{MC}}$ distribution is used for all energies. The \rmsninety of the
\mbox{$E_{\mathrm{rec}}-E_{\mathrm{MC}}$} distributions is used to obtain the energy resolutions.

\clearpage
\subsubsection{Forthcoming Studies and Improvements}
\label{sec:FutureImp}

In the near future, improvements are foreseen in terms of tracking and particle flow reconstruction,
 as already introduced in \cref{sec:flavTag}. 

Concerning the tracking, a systematic study is planned to assess the quality of the Kalman filter fit. This effort is expected to improve especially low energy tracks, which have hits correctly assigned by the pattern recognition but rejected by the fit. As a consequence, the fake rate is also expected to decrease.


Possible improvements of the particle flow reconstruction are partially linked to better track reconstruction. Fewer \emph{ghost} tracks 
and better track parameter reconstruction results in a better link between tracks and clusters. 
Improvements in the track reconstruction are therefore expected to translate in improved jet energy resolutions.
A second area of improvements for  particle flow reconstruction are the selection criteria for using tracks. 

Finally, better track and particle flow reconstruction will certainly improve the flavour tagging performances.

\section{Summary}

An overview of the recent CLIC detector model (CLICdet) has been given, together with updated results of 
beam--beam backgrounds at \SI{380}{GeV} and \SI{3}{TeV} CLIC\@. Occupancies in the CLICdet subdetectors resulting from
these backgrounds have been shown. In the case of the HCAL endcap, the high occupancies found need further studies, e.g.\ to mitigate back- or re-scattering of particles from the very forward region into the HCAL\@.

The detector performance has been illustrated with a series of results from single particle studies, 
covering momentum, position and angular resolution as well as particle identification efficiencies.
Studies using more complex events (light flavour \PZgstarToqq events, \bb, \ttbar) show the performance without and with \gghadron{} background overlaid with the physics events.
Detailed investigations of the jet energy resolution at all jet energies up to \SI{1.5}{TeV}, using software compensation, have been performed, and the separation of W and Z mass peaks has been studied.
First results for the efficiency and possible contaminations in b- and c-tagging have been obtained, and forthcoming work to improve the performance in this domain has been outlined.
Finally, the performance of the very forward electromagnetic calorimeters for electron tagging has been demonstrated.

\section*{Acknowledgements}

This project has received funding from the European Union's Horizon 2020
Research and Innovation programme under Grant Agreement no. 654168. This work
benefited from services provided by the ILC Virtual Organisation, supported by
the national resource providers of the EGI Federation. This research was done
using resources provided by the Open Science Grid, which is supported by the
National Science Foundation and the U.S. Department of Energy's Office of
Science.

\clearpage
\appendix
\numberwithin{table}{section}
\numberwithin{figure}{section}
\section{Illustration of Selected Angles}
\label{sec:appendix}

\begin{figure}[htbp]
    \includegraphics[width=1.0\textwidth]{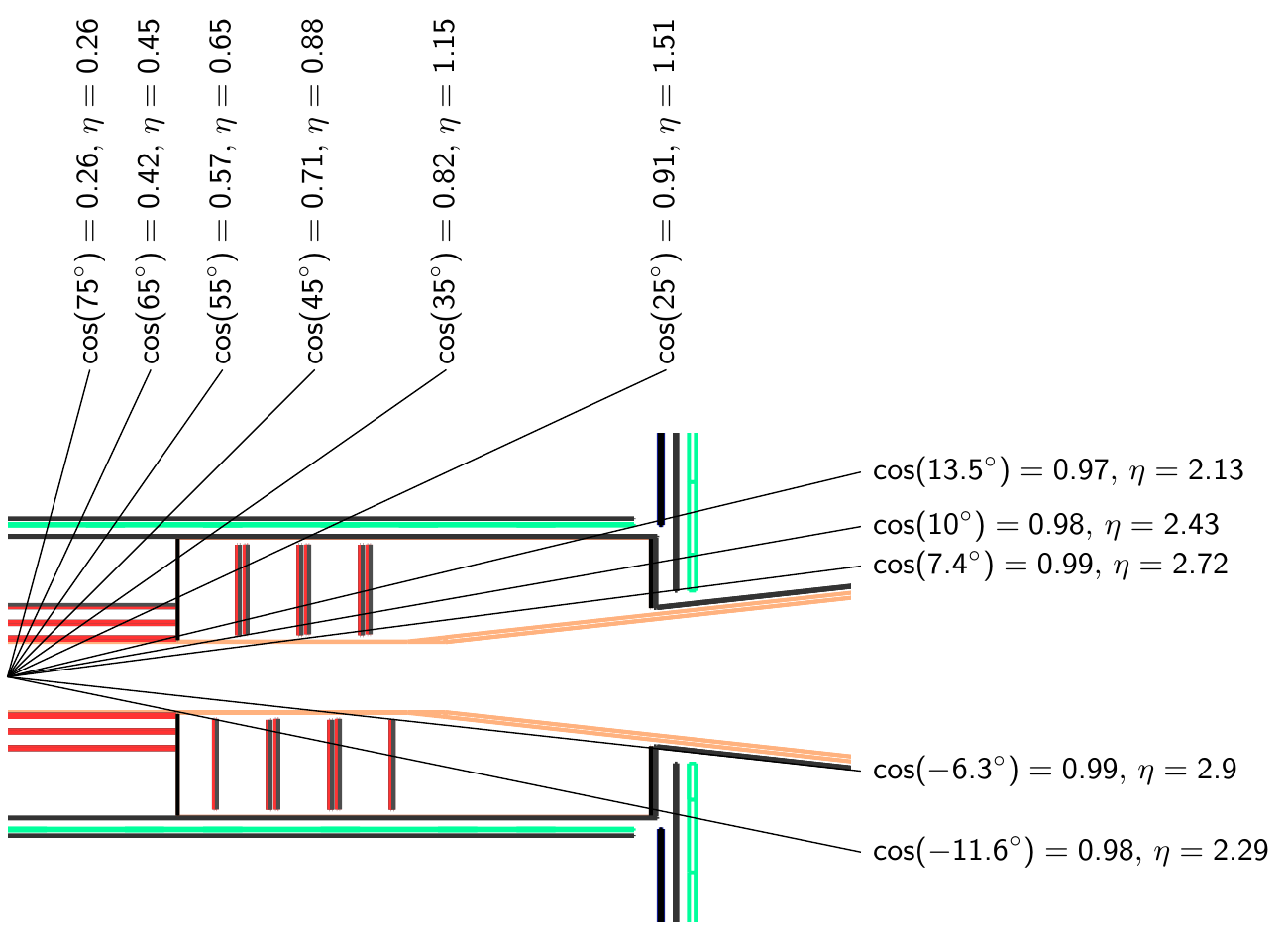}
  \caption{Illustration of selected angles in the vertex region of CLICdet.
\label{fig:vertex_angles}}
\end{figure}

\begin{figure}[htbp]
    \includegraphics[width=1.0\textwidth]{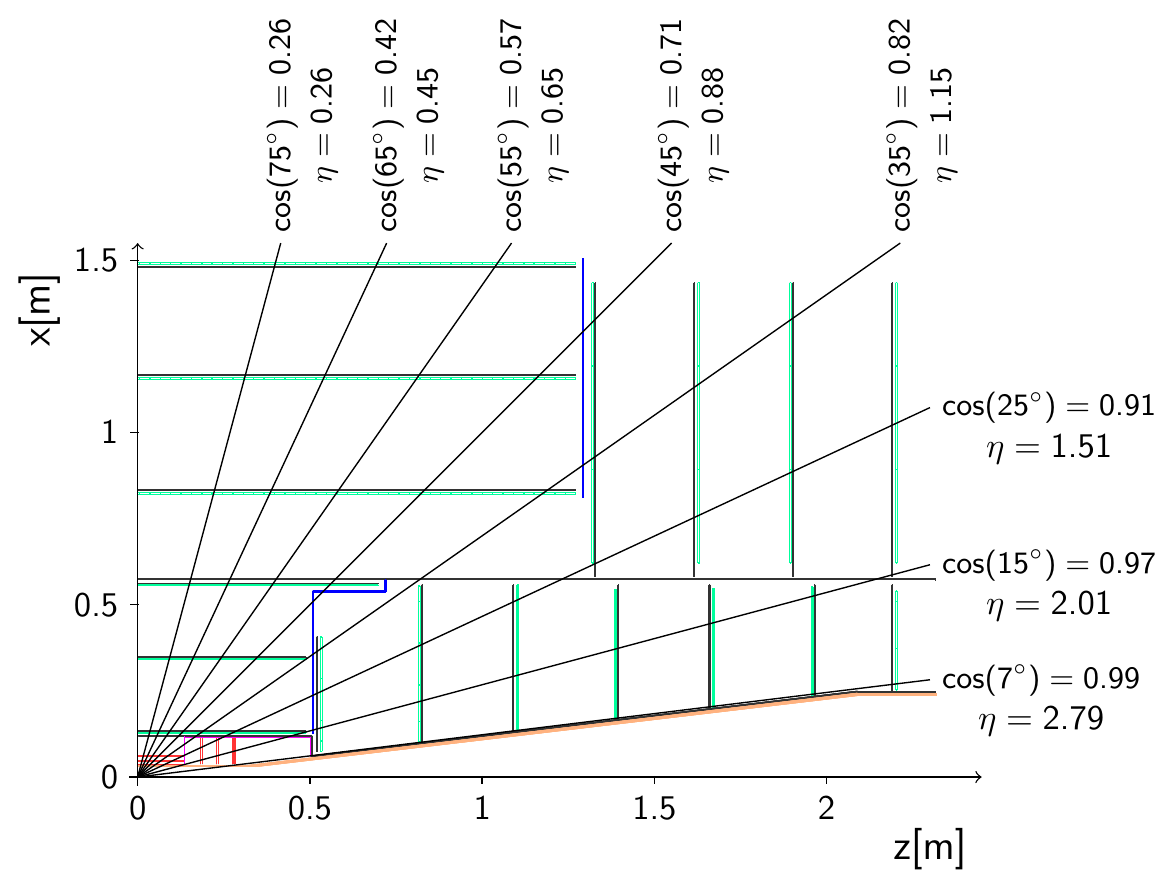}
  \caption{Illustration of selected angles in the tracker region of CLICdet.
\label{fig:tracker_angles}}
\end{figure}

\begin{figure}[htbp]
    \includegraphics[width=1.0\textwidth]{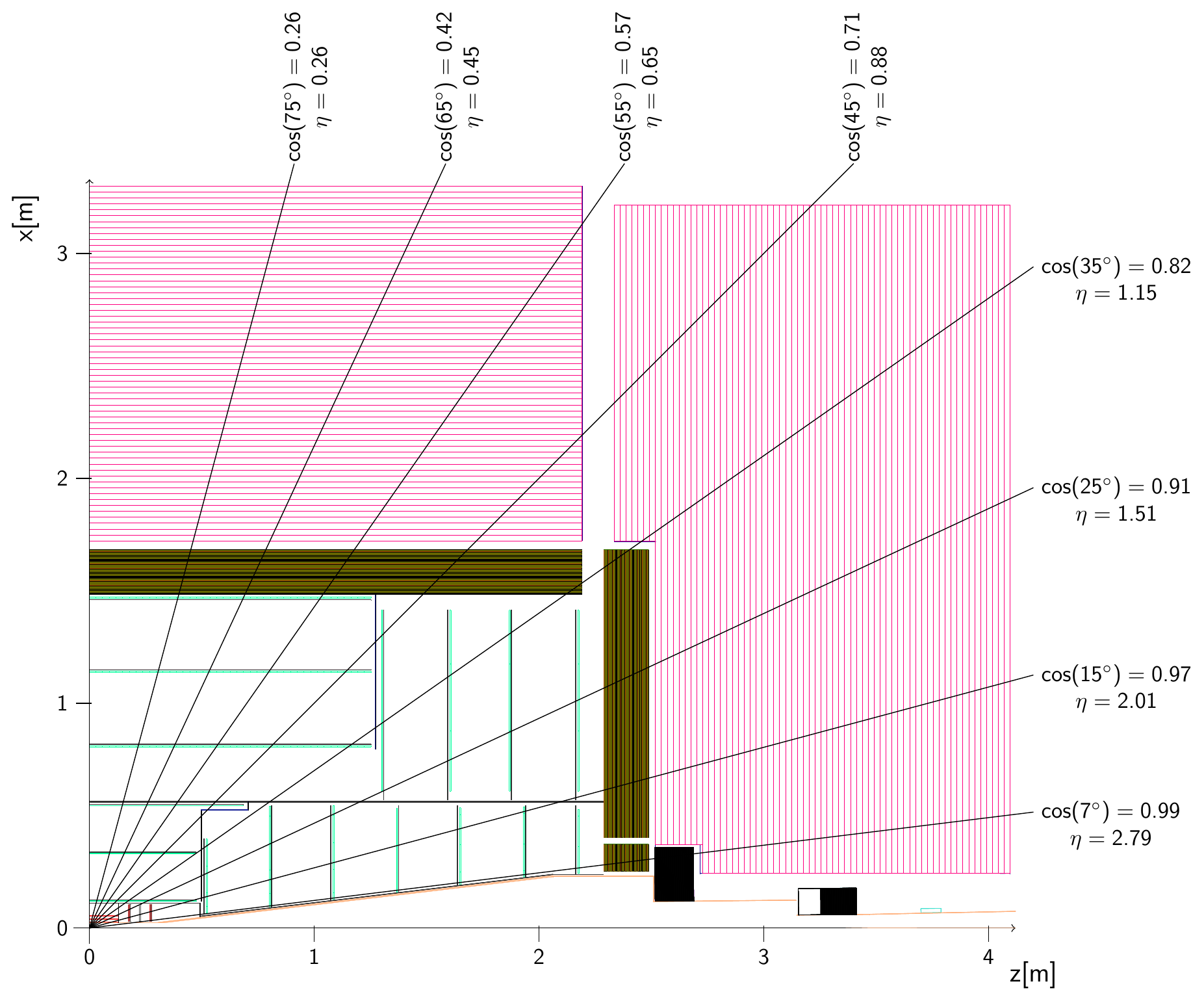}
  \caption{Illustration of selected angles in CLICdet.
\label{fig:calo_angles}}
\end{figure}

\clearpage
\section{Deposited Background Energies in the Calorimeters}\label{sec:appendixBKG}
\begin{table}[h]
\centering
  \caption{Raw deposited energy from beam-induced backgrounds in the CLICdet calorimeters. The
    numbers correspond to the background for an entire CLIC bunch train and nominal background rates.
    Safety factors representing the simulation uncertainties are not included.
    \label{tab:clicBackgroundsRaw}}
\begin{threeparttable}
  \begin{tabular}{l *4{S[table-format=5.5,round-mode=figures,round-precision=2]} }
    \toprule
    Energy stage          & \multicolumn{2}{c}{\SI{380}{GeV}} & \multicolumn{2}{c}{\SI{3}{TeV}}                             \\\cmidrule(r){2-3}\cmidrule(l){4-5}
    Subdetector           & \tabt{Incoherent pairs}           & \tabt{\gghad{}} & \tabt{Incoherent pairs} & \tabt{\gghad{}} \\
                          & \tabt{[GeV]}                      & \tabt{[GeV]}    & \tabt{[GeV]}            & \tabt{[GeV]}    \\\midrule
    ECAL barrel           &          3.5816704                &   2.1405824     &     13.7914300          &  52.206336      \\
    ECAL endcaps\tnote{a} &         11.137597                 &   9.4491021     &     39.218151           & 252.02705       \\
    HCAL barrel           &\sr{1}    0.0529183                &   0.1806999     &      0.2234475          &   4.973529      \\
    HCAL endcaps          &\sr{4} 2874.2560000                &   6.9521408     &\sr{4}11789.6380         & 311.6018900     \\\midrule
    ECAL\&HCAL            &\sr{4} 2889.0282                   & 18.722525       &\sr{4}11842.871          & 620.80881       \\\midrule
    LumiCal               &         68.5220800                &  4.5247840      &    282.6429800          & 192.6478300     \\
    BeamCal               &\sr{4}54726.4960000                &  5.6412224      &\sr{4}270566.400         & 540.4744800     \\\bottomrule
  \end{tabular}
\begin{tablenotes}
\item[a] Including the ECAL plugs
\end{tablenotes}
\end{threeparttable}
\end{table}

\clearpage
\section{Jet Energy Resolution Plots with Different Y-Axis Ranges}
\label{sec:appendixJER}

\begin{figure}[h]
  \centering
  \begin{subfigure}{\subfigwidth}
    \includegraphics[width=1.0\textwidth]{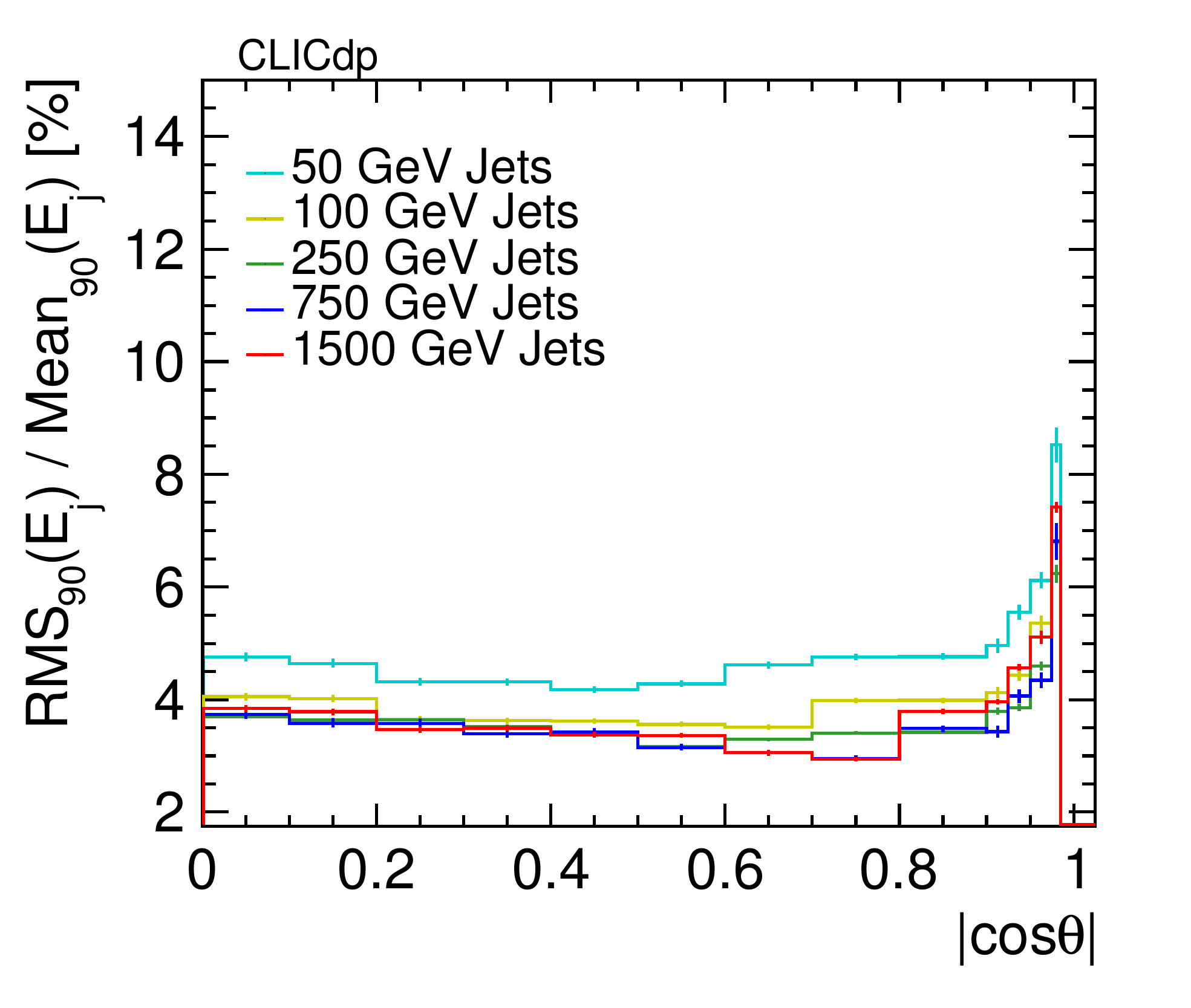}%
  \end{subfigure}%
  \begin{subfigure}{\subfigwidth}
    \includegraphics[width=1.0\textwidth]{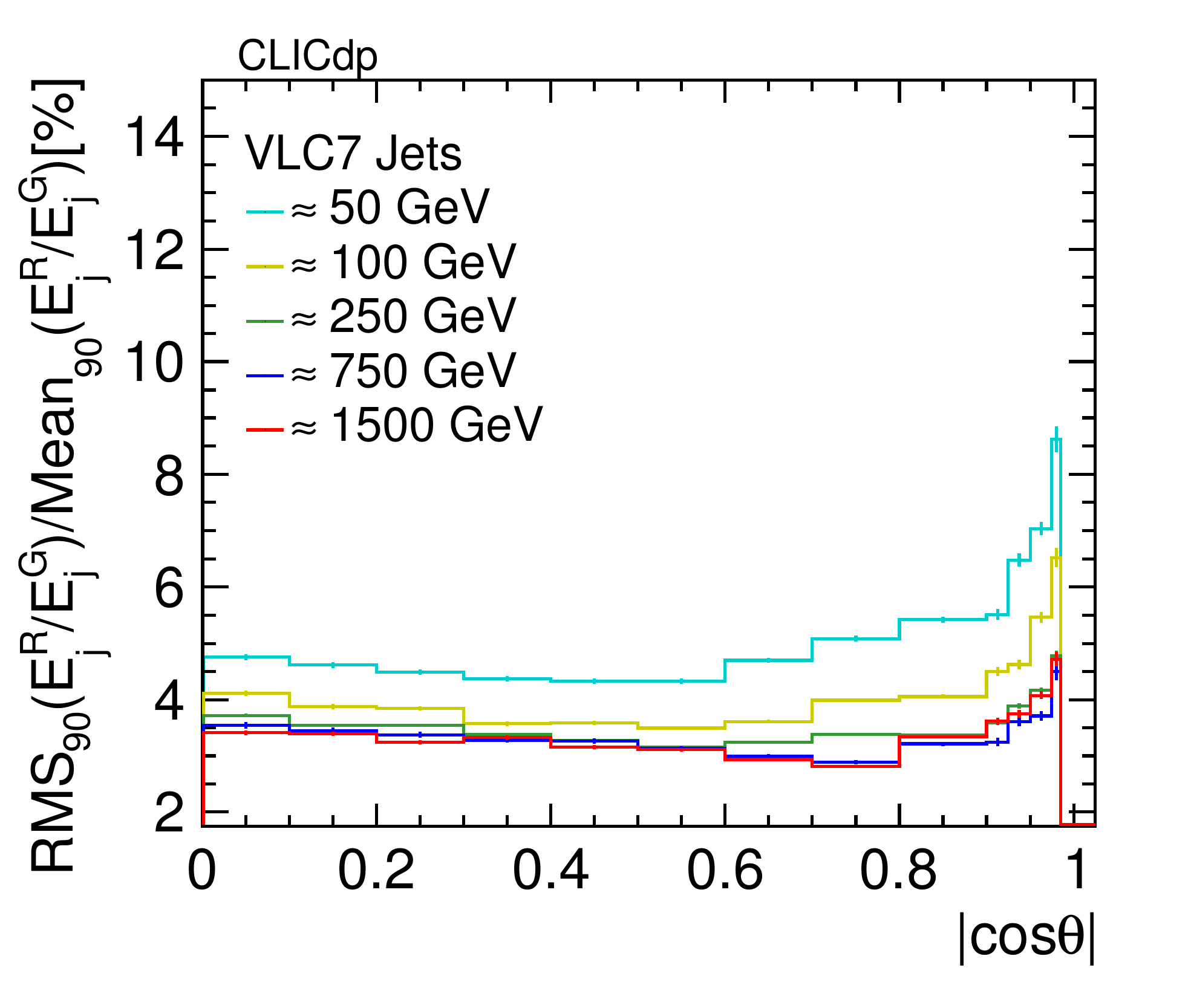}%
  \end{subfigure}
  \caption{Jet energy resolution distributions for various jet energies as a function of the $|\cos\theta|$ of the quark
    using two methods. The first method compares the total reconstructed energy with the energy sum from all visible
    particles on MC truth (left). The second method compares the jet energy of reconstructed jets and matched MC truth
    particle jets, using the VLC algorithm with an $R=0.7$ (VLC7, right)}\label{fig:JER_totE_vs_jetE_zoomedout}
\end{figure}

\begin{figure}[h]
  \centering
  \begin{subfigure}{\subfigwidth}
    \includegraphics[width=1.0\textwidth]{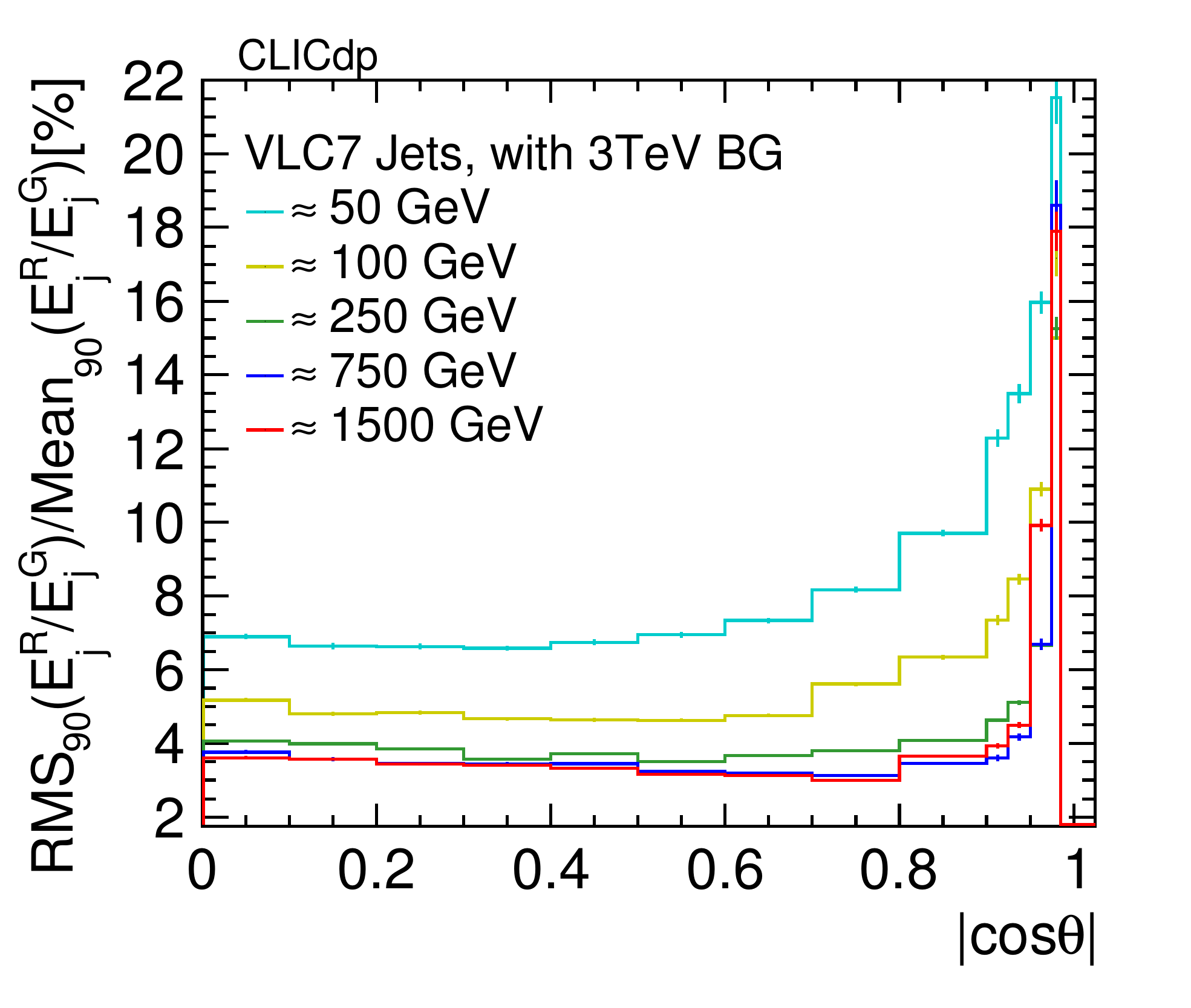}%
  \end{subfigure}%
  \begin{subfigure}{\subfigwidth}
    \includegraphics[width=1.0\textwidth]{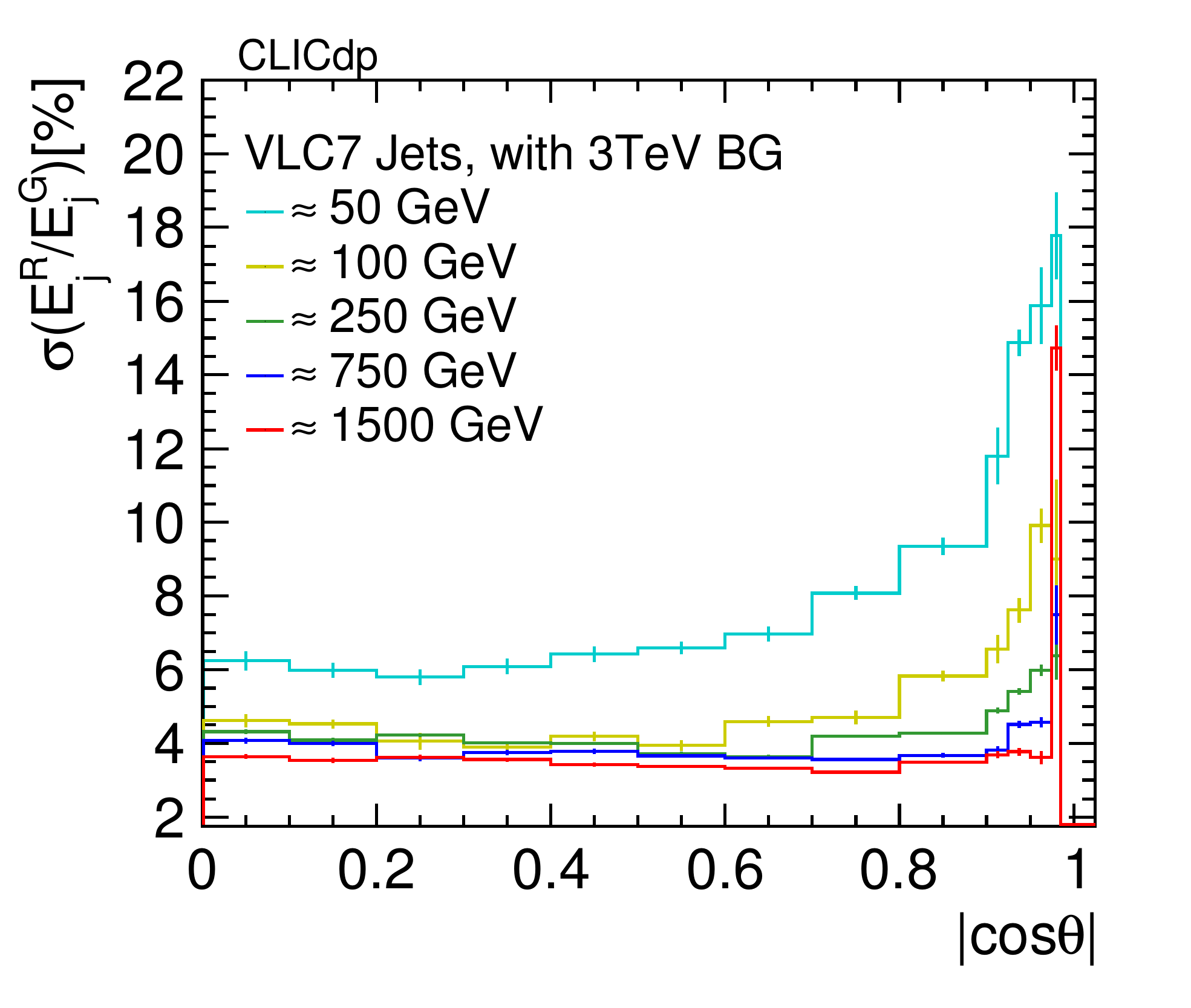}%
  \end{subfigure}
  \caption{Jet energy resolution for various jet energies as a function of the $|\cos\theta|$ of the quark with 3~TeV
    \gghadron{} background overlaid on the physics di-jet event. In the first method \rmsninety is used as measure of
    the jet energy resolution (left), the standard deviation $\sigma$ of the Gaussian core of the double-sided Crystal
    Ball fit quantifies the jet energy resolution in the second method (right). Tight PFO selection cuts are
    used.}\label{fig:jet_response_jets_wBG_zoomedout}
\end{figure}

\clearpage
\printbibliography[title=References]

\end{document}

%% file: include/packages.tex
\usepackage{threeparttable}
\usepackage{amsbsy,amsmath,amssymb}


%% file: include/definitions.tex
\newcommand{\nuclen}{\ensuremath{\lambda_{\mathrm{I}}}\xspace}

\newcommand{\ddhep}{\textsc{DD4hep}\xspace}
\newcommand{\ddg}{\textsc{DDG4}\xspace}
\newcommand{\ddrec}{\textsc{DDRec}\xspace}
\newcommand{\ilcdirac}{\textsc{iLCDirac}\xspace}
\newcommand{\PZgstar}{\ensuremath{\PZ/\gamma^{*}}\xspace}
\newcommand{\PZgstarToqq}{\ensuremath{\PZ/\gamma^{*}\to \PQq\PAQq}\xspace}

\newcommand{\microrad}{\ensuremath{\upmu\mathrm{rad}\xspace}}

\newcommand{\rmsninety}{\ensuremath{\mathrm{RMS}_{90}}\xspace}
\newcommand{\deltamc}{\ensuremath{\Delta_{\mathrm{MC}}}\xspace}
\newcommand{\ign}[1]{#1}

\crefname{section}{Chapter}{Chapters}
\Crefname{section}{Chapter}{Chapters}
\crefname{subsection}{Section}{Sections}
\Crefname{subsection}{Section}{Sections}
\crefname{subsubsection}{Section}{Sections}
\Crefname{subsubsection}{Section}{Sections}
\crefname{subsubsubsection}{Section}{Sections}
\Crefname{subsubsubsection}{Section}{Sections}

\makeatletter
\def\lr#1{\expandafter\@lr\csname c@#1\endcsname}
\def\@lr#1{%
  \ifnum#1=0%
    isZero%
  \fi
  \ifnum#1=1%
    left%
  \fi
  \ifnum#1=2%
    right%
  \fi
}
\makeatother



%% file: include/abbreviations.tex
\begin{acronym}[BX]
\acro{BX}{bunch crossing}
\acro{IP}{interaction point}
\acro{QD0}{final focusing quadrupole}
\end{acronym}

%% file: AuthorList/authorslist.tex
\addauthor{Dominik~Arominski}{\institute{1}\hcomma\institute{2}}
\addauthor{Jean-Jacques~Blaising}{\institute{3}}
\addauthor{Erica~Brondolin}{\institute{1}}
\addauthor{Dominik~Dannheim}{\institute{1}}
\addauthor{Konrad~Elsener}{\institute{1}}
\addauthor{Frank~Gaede}{\institute{4}}
\addauthor{Ignacio~Garc\'{\i}a~Garc\'{\i}a}{\institute{1}\hcomma\institute{5}}
\addauthor{Steven~Green}{\institute{6}}
\addauthor{Daniel~Hynds}{\institute{1}\hcomma\thanks{Now at Nikhef, Amsterdam, The Netherlands}}
\addauthor{Emilia~Leogrande}{\institute{1}\hcomma\editor{}}
\addauthor{Lucie~Linssen}{\institute{1}}
\addauthor{John~Marshall}{\institute{7}\hcomma\thanks{Formerly at University of Cambridge, Cambridge, United Kingdom}}
\addauthor{Nikiforos~Nikiforou}{\institute{1}\hcomma\thanks{Now at University of Texas at Austin, Austin, Texas, USA}}
\addauthor{Andreas~N\"{u}rnberg}{\institute{1}\hcomma\thanks{Now at Karlsruhe Institute of Technology, Karlsruhe, Germany}}
\addauthor{Estel~Perez-Codina}{\institute{1}}
\addauthor{Marko~Petri\v{c}}{\institute{1}}
\addauthor{Florian~Pitters}{\institute{1}\hcomma\institute{8}}
\addauthor{Aidan~Robson}{\institute{9}\hcomma\thanks{Also at CERN, Geneva, Switzerland}}
\addauthor{Philipp~Roloff}{\institute{1}}
\addauthor{Andr\'{e}~Sailer}{\institute{1}\hcomma\editor{}}
\addauthor{Ulrike~Schnoor}{\institute{1}}
\addauthor{Frank~Simon}{\institute{10}}
\addauthor{Rosa~Simoniello}{\institute{1}\hcomma\thanks{Now at Johannes Gutenberg Universit\"{a}t Mainz, Mainz, Germany}}
\addauthor{Simon~Spannagel}{\institute{1}}
\addauthor{Rickard~Stroem}{\institute{1}}
\addauthor{Oleksandr~Viazlo}{\institute{1}}
\addauthor{Matthias~Weber}{\institute{1}\hcomma\editor{}}
\addauthor{Boruo~Xu}{\institute{6}}
\addinstitute{1}{CERN, Geneva, Switzerland}
\addinstitute{2}{Warsaw University of Technology, Warsaw, Poland}
\addinstitute{3}{Laboratoire d'Annecy-le-Vieux de Physique des Particules, Annecy-le-Vieux, France}
\addinstitute{4}{DESY, Hamburg, Germany}
\addinstitute{5}{IFIC, Universitat de Valencia/CSIC, Valencia, Spain}
\addinstitute{6}{University of Cambridge, Cambridge, United Kingdom}
\addinstitute{7}{University of Warwick, Coventry, United Kingdom}
\addinstitute{8}{Technische Universit\"{a}t Wien, Vienna, Austria}
\addinstitute{9}{University of Glasgow, Glasgow, United Kingdom}
\addinstitute{10}{Max-Planck-Institut f\"{u}r Physik, Munich, Germany}

%% file: include/software.tex
\subsection{Simulation and Reconstruction}
\label{sec:simreco}

The detector simulation and reconstruction programs used for the results
presented here are developed together with the Linear Collider community. The
\ddhep~\cite{frank13:dd4hep} detector simulation and geometry framework was
developed in the AIDA and AIDA2020 projects. Large simulation and reconstruction samples
were produced with the \ilcdirac{} grid production
tool~\cite{dirac08,ilcdirac13}. The software packages of \ign{iLCSoft-2018-10-11\_gcc62} have been used throughout the study with the CLICdet geometry version CLIC\_o3\_v14.

\subsubsection{Event Generation}
\label{sec:event-generation}

The detector performance is studied with single particles or simple event
topologies. The individual particles are used to probe the track reconstruction
and the particle ID.\@ The reconstruction of particles inside jets is tested
through \PZgstar events decaying  into pairs of \PQu{}, \PQd{}, or \PQs{} quarks
 at different centre-of-mass energies. These events were created with
\pythia6.4~\cite{PYTHIAmanual}. To study the track reconstruction and particle ID
in complex events and for the flavour tagging studies \uu{}, \dd{}, \ssbar{},
\cc{}, \bb{}, and \ttbar{} events were created with
\whizard1.9~\cite{Kilian:2007gr,Moretti:2001zz}. The \gghad{} event initial
states are created in the \guineapig{}~\cite{schulte1996} simulation of the CLIC
collisions and hadronised in \pythia{}. In all cases, parton showering,
hadronisation, and fragmentation is performed in \pythia with the
fragmentation parameters tuned to the OPAL data taken at
LEP~\cite[Appendix B]{cdrvol2}.

\subsubsection{Detector Simulation}
\label{sec:detector-simulation}

The CLICdet detector geometry is described with the \ddhep{}
software framework, and simulated in \geant{}~\cite{Agostinelli2003,
  Allison2006, Allison2016186} via the
\ddg{}~\cite{frank15:ddg4} package of \ddhep{}. The \geant{} simulations are
done with the \texttt{FTFP\_BERT} physics list of \geant{} version 10.02p02.

\subsubsection{Event Reconstruction}
\label{sec:event-reconstruction}

The reconstruction software is implemented in the Linear Collider
\marlin{}-framework~\cite{MarlinLCCD}. The reconstruction algorithms take
advantage of the geometry information provided via the \ddrec{}~\cite{sailer17:ddrec}
data structures and surfaces. The reconstruction starts with the
overlay of simulated hits from \gghad{} events via the \emph{overlay} processor~\cite{LCD:overlay},
which selects only the energy deposits inside the timing windows of 10~ns
following the physics event (cf.\ \cref{sec:overv-detect-timing}). In
the next step, the hit positions in the tracking detectors are smeared with
Gaussian distributions according to the expected resolutions described in
\cref{sec:tracker}.
For the calorimeter and muon system digitisation, the hit position is taken from the centre of the cell. No smearing of the hit energy is done.
The calorimeter hits are scaled with the calibration
constants obtained from the reconstruction of mono-energetic \SI{10}{GeV} photons and \SI{50}{GeV}
\PKzL{}.

\paragraph{Tracking}

\emph{ConformalTracking} is the tracking algorithm used for reconstruction at CLICdet~\cite{Leogrande:2630512}. It is composed of a novel pattern recognition strategy followed by a Kalman-filter based fit. A detailed description is given below.

In modern pattern recognition algorithms, the use of cellular networks has been
shown to be a powerful tool, providing robustness against missing hits and the
addition of noise to the system. For a detector with solenoid field and barrel
plus endcap configuration, a cellular automaton (CA) may be applied to provide
 efficient track finding. Several aspects of CA algorithms may however
impact performance negatively: producing many possible hit combinations requires
a fit to be performed on a large number of track candidates. This may be costly
in processing time. Methods to reduce combinatorics at this stage may, in turn, compromise the final track finding performance. One way around such issues is the additional application of conformal mapping. 

Conformal mapping is a geometry transformation which has the effect of mapping
circles passing through the origin of a set of axes (in this case the global
$xy$ plane) onto straight lines in a new $uv$ co-ordinate system, with $u$ and $v$ defined as follows~\cite{hansroul1988498}:
\begin{equation}
\begin{aligned}[c]
u = \frac{x}{x^{2}+y^{2}}
\qquad \qquad
v = \frac{y}{x^{2}+y^{2}}.
\end{aligned}
\label{eq:confMappEq}
\end{equation}

\begin{figure}[tbp]
  \centering
  \includegraphics[width=0.5\linewidth]{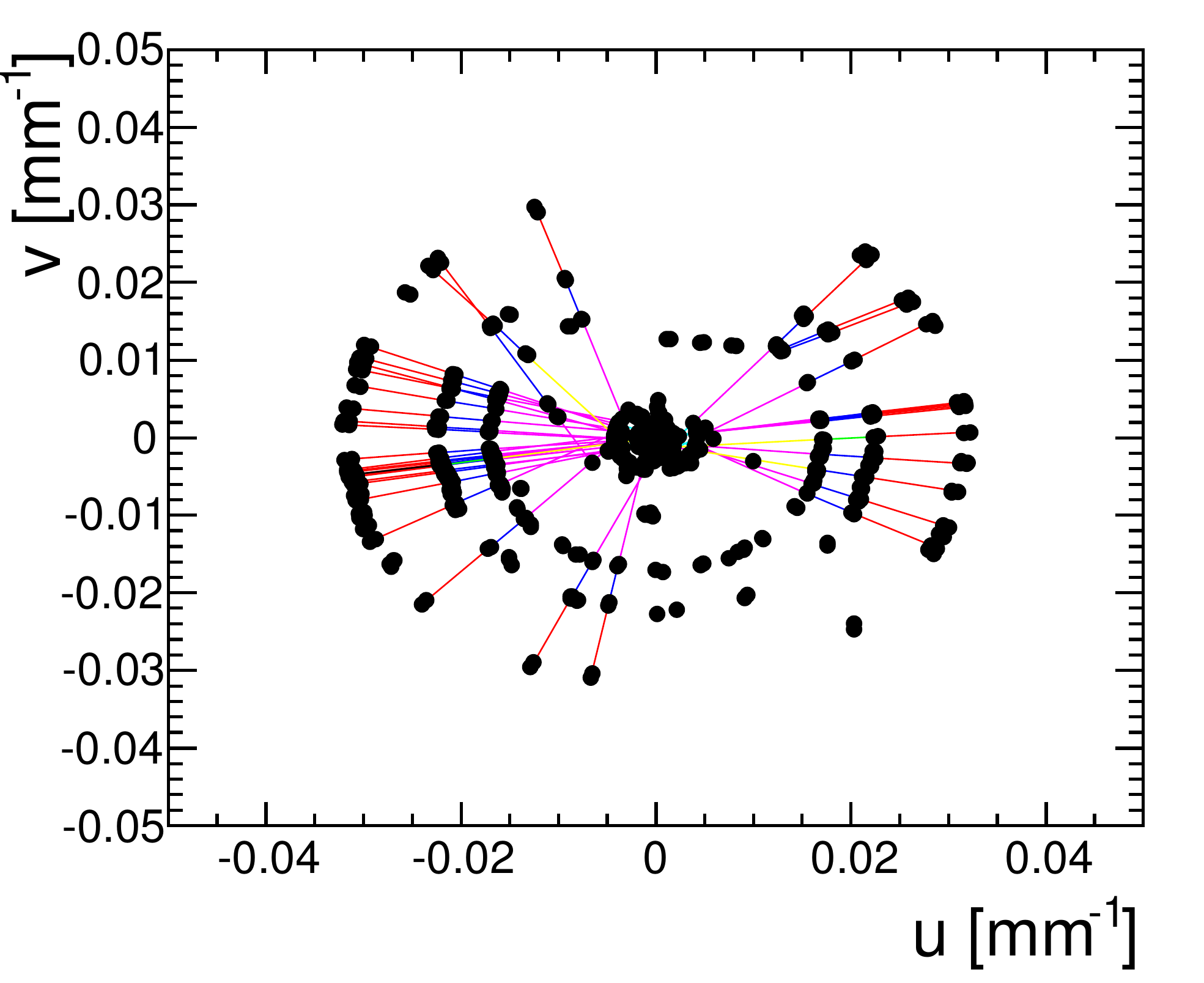}
\caption{Cellular automaton in conformal space.}
\label{fig:confMapping}
\end{figure}

By performing such a transform on an $xy$ projection of the detector (where the $xy$ plane is
the bending plane of the solenoid), the pattern recognition can be reduced to a
straight line search in two dimensions. A cellular automaton can then be applied
in this 2D space, with the use of a simple linear fit to differentiate between track
candidates. \Cref{fig:confMapping} shows an example of the cellular automaton in conformal space.

To make this approach flexible for changes in the geometry and for
applications to other detector systems, all hits in conformal space are treated
identically, regardless of sub-detector and layer. Cells between hits are
produced within a given spatial search window, employing \emph{kd}-trees~\cite{kdtree} for a fast
neighbour lookup. This provides additional robustness against missing hits in
any given detection layer. A second 2D linear fit in the $sz$ parametrisation\footnote{$s$ is the coordinate along the helix path.} of
the helix is also implemented, to recover the lost information resulting from
the 2D projection onto the $xy$ plane and reduce the number of \emph{ghost} tracks.
A minimum number of 4 hits is required to reconstruct a track.

For displaced tracks, which do not comply with the requirement of passing through the origin of the global $xy$ plane,
second-order corrections are applied to the transformation equations. Additionally, the search strategy was modified:
\begin{itemize}
\item broader angles in the search for nearest neighbours;
\item minimum number of 5 hits to reconstruct a displaced track;
\item inverted search order, from tracker to vertex hits.
\end{itemize}

Tracks reconstructed via the pattern recognition are then fitted with a Kalman filter approach. 
A preliminary helical trajectory is obtained by fitting only three hits (typically the first, last and intermediate hits on the track). The helix parameters are then given as initial input to the Kalman filter, which refits the track while adding hits one by one and progressively updating the track parameters.
The default Kalman filter starts from the innermost hits (vertex hits) and proceeds outward. The fit is complemented by a smoothing backward to the \ac{IP}.

The performance studies presented in this note assume a homogeneous magnetic field of 4 T.

\paragraph{Particle Flow Clustering}

The calorimeter clusters are reconstructed in the particle flow approach by
\pandora~\cite{Marshall:2015rfaPandoraSDK,Thomson:2009rp,Marshall:2012ryPandoraPFA}. \pandora{} uses the reconstructed tracks and
calorimeter hits as input to reconstruct all visible particles. The procedure is optimised
to achieve the best jet energy resolution. This may not be the ideal procedure for isolated particles,
which can benefit from a dedicated treatment. The output of the particle flow
reconstruction are \emph{particle flow objects} (PFOs).

\paragraph{Forward Calorimeter Reconstruction}

The high energy electrons and photons in the
forward calorimeters LumiCal and BeamCal are reconstructed with special
considerations for the large amount of incoherent pairs impacting on
them~\cite{sailer:bcalreco}. The expected average energy deposits from incoherent pair
background is subtracted from the total amount of deposited energy before a
nearest neighbour clustering is performed on the pads which have sufficient remaining
energy. LumiCal and BeamCal performances are described in \cref{sec:VF_perf}.

\subsubsection{Treatment of \gghad{} Background}
\label{sec:treatment-background}

The largest impact on the detector performance from beam-induced backgrounds
comes in the form of the \gghadron{} events as discussed in
\cref{clic_beam_and_bg}. When studying the detector performance degradation due
to these backgrounds the expected number of events $n_{\mathrm{Had}}$
(\cref{tab:clicBeam}) are overlaid for 30 bunch crossings around the physics
event, which is placed in bunch crossing 11. Accounting for the expected
detector timing resolutions and integration times, time windows of 10~ns
following the physics event are applied to the hits of the background events and
physics event. All hits inside the time window are then passed forward to the
reconstruction.

Once the particle flow clustering is finished, additional \pT{} dependent timing
cuts are applied. Depending on the particle type -- photon, neutral hadron, or
charged particle -- and the transverse momentum and based on the time of the
clusters, reconstructed particles are rejected. The time of a clusters is the
truncated energy-weighted mean time of its hits. To allow for different physics
topologies three sets of timing cuts were created, \emph{Loose},
\emph{Selected}, and \emph{Tight} timing cuts for the studies at $\roots=\SI{3}{TeV}$ and
$\roots=\SI{1.5}{TeV}$. More relaxed cuts were also created for $\roots=\SI{500}{GeV}$ and
below. Detailed information on the timing cuts are given in the
CDR~\cite[Appendix B]{cdrvol2} and in a separate study for the \SI{380}{GeV} energy stage~\cite{Brondolin:2641311}.
